%% file: main.tex
\documentclass[11pt,a4paper,oneside,openright]{book}
\usepackage[greek,english]{babel}
\usepackage[utf8]{inputenc}
\input{preamble}
\usepackage[pdftex, pdftitle={Article}, pdfauthor={Author}, linktocpage]{hyperref} 
\hypersetup{
     colorlinks=true,
     allcolors=mycolor
     }

\usepackage{hyperref}

\definecolor{mycolor}{HTML}{000067}
\definecolor{BergerColor}{HTML}{cccaca}

\newcommand{\mc}[1]{\mathcal{#1}}
\newcommand{\pd}{\partial}
\begin{document}

\input{chaps/title.tex}
\input{chaps/declaration.tex}
\input{chaps/abstract.tex}
\input{chaps/acknowledgements.tex}
\input{chaps/TOC.tex}
\input{chaps/chap01.tex}  
\input{chaps/chap03.tex}
\input{chaps/chap04.tex}
\input{chaps/chap05.tex}
\input{chaps/chap06.tex}
\input{chaps/chap07.tex}

\input{appendix.tex}

\bibliography{newRefs.bib}
\end{document}

%% file: preamble.tex
\usepackage{amsthm}
\usepackage{mathtools}
\usepackage{physics}
\usepackage[table]{xcolor}
\usepackage{graphicx}
\usepackage[left=40mm,right=25mm,top=25mm,bottom = 25mm,includehead,includefoot]{geometry} 
\usepackage{adjustbox}
\usepackage{placeins}
\usepackage[T1]{fontenc}
\usepackage{lipsum}
\usepackage{csquotes}
\usepackage{dsfont}
\usepackage{cite}
\usepackage{graphicx}
\usepackage{amsmath}
\usepackage{amssymb}
\usepackage{subcaption}
\usepackage{mathtools}
\usepackage{mathrsfs}

\usepackage{xcolor}
\usepackage[most]{tcolorbox}
\usepackage{blindtext}
\usepackage{tikz-cd}
\usepackage{wrapfig}
\usepackage{subfiles}
\usepackage{titlesec}
\usepackage{parskip}
\usepackage{tensor}

\def\beq{\begin{equation}}
\def\eeq{\end{equation}}
\allowdisplaybreaks
\newcommand{\hol}{\text{Hol}}
\newcommand{\ghol}{\mathfrak{hol}}
\newcommand{\Ft}{\tilde{F}}

\newcommand{\Indx}{\,\,\,}
\newtheorem{theorem}{Theorem}[section]
\newtheorem{exmp}{Example}
\theoremstyle{definition}
\newtheorem{definition}{Definition}[section]
\newtheorem{proposition}{Proposition}[section]

\tikzset {_f2p9j0ezo/.code = {\pgfsetadditionalshadetransform{ \pgftransformshift{\pgfpoint{0 bp } { 0 bp }  }  \pgftransformscale{1 }  }}}
\pgfdeclareradialshading{_tw3gbh7dx}{\pgfpoint{0bp}{0bp}}{rgb(0bp)=(1,1,1);
rgb(0bp)=(1,1,1);
rgb(25bp)=(0,0,0);
rgb(400bp)=(0,0,0)}
\tikzset{every picture/.style={line width=0.75pt}} 

%% file: chaps/title.tex
\newcommand{\HRule}{\rule{\linewidth}{0.5mm}}

\begin{titlepage}

\begin{centering}

\begin{figure}[h]
\centering
\includegraphics[scale=0.2]{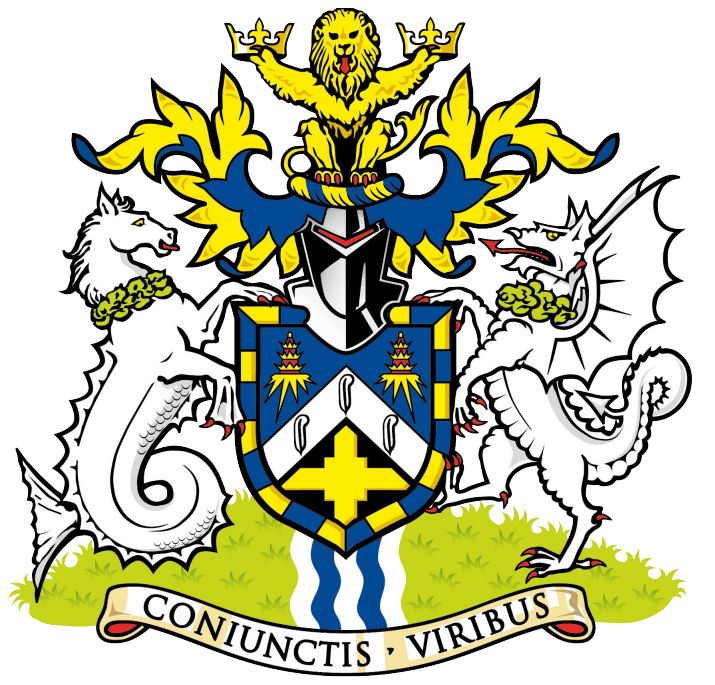}
\end{figure}\ \\

{\LARGE
\textsc{Thesis submitted in partial fulfilment for the award of the degree of \\[0.25cm] Doctor of Philosophy}
}\\[1.5cm]

{\huge
\bfseries{Aspects of the Classical Double Copy
}
}\\[1.5cm]

{\huge
\textsc{Rashid Alawadhi}
}\\[2cm]

{\Large
Supervisor \\[0.3cm]
\textsc{Prof. David S. Berman}\\[1.5cm]
\today}\\[0.5cm]

{\large
\textsf{
Centre for Theoretical Physics \\[0.1cm]
School of Physical and Chemical Sciences \\[0.1cm]
Queen Mary University of London}}

\end{centering}

\end{titlepage}

%% file: chaps/declaration.tex
\chapter*{Declaration}\label{Declaration}
\vspace{-0.5cm}
I, Rashid Alawadhi, confirm that the research included within this thesis is my own work or that where it has been carried out in collaboration with, or supported by others, that this is duly acknowledged below and my contribution indicated. Previously published material is also acknowledged below. \\

\noindent I attest that I have exercised reasonable care to ensure that the work is original, and does not to the best of my knowledge break any UK law, infringe any third party’s copyright or other Intellectual Property Right, or contain any confidential material. \\

\noindent I accept that the university has the right to use plagiarism detection software to check the electronic version of the thesis. \\

\noindent I confirm that this thesis has not been previously submitted for the award of a degree by this or any other university. \\

\noindent The copyright of this thesis rests with the author and no quotation from it or information derived from it may be published without the prior written consent of the author. \\

\noindent Signature:

\noindent \vspace{-0.1cm}Date: \today \\

\noindent Details of collaboration and publications:\\

\noindent This thesis describes research carried out with my supervisor Prof. David S. Berman and Prof. Bill J. Spence, which was published in \cite{Alawadhi:2020jrv} and \cite{Alawadhi:2019urr} with the latter being also in collaboration with David Peinador Veiga.
This thesis also describes research carried out in collaboration with Prof. David S. Berman, Dr Chris D. White, and Sam Wikeley, published in \cite{Alawadhi:2021uie}.
It also describes a paper I wrote in \cite{Alawadhi:2022gwy}.

%% file: chaps/abstract.tex
\chapter*{Abstract}\label{Abstract}
This thesis applies the Kerr-Schild and the Weyl double copy formalisms to study various concepts in the physics literature. First we apply both the Kerr-Schild and the Weyl double copy to solution generating transformations in General Relativity, where we identify Ehlers transformation as the double copy of electromagnetic duality transformation. Secondly, as a spin-off of the Weyl double copy, we use gauge fields defined on curved spacetimes to construct the Weyl tensor and study a host of solution of Einstein's equations. This study provides a test of the non-triviality of the double copy formalism.

The second half of the thesis deals with mathematical concepts of physical relevance. First we apply the Kerr-Schild double copy to the concept of holonomy groups of Riemannian manifolds. We find that the single copy of the Riemannian holonomy operator, which we dub SCH, to be a similar operator constructed from the single copy gauge-field curvature. This is followed by a study of this single copy operator on different solutions of Einstein equations and their respective single copies, where we find that the holonomy and SCH groups differ for the Taub-NUT metric, while both reducing to $\text{SU}(2)$ for self-dual solutions. Lastly, we apply the Kerr-Schild double copy to the Ricci flow equation, interpreted as the beta function of the closed string, and obtain the Yang-Mills flow equation, which is physically interpreted as the beta function of the open string coupled to a gauge field.

%% file: chaps/acknowledgements.tex
\chapter*{Acknowledgements} \label{Acknowledgements}
First and foremost, I would like to thank my supervisor Prof. David S. Berman for the encouragement, interesting discussion, and never failing to answer any questions I had. I would like to also extend my thanks to my collaborators, Prof. Bill J. Spence, David Peinador Veiga, Dr Chris D. White, and Sam Wikeley, with whom it was a pleasure to work. I would also like to extend my thanks to the rest of the faculty members for interesting discussions, and to the UAE ministry of education for sponsoring my academic endeavours.

As someone who appreciates good company and friendship, I would like to express my deep gratitude to my colleagues, Nadia Bahjat-Abbas, Graham Brown, Stefano De Angelis, Josh Gowdy, Manuel Accettulli Huber, Gergely Kántor and Sam Wikeley for the support and great time we had together throughout our studies. Of course, I cannot fail to mention my friends, Marcel Hughes, Rajath Radhakrishnan, Adrian Padellaro, David Peinador Veiga, and Shun-Qing Zhang, with whom I have developed a friendship which I hope for it to last for a long time to come. Special thanks to James Prosser for his friendship, company, and support.

Lastly, I would like to express my sincere and deepest gratitude to my mother and my late father, for providing me with a loving environment to cultivate my interests, and allowing me to explore the wonders of our world. Indeed, their encouragement and support for me to always stay curious is something I shall forever be grateful for.
\newpage
\vspace*{55pt}
\textit{\large
    I think it's a peculiarity of myself that I like to play about with equations, just looking for beautiful mathematical relations which maybe don't have any physical meaning at all. Sometimes they do.
}
--- Paul Dirac\footnote{Interview of P. A. M. Dirac by Thomas S. Kuhn on 1963 May 6}

%% file: chaps/TOC.tex
\begingroup
\hypersetup{linkcolor=mycolor, linktocpage=true}
\tableofcontents
\endgroup

%% file: chaps/chap01.tex
\chapter{Introduction}
\label{chap:Intro}

\section{Gravity and its relation to gauge theories}
The force of gravity is the oldest force known to man and the most mysterious. It does not require a great deal of imagination to think that philosophers and thinkers were occupied in the search for an explanation of this mysterious phenomenon. Throughout the history of mankind, a great number of philosophers tried to explain the nature of this force, whether by considering it as a manifestation of inherent qualities that objects affected by this force may posses \cite{AristotleRef}, or by thinking of it as an ether-like medium through which objects travel \cite{2006APS..NWS.F1001E}. Indeed, a pivotal point in the history of gravitational theories is when Galileo Galilei, thanks to his experiments, showed that objects free fall at the same rate under the influence of gravity regardless of their internal qualities, such as mass. This eventually paved the way for Isaac Newton to formulate his law of gravitation, and produce Kepler's laws\cite{principia}. However, Newton could not explain the nature of the force of gravity as he clearly stated in his \textit{Principia Mathematica}. Newton's model relied on the concept of Force, which treated gravity as some sort of instantaneous, non-mechanical interaction. Attempts at trying to explain gravity by a mechanical model saw a rise in popularity in the scientific community, most relying on the existence of an ether; however, these attempts produced unacceptable amount of drag and eventually failed as models of gravity. Eventually, in 1915, Einstein published his geometric theory of gravity which still remains the only accepted model of gravity thanks to its excellent agreement with experimental tests. Einstein's theory does not suffer from action at a distance, and does not predict an ether to explain gravity.

More relevant to the concept of the double copy is the similarity, difference, and interaction of gravitational theories with electromagnetism. Without concerning ourselves with an accurate historical treatment, let us compare some of the qualities of gravity and electromagnetism. Newton formulated his law of universal gravitation in 1687, where he states that the force of attraction is proportional to the inverse square of the distance between two massive objects\footnote{Some authors write that the assumption of an inverse relation between gravity and the square of the distance was common by the time Newton published his \textit{Principia Mathematica}\cite{HookvsNewton, InvtCele,HookNewtonCont}}; that is, in mathematical form,
\begin{equation}
    |F_{\text{G}}| = \frac{Gm_1m_2}{|r|^2},
\end{equation}
where $m_1$, $m_2$, and $G$ are the masses of the first and second objects, and Newton's gravitational constant respectively. Let us compare this equation to that of the force between two electrically charged objects, given by Coulombs law
\begin{equation}
    F_{\text{Coulomb}} = \frac{kq_1q_2}{|r|^2}.
\end{equation}
We notice a clear resemblance to the gravitational equation. The gravitational universal constant is replaced with Coulomb's constant $k$, and $q_1$, $q_2$ are the magnitude of the objects' charges. The two equations both follow the inverse square law; however, there is a clear difference between them. In Coulombs law, the charges could take negative values. This leads to the possibility of having both repulsive and attractive forces between charged objects, in clear contrast to the gravitational case, where the value of mass is always positive(in general relativity things can be quite different; in particular, when the cosmological constant is present). Almost a century later, it was discovered by Maxwell that there is such as a thing as the electromagnetic \emph{field}, wherein electrically charged objects interact, providing a non-action at a distance explanation of electromagnetism. Similarly in the case of gravity, Einstein showed that gravity is a manifestation of the curvature of spacetime acting as the gravitational field. With the advent of quantum field theory, and its formulation as a (quantum-)gauge field theory, one thinks of the gravitational and gauge fields, and their quantised versions as force carriers of the respective interactions. In the case of electromagnetism, charged particles interact by exchanging spin-1 carrier particles, called photons, which classically manifest as a Coulomb interaction in the appropriate limit. Similarly in the case of gravity, gravitons are the carrier particles which are exchanged between particles possessing mass(and/or energy). Classically, in the appropriate non-relativistic limit one obtains Newton's potential. Currently, the accepted framework for the interaction found in nature are General Relativity and gauge theories; the latter describing electromagnetism, weak force, and the strong force. The point of the author's humble historical account of gravity as a theory, and its interaction with electromagnetism is to showcase some of the similarities between the two(or more) forces, and the possibility of a dual description of them. We found considerable success in formulating the forces of nature, excluding gravity, as gauge theories; finding also a quantum description of said forces. Gravity, the oldest known force, comes back to haunt us with a lack of a satisfactory formulation of it as a gauge theory, let alone a quantum one. The point of view of the author is to study gravity and gauge theories, their similarities and differences, to better understand the underlying principles governing the interaction. This is by no means a novel idea, for in the literature a great many examples of research exist wherein gravity is studied as a gauge theory, or at least attempted to make it resemble our conventional gauge theories. Famous examples are, Poincaré gauge theory, string theory, and loop quantum gravity; both attempting to cast gravity in the language of gauge theories, where the first one gauges the Poincaré group, the second by treating it as an excitation of a closed string, while the third by the use of Ashtekar variables in hope of formulating quantum gravity canonically via commutation relations\footnote{This is an extremely simplified description of these theories that does not do justice to them. The interested reader should consult the vast amount of literature on these topics.}.

The author views the double copy as yet another mean to understand gravity and its relation to gauge theory. In this thesis only the classical aspect of the double copy is studied. This is a more recent topic of study as opposed to the amplitudal point of view which has seen great success and is ever growing.


\section{What is the double copy?}
The double copy was first investigated in a series of works~\cite{Bern:2008qj,Bern:2010ue,Bern:2010yg} as a relationship between perturbative scattering
amplitudes in gauge theory and gravity. It has been proved at tree
level~\cite{BjerrumBohr:2009rd,Stieberger:2009hq,Bern:2010yg,BjerrumBohr:2010zs,Feng:2010my,Tye:2010dd,Mafra:2011kj,Monteiro:2011pc,BjerrumBohr:2012mg},
where it has a stringy origin~\cite{Kawai:1985xq}. But there is still no non-perturbative proof of the double copy, although the evidence is mounting with a series of papers showing double copy behaviour for amplitudes at higher loop order \cite{Bern:2010ue,Bern:1998ug,Green:1982sw,Bern:1997nh,Carrasco:2011mn,Carrasco:2012ca,Mafra:2012kh,Boels:2013bi,Bjerrum-Bohr:2013iza,Bern:2013yya,Bern:2013qca,Nohle:2013bfa,
  Bern:2013uka,Naculich:2013xa,Du:2014uua,Mafra:2014gja,Bern:2014sna,
  Mafra:2015mja,He:2015wgf,Bern:2015ooa,
  Mogull:2015adi,Chiodaroli:2015rdg,Bern:2017ucb,Johansson:2015oia,Oxburgh:2012zr,White:2011yy,Melville:2013qca,Luna:2016idw,Saotome:2012vy,Vera:2012ds,Johansson:2013nsa,Johansson:2013aca}.

More recently, the double copy/single copy was applied to a class of exact classical solutions. {\it{Double copy}} refers to moving from gauge theory to gravity while {\it{single copy}} is the inverse map from gravity to gauge theory. The Schwarzschild solution was shown to single copy to an electric charge\footnote{More recently, using the double field theoretic formulation of the Kerr-Schild double copy, the point charge in gauge theory is shown to double copy to JNW solution in gravity which includes a dilaton. } \cite{Monteiro:2014cda} and the Taub-NUT solution single copy to a magnetic monopole \cite{Luna:2015paa}. Subsequent to that the single copy of the Eguchi-Hanson solution has been mapped to a self-dual gauge field \cite{Berman:2018hwd,Luna:2018dpt,Armstrong-Williams:2022apo}. More general topologically non-trivial solutions have been double copied in \cite{Sabharwal:2019ngs}. Other work examining symmetries of the perturbative double copy have also appeared in a series of works \cite{Anastasiou:2014qba,Borsten:2015pla,Anastasiou:2016csv,Anastasiou:2017nsz,Cardoso:2016ngt,Borsten:2017jpt,Anastasiou:2017taf,Anastasiou:2018rdx,LopesCardoso:2018xes}.

To make the thesis as self contained as possible and make clear our conventions for the double/single copy, in Chapter \ref{chap:s-duality}, we begin with a description of the Kerr-Schild form in GR and then the single copy prescription.{\footnote{Other conventions and methods for the double copy are also available, see  for example \cite{Anastasiou:2018rdx,LopesCardoso:2018xes}.}} We give the detailed examples of the Schwarzschild black hole and the Taub-NUT solution and their single copies as described first in \cite{Monteiro:2014cda,Luna:2015paa}. Then we describe the Ehlers transform in general before moving on to its application to Schwarzschild and its single copy. We will show  in a number of cases that the duality transformations in electromagnetism  correspond to solution-generating transformations in general relativity, and together preserve the form of the double copy.
\section{The double copy and solution generating transformations}
In Chapter \ref{chap:s-duality}, to further investigate the double copy formalism beyond the perturbative regime, we examine how non-perturbative symmetries in the gauge theory are double copied to gravity. In particular, gauge theories exhibit electromagnetic duality which exchanges electric and magnetic charges. This symmetry, often also called S-duality, was first discovered for classical Maxwell theory but then became a crucial ingredient in the study of properties of Yang-Mills theory \cite{Montonen:1977sn}. General relativity is not known to exhibit S-duality which leads to the question of what S-duality could double copy to. We will address this at the non-perturbative classical level by identifying the solution generating symmetry in general relativity that single copies to electromagnetic duality.

We will show that the double copy of electromagnetic duality is identified as the Ehlers transformation in general relativity. It is worth commenting further that we are working with exact and thus classically non-perturbative solutions in the double copy and the metrics corresponding to the {\it{electric}} and {\it{magnetic}} solutions are mapped between each other using an exact non-local transformation. We will then demonstrate how another solution generating symmetry discovered by Buchdahl \cite{PhysRev.115.1325}, is single copied to the charge conjugation.

We adopt two complementary approaches. First we will use the Kerr-Schild form of the metric where the metric in this form is related to gauge fields in the single copy. Second, in Chapter \ref{chap:Weyl}, we will develop the correspondence introduced in \cite{Luna:2018dpt} where the Weyl curvature in gravity is related to a combination of field strengths in the single copy. In each case we will examine how the electromagnetic transformation in the single copy is related to a transform in relativity.
\section{The Weyl formulation}
In \cite{Luna:2018dpt}  four-dimensional type D spacetimes were investigated, using a double copy formula for the Weyl curvature spinor in terms of a Maxwell spinor. The earlier work of \cite{Mars:2001gd} had used a self-dual Maxwell field, defined in terms of a Killing vector on the spacetime, in order to study how the Weyl tensor transformed under $\text{SL}(2,\mathbb{R})$, and noted in particular that if the Weyl tensor was given by a suitable function quadratic in the Maxwell field, then the $\text{SL}(2,\mathbb{R})$-transformed metric also had a Weyl tensor satisfying this property. In \cite{Alawadhi:2019urr} we, with Peinador Veiga,  studied various metrics for which this is the case, showing how they transform under duality. This work, and that of  \cite{Luna:2018dpt}, suggested that it would be interesting to study cases where the Weyl tensor is given in terms of an Abelian gauge field, by what we will call `Weyl doubling'. 

In Chapter \ref{chap:Weyl} we would thus like to explore classes of gravity and gravity-gauge field systems where the Weyl curvature of a spacetime is given as a quadratic function of an Abelian field strength which is also defined on the spacetime. Crucially we differ from the normal double copy scheme as our gauge field will be taken to be in the curved space time rather than some auxiliary flat background.
We will investigate two such classes - the first where the gauge  field is defined using intrinsic geometric properties of the spacetime, and the second where it is an additional field in the theory.  

\section{The double copy and topology}
An important focus of ongoing research is to ascertain how
general the double copy is, and in particular whether it applies to
non-perturbative information. A successful understanding of the latter
may elucidate the underlying origin of the double copy, or reveal new
ways of thinking about different field theories, that make their
common structures manifest. Previous attempts to study non-perturbative effects include analysing
strong coupling solutions of equations of
motion~\cite{White:2016jzc,DeSmet:2017rve,Bahjat-Abbas:2018vgo,Bahjat-Abbas:2020cyb,Berman:2020xvs},
examining exact algebras underlying the kinematic sectors of different
theories~\cite{Monteiro:2011pc,Borsten:2021hua,Chacon:2020fmr}, using
twistor
methods~\cite{White:2020sfn,Chacon:2021wbr,Farnsworth:2021wvs},
matching solution-generating transformations between different
theories~\cite{Alawadhi:2019urr,Huang:2019cja}, and studying whether
topological information (such as characteristic classes) can be
identified in gauge and gravity
solutions~\cite{Berman:2018hwd,Alfonsi:2020lub}. These studies suggest
that it is worthwhile to consider other global properties of gauge or
gravity solutions, and to ascertain whether or not they can be matched
according to a double copy prescription. In Chapter \ref{chap:Holonomy}, we study the
notion of {\it holonomy} which, loosely speaking, describes how a
vector is transformed after parallel transport around a closed
curve. The set of all such transformations forms a {\it holonomy
  group} and particular elements of the holonomy group are described
by path ordered integrals of the Christoffel connection along the
curve. In gauge theories, the equivalent concept is how the phase of a
charged particle transforms as it moves along a path. This is then
described by {\it Wilson line} operators, which involve integrating
the gauge field along a curve.

It has long been known that holonomy properties of gauge and gravity theories are mathematically analogous. Gauge theories can be thought of in terms of principal fibre bundles, where a base space (corresponding to spacetime) is dressed by fibres acted on by the gauge group. The gauge field itself is then associated with a connection on the fibre bundle. The description in gravity is
similar: one considers the tangent bundle obtained by dressing
spacetime with its tangent space at each point such that the tangent space is now the fibre. The connection on this
bundle corresponds to the connection in gravity. Thus, the holonomy in gauge and gravity theories share a common geometrical origin in that they both arise due to the fact that parallel transport in the base manifold induces transport in the fibre. It is therefore tempting to conclude
that the holonomy groups of gauge and gravity theories are directly
related by the double copy. As we will discuss in detail in this
Chapter \ref{chap:Holonomy}, this assumption is false.

Attempts to explicitly relate holonomy properties and/or Wilson
lines in gauge and gravity theories have been made before. In
particular, refs.~\cite{Modanese:1991nh,Modanese:1993zh} studied the
holonomy properties of gravity solutions using a perturbative field
theory approach, based on Wilson line operators involving the
Christoffel symbol.\footnote{Similar gravitational Wilson lines have
  been used in lattice studies of quantum
  gravity~\cite{Hamber:2007sz,Hamber:2009uz}, and even date back to
  much earlier work~\cite{Mandelstam:1962us} that attempted to recast
  General Relativity in a manifestly coordinate-independent
  form.} The behaviour of these operators in perturbation theory was
found to be in striking contrast with the situation in gauge theory,
an observation that has arisen more recently in the study of
perturbative Wilson loops~\cite{Brandhuber:2008tf}. Given that the
double copy relates scattering amplitudes in perturbation theory, this
already suggests that the traditional holonomy operator in gravity is
not a double copy of its gauge theory counterpart. Indeed, there is a
second operator that one may write down in gravity, involving the
path-length of a particle traversing a curve, which has also been
called a Wilson
line~\cite{Hamber:1994jh,Brandhuber:2008tf,Naculich:2011ry,White:2011yy,Miller:2012an}. It
represents the phase experienced by a scalar particle that travels
around a closed loop (see also the earlier work of
ref.~\cite{Dowker:1967zz}), and in this sense is the correct physical
analogue of the holonomy operator in gauge theory, which arises in the
description of the Aharonov-Bohm effect. It has also been used in the
description of soft radiation~\cite{White:2011yy}, and high-energy
scattering~\cite{Melville:2013qca,Luna:2016idw}, in both cases
overlapping with results that can also be obtained via the double copy
of scattering
amplitudes~\cite{Oxburgh:2012zr,Vera:2012ds,Johansson:2013aca}. Recently,
ref.~\cite{Alfonsi:2020lub} argued that the single copy of this second
operator is indeed the (unique) gauge theory Wilson line, and that it can be
used to express non-trivial topological information of gauge and
gravity solutions in a common language (see also
refs.~\cite{Plefka:2018dpa} for a related study from a different point
of view).\\

The above discussion begs the following question: does the holonomy operator in gravity have a single copy?  The aim of this chapter is to systematically
explore this question, and thus construct a square of four
operators, such that we have a pair of gauge theory operators which
are a meaningful single copy of the two gravity operators mentioned
above. Doing so will allow us to clear up some confusions in
the literature regarding the nature of Wilson lines in gravity, as well as provide yet more glimpses of potential
non-perturbative aspects of the double copy.
\section{The Ricci flow and the double copy}
In Chapter \ref{chap:RicciFlow}, we introduce the concept of the Ricci flow as first introduced by Hamilton \cite{HamiltonRicciFlow} as a way to evolve the metric on a Riemannian manifold. Following the spirit of Chapter \ref{chap:Holonomy}, we attempt to use the Kerr-Schild formulation with the Ricci flow to obtain the Yang-Mills flow, which is a similar evolution equation for a Yang-Mills connection, or in the double copy context to the single copy gauge field.

%% file: chaps/chap03.tex
\chapter{S-duality and the Double Copy}\label{chap:s-duality}
This chapter is based on the first half of \cite{Alawadhi:2019urr}. The second half, which deals with the Weyl double copy, is contained in Chapter \ref{chap:Weyl}.

\section{Introduction and conclusion}
In this chapter we consider the notion of electromagnetic duality in the context of the double copy. More specifically, we pose the following question: Given that the double copy relates local quantities, such as solutions to the equation of motions, on the gauge theory and gravity sides to each other; are non-perturbative properties that these solutions exhibit also related by the double copy map? We know that Maxwell's theory of electromagnetism enjoys a duality, namely, the electromagnetic (EM) duality between the electric and magnetic fields described by the theory as vector fields (pseudo-vector for the magnetic field). For example, if we label the electric field by $E$ and the magnetic field by $B$, then any given solution to the equations of motion is also a solution upon the substitution $(E,B)\rightarrow (B,-E)$. General Relativity is not known to exhibit such a symmetry. Using the double copy map, we will show that in a classical and non-perturbative regime, the corresponding transformation in General Relativity is called Ehlers transformation, and we shall do so using the classical Kerr-Schild double copy map.

First we introduce the Kerr-Schild class of solutions of Einstein's equations, write down the linearised equations, and see how they imply the Abelian Yang-Mills equations. Then we explain how to take the single copy of Kerr-Schild metrics. Next, we examine two relevant solutions to Einstein equations, namely the Schwarzschild and Taub-NUT solutions, and show that they reduce to the Coulomb charge and dyon respectively. In the final part of this chapter, we examine solution generating transformations in General Relativity. First we show how to apply Ehlers transformation as described by Ehlers on the Schwarzschild solution to obtain the Taub-NUT solution. Next we examine another example of a solution generating transformation first discovered by Buchdahl which multiplies the mass parameter in the Schwarzschild metric by a minus sign. Finally, we apply the above transformations to the Taub-NUT solution and show that Ehlers transformation is the gravitational double copy of electromagnetic duality.
\section{The Kerr-Schild classical double copy}\label{sec:DC}

The Kerr-Schild classical double copy was first introduced in \cite{Monteiro:2014cda} as a map between gravity and gauge theory. More specifically, the map requires the existence of a coordinate system such that the line element takes on a special form. Remarkably, in this coordinate system the vacuum Einstein field equations exactly linearise. Here, exact means that no small parameter expansion is performed. Thereby, the Kerr-Schild class of solutions simplifies the study of various spacetimes, a welcoming change from the notoriously difficult to solve second order partial differential equations of Einstein. Indeed, using the Kerr-Schild metric, gauge theory counterparts to the gravitational solutions have been found under the double/single copy map. Following the paper by Monteiro et al. \cite{Monteiro:2014cda}, where the Schwarzschild and Kerr solutions where found to single copy to a Coulomb charge and a charged rotating respectively, follow-up papers started to appear examining more interesting solutions including Taub-NUT \cite{Luna:2015paa} and the topologically non-trivial Eguchi-Hanson space \cite{Berman:2018hwd}. Indeed, the classical Kerr-Schild double copy has grown to be an active area of research giving rise to many interesting avenues of research, ranging from single copies of curved background spaces of AdS/dS; single copies of stress-energy tensors and sources; examining topological properties of gravity/gauge theory solutions, and more \cite{Bahjat-Abbas:2017htu,Carrillo-Gonzalez:2017iyj,Bah:2019sda,Alfonsi:2020lub,Alawadhi:2021uie,CarrilloGonzalez:2019gof}. More recently, the double copy has also seen a Double Field theoretic formulation in \cite{Lee:2018gxc,Cho:2019ype,Kim:2019jwm} by which the Coulomb solution was revisited under a new light, and was found to double copy to a spherically symmetric solution of Einstein-Dilaton gravity, resembling the situation on the scattering amplitude formulation of the double copy. The Kerr-Schild double copy formulation has also been used in the study of non-singular black holes \cite{Easson:2020esh,Pasarin:2020qoa}. The perturbative double copy has also been applied to massive gravity in \cite{Momeni:2020vvr,Momeni:2020hmc,CarrilloGonzalez:2022mxx,Gonzalez:2021bes,Gonzalez:2021ztm}. However, in this thesis, only the original Einsteinian formulation of the Kerr-Schild double copy will be utilised. We now review the basics of Kerr-Schild class of solutions and its double/single copy.
\subsection{A review of the Kerr-Schild double copy}\label{KSReview}
First let use introduce the Kerr-Schild form of the metric and examine the behaviour of Einstein's field equations. Writing  the metric in this form is a crucial step in making the double/single copy  manifest. In what follows we shall use the procedure outlined in \cite{Monteiro:2014cda}. We take the metric $\eta_{\mu\nu}={\rm diag}(-1,+1,+1,+1)$ throughout.

Suppose one has a 4-dimensional manifold $\mc{ M}$ endowed with a Lorentz metric $g_{\mu\nu}$ and a Levi-Civita connection $\nabla$ associated to $g$. A solution to the Einstein's field equations is `Kerr-Schild'  if a set of coordinates $x^\alpha: p\in \mc M \rightarrow \mathbb{R}^4$ on an open subset $U$ of $\mc M$ may be found such that the spacetime metric $g_{\mu\nu}$ may be put in the form
\begin{equation}\label{KSFORM}
  g_{\mu\nu}=\eta_{\mu\nu}+\phi \,k_\mu k_{\nu}\, ,
\end{equation}
where $\eta_{\mu\nu}$ is the metric on Minkowski spacetime $\mathbb{R}^{1,3} = (\mathbb{R}^4, \eta)$, $\phi$ is a scalar field $\phi : U \subseteq M \rightarrow \mathbb{R}$ and $k_{\mu}$ is a co-vector $k: T_p\mc M\rightarrow \mathbb{R} $ satisfying
\begin{equation}\label{KSrelations}
  \eta_{\mu\nu}k^{\mu}k^{\nu}=0=g_{\mu\nu}k^{\mu}k^{\nu}, \quad k^\nu\nabla_{\nu}k_{\mu} =k^\nu\pd_{\nu}k_{\mu}= 0, \quad \text{det}(g) = -1,
\end{equation}
i.e., it is null and geodesic with respect to both the full and background Minkowski metric. The background metric could in principle be curved(see \cite{Bahjat-Abbas:2017htu}). However, for the purposes of the work presented in this thesis, the background metric is taken to be Minkowski unless stated otherwise. The second equation in \eqref{KSrelations} follows from the Einstein field equations
\begin{equation}
R_{\mu\nu}k^\mu k^\nu = 0.
\end{equation}
More generally, in the presence of a matter stress-energy tensor, the condition for $k^{\mu}$ to be geodesic with respect to both the background and the full metric is
\begin{equation}
  T_{\mu\nu}k^{\mu}k^{\nu} = 0.
\end{equation}
For a detailed discussion and derivation see \cite{Kerr-Schild-cond} and \cite{Stephani:2003tm}. The third property can be obtained by using the identity \cite{bernstein11},
\begin{equation}
  \text{det}(I + xy^T) = 1 + y^T I^{-1}x \text{det}(I),
\end{equation}
where $I$ is the identity matrix and $x,y\in \mathbb{R}^n$. Here $xy^T$ inside the determinant is equivalent to the outer product of two vectors. In our case we have
\begin{equation}
  \begin{split}
  \text{det}(\eta + \phi\, k \otimes k)&=(1 + k^{\mu}\eta_{\mu\nu}k^\nu)\text{det}(\eta)\\
  & = \text{det}(\eta),
\end{split}
\end{equation}
which is equal to minus one. The inverse metric can be found by noting that the variation of the Kronecker delta is zero,
\begin{equation}
 \delta (g g^{-1}) = 0,
\end{equation}
from which we find the variation of the inverse metric,
\begin{equation}
  \delta(g^{-1}) = -g^{-1}\delta g g^{-1}.
\end{equation}
Writing it in index notation we have,
\begin{equation}
  \begin{split}
\delta g^{\beta \nu} &= - g^{\mu\beta} \delta g_{\mu\alpha}g^{\alpha \nu}\\
& = - \phi\, k^{\beta}k^{\alpha},
  \end{split}
\end{equation}
where we have used $\delta g_{\mu\nu} = \phi k_{\mu}k_{\nu}$ and raised the indices of the Kerr-Schild vectors with the inverse metrics. Therefore, if we assume our inverse metric to have the form $g^{\mu\nu} = \eta^{\mu\nu} + \delta g^{\mu\nu}$, then we have
\begin{equation}
  g^{\mu\nu}=\eta^{\mu\nu}-\phi \,k^\mu k^{\nu} \, .
\end{equation}
In fact, this is the exact inverse metric because if we assume there are higher order terms of the form,
\begin{equation}\begin{split}
  g^{\mu\nu} &= \eta^{\mu\nu} - \sum_{n= 1}^{\infty} \phi^n( k \otimes_n k)^{\mu\nu}\\
  & = \eta^{\mu\nu} - \phi\, k^\mu k^\nu - \phi^2 k^\mu k^\alpha k_\alpha k^\nu \ldots,
\end{split}
\end{equation}
one is bound to have the contraction of $k_\mu$ with itself in higher order terms, which is null by definition, and hence the series stops at order one in $\phi$. The metric is exact and there is no requirement for the `perturbation' $\delta g$ to be small. The connection components associated with the Kerr-Schild metric simplify greatly \cite{Stephani:2003tm},
\begin{equation}
  \Gamma^\mu_{\alpha\beta} k^\alpha k^{\beta} =0, \quad \Gamma^{\alpha}_{\mu\beta}k_\alpha k^\beta = 0.
\end{equation}

In terms of the scalar field $\phi$ and co-vector $k_\mu$, the Ricci tensor and Ricci scalar in Kerr-Schild coordinates take the form
\begin{equation}
  \begin{split}
    &R^\mu{}_\nu=\frac{1}{2}(\partial^\mu\partial_\alpha(\phi\, k^\alpha k_\nu)+\partial_\nu\partial^\alpha(\phi\, k_\alpha k^\mu)-\partial^2(\phi\, k^\mu k_\nu)) \, ,\\
    &R=\partial_\mu\partial_\nu(\phi \,k^\mu k^\nu)\, ,
\end{split}
\end{equation}
where $\partial^\mu=\eta^{\mu\nu}\partial_\nu$. In the stationary spacetime case ($\partial_0 \phi=\partial_0 k^\mu=0$) one may take the time component of the Kerr-Schild vector as $k^0=1$, with the dynamics in the time component contained in $\phi$. Demanding stationarity further simplifies the components of the Kerr-Schild Ricci tensor as follows 
\begin{equation}\label{KSNabla}
  R^{0}{}_{0}=\frac{1}{2}\nabla^{2}\phi\, ,
\end{equation}
\begin{equation}\label{RicciSingle}
  R^{i}{}_{0}=-\frac{1}{2}\partial_{j}[\partial^{i}(\phi\, k^j)-\partial^j(\phi \,k^i)]\, ,
\end{equation}
\begin{equation}
  R^{i}{}_{j}=\frac{1}{2}\partial_{l}[\partial^{i}(\phi\, k^{l}k_{j})+\partial_{j}(\phi\, k^{l}k^{i})-\partial^{l}(\phi \,k^{i}k_{j})]\, ,
\end{equation}
\begin{equation}
  R=\partial_{i}\partial_{j}(\phi\, k^{i}k^{j})\, ,
\end{equation}
where Latin indices indicate the spatial components. Demanding the spacetime to be stationary is not required to put the metric in Kerr-Schild form; however, it is needed for the double copy procedure. Nevertheless, some non-stationary Kerr-Schild metric do allow a double copy prescription like the plane-wave solution \cite{Monteiro:2014cda}.

Now  define a local gauge field $A_{\mu}=\phi \,k_\mu$, with the Maxwell field strength $F_{\mu\nu}=\partial_{\mu} A_{\nu}-\partial_{\mu}A_\nu$. Taking the stationary case of the vacuum Einstein equations $R^\mu_\nu=0$, namely equation \eqref{RicciSingle}, one finds that the gauge field satisfies the stationary Abelian Maxwell equations
\begin{equation}\label{RicciField}
  \partial_{\mu}F^{\mu\nu}=\partial_{\mu}(\partial^{\mu}(\phi\, k^{\nu})-\partial^{\nu}(\phi\, k^{\mu}))=0 \, ,
\end{equation}
where the $\mu = 0, \nu = 0$ component follows from \eqref{KSNabla} and the rest from \eqref{RicciSingle} . The remarkable thing about the double copy is that if now one considers a non-Abelian gauge field $A^a_{\mu}$ with the gauge group index $a$, there is still a single copy/double copy relationship. The recipe is to take the quantity $\phi \,k_\mu k_\nu$ of a given gravity solution and strip off one of the Kerr-Schild vectors and dress with a gauge group index to get the corresponding gauge field $A^a_\mu=c^a \phi \,k_\mu$. There is no derivation as such for this procedure but by now there is a compelling amount of evidence as listed in the introduction.
Thus, the basic statement of the double/single copy we will be applying is the following:

\textit{If $g_{\mu\nu}=\eta_{\mu\nu}+\phi\, k_\mu k_{\nu}$ is a stationary solution of Einstein's equations, then $A_{\mu}^a=c^a\phi\, k_\mu$ is a solution of the Yang-Mills equations which is linearised by the Kerr-Schild coordinates allowing an arbitrary choice for the constant $c^a$.}

Essentially, the above amounts to the replacement of one Kerr-Schild vector with the trivial colour vector.

\subsection{Schwarzschild and NUT spacetimes}

We shall demonstrate the double/single procedure on two important gravity solutions, the Schwarzschild black hole and the Taub NUT spacetime.

\subsubsection{Schwarzschild spacetime}
The Schwarzschild solution defined on $\mathbb{R}\times (2M, \infty)\times S^2$ is an asymptotically flat, static and spherically symmetric solution of Einstein's field equations. In Schwarzschild coordinates $(t, r, \theta, \phi)$, the line element takes the form
\begin{equation}
  ds^2=-\Big(1-\frac{2GM}{r}\Big)dt^2+\Big(1-\frac{2GM}{r}\Big)^{-1}dr^2+r^2d\Omega^2
  \end{equation}

where $d\Omega^2=d\theta^2+\sin^2\! \theta\, d\phi^2$ is the line element on the unit sphere, $G$ is Newton's constant, and $r^2=x^2+y^2+z^2$ is the radial distance from the origin, and constant $M\in \mathbb{R}$ is the `mass' of the gravitating body.

In order to apply the single copy procedure, one must write this metric in Kerr-Schild form. To do this, we apply the coordinate transformation $l=t+\bar{r}$ with $d\bar{r}=(1-\frac{2GM}{r})^{-1}dr$, so that the metric takes the form
\begin{equation}
  ds^2=-dl^2+2dldr+r^2d\Omega^2+\frac{2GM}{r}dl^2 \, .
\end{equation}
A further coordinate transformation of the form $l=\bar{t}+r$ is then applied so that the metric becomes
\begin{equation}
  ds^2=-d\bar{t}^2+dr^2+r^2d\Omega^2+\frac{2GM}{r}(d\bar{t}^2+dr^2+2d\bar{t}dr) \, .
\end{equation}
Notice that the first three terms are just the flat Minkowski metric in spherical coordinates. Using the definition of $r$ given above to transform into `Cartesian' coordinates the metric becomes
\begin{equation}
  ds^2=\eta_{\mu\nu}dx^\mu dx^\nu+\frac{2GM}{r}k_\mu k_\nu dx^\mu dx^\nu\, ,
\end{equation}
where the null vector $k^\mu$ is defined by
\begin{equation}
  k^\mu=\left(1,\frac{x^i}{r}\right),\qquad i=1...3.
\end{equation}
The metric is now in Kerr-Schild form:
\begin{equation}\label{kerrsh}
  g_{\mu\nu}=\eta_{\mu\nu}+\frac{2GM}{r}k_\mu k_\nu \, .
\end{equation}
Note that in this coordinate system, the radial distance $r$ has been continued past the point $r = 2M$ such that the range is now $0 < r < \infty $
In terms of a metric
\begin{equation}
 g_{\mu\nu}=\eta_{\mu\nu}+\kappa\, h_{\mu\nu}\,
\end{equation}
we have
\begin{equation}
  h_{\mu\nu}=\frac{\kappa}{2}\phi k_\mu k_\nu, \qquad \phi=\frac{M}{4\pi r},\quad \kappa^2 = 16\pi G\, .
\end{equation}
The single copy then yields
\begin{equation}
  A^\mu=\frac{gc_aT^a}{4\pi r}k^\mu \, ,
 \end{equation}
where, motivated by the prescription for amplitudes, we have made the replacements
\begin{equation}\label{DCr}
  \frac{\kappa}{2}\rightarrow g, \quad M\rightarrow c_aT^a, \quad k_\mu k_\nu\rightarrow k_\mu \, .
\end{equation}
The first one is just choosing the right coupling constant, the second one being the correspondence between the charges of  the two theories. 

It is interesting to note that we have single copied a spacetime that includes the event horizon, i.e., all the points $p\in \mc M$ such that $r > 0$ are included, and not just the exterior solution of Schwarzschild. However, we lose all information about the event horizon at $r = 2GM$ on the gauge theory side upon single copying. In other words, what happens to the event horizon once single copied?

\subsubsection{Taub-NUT spacetime}
\textit{This discussion makes use of fibre bundle theory. Should the reader require a refresher for the topic, Chapter \ref{chap:Holonomy} provides a short review.}\\

The solution known as Taub-NUT was first derived by Taub \cite{Taub:1950ez} and generalised by Newman, Tamburino, and Unti \cite{Newman:1963yy} in 1963. The NUT solution is a non-asymptotically flat, and stationary solution with isometry group $\text{U}(1)\otimes \text{SU}(2)$ \cite{hawkingInstanton, Yohannes:2021iwc, Cotaescu:2003gx,Ortin:2015hya}.
Following Ort\'in \cite{Ortin:2015hya}, the Taub-NUT metric can be written in polar coordinates $(t, r , \theta, \phi)$ as
\begin{equation}
  ds^2=-f(r)(dt-2N\cos{\theta}d\phi)^2+f(r)^{-1}dr^2+(r^2+N^2)d\Omega^2,\label{Taub-NUT metric}
\end{equation}
where
\begin{equation}
  f(r)=\frac{(r-r_{+})(r-r_{-})}{r^2+N^2},\qquad r_{\pm}=M\pm r_0,\qquad r_{0}^2=M^2+N^2.
\end{equation}
This can be thought of as a generalisation of the Schwarzschild solution with an additional topological  charge that is called the \emph{NUT charge} $N\in\mathbb{R}$. This solution exhibits the following interesting properties:
\begin{itemize}
  \item The solution reduces to the Schwarzschild metric for $N=0$ and $M\neq 0$, but is not trivial in the limit $M\rightarrow 0$ and $N \neq 0$.
  \item Taking the Newtonian limit shows that $M$ is indeed the mass of the source. The NUT charge has no Newtonian analogue. In fact, if we look at the diagonal component of the metric tensor, we find that in the limit $r\rightarrow \infty$ the metric component behaves as $g_{t\phi}|_{r\rightarrow\infty}\sim 2N\cos{\theta}$, which is the gauge field describing a \textit{magnetic monopole} with charge $N$.
  \item The solution defines its own class of asymptotic behaviour labelled by $N$ and is associated with the non-vanishing at infinity of the $g_{t\phi}$ component of the metric.
  \item The solution admits Dirac-like singularities at $\theta=0,\pi$, requiring the introduction of two coordinate patches with different time coordinates given by $t_N = t_S + 4N\phi$. Since $\phi$ is compact with period $2 \pi$, $t_{N,S}$ is then compact with period $8\pi N$. This changes the topology of the solution such that it acquires spherical symmetry which can be viewed as a Hopf fibration over $S^2$ along the time direction. More succinctly, the metric is defined on $\mathbb{R}\times S^3$:
\end{itemize}

\begin{equation}\label{Taub-NUTBundle}
  ds^2_{N,S}=f(r)^{-1}dr^2 \underbrace{-f(r)\big(dt_{N,S}-2N(\mp 1 +\cos{\theta})d\phi\big)^2+(r^2+N^2)d\Omega^2}_{S^3},
\end{equation}
where $t, \theta, \phi$ are Euler coordinates on $S^3$ with ranges $0\leq t \leq 8\pi N,\, 0\leq \theta \leq \pi,\, 0\leq \phi\leq 2\pi$ . In fact, the last point above implies that the NUT charge is quantised much like the case of a magnetic monopole defined on $S^2$(See \ref{subsec: DiracMonopole}). To see this, let us view the Taub-NUT space as a product $\mathbb{R}\times S^3$ where the 3-sphere part is a Hopf fibration treated as a non-trivial $\text{U}(1)$-principal bundle. The bundle is constructed by taking the base space to be $S^2$ and fibering along the time direction which is a circle $\text{U}(1) \cong S^1$. In terms of coordinates, the projection map is given by $(t, \theta,\phi) \xrightarrow{\pi} (\theta, \phi)$. The bundle induces a connection on the base space given by taking local sections on each side of the Hemisphere, $A_N=\sigma_{N}^*\omega$ and $A_S=\sigma_S^*\omega$. The transition function, $t_{NS}: S^1 \rightarrow \text{U}(1)$, tells us how to transform from one side of the $S^2$ to the other and is given by $t_{NS} = \exp[i\varphi(\phi)]$. This induces a relation between the local connections $A_N$ and $A_S$ on the equator, $U_N\cup U_S$, given by a gauge transformation,
\begin{equation}
  A_N = t^{-1}_{NS}A_St_{NS}  - i t^{-1}_{NS}dt_{NS} = A_S + d\varphi.
\end{equation}
More explicitly, upon viewing eq.~\eqref{Taub-NUTBundle} as a bundle metric, the local connections are just the $t\phi$ component of the metric,
\begin{equation}\label{NUTConnection}
  A_{N,S} = -2N(\mp 1+ \cos{\theta})d\phi.
\end{equation}
To see how the quantisation of the NUT charge arises, one requires that the transition function to be single valued around the equator, meaning,
\begin{equation}\begin{split}
  \Delta\varphi = \oint_{S^1}d\varphi &= \oint_{S^1} A_N - \oint_{S^1}A_S\\
  & = 4N\oint_{S^1}d\phi,
\end{split}
\end{equation}
using eq.~\eqref{NUTConnection}. Now as $\phi$ runs from $0$ to $2\pi$, for $t_{N,S}$ to be uniquely defined, $\varphi$ must be a multiple of $2\pi$, i.e.,
\begin{equation}
  \Delta\varphi = 4N (2\pi) = 2\pi m,\quad m\in \mathbb{Z}.
\end{equation}
This implies that the NUT charge is quantised(see \cite{Emond:2021lfy} for related results),
\begin{equation}\label{QuantNUT}
  N = \frac{m}{4}.
\end{equation}
This result is analogous to the Dirac quantisation of the magnetic monopole charge. In the next section we will see that this quantisation of the NUT charge forces the Ehlers group, thought of as $\text{SL}(2,\mathbb{R})$, to reduce to $\text{SL}(2,\mathbb{Z})$. Indeed, similar to what happens to the electromagnetic duality upon quantisation; more on this later on in this chapter. 

We saw earlier that the mass in gravity single copies to an electric or colour charge. In the case of the Taub-NUT solution, the NUT charge $N$ is analogous to the magnetic monopole charge. The single copy provides a way to make this statement precise. Under the single copy, the NUT charge $N$ single copies to a magnetic monopole on the gauge theory side as follows.

The Taub-NUT metric written in coordinates introduced by Plebanski \cite{PLEBANSKI1975196} exhibits a double Kerr-Schild form, as shown in \cite{Chong:2004hw}. This means the Taub-NUT metric may be written as:
\begin{equation}
    g_{\mu\nu}=\eta_{\mu\nu}+\kappa h_{\mu\nu}
    =\eta_{\mu\nu}+\kappa(\phi\, k_{\mu}k_{\nu}+\psi\, l_{\mu}l_{\nu}) \, ,
\end{equation}
where we restated the gravitational coupling constant $\kappa$, and the vectors $k^\mu$ and $l^\mu$ must satisfy the following conditions
\begin{equation}
  k^2=l^2=k\cdot l=0\, ,\qquad (k\cdot \nabla)k_\mu=0,\qquad (l\cdot \nabla)l_\mu=0 \, ,
\end{equation}
where the geodesic condition is valid with respect to both the background and full metric. The Plebanski coordinate system linearises the Einstein equations as with the single Kerr-Schild form \cite{PLEBANSKI1975196,Luna:2015paa}. 
The generalisation of the single copy prescription for the gauge field, for our stationary Taub-NUT spacetime, is then
\begin{equation}
 A_{\mu}^a=c^a(\phi\, k_{\mu}+\psi\, l_{\mu}) \, .
\end{equation}
Following \cite{Luna:2015paa}, we have made the following substitutions
\begin{equation}
  \frac{M\kappa}{2}\rightarrow (c_aT^a)g_s\, ,\qquad \frac{N\kappa}{2}\rightarrow (c_aT^a)\tilde{g}_s\, .
\end{equation}
The field strength is then
\begin{equation}
  F=\frac{1}{2}F_{\mu\nu}\,dx^\mu \wedge dx^\nu=-\frac{c_aT^a}{8\pi}\Big(\frac{g_{s}}{r^2}dt\wedge dr+\tilde{g}_{s}\sin{\theta}d\theta\wedge d\phi\Big) \, .\label{Taub-NUT single copy}
\end{equation}
The first term on the right-hand side above is  the pure electric charge corresponding to a Coulomb solution that was derived in the Schwarzschild case. The second term is a magnetic monopole charge which is single copied from the NUT contribution on the gravity side. In summary, the Schwarzschild single copies to a Coulomb solution and the NUT charge to a magnetic monopole. Recently, generalisations of the Kerr-Schild double copy have appeared in the literature. Starting with the work of K. Lee \cite{Lee:2018gxc} which introduced a double field theoretic(DFT) formulation of the double copy, and subsequently generalised to heterotic DFT \cite{Cho:2019ype}. Moreover, by applying the DFT formulation to the point charge case, it has been found that the Coulomb charge double copies to the two parameter Janis-Newman-Winicour solution in gravity. The latter being a static, spherically symmetric, asymptotically flat solution that generically includes a dilation field \cite{Kim:2019jwm}. Indeed, this generalisation of the Coulomb solution is more faithful to the string theory origin of the double copy for scattering amplitudes.

\section{Solution generating transformations in general relativity}\label{ehlerss}

A solution generating transformation is simply a recipe for obtaining a new solution to Einstein's equations from a known one. Their existence is remarkable given that Einstein's equations are a set of ten  second order coupled non-linear differential equations that are notoriously difficult to solve. Whilst explicit symmetries in Einstein's equations make some transformations straightforward, there are  ``hidden symmetries" that allow us to generate very non-trivial solutions through this transformation technique. This approach to solving Einstein's equations was studied by Buchdahl, Ehlers, Geroch and  Ernst \cite{Ernst:1967wx}. The transformations require a Killing symmetry in the spacetime. When such a symmetry is present then there are a set of solution-generating techniques which we describe below.

\subsection{Ehlers transformation}

The Ehlers transformation is a transformation that acts on the parameters of a static solution of Einstein gravitational field equations and generates other solutions of the field equations that are in general stationary. In what follows we describe how the Ehlers transformation works \cite{Ehlers:1961zza}, following \cite{Momeni:2005uc}.
It is assumed that the spacetime possesses a time-like Killing vector $\xi$ that generates an isometry. We would like to mention that while this formulation relies on the existence of a time-like killing vector field, the formulation of Ehlers transformation as introduced by Mars in \cite{Mars:2001gd}, which we will see in the next chapter, does not; moreover, it works even for null and space-like killing fields. Now back to the formulation as introduced by Ehlers. First perform a (1+3) decomposition of the metric and choose coordinates $x^{\mu}=\{x^0,x^i\}$ to put the line element in the following form
\begin{equation}
ds^2=-e^{2U}(dx^0+A_{i}dx^{i})^2+d\ell^2\, ,
\end{equation}
where  we define
\begin{equation}
A_{i}=\frac{-g_{0i}}{g_{00}} \, , \quad 
e^{2U}= -g_{00}\, , \quad 
dl^2 = \gamma_{ij}dx^i dx^j
\end{equation}
and
\begin{equation}
  \gamma_{\alpha\beta}=(-g_{\alpha\beta}+\frac{g_{0\alpha}g_{0\beta}}{g_{00}})  \, .
\end{equation}
The Ehlers transformation states that if
\begin{equation}\label{Ehler form}
  g_{\mu \nu}dx^\mu dx^\nu=-e^{2U}(dx^0)^2+e^{-2U}d\tilde{\ell}^2\, ,
\end{equation}
with $d\tilde{\ell}^2=e^{2U}d\ell^2$, is the metric of a static spacetime, then
\begin{equation}\label{result form}
  \bar{g}_{\mu \nu}dx^\mu dx^\nu=-(\alpha\cosh(2U))^{-1}(dx^0+A_{i}dx^{i})^2-\alpha\cosh(2U)d\tilde{\ell}^2\, ,
\end{equation}
where $\alpha$ is a positive constant, $U=U(x^i)$ and $A_i=A_{i}(x^j)$, is the metric of a stationary spacetime provided that $A_{i}$ satisfies Ehlers equation
\begin{equation}\label{Ehlers eq2}
 -\alpha \sqrt{\tilde{\gamma}}\,\epsilon_{ijk}\, U^{,k}=A_{[i,j]}\, ,
\end{equation}
where $\tilde{\gamma}_{ab}$ is the conformal spatial metric. Using this method, one generates a stationary solution from a static one by finding the potential $A_{a}$ which is related to a given static potential $U$. Let us see how this comes about by applying this procedure to the Schwarzschild metric.

Start with the Schwarzschild metric in the form \eqref{Ehler form}, with the potential taking the form
\begin{equation}\label{Pot}
U_c=\frac{1}{2}\ln{(1-\frac{2m}{r})}+C
\end{equation}
and
\begin{equation}
  d\tilde{\ell}^2=dr^2+r^2e^{2U}(d\theta^2+\sin^2{\theta}d\phi^2)\, ,
\end{equation}
where $C$ is a constant and $U(x^a)$ only depends on the spatial coordinates. Then the Ehlers equation \eqref{Ehlers eq2} reads
\begin{equation}
  -\alpha r^2e^{2U}\sin{\theta}\,\tilde{\gamma}^{rr}U_{,r}=\frac{1}{2}(A_{\theta,\phi}-A_{\phi,\theta}) \, .
\end{equation}
One may choose a gauge such that $A_{\theta,\phi}=0$. Then substituting \eqref{Pot} in the above equation and solving for the field $A_{\phi}$ we find
\begin{equation}\label{gravmag}
  A_{\phi}(\theta)=-2\,\alpha\, m \cos{\theta}\, .
\end{equation}
The resulting metric is then given by  \eqref{result form} as
\begin{equation}\label{momenehler}
ds^2=-\frac{r(r-2m)}{\alpha f(r)}(dt-2\alpha m\cos{\theta}d\phi)^2+\frac{\alpha f(r)}{r(r-2m)}dr^2+\alpha f(r)d\Omega^2,
\end{equation}
where
\begin{equation}
  f(r)=r^2\left(\frac{1+c_1^2}{2c_1}\right)+2m^2c_1-2mrc_1\, , \quad c_1=e^{2C} \, .
\end{equation}

Recovering the Schwarzschild metric as $\alpha\rightarrow 0$ requires us to set
\begin{equation}
  \alpha=\frac{2c_1}{1+c_1^2}, \quad |c_1|\leq 1.
\end{equation}
Now apply the following changes of variables and redefinitions of constants;
\begin{equation}\label{massc}
  M=m\,\frac{1-c_1^2}{1+c_1^2}\, , \quad \alpha\, m=N, \quad r-N c_1=R\, ,
\end{equation}
to get
\begin{equation}
  ds^2=-\frac{R^2-2MR-N^2}{R^2+N^2}(dt-2N\cos{\theta}d\phi)^2
  +\frac{R^2+N^2}{R^2-2MR-N^2}dR^2
  +(R^2+N^2)d\Omega^2 \, ,
\end{equation}
which is the metric for the NUT spacetime with NUT charge $N$. Remarkably, the constant $C$ in the potential $U$, while having no effect on the Schwarzschild spacetime, plays an important role in generating solutions when the Ehlers transformation is applied. Had we started with $C=0$ we would have ended up with the pure NUT spacetime ($M=0$). This means that the seed metric for both the NUT with mass and pure NUT metric is the Schwarzschild metric.


\subsection{Buchdahl's reciprocal transformation}
In \cite{PhysRev.115.1325} Buchdahl showed that if a solution of Einstein's equations admits a Killing vector field one can obtain a new solution by applying the so called ``reciprocal transformation''. For coordinates adapted to the Killing direction such that the Killing vector is given by $\frac{\partial}{\partial x^\alpha}$, then the line element may be written, where there is no summation on the index $\alpha$,
\begin{equation}
  ds^2=g_{\beta\gamma}dx^{\beta}dx^{\gamma}+g_{\alpha\alpha}(dx^\alpha)^2 \, .
\end{equation}
The reciprocal transformation then generates the following line element
\begin{equation}
  ds^2=g_{\alpha\alpha}^{\frac{2}{d-3}}g_{\beta\gamma}dx^{\beta}dx^{\gamma}+g_{\alpha\alpha}^{-1}(dx^\alpha)^2 \, .
\end{equation}
This transformation may be written acting on metric components in this adapted coordinate system as
\begin{equation}
  (g_{\alpha\alpha},g_{\beta\gamma})\rightarrow (g_{\alpha\alpha}^{-1},g_{\alpha\alpha}^{\frac{2}{d-3}}g_{\beta\gamma})\, .
\end{equation}
The reader familiar with string theory will note that Buchdahl's reciprocal transformation is essentially T-duality  as a solution generating transformation,  predating it by some thirty years.
We shall apply this to the Schwarzschild metric. As noted above, the $d=4$ Schwarzschild metric admits a time-like Killing vector $\partial/\partial t$ so that the transformation yields
\begin{equation}
  (g_{tt},g_{ij})\rightarrow (g_{tt}^{-1},g_{tt}^2g_{ij})\, ,
\end{equation}
which upon substitution into the metric gives
\begin{equation}\label{timelikebuch}
  ds^2=-\frac{dt^2}{1-\frac{2M}{r}}+\Big(1-\frac{2M}{r}\Big) dr^2+r^2\Big(1-\frac{2M}{r}\Big)^2(d\theta^2+\sin{\theta}^2d\phi^2).
\end{equation}
The Schwarzschild metric also admits the spatial Killing vector $\partial/\partial\phi$. Applying the same transformation, but now with $\alpha=\phi$, we get the following metric
\begin{equation}
  ds^2=-r^4\sin^4{\theta}\Big(1-\frac{2M}{r}\Big)  dt^2+r^4\sin^4{\theta}\Big(1-\frac{2M}{r}\Big) ^{-1} dr^2  +r^6\sin^4{\theta}d\theta^2
  +\frac{1}{r^2\sin^2{\theta}}d\phi^2\, .
\end{equation}
This new solution is completely unrelated to the original seed metric, unlike \eqref{timelikebuch} which was obtained using the time-like Killing symmetry and is indeed related to the Schwarzschild metric by a simple coordinate transformation  $\bar{r}=r-2M$ whereby one obtains the Schwarzschild metric but with negative mass parameter
\begin{equation}
  ds^2=-\Big(1-\frac{2M}{\bar{r}}\Big) dt^2+\Big(1+\frac{2M}{\bar{r}}\Big)^{-1}d\bar{r}^2 +\bar{r}^2(d\theta^2+\sin{\theta}^2d\phi^2).
\end{equation}

Now if we drop the bar from $\bar{r}$ and write the negative mass Schwarzschild metric in Kerr-Schild form,
\begin{equation}
g_{\mu\nu}=\eta_{\mu\nu}-\frac{2GM}{r}k_\mu k_\nu
\end{equation}
and then do the usual single copy procedure where $-M\rightarrow -c_aT^a$, we arrive at
\begin{equation}
  \tilde{A}_\mu=(-\frac{gc_aT^a}{4\pi r},0,0,0)\, ,
\end{equation}
which has the opposite sign of the gauge field $A_{\mu}$ compared to the single copy of the positive mass Schwarzschild solution \cite{Monteiro:2014cda}, {i.e.},
\begin{equation}
  \textrm{`Single copied 4D Buchdahl'}: A_{\mu} \rightarrow \hat{C}A_\mu = -{A_\mu}\, .
\end{equation}
This shows that the Buchdahl reciprocal transformation associated with the time-like Killing vector in the Schwarzschild spacetime is the gravitational analogue of charge conjugation on the gauge theory side acting as a $\mathds{Z}_2$ transformation on the charge.

We now examine this transformation acting on the Taub-NUT solution by first using the Schwarzschild metric as our seed metric on which we act with the Buchdahl transform, followed by the Ehler transformation. Firstly, note that Buchdahl transformation corresponds to sending $U\rightarrow -U$ when one writes the Schwarzschild metric in the form
\begin{equation}
  ds^2=-e^{2U}(dx^0)^2+e^{-2U}d\tilde{\ell}^2, \quad U=\frac{1}{2}\ln(1-\frac{2M}{r})\, .
\end{equation}
Thus the Buchdahl-transformed Schwarzschild metric reads
\begin{equation}
    ds^2=-e^{-2U}(dx^0)^2+e^{2U}d\hat{\ell}^2, \quad d\hat{\ell}^2=dr^2+e^{-2U}r^2d\Omega^2.
\end{equation}
which as before is just the Schwarzschild metric with negative mass as shown earlier. Now construct the Ehlers transformed metric as before. Considering the Ehlers equation for the reciprocal solution
\begin{equation}
  -\alpha \sqrt{\gamma}\epsilon_{\alpha\beta\gamma}\hat{U}^{,\gamma}=\alpha \sqrt{\gamma}\epsilon_{\alpha\beta\gamma}U^{,\gamma}=\hat{A}_{[\alpha,\beta]},
\end{equation}
and comparing it to the standard one \eqref{Ehlers eq2} we find that
\begin{equation}
\hat{A}_\mu=-A_\mu\, .
\end{equation}
Then, following exactly the same steps as before, i.e., solving for $A_\mu$ while imposing axial symmetry, we find
\begin{equation}
\hat{A}_\phi(\theta)=2\alpha M \cos{\theta}
\end{equation}
up to a constant term, and as before the NUT charge is given by
\begin{equation}
  \hat{N}=\alpha M=-N \, .
\end{equation}
Thus the NUT charge has acquired a negative sign. Comparing the field $A_\mu$ to Eq. \eqref{gravmag}, the reciprocal Taub-NUT metric then reads
\begin{equation}\begin{split}
  &ds^2=-\big(\alpha\cosh(2U)\big)^{-1}(dx^0+\hat{A}_{\beta}dx^{\beta})^2+\alpha\cosh(2U)d\hat{\ell}^2\, , \\
  &d\hat{\ell}^2=dr^2+e^{-2U}r^2d\Omega^2.
\end{split}
\end{equation}
Upon applying the single copy procedure \cite{Luna:2015paa} as in the previous sections using the multi-Kerr-Schild form of the metric, one finds again that the resulting gauge field has negative electric and magnetic monopole charges.

In summary, we show below diagrams of the transformations and their effects on the parameters of both the Schwarzschild and Taub-NUT solutions. The vertical lines indicate the application of the double/single copy (`S.D.') procedure.
\begin{center}
\begin{tikzpicture}[scale=3]
\node (A) at (0,1) {$M$};
\node (B) at (1,1) {$-M$};
\node (C) at (0,0) {$Q_e$};
\node (D) at (1,0) {$-Q_e$};
\path[<->,font=\tiny]
(A) edge node[above]{$Buchdahl$} (B)
(A) edge node[left]{$S.D.C$} (C)
(B) edge node[right]{$S.D.C$} (D)
(C) edge node[above]{$\hat{C}$} (D);
\end{tikzpicture}
\hspace{1cm}
\begin{tikzpicture}[scale=3]
\node (A) at (0,1) {$M$};
\node (B) at (1,1) {$(M,N)$};
\node (C) at (0,0) {$Q_e$};
\node (D) at (1,0) {$(Q_e,Q_M)$};
\path[<->,font=\tiny]
(A) edge node[above]{$Ehlers$} (B)
(A) edge node[left]{$S.D.C$} (C)
(B) edge node[right]{$S.D.C$} (D)
(C) edge node[above]{$E-M duality$} (D);
\end{tikzpicture}
\end{center}

From the action of the Ehlers transformation on the charges in the double copy one sees that the single copy of the Ehlers transformation is electromagnetic duality in the gauge theory;  alternatively the double copy of electromagnetic duality is generated by the Ehlers transformation in gravity.

More formally, the Ehlers transformation is an element of $\text{SL}(2,\mathbb{R})$ as introduced by Geroch \cite{doi:10.1063/1.1665681}. However when this acts on the Taub NUT solution the quantisation of NUT charge, as shown in the argument preceding eq.~\eqref{QuantNUT}, means that the group must be broken to $\text{SL}(2,\mathbb{Z})$ so as to preserve the quantisation condition. This exactly follows what happens with the electromagnetic duality group acting on the dyon spectrum. Classically the group  is $\text{SL}(2,\mathbb{R})$ but this reduces to $\text{SL}(2,\mathbb{Z})$ in order to maintain Dirac quantisation for the magnetic charges.

\section{From Hodge duality to $\text{SL}(2,\mathbb{Z})$}\label{sec:HodgeMag}
It is a good idea at this point to remind the reader of the duality property of electromagnetism and how the ideas discussed above tie together. We will examine the duality operations of Maxwell's electromagnetism as a duality transformation on the field strength tensor. Our playground here is a four dimensional Minkowksi spacetime $(\mathbb{R}^4, \eta)$ with $\text{diag}(-1,1,1,1)$. Quantisation will also be taken into account throughout the discussion. In fact, it is quantisation that forces the duality group to reduce to its discrete subgroup.

Let us begin by writing the vacuum Maxwell's equations in their non-Lorentz covariant form,

\begin{align}
    \vec{\partial}\cdot \vec{E} &= 0 &  \vec{\partial}\cdot \vec{B}&= 0\\ 
    \vec{\pd}\times \vec{E} &= -\frac{\pd\vec{B}}{\pd t} &  \vec{\pd}\times \vec{B} &= \frac{\pd \vec{E}}{\pd t},
\end{align}
one notices that the equations are invariant under the map $(\vec{E},\vec{B}) \rightarrow (\vec{B},-\vec{E})$. Indeed, if applied again we have $(\vec{B},-\vec{E}) \rightarrow (-\vec{E},-\vec{B})$ which is equivalent to the pair $(\vec{E}, \vec{B})$. Hence the duality group is the cyclic group ${\mathds Z}_2$. However, once the equations are written in a Lorentz covariant form using the field strength, $F_{\mu\nu}$ with the electric and magnetic fields corresponding to the components $(F^{0i}, F^{ij}) = (-E^i, -\epsilon_{ijk}B^k)$, where the latin indices are understood to be spatial and $\epsilon_{ijk}$ is the totally symmetric Levi-Civita symbol. In this description, the duality transformation is nothing but the application of the Hodge dual on a the two-form $F_{\mu\nu}$ defined by $\star F^{\mu\nu}  = \frac{1}{2}\epsilon^{\mu\nu\lambda\rho} F_{\lambda\rho} $. For Maxwell's equations are now,
\begin{equation}
  \pd_{\nu}F^{\mu\nu} = 0,\quad \pd_\nu \star F^{\mu\nu} = 0,
\end{equation}
and the duality transformation is given by,
\begin{equation}
  (F^{\mu\nu} ,\star F^{\mu\nu}) \rightarrow (\star F^{\mu\nu} ,- F^{\mu\nu}),
\end{equation}
where the minus sign is due to the fact that $\star\star = -1$ on a 4-dimensional Lorentzian manifold. Indeed, the Lorentz covariant Maxwell's equations are invariant under the duality transformation given by the action of the hodge dual on $F_{\mu\nu}$. In the presence of sources, the Maxwell's equations are modified by the addition of the current term,
\begin{equation}
  \pd_{\nu}F^{\mu\nu} = j_e^\mu,\quad \pd_\nu \star F^{\mu\nu} = j_m^\mu,
\end{equation}
where $j_e^\mu$ and $j_m^\mu$ are the electric and magnetic sources. Now remarkably, the electromagnetic duality still holds provided one also map the sources into each other via the map $(j_e^\mu, j_m^\mu) \rightarrow (j_m^\mu, - j_e^\mu) $. In fact we can go even further, under further scrutiny, the reader should notice that the equations of motion are invariant under a continuous transformation given by
\begin{equation}\begin{split}
  F^{\mu\nu}& \longrightarrow \cos{\phi} F^{\mu\nu} + \sin{\phi} \star F^{\mu\nu},\\
  \star F^{\mu\nu}& \longrightarrow \cos{\phi} \star F^{\mu\nu} - \sin{\phi} F^{\mu\nu},
\end{split}
\end{equation}
and similarly for the currents,
\begin{equation}\begin{split}
  j_e^{\mu}& \longrightarrow \cos{\phi}  j_e^{\mu} + \sin{\phi}  j_m^{\mu},\\
  j_m^{\mu}& \longrightarrow \cos{\phi}  j_m^{\mu} - \sin{\phi}  j_e^{\mu}.
\end{split}
\end{equation}
This further augments the duality group from $\mathds{Z}_2$ to $SO(2)$ since we are effectively rotating the charges into each other,
\begin{equation}\label{ChargeRot}
\begin{pmatrix}
  \cos{\phi}& \sin{\phi}\\
  -\sin{\phi}& \cos{\phi}
\end{pmatrix}
\begin{pmatrix}
  e \\
  g   
\end{pmatrix}
=
\begin{pmatrix}
  e\cos{\phi}  + g\sin{\phi} \\
  g\cos{\phi} - e\sin{\phi}  
\end{pmatrix},
\end{equation}
where $e$ and $g$ are the electric and magnetic charges respectively.Therefore, so far we have established the relationship between electromagnetic duality, hodge duality, $\text{SO(2)}$, and $\mathds{Z}_2$. Where does the $\text{SL}(2,\mathbb{R})$ come into play? First, let us go back to Dirac and his quantisation condition to see what kind of effect does it have on electromagnetic duality. We briefly review the setup and results of the magnetic monopole and its effect on charge quantisation.
\subsection{Dirac quantisation and the reduction of $\text{SO}(2)$ duality group}
\label{subsec: DiracMonopole}
A magnetic monopole is a point-like solution of Maxwell's equations, i.e., a magnetic field with a point-like source placed at the origin of $\mathbb{R}^{3}$ of the form,
\begin{equation}\label{PointMono}
  \vec{B}(r) = \frac{g}{4\pi}\frac{\vec{r}}{r^3},
\end{equation}

  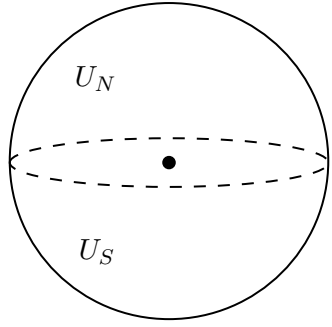
\begin{figure}
    \centering
    \begin{tikzpicture}[x=0.75pt,y=0.75pt,yscale=-1,xscale=1]

      \draw   (244,145.5) .. controls (244,101.59) and (279.59,66) .. (323.5,66) .. controls (367.41,66) and (403,101.59) .. (403,145.5) .. controls (403,189.41) and (367.41,225) .. (323.5,225) .. controls (279.59,225) and (244,189.41) .. (244,145.5) -- cycle ;
      \draw  [dash pattern={on 4.5pt off 4.5pt}] (244,146.25) .. controls (244,139.48) and (279.59,134) .. (323.5,134) .. controls (367.41,134) and (403,139.48) .. (403,146.25) .. controls (403,153.02) and (367.41,158.5) .. (323.5,158.5) .. controls (279.59,158.5) and (244,153.02) .. (244,146.25) -- cycle ;
      \draw  [fill={rgb, 255:red, 0; green, 0; blue, 0 }  ,fill opacity=1 ] (326.38,146.25) .. controls (326.38,144.66) and (325.09,143.38) .. (323.5,143.38) .. controls (321.91,143.38) and (320.63,144.66) .. (320.63,146.25) .. controls (320.63,147.84) and (321.91,149.13) .. (323.5,149.13) .. controls (325.09,149.13) and (326.38,147.84) .. (326.38,146.25) -- cycle ;

      \draw (275,96.4) node [anchor=north west][inner sep=0.75pt]    {$U_{N}$};
      \draw (277,183.4) node [anchor=north west][inner sep=0.75pt]    {$U_{S}$};
      
    \end{tikzpicture}
      
  \caption{Covering $S^2$ with two charts. The equator is the intersection $U_N\cap U_S$}
  \label{fig:S^2cover}
  \end{figure}
with the radial coordinate $r\in (0, \infty)$. Since the solution is singular at the origin, $r = 0$, it is only valid over $\mathbb{R}^3 -\{0\}$, $\mathbb{R}^3$ with the origin removed. And indeed, $\mathbb{R}^3-\{ 0\}$ has the same homotopy group as $S^2$ and therefore $\mathbb{R}^3-\{0\} \simeq S^2$. Hence the appropriate formulation of the problem is in terms of a non-trivial $\text{U}(1)$-line bundle over $S^2$ as the base space. In complete analogy to the Taub-NUT case of section \ref{sec:DC}, this forces us to define two local connections on the base space since a global section does not exist on $S^2$. On the northern and southern hemispheres we define the connections $A_N$ and $A_S$ respectively, where they are patched together at the equator $\gamma := U_N\cap U_S$ via a gauge transformation,
\begin{equation}\label{ConnectionGauge}
  A_N = A_S + df,
\end{equation}
where $f$ is a function on the equator. First notice that the magnetic flux through $S^2$, in our convention, is given by
\begin{equation}\label{MagCharge}
  \int_{S^2} F = \int_{U_N} dA_N + \int_{U_S}dA_S = g,
\end{equation}
where $F$ is defined by $F_{ij} = \epsilon_{ijk}B^k$ for $B^k$ as defined in \eqref{PointMono}. Now suppose one is describing a charge particle with charge $e$ using the wave function $\psi(x)$. Suppose that we choose a wavefunction at a point $p\in U_N\cap U_S$ such that $\psi_N|_p=\psi_S|_p = \psi$. Now If we parallel transport the wavefunction around the equator, the wavefunction will experience a phase-shift given by the holonomy of the connection around the equator, namely,
\begin{equation}\label{HoloPhase}
  \exp\Big[ i e \oint_\gamma A_N \Big]\psi.
\end{equation}
However, we could have equally used the connection defined on $U_S$ instead and the result should not depend on the choice on connection. The two connections are related by the gauge transformation \eqref{ConnectionGauge},
\begin{equation}
  \exp\Big[ i e \oint_\gamma A^S \Big]\psi
  = \exp\Big[ i e \oint_\gamma A^N \Big] \exp\Big[ -i e \oint_\gamma df \Big] \psi.
\end{equation}
Comparing this equation to eq. \eqref{HoloPhase}, we see that for the phase to be single valued we require that the integrand is an integer multiple of $2\pi$,
\begin{equation}\label{DiracQuant}
  e\oint df = 2\pi n, \quad n \in \mathbb{Z}.
\end{equation}
Now to find an expression for the above integral. Recalled that from eq. \eqref{MagCharge}
\begin{equation}
  \int_{S^2} F = \oint_\gamma A_N - \oint_\gamma A_S = \oint_{\gamma}df = g,
\end{equation}
where we have used stokes theorem to convert the surface integral on $U_N(U_S)$ into a loop integral around the equator, taking into account the orientation of the loop $\gamma$. Therefore, the we see that $\oint_{\gamma} df$ is the magnetic charge of the monopole. Therefore, by eq. \eqref{DiracQuant} we find that the electric charge is quantised,
\begin{equation}\label{DiracInt}
  eg = 2\pi n \in \mathbb{Z},
\end{equation}
which is the famous \emph{Dirac quantisation}
condition. Notice that this is also a manifestation of S-duality, where one coupling constant is the inverse of the other, $e\rightarrow g^{-1}$, relating strong and weak regimes of theories. Moreover, quantisation of the charges require the classical $\text{SO}(2)$ duality group from eq. \eqref{ChargeRot} to reduce to the group of integers $\mathbb{Z}$. This is because the charges now can only take values that are integer multiples of some number as given by eq. \eqref{DiracInt}. In fact, equation \eqref{MagCharge} is nothing but the first \emph{chern number} defined as the integral of the first \emph{Chern class},
\begin{equation}
  c := \text{det} \Big(I + \frac{iF}{2\pi}\Big) = 1 + c_1(F) + c_2(F)+\mathellipsis,
\end{equation}
where each $c_j\in \Omega^{2j}(M)$ is the $j\text{th}$ Chern class. Since in the Dirac monopole case the base space is a 2-dimensional manifold, only the first chern class survives since all forms in $\Omega^{2j}(S^2)$ vanish for $j > 1$. The Chern class measures the non-triviality of our bundle. More specifically, it classifies the principal bundle by identifying the ways one can construct the bundle by gluing the patches $U_N$ and $U_S$.
\subsection{The theta term, $\text{SL}(2,\mathbb{R})$, and $\text{SL}(2,\mathbb{Z})$}
Another way to see an S-duality transformation in the context of Maxwell's theory is to consider the following action on $\mathbb{R}^{4}$ with metric $\text{diag}(+1,+1,+1,+1)$
\begin{equation}\label{MaxAction}
  S = \int d^4x \big(-\frac{1}{4g^2}F^{\mu\nu}F_{\mu\nu} + \pd_\mu \tilde{A}_\nu (\star F)^{\mu\nu}\big),
\end{equation}
where at the present we do not give $F_{\mu\nu}$ the interpretation of the field strength of some gauge field, and $\tilde{A}_\mu$ is some one-form independent of $F_{\mu\nu}$ acting as a Lagrange multiplier. If one varies with respect $\tilde{A}_\mu$, it is easy to see that the Bianchi identity  is found
\begin{equation}
  \epsilon^{\mu\nu\rho\sigma}\pd_\mu F_{\rho\sigma} = 0 \implies F_{\mu\nu} = \pd_{[\mu}A_{\nu]}\quad  \text{locally} .
\end{equation}
On the other hand, varying with respect to $F_{\mu\nu}$ yields
\begin{equation}
  \begin{split}
 F_{\mu\nu} &= 2 g^2 \pd_\rho \tilde{A}_{\sigma}\epsilon^{\rho\sigma}{}_{\mu\nu}\\
 & :=  g^2 (\star \tilde{F})_{\mu\nu}.
\end{split}
\end{equation}
Now plugging our newfound expression for $F_{\mu\nu}$ into the action eq. \eqref{MaxAction} we find
\begin{equation}
  S = \int d^4x \big(-\frac{g^2}{4} \tilde{F}^{\mu\nu}\tilde{F}_{\mu\nu} + g^2 \pd_\mu \tilde{A}_\nu \tilde{F}^{\mu\nu}\big),
\end{equation}
where we have used the fact $\star\star\alpha = (-1)^{r(m-r)}\alpha$ for an r-form $\alpha$ on an $m$-dimensional Riemannian manifold. Notice the inverted coupling constant in comparison to the initial action \eqref{MaxAction} we started with. This exercise shows that the electromagnetic duality is a strong-weak one, justifying the name S-duality. In fact, the second term we added to Maxwell action is in fact the well-known theta term; the second term in the following action:
\begin{equation}\label{MaxthetaAction}
S = -\frac{1}{4g^2}\int d^4x\, F^{\mu\nu}F_{\mu\nu} +  \frac{\theta }{4\pi^2 \hbar}\int d^4x \,\frac{1}{4}(\star F)^{\mu\nu}F_{\mu\nu},
\end{equation}
which is a surface term that does not contribute to the equations of motion, which can be easily shown by taking the hodge dual, and then integrating by parts. The quadratic curvature integrand in the theta term is, in fact, the second \emph{Chern class} which is known to be an integer. Taking this into account and the fact that the partition function, given the action with the theta term, receives a $e^{i\theta n}$ factor where $n\in \mathbb{Z}$ from the integral of the second Chern class. Now to make the partition function invariant, we demand that $\theta$ is an angle with periodicity $2\pi$(See \cite{Verlinde:1995mz, Witten:1995gf} for more on this). Equipped with this information, now it is easy to see that the modified Maxwell action \eqref{MaxthetaAction} is invariant under $\theta \rightarrow \theta + 2\pi$, since the theta term is a surface term which does not affect the equation of motion. On the other hand we also have S-duality, as we saw in the argument under eq. \eqref{MaxAction}, which takes the coupling constant $g^2$ to $1/g^2$. Indeed, one can rewrite the action by defining a complex curvature and a new complex coupling constant given by
\begin{equation}
  \tau = \frac{\theta}{2\pi} + i\frac{4\pi}{g^2}.
\end{equation}
Now notice that the periodicity of $\theta$ implies that the action is invariant under $\tau \rightarrow \tau +1$, and that for $\theta = 0$, S-duality as shown above implies that the action is invariant under $\tau \rightarrow -1/\tau$. Together the set of transformations
\begin{equation}\begin{split}
  T &: \tau \rightarrow \tau +1\\
  S &: \tau \rightarrow -\frac{1}{\tau}
\end{split}
\end{equation}
generate the group $\text{SL}(2,\mathbb{Z})$. Classically, in absence of charge quantisation, the duality group is instead $\text{SL}(2,\mathbb{R})$. This is because the theta angle can take any real value. It is in this sense we have identified Ehlers transformation, an $\text{SL}(2,\mathbb{R})$ action(see Chapter \ref{chap:Weyl}), with the classical electromagnetic duality transformation via the machinery of the double copy.
\section{Conclusion}
In this chapter we introduced the Kerr-Schild class of solutions in General Relativity wherein the Field equations are linearised. Using the form of the metric, we showed how to take the single copy to obtain a gauge field as a solution to the Abelian Yang-Mills equation. We demonstrated the single copy procedure using two canonical examples, the Taub-NUT and Schwarzschild solutions. The latter, being a static and spherically symmetric, single copies to a Coulomb charge, while the former, known to have a dyon-like character to it single copies to a dyon with electric and magnetic charges. The double copy provides a concrete evidence of the Taub-NUT spacetime's magnetic character and makes the analogy exact. In the second half of the chapter, we introduced the idea of solution generating transformation as a way of obtaining new solutions to Einstein's equations from old ones. This was demonstrated using two known transformations, one being Ehlers transformation and the other known as Buchdahl transformation. The former, once applied to the Schwarzschild metric generates the Taub-NUT metric, while the latter, after choosing the appropriate killing vector produces the negative mass Schwarzschild. These transformations resemble electromagnetic duality and charge conjugation. Indeed, the double copy makes the analogy exact by single copying Ehlers transformation to electromagnetic duality and Buchdahl transformation to charge conjugation. Viewed as an element of the group $\text{SL}(2,\mathbb{R})$, Ehlers transformation reduces to $\text{SL}(2,\mathbb{Z})$ in the Taub-NUT case upon the introduction of the two patches in order to avoid Dirac singularities. This is similar to what happens when the electromagnetic duality group acts on the dyon spectrum. To maintain the dirac quantisation of the magnetic charges, the classical $\text{SL}(2,\mathbb{R})$ group is forced to reduce to $\text{SL}(2,\mathbb{Z})$.

In the next chapter, the relation of Ehlers transformation to electromagnetic duality on the gauge theory side will be more apparent under the lens of the Weyl double copy.

%% file: chaps/chap04.tex
\chapter{The Weyl Double Copy}\label{chap:Weyl}

This chapter is based on the second half of \cite{Alawadhi:2019urr}, and \cite{Alawadhi:2020jrv}.
\section{Introduction and conclusion}
In the previous chapter we have introduced the Kerr-Schild double copy as relation between solutions of Einstein's field equations and those of the linearised Yang-Mills equations, effectively the well-known Maxwell's electromagnetism. While the Kerr-Schild formalism provides a powerful non-perturbative way to study famous and physically relevant solutions of Einstein's equations and their respective single copies, one can not overlook some of its shortcomings. The first one being the requirement of metrics of a very particular and special form, i.e., the algebraically special Kerr-Schild form. Not all solutions of General Relativity admit such a form of the metric. Therefore, in our study of the single copy, we are restricted to this particular class of metrics. As for the second shortcoming of the Kerr-Schild double copy, the fact that we are mapping metrics to the gauge connection on the gauge theory side, which is a non-gauge invariant object. Generally, mapping to objects that are gauge invariant is more desirable. As a small example to illustrate this, take the Abelian gauge connection Dirac used to describe a magnetic monopole \cite{Gockeler:1987an},
\begin{equation}
  A = i \frac{m}{2r}\frac{1}{x^3 - r}(x^1 dx^2 - x^2 dx^1),
\end{equation}
defined on Minkowski spacetime $\mathbb{R}^{1,3}$ with coordinates $(x^0, x^1, x^2, x^3)$ corresponding to the time, $x$, $y$, and $z$ axes; with $r^2 = \sum_{i=1}^{3} x^i x_i$ and $m>0$ constant. Now observe that the connection is not defined on the $x^3$-axis and is singular. The curvature of the connection however does not suffer from such a singularity,
\begin{equation}
  F = dA = i \frac{m}{2r^3}(x^1 dx^2\wedge dx^3 + x^2 dx^3\wedge dx^1 + x^3 dx^1\wedge dx^2 ).
\end{equation}
This has to do with the fact that the curvature is invariant under a gauge transformations of the form,
\begin{equation}
  A \rightarrow A + f(x), \quad f : \mathbb{R}^{1,3} \rightarrow \mathbb{R}.
\end{equation}
Indeed, it is to be able to map solutions of General Relativity via the double copy to gauge invariant objects on the gauge theory side. This takes us the double copy formalism known as the Weyl double copy, first appearing in \cite{Luna:2018dpt}. It relates the Weyl tensor on the gravity side to the curvature of the gauge field on the gauge theory side upon single copying. The latter is gauge invariant for Abelian fields and covariant for non-Abelian fields. This is a powerful way to non-perturbatively study solutions of the four dimensional Einstein's equations. This time however, one is not restricted to work in the Kerr-Schild class of metrics. The metric can in principle be in any coordinate system as long as it is in type D class of solutions of Einstein's equations as classified\footnote{For more details on this classification see \cite{Stephani:2003tm}} by A. Z. Petrov in 1954 \cite{Petrov:2000bs}. The idea of relating the Weyl tensor to the gauge field comes from an observation in the perturbative scattering amplitude side of the pond. Consider linear waves in gauge theory and gravity. Their fields and field strengths are given by:
\begin{equation}
  \begin{split}
A_\mu &= \epsilon_\mu e^{ik\cdot x}, \quad F_{\mu\nu} = i ( k_\mu \epsilon_\nu - k_\nu \epsilon_\mu )e^{ik\cdot x}\\ 
h_{\mu\nu} &= \varepsilon_{\mu\nu}e^{ik\cdot x}, \quad R_{\mu\nu\rho\lambda} = \frac{1}{2}( k_\mu \epsilon_\nu - k_\nu \epsilon_\mu )( k_\rho \lambda_\nu - k_\lambda \rho_\mu )e^{ik\cdot x},\quad \varepsilon_{\mu\nu} = \epsilon_\mu \epsilon_\nu,
  \end{split}
\end{equation}
where $h_{\mu\nu}$ is the graviton field described as a linearised perturbation around the flat Minkowski background, $R_{\mu\nu\rho\lambda}(h)$ is the linearised Riemann tensor. One sees that there is a doubling-like relationship between gauge and gravity theory:
\begin{equation}
  e^{k\cdot x}R_{\mu\nu\rho\lambda} \sim F_{\mu\nu}F_{\rho\lambda}.
\end{equation}
The idea of the Weyl double copy is to find and extend such a relationship to non-perturbative and exact solutions. It turned out that such a relationship exists for the Weyl tensor if one goes to spinor space. Once one writes the Weyl tensor in spinorial basis, a coordinate change in spinor space can be done such that the Weyl spinor is written in a basis where it is linearly proportional to the mass of the gravitating body. In this basis, the Weyl spinor equals its linear approximation, the integral of the 3-point amplitude. However, in this basis the relationship is exact. While it is indeed true that this relationship is valid for the Weyl tensor as opposed to the Riemann tensor, this does not pose a problem if we consider vacuum solutions of Einstein's equation. Because the decomposition of the Riemann tensor into its trace-free part, Ricci tensor and Ricci scalar is given by,
\begin{equation}
    \label{RieDecomp}
    R_{\mu\nu\rho\lambda} = C_{\mu\nu\rho\lambda} + \frac{2}{n-2}( g_{\mu[\rho} R_{\lambda]\nu} - g_{\nu[\rho} R_{\lambda]\mu})
    - \frac{2}{(n-1)(n-2)}R g_{\mu[\rho} g_{\lambda]\nu} \,,
\end{equation}
where $C_{\mu\nu\rho\lambda}$ is the Weyl tensor which is trace-free, i.e., metric contraction on any pair of indices yields zero. Therefore, in the vacuum case, the Riemann tensor is equal to the Weyl tensor. All propagating degrees of freedom which have nothing to do with the direct presence of matter are encoded in the Weyl tensor. As a result, solutions of Einstein's equations are completely characterised by the Weyl tensor, and the Weyl double copy exploits this alongside the spinorial representation to obtain solutions of gauge theories.

The Weyl double copy as it was originally introduced in \cite{Luna:2018dpt} has been shown to work for all type D and non-twisting type N spacetimes \cite{Godazgar:2020zbv} which describe gravitational waves. For algebraically general solutions, the Weyl double copy was found to hold asymptotically for a number of spacetimes \cite{Godazgar:2021iae, Adamo:2021dfg}. A twistorial interpretation of the double copy has been recently discovered \cite{White:2020sfn, Chacon:2021wbr } where techniques from the twistor formalism was used to shed light and explain the origin of the Weyl double copy.

In this chapter however, we shall utilise the tensorial form of the Weyl double copy. We do this by writing the Weyl tensor in terms of the Maxwell field strength constructed from a killing vector field associated to the metric. The manner in which this field strength is related to the usual single copy field strength will be explained in the following section. But generally, the killing field strength is generally interpreted as the single copy field strength but living on the same spacetime as that on which the killing vector is defined on. In some cases, the killing field strength will coincide with the flat space single copy field strength. We will see that it is indeed the case for the Schwarzschild and Taub-NUT spacetimes, which allowed us to verify and further study the Ehlers transformation and its single copy as presented in the previous chapter. We will find that the Ehlers transformation, acting as a $\text{SL}(2, \mathbb{R})$ transformation on solutions of Einstein's equations, or more specifically, on their charges.

In the second part of this chapter, namely in \ref{sec:WeylSpinOff}, we will study a wide range of solutions of Einstein's equations in various dimensions including those with special holonomy groups such as $\text{Spin}(7)$ and $\text{G}_2$. Here however, we are fully embracing the tensorial version of the Weyl double copy and drop the requirement of our `single copy' field strengths to be living on flat spacetime. We will find that while numerous solutions has a Weyl doubling form, some require derivative corrections showcasing the non-triviality of writing the Weyl tensor in terms of the field strength. We will also apply our analysis to spacetimes with an additionally defined gauge field on them such as the Reissner-Nordstrom solution. Again, we will see that using the predefined gauge field, it is possible to construct the Weyl tensor from it. This generalises to higher dimensions and also to the more general Born-Infeld theory.

\section{The Weyl tensor via the field strength $F$}
The double copy as described in the previous chapter is defined in terms of the gauge connection and the metric. It is a natural question to ask whether one might study a double copy directly in terms of field strengths and curvatures. This was investigated in \cite{Luna:2018dpt}, where a particularly nice form of the double copy was obtained using spinors -- writing the spinor corresponding to the Weyl tensor in terms of the spinor for the Maxwell tensor as
\begin{equation}
\label{spinorDC}
C_{ABCD} = \frac{1}{S} f_{(AB}f_{CD)} \, ,
\end{equation}
where $S$ is a suitable function related to the `zeroth copy'. For the benefit of the reader and to make sense of \eqref{spinorDC}, we quote the following from \cite{Luna:2018dpt}. One starts with the `spinorial vielbein' $\sigma^{\mu}_{A\dot{A}}$, which is related to the metric tensor of spacetime according to
\begin{equation}
  \left(\sigma^\mu_{A\dot{A}} \sigma^\nu_{B\dot{B}}+\sigma^\nu_{A\dot{A}} \sigma^\mu_{B\dot{B}}\right) \epsilon^{\dot{A}\dot{B}} = g^{\mu\nu} \epsilon_{AB},
\end{equation}
where $\epsilon_{A\dot{A}}$ is an antisymmetric `metric' spinor for the 2-dimensional complex spin space with $\epsilon^{12} = 1$. Defining an orthonormal frame, and using a vielbein and its inverse as a map from spacetime to the tangent space such that
\begin{equation}
  g^{\mu\nu} = (e^{-1})^\mu_a\, (e^{-1})^\nu_b\, \eta^{ab}, \qquad \eta^{ab}=\eta_{ab}=\text{diag}(-1,1,1,1),
\end{equation}
where $g_{\mu}$ is the metric in coordinate basis and $\eta_{ab}$ is the flat minkowski metric. Using the vielbein, one can convert the spacetime index of the spinorial vielbein to the orthonromal basis index
\begin{equation}
  \sigma^\mu_{A\dot{A}} = (e^{-1})^\mu_a \,\sigma^a_{A\dot{A}}, \qquad \sigma^a=\frac{1}{\sqrt{2}}({I},\sigma^i),
\end{equation}
where $\sigma^i$ are the three Pauli matrices. The spinorial vielbein and its inverse satisfy the relations
\begin{equation}
  \sigma_\mu^{A\dot{A}} = g_{\mu\nu} \,\epsilon^{AB}\,\sigma^\nu_{B\dot{B}}\,\epsilon^{\dot{B}\dot{A}},
\qquad \sigma^\mu_{A\dot{A}} \sigma_\nu^{A\dot{A}} =\delta^\mu_\nu,
\qquad \sigma^\mu_{A\dot{A}} \sigma_\mu^{B\dot{B}} = \delta^B_A \delta^{\dot{B}}_{\dot{A}},
\end{equation}
which we can use to write any tensorial object in spinorial basis, i.e., $V_{A\dot{A}} = \sigma^\mu_{A\dot{A}}V_\mu$. Since the Weyl double copy as originally presented describes a relation between the Weyl tensor and the Maxwell tensor, which we will define shortly, we proceed to write the Weyl tensor in spinorial form
\begin{equation}
  \label{eq:Wspinor}
W_{A\dot{A} B\dot{B} C\dot{C} D\dot{D}} = C_{ABCD} \,\epsilon_{\dot{A}\dot{B}} \epsilon_{\dot{C}\dot{D}} + \bar{C}_{\dot{A}\dot{B}\dot{C}\dot{D}} \,\epsilon_{AB} \epsilon_{CD},
\end{equation}
where the two quantities on the right side labelled as $C$ are completely symmetric and related by complex conjugation for a real spacetime. In fact, they are the anti-self-dual and self-dual parts of the Weyl tensor. Once can wrote $C_{ABCD}$ from the Weyl tensor by
\begin{equation}
  C_{ABCD} = \frac{1}{4} W_{\mu\nu\rho\lambda}\,\sigma^{\mu\nu}_{AB}\,\sigma^{\rho\lambda}_{CD},
\end{equation}
using 
\begin{equation}
  \sigma^{\mu\nu}_{AB} = \sigma^{[\mu}_{A\dot{A}} \,{\tilde{\sigma}}^{\nu]\; \dot{A} C}\, \epsilon_{CB},
\qquad {\tilde{\sigma}}^{\mu\; \dot{A} A} =(e^{-1})^\mu_a\, {\tilde\sigma}^{a\;\dot{A} A}\,,
\qquad {\tilde\sigma}^a=\frac{1}{\sqrt{2}}({I},-\sigma^i).
\end{equation}
For a gauge theory field strength $F_{\mu\nu}$, we have analogously
\begin{equation}
\label{eq:Fspinor}
F_{A\dot{A} B\dot{B}} = f_{AB} \,\epsilon_{\dot{A}\dot{B}} + \bar{f}_{\dot{A}\dot{B}} \,\epsilon_{AB}\,, \qquad \text{with} \qquad
f_{AB} = \frac{1}{2} F_{\mu\nu}\, \sigma^{\mu\nu}_{AB}
\end{equation}
and $f_{AB}=f_{BA}$, as well as $f_{AB}=(\bar{f}_{\dot{A}\dot{B}})^*$ for a real field strength.

This `Weyl double copy' was shown to be consistent with the previously known Kerr-Schild double copy, and resolved some of ambiguities in that formulation. It also presented new double copy interpretations of the Eguchi-Hanson instanton, and the C-metric, relating the latter to the Li\'enard-Wiechert potential for a pair of uniformly accelerated charges. Extending this four-dimensional result to higher dimensions requires an appropriate study of spinors and curvature invariants in higher dimensions, and the latter has been explored recently in \cite{Monteiro:2018xev}.
\\

A higher-dimensional Weyl double copy might also be investigated in terms of a tensor version of the spinor Weyl double copy. One can obtain this by translation from the spinor equations of course;  more directly one can note that this must involve writing the Weyl tensor as a sum of terms quadratic in the Maxwell tensor. Keeping in mind the algebraic symmetries of the Weyl tensor, in four dimensions there are two independent expressions  that one can write down:
\begin{align}
\label{CDdefns}
C_{\mu\nu\rho\sigma} [F] &= F_{\mu\nu}F_{\rho\sigma} - F_{\rho\mu}F_{\nu\sigma}-3g_{\mu\rho}{F_\nu}^\lambda F_{\sigma\lambda} + \frac{1}{2}g_{\mu\rho}g_{\nu\sigma}F^2   \; \Big\vert_s  \, ,
\\
D_{\mu\nu\rho\sigma} [F] &= \frac{1}{2}\left( F_{\mu\nu}\Ft_{\rho\sigma} - F_{\rho\mu}\Ft_{\nu\sigma}-3g_{\mu\rho}{F_\nu}^\lambda \Ft_{\sigma\lambda} + \frac{1}{2} g_{\mu\rho}g_{\nu\sigma}F.\Ft \right) + (F \leftrightarrow \Ft)   \;  \Big\vert_s  \, ,
\end{align}
where $F^2= F^{\lambda\delta}F_{\lambda\delta}$,   $F.\Ft =F^{\lambda\delta}\Ft_{\lambda\delta}$, and $\Ft_{\mu\nu} =\frac{1}{2}\sqrt{g}\, \epsilon_{\mu\nu\rho\sigma}F^{\rho\sigma}$ with  $\epsilon_{\mu\nu\rho\sigma}$ the numerical alternating symbol.
In equations like those above, the symbol ``$\,\vert_s$'' applies to the expression on the right-hand side of the equation, and it means to anti-symmetrise in the indices $\mu,\nu$ and in $\rho, \sigma$, with unit weight.
In $D$  dimensions there is no equivalent of  $ D_{\mu\nu\rho\sigma} [F]$ and one just has
\begin{equation}
\label{Dhigher}
C^{(D)}_{\mu\nu\rho\sigma} [F] = F_{\mu\nu}F_{\rho\sigma} - F_{\rho\mu}F_{\nu\sigma}-\frac{6}{D-2}g_{\mu\rho}{F_\nu}^\lambda F_{\sigma\lambda} + \frac{3}{(D-1)(D-2)}g_{\mu\rho}g_{\nu\sigma}F^2   \; \Big\vert_s  \, .
\end{equation}
\\
We make some comments on the higher-dimensional double copy at the end of the chapter and hereon consider four dimensions. A four-dimensional Weyl tensor double copy must  involve a linear sum of the two expressions $C_{\mu\nu\rho\sigma} $ and $D_{\mu\nu\rho\sigma} $, with suitable coefficients which will in general be functions of the relevant variables and constants.
We note some useful properties of these expressions:
 $C_{\mu\nu\rho\sigma} [\Ft] = C_{\mu\nu\rho\sigma} [F]$, $\tilde C_{\mu\nu\rho\sigma} [F] = D_{\mu\nu\rho\sigma} [F]$, $D_{\mu\nu\rho\sigma} [\Ft] = -D_{\mu\nu\rho\sigma} [F] $ and
\begin{equation}
C_{\mu\nu\rho\sigma} [aF+b\Ft] = (a^2+b^2)\,C_{\mu\nu\rho\sigma} [F] + 2ab\, D_{\mu\nu\rho\sigma} [F]
\end{equation}
for any coefficients $a,b$. Define the self-dual and anti-self-dual parts of a two form $F$ via $F^\pm =\frac{1}{2}(F \pm \Ft)$ with a similar formula for $C^\pm$, with the action either on the left or right pair of indices. Then
\begin{equation}
C_{\mu\nu\rho\sigma} ^\pm[a F+b\Ft] = C_{\mu\nu\rho\sigma} [(a\pm b)F^\pm] \, .
\end{equation}
\section{Ehlers transformation take two}
\subsection{Ehlers and the double copy: the Schwarzschild case}

We would like to explore how the Ehlers transformation described earlier may be understood in terms of the double copy. Consider as starting point the Schwarzschild metric. We will find it useful to work in coordinates $(u,v,p,q)$ with the metric
\begin{align}
\label{Schw_metric}
ds^2 &=  \frac{1}{(1-pq)^2} \Bigg[2 i (du+q^2 dv) dp -2(du-p^2 dv) dq   + \frac{2mp^3}{(p^2+q^2)} (du+q^2 dv)^2   \nonumber \\
&\qquad\qquad\qquad+ \frac{2mq}{(p^2+q^2)}(du-p^2 dv)^2 \Bigg]   \, .
\end{align}
The single copy Maxwell tensor, which satisfies Maxwell's equations on Minkowski spacetime, is then \cite{Luna:2018dpt}
\begin{equation}
F_S = {\frac{e}{ (p^2+q^2)^2}} \Big[ 2 pq (du+q^2dv)dp - (p^2-q^2) (du-p^2dv)dq \Big]  \, .
\end{equation}
It is then straightforward to show that the Weyl tensor $C_S$ for the Schwarzschild metric is given by
\begin{equation}
C^S_{\mu\nu\rho\sigma}  = -{\frac{4}{(1-pq)e}} \Big( q \,C_{\mu\nu\rho\sigma} [F_S] -i p\,D_{\mu\nu\rho\sigma} [F_S] \Big) \,\Big\vert_{e\rightarrow m}  \, ,
\end{equation}
where $\vert_{e\rightarrow m}$ means to replace $e$ by $m$ on the right-hand side of the equation. This result may be written simply as
\begin{equation}
\label{SchwWeyl}
C^S_{\mu\nu\rho\sigma} = -{\frac{4}{ (1-pq)}} \Big( C_{\mu\nu\rho\sigma} [ \alpha_S F_S^+ + (c.c)]\Big) \,\Big\vert_{e\rightarrow m}  \, ,
\end{equation}
where $\alpha_S = \sqrt{{\frac{-i(p+iq)}{ e}}}$.

Now let us consider the Taub-NUT metric in the corresponding coordinate system:
\begin{align}
\label{TNmetric}
ds^2 &=  \frac{1}{(1-pq)^2} \Bigg[2 i (du+q^2 dv) dp -2(du-p^2 dv) dq   +2p \frac{mp^2+n}{(p^2+q^2)} (du+q^2 dv)^2   \nonumber \\
&\qquad\qquad\qquad+ 2q\frac{m+nq^2}{(p^2+q^2)}(du-p^2 dv)^2 \Bigg]   \, ,
\end{align}
where $n$ is the Taub-NUT charge. The known single copy Maxwell tensor on Minkowski space in this case is \cite{Luna:2015paa}
\begin{equation}
\label{MaxwellTN}
F_T = {\frac{1}{(p^2+q^2)^2}} \Big[ \big(2 e p q +g(p^2-q^2) \big)(du +q^2 dv) dp
         +  \big(2 g p q -e(p^2-q^2) \big) (du-p^2dv) dq    )   \Big]  \, .
\end{equation}
This can be  expressed simply in terms of the Schwarzschild Maxwell single copy tensor as
\begin{equation}
F_T = F_S - {\frac{i g}{ e}}\Ft_S  \, .
\end{equation}
Now, if we  make the replacements
\begin{equation}
\label{Ehlers}
F_S \rightarrow  F_S - \frac{ ig}{e} \Ft_S, \qquad e\rightarrow e-i g\, ,
\end{equation}
on the right-hand side of
\eqref{SchwWeyl}, and then make the replacements
$e\rightarrow m, g\rightarrow n$ then we find a tensor that we will call $C^T$ which is
\begin{equation}
\label{TNWeyl}
C_{\mu\nu\rho\sigma} ^T := -{\frac{4}{(1-pq)}} \Big( C_{\mu\nu\rho\sigma} [ \alpha _T F_T^+ + (c.c)]\Big) \Big\vert_{e\rightarrow m, g\rightarrow n} \, ,
\end{equation}
where $\alpha_T = \sqrt{{\frac{-i(p+iq)}{ (e-ig)}}}$.
It can be checked that $C^T$ is the Weyl tensor for the Taub-NUT metric   \eqref{TNmetric}. (Note that we implicitly assumed that the charge $e$ is complex prior to the shift, and that in going from \eqref{SchwWeyl} to \eqref{TNWeyl} the metric dependence in $C_{\mu\nu\rho\sigma} [F]$ also needs to shift from \eqref{Schw_metric} to  \eqref{TNmetric}.)
\\

Thus we see that the Ehlers transformation which takes one from the Schwarzschild to the Taub-NUT spacetime can be seen via the Weyl double copy as a simple duality transformation \eqref{Ehlers} (combined with identifying $(e,g)$ with $(m,n)$) which maps between the two Weyl double copy curvatures. It is instructive to return to the spinor form of the Weyl double copy \eqref{spinorDC} in the light of this (see Section 4 of \cite{Luna:2018dpt}). The transformation $eF\rightarrow (e-ig)F$ induces the shifts $eF^\pm\rightarrow (e\mp ig)F^\pm$. The Maxwell field strength spinor $f_{AB}$ depends only on the self-dual part of the Maxwell tensor and thus transforms according to this formula. The scale function $e\, S$ in \eqref{spinorDC}  transforms to $(e-ig)S$ and hence the double copy formula yields
\begin{equation}
\label{WeylSpinorTransform}
m\, C_ {ABCD}\rightarrow  (m-in)\, C_ {ABCD}\, ,
\end{equation}
correctly mapping the Schwarzschild Weyl spinor to the Taub-NUT one.


\subsection{Type D metrics}\label{subsec: Type D metrics}

The Taub-NUT example considered above is a special case of the general vacuum type D solution with vanishing cosmological constant \cite{Plebanski:1976gy}, as given in  \cite{Luna:2018dpt}:
\begin{equation}
\label{typeDmetric}
ds^2 = {\frac{1}{(1-pq)^2}}  \Bigg[2 i (du+q^2dv) dp - 2(du-p^2dv)dq +  \frac{P(p)}{p^2+q^2}\,(du+q^2dv)^2
-  \frac{Q(q)}{p^2+q^2}\,(du-p^2dv)^2  \Bigg] \, ,
\end{equation}
with
\begin{align}
P(p) =  \gamma (1-p^4) +2 n p- \epsilon p^2 +2 m p^3 \,, \nonumber \\
Q(q) =  \gamma (1-q^4) - 2m q + \epsilon q^2 -2 n q^3 \,,
\end{align}
where the parameters $m,n,\gamma,\epsilon$ are related to the mass, NUT charge, angular momentum and acceleration (see \cite{Griffiths:2005qp} for a discussion of the various limits and definitions which enable the identifications in different cases).

The single copy Maxwell tensor in this case is the same as the one for the Taub-NUT metric \eqref{TNmetric}. It is then natural to investigate the Weyl double copy in this case and, indeed, one finds that the Weyl tensor $C^D$ for the type D metric \eqref{typeDmetric} is  given by the same formula as that for the TN case:

\begin{equation}
\label{TypeDWeyl}
C^D_{\mu\nu\rho\sigma} := -{\frac{4}{ (1-pq)}} \Big(C_{\mu\nu\rho\sigma} [ \alpha _T F_T^+ + (c.c)]\Big) \, \Big\vert_{e\rightarrow m, g\rightarrow n}  \, ,
\end{equation}
with $\alpha_T = \sqrt{{\frac{-i(p+iq)}{ (e-ig)}}}$. Note that the metric \eqref{typeDmetric}  enters the right-hand side of  \eqref{TypeDWeyl} so that this doesn't simply reproduce $C_{\mu\nu\rho\sigma} ^T$.

One can then ask if an Ehlers transformation will take one from the spacetime with Type D metric \eqref{typeDmetric} with $n = 0$, to that with nonzero $n$. To see this, consider the type D metric $g_{D_0}$ with vanishing NUT charge. This satisfies
\begin{equation}
\label{eq::TypeDWeylNoNUT}
C^{D_0}_{\mu\nu\rho\sigma} = -{\frac{4}{(1-pq)}} \Big(C_{\mu\nu\rho\sigma} [ \alpha _S F_S^+ + (c.c)]\Big) \, \Big\vert_{e\rightarrow m}  \, .
\end{equation}
Then if we make the replacements $F_S \rightarrow  F_ S- {\frac{i g}{ e}}\Ft_ S$ and $e\rightarrow e-i g$ in the right-hand side of the above, and shift the metric from $g_{D_0}$ to \eqref{typeDmetric}, we find that we reproduce  \eqref{TypeDWeyl}.


\section{The $\text{SL}(2,\mathbb{R})$ transformations}

\subsection{The spacetime Ehlers group}

We would now like to  discuss how   $\text{SL}(2,\mathbb{R})$ transformations  act more generally in the context of the double copy.
For this purpose, it will be useful to use a generalisation of the Ehlers procedure for  more general Killing vector fields: the \textit{spacetime Ehlers group} \cite{Mars:2001gd}, which we now summarise briefly.\footnote{We comment that this paper and  work following from it (see for example \cite{Mars:2016ynw} and references therein) anticipate some of the formul\ae\ of the double copy -  e.g., the vanishing of the Mars-Simons tensor defined below corresponds to the self-dual part of the tensor double copy, and the spinor form of this can  be found in \cite{Gomez-Lobo:2016ykv}.}
Given a Killing vector field $\xi=\xi^\mu\partial_\mu$ and one-form $W=W_\mu dx^\mu$ on a Lorentzian manifold with metric $g_{\mu\nu}$, satisfying the vacuum Einstein equations, the spacetime Ehlers group is defined in \cite{Mars:2001gd} by the transformation
\begin{equation}
\label{newmetric}
g_{\mu\nu}\rightarrow \Omega^2 g_{\mu\nu}-2\xi_{(\mu}W_{\nu)}-\frac{\lambda}{\Omega^2}W_\mu W_\nu,
\end{equation}
where $\lambda=-\xi^\mu\xi_\mu$ and $\Omega^2\equiv\xi^\mu W_\mu+1\geq 1$ with the inequality holding over the whole geometry.
Define  the twist potential $\omega_\mu=\sqrt{-\operatorname{det}(g)}\,\epsilon_{\mu\nu\sigma\rho}\xi^\nu\nabla^\sigma\xi^\rho$
and the Killing tensor two form $F_{\mu\nu}=2\partial_{[\mu}\xi_{\nu]}$ (note that we have a factor of $2$ here, and a factor of $1/2$ in the definitions of the (anti-)self-dual parts of $F$, in comparison with \cite{Mars:2001gd})\footnote{The field strength $F_{\mu\nu}$ constructed with a Killing vector is guaranteed to satisfy Maxwell's equations on the spacetime on which the killing vector $\xi_\mu$ is defined. See Appendix \ref{Maxapp} for a derivation}.
Then the Ernst one-form
\begin{equation}
\sigma_\mu :=  2 \xi^\nu F^+_{\nu\mu} =\nabla_\mu\lambda-i\omega_\mu \,
\end{equation}
is closed, following from the vanishing of the Ricci tensor, and so locally
$\sigma_\mu =   \nabla_\mu\sigma \label{ernstoneform2}$
for some complex function $\sigma$.\footnote{This satisfies $\nabla^2\sigma=-(F^+_{\mu\nu})^2$. In this construction we do not see a direct emergence of the standard harmonic \lq\lq zeroth copy" function, as in the Kerr-Schild formulation. }
\\
The spacetime Ehlers group is then defined for $W$ satisfying
\begin{subequations}\label{Wform}
\begin{gather}
2\nabla_{[\mu}W_{\nu]}=-4\gamma\,{\rm Re}[(\gamma \bar\sigma+i\delta)\,F^+_{\mu\nu}]\label{Wforma}\, ,\\
\xi^\mu W_\mu+1 = (i\gamma\sigma+\delta)(-i\gamma \bar{\sigma}+\delta) \label{Wformb}\, ,
\end{gather}\end{subequations}
where a bar denotes complex conjugation, and $\gamma$ and $\delta$ are non-simultaneously vanishing real constants, which as a pair fix the gauge of $W$. After repeated action, the transformation defines an $\text{SL}(2,\mathbb{R})$ group action on the Ernst scalar by the M\"obius map
\begin{equation}
\label{sl2r}
\sigma\rightarrow\frac{\alpha\sigma+i\beta}{i\gamma\sigma+\delta},\qquad\text{where}\quad \beta\gamma+\alpha\delta=1.
\end{equation}
The self-dual part of the Killing tensor transforms as
\begin{equation}
\label{Ftransform}
F^+_{\mu\nu} \rightarrow \frac{1}{(i\gamma\sigma+\delta)^2}\Big(\Omega^2 F^+_{\mu\nu} -W_{[\mu} \sigma_{\nu]}  \Big)\, ,
\end{equation}
where $W, \sigma$ are the one-forms defined above. The self-dual part of the Weyl tensor transforms as
\begin{equation}
\label{Weyltransform}
C^+_{\mu\nu\rho\sigma} \rightarrow     \frac{1}{(i\gamma\sigma+\delta)^2} P^{\alpha\beta}_{\mu\nu}
P^{\gamma\delta}_{\rho\sigma}  \left( C^+_{\alpha\beta\gamma\delta}   -   \frac{6i\gamma}{i\gamma\sigma+\delta} \Big( F^+_{\alpha\beta}F^+_{\gamma\delta} -\frac{1}{3}I_{\alpha\beta\gamma\delta} (F^+)^2\Big) \right) \,  ,
\end{equation}
where in our conventions
 $I_{\mu\nu\rho\sigma}= \frac{1}{4}(g_{\mu\rho}g_{\nu\sigma} - g_{\nu\rho}g_{\mu\sigma} + \epsilon_{\mu\nu\rho\sigma})$
is the canonical metric in the space of self-dual two-forms and $P^{\alpha\beta}_{\mu\nu}   =\Omega^2\delta^\alpha_\mu \delta^\beta_\nu -
\delta^\alpha_\mu \xi^\beta W_\nu - \xi^\alpha W_\mu \delta^\beta_\nu$.
Notice that
\begin{equation}
 F^+_{\mu\nu}F^+_{\rho\sigma} -\frac{1}{3}I_{\mu\nu\rho\sigma} (F^+)^2 = \frac{2}{3}  C_{\mu\nu\rho\sigma}  [F^+]  \,
\end{equation}
in terms of the definition in \eqref{CDdefns}.
The Mars-Simons tensor is then defined as
\begin{equation}
   S_{\mu\nu\rho\sigma} =  C^+_{\mu\nu\rho\sigma}     - \frac{2}{3}Q\, C_{\mu\nu\rho\sigma}  [F^+] \, ,
\end{equation}
for a suitable function $Q$.
 Finally,  as discussed in Section 6 of  \cite{Mars:2001gd},  we note that the vanishing of the Mars-Simons tensor is maintained under the Ehlers transformation, with $Q$ transforming appropriately.
%


\subsection{The Taub-NUT case}
Let us now consider applying these arguments in the context of the Weyl double copy  described earlier.
Starting from the Taub-NUT metric\footnote{Much like how one obtains Taub-NUT from the Schwarzschild spacetime, applying Ehlers transformation to the Kerr solutions yields the Kerr-NUT solution \cite{Mars:2001gd}}, in the real form \eqref{Taub-NUT metric}, consider the Killing vector $\xi=\partial_t$. Its associated two-form $F_{\mu\nu}=2\,\partial_{[\mu} \xi_{\nu]}$ is
\begin{align}
\label{F TN}
F&=\frac{2 M \left(r^2-N^2\right)+4 N^2 r}{\left(N^2+r^2\right)^2}\,dt\wedge dr
+\frac{4 N \cos (\theta ) \left(M \left(r^2-N^2\right)+2 N^2 r\right)}{\left(N^2+r^2\right)^2}\,dr\wedge d\phi \\ \nonumber
&\qquad +\frac{2\,N \sin (\theta ) \left(r (2 M-r)+N^2\right)}{N^2+r^2}\,d\theta\wedge d\phi
~.
\end{align}
This solves the Maxwell equations on the Taub-NUT background. The single copy of Taub-NUT was found in \cite{Luna:2015paa,Luna:2018dpt} and solves the flat-background Maxwell equations. The Ernst one-form is obtained from its definition
\begin{equation}
\sigma_\mu:=2\xi^\nu F^+_{\nu\mu}=\frac{2 (M-i\, N)}{( r-i\,N)^2}\,\delta^r_\mu~.
\end{equation}
In \cite{Mars:2001gd} it was proved that the Ernst one-form is exact, $\sigma_\mu=\partial_\mu \sigma$, and the integration constant can be chosen such that ${\rm Re}(\sigma)=-\xi^\mu\xi_\mu$, giving
\begin{equation}
\sigma=1-\frac{2 (N+i M)}{N+i\, r}~.
\end{equation}
Additionally, the fact that \eqref{Taub-NUT metric} has a Weyl double copy structure implies that
\begin{equation}
\begin{split}
C^+_{\alpha\beta\gamma\delta}&=-\frac{6}{c-\sigma}\left(F^+_{\alpha\beta}F^+_{\gamma\delta}-\frac{(F^{+})^{2}}{3}I_{\alpha\beta\gamma\delta}\right)~,\\
(F^{+})^{2}&= A(c-\sigma)^4~,\label{C+ as FF+}
\end{split}
\end{equation}
with $c=1$ and $A=-(4(M-iN))^{-1}$. Next, $W$ is found  by solving  \eqref{Wform}.
After this,  we can transform the original metric into \eqref{newmetric}
\begin{equation}
g^{\prime }_{\mu\nu}=\Omega^2\,g_{\mu\nu}-2\xi_{(\mu}W_{\nu)}+\frac{\xi^\sigma\xi_{\sigma}}{\Omega^2}W_{\mu}W_\nu~.
\end{equation}
In order to interpret this new metric, it is convenient to define polar coordinates in the parameter space
\begin{equation}
\rho=\sqrt{\delta^2+\gamma^2}~,\qquad \tan\zeta=\frac{\delta}{\gamma}~.
\end{equation}
Performing a charge redefinition and a change of coordinates 
\begin{equation}
\begin{gathered}
\begin{pmatrix}
M^\prime\\
N^\prime
\end{pmatrix}=\begin{pmatrix}
\cos 2\zeta& -\sin2\zeta \\
\sin2\zeta & \cos 2\zeta
\end{pmatrix}\begin{pmatrix}
\rho \,M\\
\rho\, N
\end{pmatrix}~,\label{New Charges}\\
t^\prime=\frac{ t}{\rho}~,\qquad r^\prime=\rho\,r+M^\prime(1-\cos 2\zeta)-N^\prime \,\sin 2\zeta~,
\end{gathered}
\end{equation}
the metric simplifies to
\begin{equation}
\begin{gathered}
ds^{\prime 2}=-f(r^\prime)(dt^\prime-2N^\prime\,\cos\theta\,d\phi)^2+\frac{dr^{\prime2}}{f(r^\prime)}+(r^{\prime 2}+N^{\prime 2})d\Omega^2_2~,\\
 f(r^\prime)=\frac{r^{\prime 2}-2M^\prime\,r^\prime-N^{\prime 2}}{r^{\prime 2}+N^{\prime 2}}~.\label{TN  transformed}
\end{gathered}
\end{equation}
Hence, it is still a member of the Taub-NUT family. The self-dual part of $F_{\mu\nu}$ transforms as \eqref{Ftransform}.
The integrated Ernst one-form transforms as
\begin{equation}
\label{sl2rTN}
\sigma^\prime=\frac{1}{\delta^2+\gamma^2}\,\frac{\delta \sigma+i\,\gamma}{i\gamma\sigma+\delta}~.
\end{equation}
After the transformation, \eqref{C+ as FF+} also holds with
\begin{equation}
c^\prime=\frac{1}{\gamma^2+\delta^2}~\qquad A^\prime=-\frac{(\delta+i\,\gamma)^4}{4(M-i\,N)}~,
\end{equation}
in agreement with (54) in \cite{Mars:2001gd}.
\\

Let us now study the implications for the single copy. The single copy of \eqref{Taub-NUT metric} can be written in flat spherical coordinates ($\tilde{t},\tilde{r},\theta,\phi$) \eqref{Taub-NUT single copy}
\begin{equation}
F_{T}=-\frac{M}{\tilde{r}^2}\,d\tilde{t}\wedge d\tilde{r}-N\,\sin\theta\,d\theta\wedge d\phi~.
\end{equation}
Hence, the single copy of the transformed space-time, on the same background reads
\begin{equation}
F_{T}^\prime=-\frac{M^\prime}{\tilde{r}^2}\,d\tilde{t}\wedge d\tilde{r}-N^\prime\,\sin\theta\,d\theta\wedge d\phi~.
\end{equation}
Using \eqref{New Charges}, it can be checked that the transformation in terms of the Ehlers group parameters is
\begin{equation}
F_{T}^\prime=\rho \cos(2\zeta)\, F_{T}+\rho \sin(2\zeta)\tilde{F}_{T}~.
\end{equation}
This corresponds to an electromagnetic duality rotation and a rescaling by $\rho$. Both transformations are contained in the electromagnetic duality, where the rescaling can be interpreted as the transformation of the gauge coupling \cite{Ortin:2015hya}. The zeroth copy is affected similarly, transforming using $M\rightarrow M^\prime$, $N\rightarrow N^\prime$, leaving the double copy structure intact.
The Weyl double copy
\begin{equation}
\label{WeylDCTN}
C^+_{\mu\nu\rho\sigma} = \frac{2}{\sigma^+}C_{\mu\nu\rho\sigma}[F^+]\, ,
\end{equation}
where $F^{'+}$ is the self-dual part of the transformed single-copy Killing tensor and  $\sigma^+ =\sigma-c= -2(N+iM)/(N+ir)$, is preserved: \eqref{WeylDCTN} transforms directly to the double copy in the transformed spacetime
\begin{equation}
\label{WeylDCTNnew}
C^{'+}_{\mu\nu\rho\sigma} = \frac{2}{\sigma^{+'}} C_{\mu\nu\rho\sigma}[F^{'+}]\, ,
\end{equation}
where $F^+$ is the self-dual part of the single-copy Killing tensor, now defined using the shifted metric (and the same Killing vector, although note of course that the co-vector differs in the new spacetime), and the transformed Ernst scalar is
\begin{equation}
\label{ErnstTNnew}
\sigma^{'+} = -\frac{2 (M-i N)}{(\gamma -i \delta ) (2 \gamma  M+N(\delta -i \gamma)  +r(i\delta-\gamma) 
  )} .
\end{equation}
\\
We see that in terms of the action on the fields, a restricted set of the $\text{SL}(2,\mathbb{R})$ transformations act in this case, and the orbit is within the Taub-NUT class of metrics. The two degrees of freedom are realised by the rotation parameter $\zeta$ and scaling $\rho$. 

\subsection{The Eguchi-Hanson metric}

It is of interest to consider a Riemannian metric example and we turn to the Eguchi-Hanson metric
\begin{equation}
\label{EHdoubleKS}
ds^2= 2dudv-2dXdY+\frac{\lambda}{(uv-XY)^3}(vdu-XdY)^2\, ,
\end{equation}
with coordinates $(u,v,X,Y)$ and constant $\lambda$. The single-copy (self-dual) Maxwell tensor is \cite{Berman:2018hwd,Luna:2018dpt}
\begin{equation}
\label{EHMax}
F =  \frac{2 \lambda}{(uv-XY)^3} \Big( (u v+XY)(du\wedge dv - dX\wedge dY) - 2vY du\wedge dX + 2uX dv\wedge dY\Big) \, .
\end{equation}
Consider the Killing vector
\begin{equation}
\label{EHKV}
K^\mu = (u,-v,-X,Y)\,
\end{equation}
and Killing two-form
\begin{equation}
\label{Killingtwoform}
K_{\mu\nu} = 2\partial_{[\mu} K_{\nu]}\, .
\end{equation}
The single-copy Maxwell tensor is then given by
\begin{equation}
\label{EHsc}
F_{\mu\nu}= K_{\mu\nu}^+\, .
\end{equation}
We have the relations
\begin{equation}
\label{EHErnstvecs}
\sigma_\mu^+:= 2K^\nu K_{\nu\mu}^+    = \partial_\mu \sigma^+\, , \qquad  \sigma_\mu^-:= K^\nu K_{\nu\mu}^-   = \partial_\mu \sigma^- \, ,
\end{equation}
with
\begin{equation}
\label{EHErnsts}
\sigma^+ = -\frac{2\lambda}{(uv-XY)} \, ,\qquad \sigma^- = 4(uv-XY) \, .
\end{equation}

We now consider the equations \eqref{Wform} with $\sigma\rightarrow \sigma^+, \bar\sigma\rightarrow\sigma^-$ and $\xi$ the Killing vector \eqref{EHKV}.
These are solved by
\begin{align}
\label{Wshifted}
W_\mu =  -\Big( \frac{8 \lambda \gamma^2}{u v - X Y} + 2 i \gamma\delta\Big) & (v,-u,Y,-X)  -\frac{2 i \lambda\gamma\delta}{(u v - X Y)^2} (v,0,0,-X)
\nonumber\\
&-(1-8 \lambda\gamma^2-\delta^2) \left(\frac{1}{u},0,0,0\right) \, .
\end{align}
The new metric is given by \eqref{newmetric} with $W$ given by the expression above. This is a complicated expression which we will not reproduce here.
We have checked that this is Ricci-flat.
The single-copy Maxwell tensor $K_{\mu\nu}^+$ transforms in the same way as $F^+$ in \eqref{Ftransform}, and the transformation of its dual is the conjugate of this. It can be checked that the transformed tensors are (anti-)self-dual with respect to the transformed metric \eqref{newmetric}, and agree with the new Killing two form obtained from \eqref{Killingtwoform}  using the same Killing vector \eqref{EHKV} but with the index lowered with the new metric.
\\
Considering the Weyl tensor, we have the double copy relation for the Eguchi-Hanson metric
\begin{equation}
\label{WeylEH}
C^{EH}_{\mu\nu\rho\sigma} =  -\frac{uv-XY}{\lambda} \, C_{\mu\nu\rho\sigma} [K^+ ]   \,  ,
\end{equation}
where $K^+$ is the single copy Maxwell tensor \eqref{EHsc}. Note that the Weyl tensor for the Eguchi-Hanson metric is  self-dual. We find that the equivalents of  \eqref{EHErnsts} in the transformed metric are
\begin{equation}
\label{EHErnstsNew}
\sigma^{'+}= -\frac{2i\lambda}{(2\gamma\delta\lambda+i\delta^2(uv-XY))} \, ,\qquad \sigma^{'-} = \frac{1}{(4\gamma^2(uv-XY)+i\gamma\delta)} \, ,
\end{equation}
and that the double copy relationship is preserved by a general transformation, with
\begin{equation}
\label{WeylEHshiftedFull}
C^{'\pm}_{\mu\nu\rho\sigma} =  \frac{2}{\sigma^{'\pm}}\, C_{\mu\nu\rho\sigma} [K^{'\pm} ] \,  .
\end{equation}
\\
To gain some insight into the action of  $SL(2,\mathbb{R})$ in this example, consider the transformations with $\delta=0$. The shifted Maxwell fields in this case are given by
\begin{equation}
\label{KshiftedSD}
K^{'+}= \frac{1}{2  \lambda  \gamma^2} \Big( - du\wedge dv + \frac{(1- 8 \lambda \gamma^2)}{u} du\wedge d(XY) \Big) - 4 dX\wedge dY\,
\end{equation}
and
\begin{align}
\label{KshiftedASD}
K^{'-}=&  \frac{-uv+XY(1-16 \lambda\gamma^2)}{4(uv-XY)^3\gamma^2} du\wedge dv
 +\frac{Y(uv-XY + 8 \lambda \gamma^2(uv+XY))}{4 u (uv-XY)^3\gamma^2}du\wedge dX   \\     \nonumber
& +\frac{X(1- 8 \lambda \gamma^2)}{4u(uv-XY)^2\gamma^2} du \wedge dY
-\frac{4u \lambda X}{(uv-XY)^3} dv\wedge dY
+\frac{2\lambda(uv+XY)}{(uv-XY)^3} dX\wedge dY\, .
\end{align}
\\
For the transformed metric in this case, from  \eqref{WeylEHshiftedFull} the new Weyl tensor is now anti-self-dual and obeys the double copy relationship
$C^{'-}_{\mu\nu\rho\sigma} =  8 \gamma^2(uv-XY)\, C_{\mu\nu\rho\sigma} [K^{-'} ]
$, where $K^{-'}_{\mu\nu}$ is the shifted Maxwell tensor given in \eqref{KshiftedASD}. When the parameter $\gamma$ satisfies $8\lambda\gamma^2=1$ we find that $K_{\mu\nu}^{'-}= -K_{\mu\nu}^+$  (note that the dual of a tensor is defined with respect to different metrics on the two sides of this equation). Thus the roles of the electromagnetic fields $E$ and $B$ are exchanged by this transformation. The new metric is given simply in terms of the original Eguchi-Hanson metric by the interchange of $(u,v)\leftrightarrow(X,Y)$. The double copy relations are consistent with this. It would be interesting to follow up with a full study of the action of the transformations when $\delta\not= 0$.

\section*{Discussion}
We have explored the double copy formalism in the context of solution generating symmetries in relativity, and in particular how these symmetries in general relativity are related to hidden or duality symmetries in the single copy gauge theory. We have used two complementary techniques, the Kerr-Schild and the Weyl double copy formalisms, and have seen in a number of cases that the transformations in the gauge theory and the corresponding transformations on the gravity side are such that  the double copy structure is preserved. This seems to us further evidence that the double copy is beyond just a perturbative symmetry for amplitudes but a fascinating relation between gravity and gauge theory.

\section{A spin-off of the Weyl double copy}\label{sec:WeylSpinOff}
\subsection{Spacetime classification}
The central object of this section will be the `doubling' formula, where a tensor, $C[F]$ with the algebraic symmetries of the Weyl tensor in $D$ dimensions can be constructed from an $n$-form $F=dA$ (with the gauge field $A$ an $n-1$ form) as follows. First, we quote again eq. \eqref{Dhigher},
\begin{equation}
C_{\mu\nu\rho\sigma} [F] = F_{\mu\nu}\cdot F_{\rho\sigma} - F_{\rho\mu}\cdot F_{\nu\sigma}-\frac{6}{D-2}g_{\mu\rho}{F_\nu} \cdot F_{\sigma} + \frac{3}{(D-1)(D-2)}g_{\mu\rho}g_{\nu\sigma}F\cdot F   \; \Big\vert_s  \, ,
\end{equation}
where a dot product  means to contract all non-visible indices between the two terms, for example for an $n$-form,
$F_{\mu\nu}\cdot F_{\rho\sigma} = F_{\mu\nu{\lambda_3}\dots{\lambda_n}} {F_{\mu\nu}}^{{\lambda_3}\dots{\lambda_n}}$.
The symbol ``$\vert_s$" above applies to the expression on the right-hand side of the equation, and it means to anti-symmetrise in the indices $\mu,\nu$ and in $\rho, \sigma$, with unit weight. 

We wish to study cases where the spacetime Weyl tensor $C_{\mu\nu\rho\pi}[g]$  is proportional to this expression - 
\begin{equation}
\label{CeqCF}
C_{\mu\nu\rho\pi}[g] = \frac{1}{\sigma}\, C_{\mu\nu\rho\pi} [F]  \, 
\end{equation}
for some function $\sigma$.  In a later example we will exhibit a derivative corrected expression where the right-hand side of \eqref{CeqCF} contains further terms determined by derivatives of $F$.
The field strength $F$ is closed and required to be divergence-free. (Note that the field is defined on the spacetime with metric $g_{\mu \nu}$ thus the divergence equation is non-trivial.)
\par

Let us first consider the case $n=2$ where we have a normal Abelian two form field strength. (The cases with $n>2$ will be relevant when we discuss supergravity brane solutions.) One can identify two cases where the divergence equation follows automatically. The first is where
\begin{equation}\label{FKill}
F_{\mu\nu} = 2\partial_{[\mu} K_{\nu]}
\end{equation}
for a Killing vector $K_\mu$, and the second is where there is a self-duality condition
\begin{equation}
F_{\mu\nu} = {\phi_{\mu\nu}}^{\rho\sigma}F_{\rho\sigma}
\end{equation}
 for some covariantly constant four form $\phi$ on the manifold. (The field strength defined via a Killing vector field may also satisfy a self-duality condition of course, or have this imposed.)The field strength as defined by \eqref{FKill} is to be interpreted as generally defined on the possibly curved geometry on which the Killing vector is defined; see appendix \ref{Maxapp}. We will study examples of both situations below.  

In the first case, where we construct $F$ from the Killing vector $K$, we note that $K^\mu F_{\mu\nu}$ is closed. This follows from application of Cartan's formula for the Lie derivative:
\begin{equation}
 \mathcal{L}_K F = i_K d F + d i_K F  \,
\end{equation}
which means for closed $F$ and $K$ Killing then $d i_K  F=0$.
Locally we can solve this condition to write  $K^\mu F_{\mu\nu}$ an exact form so that
\begin{equation}
\label{sigma}
 K^\mu F_{\mu\nu}= \partial_\nu \sigma\, .
\end{equation}
A solution to this (up to the addition of an arbitrary constant) is
\begin{equation}
\label{sigmaKsq}
\sigma= K^\mu K_\mu\, .
\end{equation}

In four dimensions, an analysis of the action of duality transformations \cite{Alawadhi:2019urr,Mars:2001gd}  shows that this is the same $\sigma$ as in equation \eqref{CeqCF}. See also  \cite{Walker:1970abc,Hughston:1972qf}  where this was seen earlier using the spinor formalism in Type D spacetimes and more recently  \cite{Frolov:2017kze,Mason:2010zzc} for related work in higher dimensions.

The formula \eqref{CeqCF} naturally implies some conditions on the spacetime. Write the eigenvector equation for 
$F$ as
\begin{equation}
\label{evaF}
{F_\mu}^\nu k_\nu= \lambda_k k_\mu \, ,
\end{equation}
for eigenvector $k_\mu$ and eigenvalue $\lambda_k$. Note that the eigenvectors are necessarily null. Since $F$ is antisymmetric, its eigenvalues form into pairs with opposite signs (and one zero eigenvalue if $D$ is odd). Then  \eqref{CeqCF} implies that 
\begin{equation}
\label{Ceva}
C_{\mu\nu\rho\sigma} k^\nu  k^\sigma =\Lambda k_\mu k_\rho\, 
\end{equation}
with 
\begin{equation}
\Lambda = \frac{3}{ 2(D-1)(D-2)}\Big[ (D-1)(D-4)\lambda_k^2 + F_{\alpha\beta}F^{\beta\alpha}\Big]\, . 
\end{equation}
This implies that the eigenvector $k_\mu$ is a principal null direction of the Weyl tensor.  
Equation \eqref{CeqCF} also implies that 
\begin{equation}
\label{typeD_moreD}
k_{[\alpha}C_{\mu]\nu[\rho\sigma} k_{\pi]}k^\nu =0 \, .
\end{equation}
and similarly for a second eigenvector $l_\mu$ with a different eigenvalue. Thus generically there are two principal null directions satisfying \eqref{typeD_moreD} which implies that the spacetime satisfies the conditions for falling within the type D  class in the appropriate higher-dimensional classification  \cite{Ortaggio:2012jd} (see also the overview \cite{Reall:2011ys}). A special case occurs when the eigenvectors are not independent and so there is only one principal null direction. This occurs when $F$ obeys a self-duality relation in which case the two eigenvectors are identical. The spacetime is then said to fall into the type II class.


\subsection{Invariants}
Equation \eqref{CeqCF} implies that scalar invariants constructed from the Weyl tensor (and its dual where there is a suitable four-form on the manifold) are functions of the traces
\begin{equation}
(F^n) = {F^{\mu_1}}_{\mu_2} {F^{\mu_2}}_{\mu_3}\dots {F^{\mu_n}}_{\mu_1} \, .
\end{equation}
We will use a similar notation for traces of products of Weyl tensors (or their duals), for example
\begin{equation}
(C^3) ={C^{\mu\nu}}_{\rho\pi} {C^{\rho\pi} }_{\alpha\beta}{C^{\alpha\beta} }_{\mu\nu}  \, .
\end{equation}
In the four-dimensional case, then,
\begin{align}
\label{trExs}
(C^2)  &= {1\over \sigma^2}\Big(\,  {9\over 4} (F^2)^2 - 3 (F^4) \Big)\,  , \nonumber     \\ 
(CC^*)  &= {1\over \sigma^2}\Big(\,  (F^2)(FF^*) +2 (F^3F^*) \Big)\, ,    \nonumber \\ 
(C^3) &= {1\over \sigma^3}\Big(  -{61\over 2}(F^6) +{201\over 8}(F^2)(F^4) - {41\over 8}(F^2)^3  \Big) \,  \nonumber\\ 
(C^2C^*) &= {1\over \sigma^3}\Big(  {21\over 2}(F^3(F^3)^*) -(F^2)(F^3F^*) +\big(- {5\over 2}(F^2)^2+2(F^4)\big)(FF^*)  +{11\over 2}(F^5F^*)\Big) \, ,
\end{align}
and so on. We define $F^*_{\mu\nu}={1\over 2}\epsilon_{\mu\nu\rho\sigma}F^{\rho\sigma}$, and ${(F^3)^*}_{\mu\nu}={1\over 2} \epsilon_{\mu\nu\rho\sigma}F^{\rho\alpha}F_{\alpha\beta}F^{\beta\sigma}$ in the above.  For four-dimensional vacuum spacetimes, the invariants \eqref{trExs} form an  independent basis (under algebraic relationships) for the set of scalar invariants \cite{Lim:2004xx}.
In addition, traces of higher powers of $F$ are related to those of lower powers by the recursion relation
\begin{equation}
\label{4Drecrel}
(F^{2n}) ={1\over 2} (F^{2n-2})(F^2) -{1\over 8}\Big( (F^2)^2 - 2 (F^4)\Big) (F^{2n-4}) \, .
\end{equation}
Using this, one can express any $(F^n)$ in terms of  $(F^2)$ and $(F^4)$, and hence all scalar curvature invariants in the vacuum case are functions of these two traces.  There are analogous results in $D> 4$.

If one block-diagonalises $F$ as
\begin{equation}
\label{Fmatrix}
F = \begin{pmatrix}
0& x & 0 & 0 \\
-x & 0 & 0 &0 \\
0& 0 & 0 & y \\
0 & 0 &-y &0 
\end{pmatrix}
 \, .
\end{equation}
then $(F^2) = -2(x^2+y^2)$ and $(F^4) = 2(x^4+y^4)$ and the invariants are functions of these combinations. If $x$ and $y$ are proportional then $(F^2)$ is the only independent function. 

Similar arguments to those above apply to express the Weyl-NP scalars in terms of the independent Maxwell-NP scalars, depending on the dimension. We now give some examples to illustrate the above.


 \subsubsection{Taub-NUT}\label{Taub-NUT}

As we have seen earlier in the chapter, the Taub-NUT metric can be written in the form \cite{Ortin:2015hya}
\begin{equation}
  ds^2=-f(r)(dt-2n\cos{\theta}d\phi)^2+f(r)^{-1}dr^2+(r^2+n^2)d\Omega^2_2\, 
 \end{equation}
with
\begin{equation}
  f(r)=\frac{r^2-2mr-n^2}{r^2+n^2}\, .
\end{equation}
From the Killing vector $K^\mu=(1,0,0,0)$ define the Maxwell field $F_{\mu\nu}=2\partial_{[\mu}K_{\nu]}$. Then we have the Weyl doubling formula \cite{Alawadhi:2019urr}
\begin{equation}
\label{TNWeyl1}
C^+_{\mu\nu\rho\pi} = {1\over\sigma^+}\, C_{\mu\nu\rho\pi} [F^+] \, ,
\end{equation}
with $\sigma^+ =-\frac{m+in}{r+in}=:-\frac{m_+}{r_+}$.
The eigenvalues of $F^+$ are $\frac{m_+}{r_+^2}(1,1,-1,-1)$ with corresponding eigenvectors
\begin{align}
\label{TNnulltetrad}
m^\mu &= \frac{1}{\sin{\theta}\sqrt{2(r^2+n^2}} \Big(-2n\cos{\theta},0,-i,0 \Big)\, , \quad 
l^\mu = \frac{1}{\sqrt{\lambda}}   \Big(\lambda,1,0,0 \Big)\, ,   \nonumber \\
\bar m^\mu &= \frac{1}{\sin{\theta}\sqrt{2(r^2+n^2}}  \Big(-2n\cos{\theta},0,i,0 \Big)\, ,  \quad
n^\mu =\frac{1}{\sqrt{\lambda}}   \Big(-\lambda,1,0,0 \Big)\, , 
\end{align}
with $\lambda=(r^2+n^2)/(n^2+2mr-r^2)$. These form a null tetrad, with $l_\mu n^\mu=-1,m_\mu\bar m^\mu=1$ and the remaining inner products zero.

As $F^+$ has repeated eigenvalues, the only independent trace is
\begin{equation}
\label{TNMaxn}
(F^+F^+)= -4\,\frac{m_+^2}{r_+^4}   \,   
\end{equation}
leading to
\begin{equation}
\label{TNWeyln}
(C^+C^+)= 24{m_+^2\over r_+^6}\, .
\end{equation}
All other scalar invariants are functions of this expression and its conjugate. 

The formula \eqref{TNWeyl1} implies corresponding relationships amongst the NP scalars. In this case, for example, the Maxwell-NP scalars are $\phi_0= F^+_{\mu\nu}l^\mu m^\nu= 0, \phi_2 = F^+_{\mu\nu}\bar m^\mu n^\nu=0$ and 
\begin{equation}
\label{TNspinor}
 \phi_1= \frac{1}{2}F^+_{\mu\nu}( l^\mu n^\nu + \bar m^\mu m^\nu)= \frac{m_+}{r_+^2}    \, .
\end{equation}
Correspondingly, the only non-vanishing Weyl-NP scalar is 
\begin{equation}
\label{TNspinor2}
 \Psi_2= C^+_{\mu\nu\rho\pi}  l^\mu m^\nu \bar m^\rho n^\pi=    -\frac{m_+}{r_+^3} =\frac{1}{\sigma_+}\phi_1^2   \, .
\end{equation}
This simple form of the Weyl doubling formula can also be seen directly from the spinor formulation (c.f. \cite{Luna:2018dpt}).


\subsubsection{Plebanski-Demianski}
The  general vacuum type D solution with vanishing cosmological constant \cite{Plebanski:1976gy}, as given in  \cite{Luna:2018dpt}, is
\begin{equation}
ds^2 = {1\over(1-pq)^2}  \Bigg[2 i (du+q^2dv) dp - 2(du-p^2dv)dq +  \frac{P(p)}{p^2+q^2}\,(du+q^2dv)^2
-  \frac{Q(q)}{p^2+q^2}\,(du-p^2dv)^2  \Bigg] \, ,
\end{equation}
with
\begin{align}
P(p) =  \gamma (1-p^4) +2 n p- \epsilon p^2 +2 m p^3 \,, \nonumber \\
Q(q) =  \gamma (1-q^4) - 2m q + \epsilon q^2 -2 n q^3 \,,
\end{align}
where the parameters $m,n,\gamma,\epsilon$ are related to the mass, NUT charge, angular momentum and acceleration (c.f. \cite{Griffiths:2005qp}).
The self-dual part of the Maxwell two form is given by
\begin{equation}
\label{PDMax}
F^+ =      {(m-in)\over2(p+iq)^2}  \Big(  i (du+q^2dv) dp + (-du+p^2dv)dq \Big), 
\end{equation}
with the anti-self dual part given by the complex conjugate.
One has the Weyl doubling formula
\begin{equation}
\label{PDWeyl}
C^+_{\mu\nu\rho\pi} = {1\over\sigma^+}\, C_{\mu\nu\rho\pi} [F^+] \, ,
\end{equation}
with $\sigma^+ =(m-in)(1-pq)^4/(4i(p+iq)$.
Then
\begin{equation}
\label{PDMaxn}
(F^2)= \Big({(m-in)(1-pq)^2\over (p+iq)^2}\Big)^2 , \quad  (F^4)=  {1\over 4} \Big({(m-in)(1-pq)^2\over (p+iq)^2}\Big)^4    \,   .
\end{equation}
These are not independent as the eigenvalues of $F^+$ are repeated - they are $\pm (m-in)(1-pq)^2/2(p+iq)^2$ twice, and hence 
\begin{equation}
\label{PDWeyln}
(C^+C^+)= -{24(m-in)^2(1-pq)^6\over(p+iq)^6}
\end{equation}
is the only independent curvature invariant involving $C^+$. Invariants involving $C$ and $C^*$ can be written in terms of this and its conjugate.

 
 \subsubsection{Eguchi-Hanson}
The Eguchi-Hanson metric is a vacuum solution with self-dual Weyl curvature. It is given by
\begin{equation}
ds^2= 2dudv-2dXdY+\frac{\lambda}{(uv-XY)^3}(vdu-XdY)^2\, ,
\end{equation}
with coordinates $(u,v,X,Y)$ and constant $\lambda$. The single-copy (self-dual) Maxwell tensor is 
\begin{equation}
\label{EHMax1}
F =  \frac{2 \lambda}{(uv-XY)^3} \Big( (u v+XY)(du\wedge dv - dX\wedge dY) - 2vY du\wedge dX + 2uX dv\wedge dY\Big) \, .
\end{equation}
This is the single copy tensor discussed in \cite{Luna:2018dpt}, rather than the ``mixed'' version of  \cite{Berman:2018hwd}.   With the Killing vector $K^\mu = (u,-v,-X,Y)$ this is given by $F_{\mu\nu}=(2\partial_{[\mu} K_{\nu]})^+$.
The Weyl curvature is then given by
\begin{equation}
\label{WeylEH1}
C_{\mu\nu\rho\sigma} =  \frac{1}{\sigma^+} \, C_{\mu\nu\rho\sigma} [F ]   \,  ,
\end{equation}
with $\sigma^+ = -\frac{\lambda}{(uv-XY)}$ and $F$ the Maxwell field \eqref{EHMax1}.

As in the case above, the eigenvalues of $F$ are repeated - here they are $\pm m/(uv-XY)^2$ twice. Thus the only independent trace of $F^n$ is
\begin{equation}
\label{EHMaxn}
(F^2)=  {4\lambda^2\over (uv-XY)^4} 
\end{equation}
and hence the only independent curvature invariant is
\begin{equation}
\label{PDWeyln2}
(C^2)= {24\lambda^2 \over(uv-XY)^6}\, .
\end{equation}
%

 
 \subsubsection{Singly rotating Myers-Perry}
 \label{MPsubsection}
 
As an example in $D>4$, consider the singly-rotating Myers-Perry metric in the form \cite{Elvang:2003mj}
\begin{align}
\label{MP}
  ds^2 =& -\frac{f(x)}{f(y)} \left(dt+
     R\sqrt{\nu} (1 + y) d\psi\right)^2       \nonumber \\
   & +\frac{R^2}{(x-y)^2}
   \left[ -f(x) \left( g(y) d\psi^2 +
   \frac{f(y)}{g(y)} dy^2 \right)
   + f(y)^2 \left( \frac{dx^2}{g(x)} 
   + \frac{g(x)}{f(x)}d\phi^2\right)\right]  \,, 
\end{align}
with
\begin{equation}
  f(\xi) = 1 - \xi \, ,
\qquad  g(\xi) = (1 - \xi^2)(1-\nu \xi) \, .
\end{equation}
From the Killing vector $K^\mu=\partial/\partial t$ form the Maxwell field strength $F_{\mu\nu} = 2\partial_{[\mu}K_{\nu]}$. Then one can check that
the Weyl tensor is given by the doubling formula
\begin{equation}
\label{MPWeyl}
C_{\mu\nu\rho\pi} = {1\over\sigma}\, C_{\mu\nu\rho\pi} [F] \, ,
\end{equation}
with $\sigma= (y-1)/(x-y)$. This satisfies \eqref{sigma}. One can ask how $K_\mu$ might be related to the standard single-copy gauge potential that arises from the Kerr-Schild form of the metric. In the notation of \cite{Frolov:2017kze}, section (E.3), the single-copy potential for the Myers-Perry metrics is $H l_\mu$, and it differs from the contravariant vector $K_\mu$ obtained from the Killing vector $\partial/\partial\tau$ by a gauge transformation with parameter $\tau$. Thus the two potentials give the same single-copy Maxwell tensor in  Kerr-Schild coordinates. When one transforms to the coordinates used in \eqref{MP} above, one obtains that metric, but with the function $g(\xi)$ having no term linear in $\xi$, i.e.,  with $g(\xi) = 1-\xi^2+\nu \xi^3$ \cite{Emparan:2001wn}. The coefficient of the linear term is purely kinematical \cite{Plebanski:1976gy} and does not affect the vanishing of the Ricci tensor. However, it does appear in the Weyl tensor, and its presence thus affects whether there is a Weyl doubling formula or not. For the function $g[\xi)=1-A \nu\xi -\xi^2+\nu \xi^3$ it can be checked that only for $A=1$ is there a Weyl doubling formula, which is \eqref{MPWeyl}.

From \eqref{Dhigher} with $D=5$, the $(C^2)$ and $(C^3)$  invariants are given by the formul\ae
\begin{align}
\label{trMP}
(C^2)  &= {1\over \sigma^2}\Big(\,  {15\over 8} (F^2)^2 - {3\over 2} (F^4) \Big)\,  ,  \nonumber    \\ 
(C^3) &= {1\over \sigma^3}\Big(  -{23\over 2}(F^6) +{81\over 8}(F^2)(F^4) - {41\over 16}(F^2)^3  \Big) \, .
\end{align}
The recursion relation \eqref{4Drecrel} also applies in five dimensions  so that the only independent traces are again $(F^2)$ and $(F^4)$. Here they are
\begin{align}
\label{trMPMax}
(F^2)  &= {2(-1+L^2(1+2x))(x-y)^3\over R^2(-1+y)^3}\,  ,      \\ \nonumber
(F^3) &={2(1-2L^2x+L^4(1+2x+2x^2))(x-y)^6\over R^4(-1+y)^6}\, .
\end{align}
The Maxwell field strength here has four distinct eigenvalues - two different pairs with opposite signs, and one zero.

Turning to the Weyl-NP scalars, the classification of  spacetimes in general dimensions via properties of the Weyl tensor has been discussed by Coley, Milson, Pravda and Pravdova (CMPP) in
\cite{Milson:2004jx,Coley:2004jv} (see also \cite{DeSmet:2002fv,Godazgar:2010ks}). More recently, the relationship between these classifications has been discussed and compared to a  spinor-based analysis  in \cite{Monteiro:2018xev}, who  show that in five dimensions the CMPP and spinor approaches are equivalent. The classification of Maxwell fields is also described there. 

We first define a null pentad of five  vectors $n^\mu, l^\mu, m_i^\mu$, ($i=1,2,3$), with
\begin{equation}
\label{nullpentad}
n^\mu l_\mu=-1, \;\; m_i^\mu m_{j\mu}= \delta_{ij}\, 
\end{equation}
and all other inner products zero. A convenient set is defined in the Appendix.
Now expand a five-dimensional Maxwell field as
\begin{equation}
\label{MaxExp5D}
F_{\mu\nu} =F_{01} n_{[\mu}l_{\nu]}   + \hat F_{0i} n_{[\mu}m_{\nu]i} + \tilde F_{1i} l_{[\mu}m_{\nu]i} + F_{ij} m_{i[\mu}m_{\nu]j} \, .
\end{equation}
There is an analogous expansion for the Weyl tensor. In this type D case the terms with non-zero weights vanish using our pentad choice and we have \cite{Milson:2004jx,Coley:2004jv}
\begin{equation}
\label{WeylExp5D}
C_{\mu\nu\rho\pi} =4C_{0101} n_{\{\mu}l_{\nu} n_{\rho}l_{\pi\}}  +  C_{01ij} n_{\{\mu}l_{\nu} m^i_{\rho}m^j_{\pi\}}  +  8C_{0i1j} n_{\{\mu}m^i_{\nu} l_{\rho}m^j_{\pi\}}  +  C_{ijkl} n_{\{\mu}m^i_{\nu} l_{\rho}m^j_{\pi\}} \, ,
\end{equation}
with the notation $T_{\{\mu\nu\rho\pi\}}:= {1\over 2} (T_{[\mu\nu][\rho\pi]} + T_{[\rho\pi][\mu\nu]} )$ for a tensor $T_{\mu\nu\rho\pi}$.

The Weyl doubling relationship \eqref{CeqCF} implies that the Weyl coefficients, such as those in \eqref{WeylExp5D}, are given in terms of the analogous Maxwell coefficients, which are given in five dimensions  by \eqref{MaxExp5D}. We also saw a simple  four-dimensional example of this for the Taub-NUT metric in subsection  \ref{Taub-NUT} above. In five dimensions we find, for example,
\begin{equation}
\label{NPmap}
C_{0101} ={1\over\sigma} \Big(  {3\over 16}F_{01}^2  - {1\over 8} \hat F_{0i}\tilde F_{1i} - {1\over 8} F_{ij}F^{ij}\Big) \, .
\end{equation}
For the Myers-Perry metric \eqref{MP} we find the non-zero Maxwell-NP scalars
\begin{equation}
\label{MPMaxScalars}
F_{01} = 2 {(x-y)\over R(-1+y)}\sqrt{{(-1+L^2 x)(x-y)\over (-1+y)}  }\, , \;\; 
F_{31}=-F_{13}= {L(x-y)\over R(1-y)}\sqrt{{(1+x)(x-y)\over(1-y)}}\, 
\end{equation}
and the non-zero independent Weyl-NP scalars
\begin{equation}
\label{MPWeylScalars}
C_{0101} = {(-3+L^2(1+4x))(x-y)^2\over 4R^2(1-y)^2}\, , \;\; 
C_{0131} = -C_{0113}= -{L^2(x-y)^4(1+x)(-1+L^2x) \over R^4(1-y)^4}\, .
\end{equation}
(The $C_{ijkl}$ are not independent quantities in five dimensions (c.f. also \cite{Monteiro:2018xev} eqn. (5.30)). One can then check that for the Myers-Perry metric \eqref{NPmap} is satisfied using \eqref{MPMaxScalars} and \eqref{MPWeylScalars} and the vanishing of $\hat F_{0i}$ and $\tilde F_{1i}$.


\section{The Gibbons-Hawking metrics}
\label{sec:GH}
 
We will now consider the extension of the Eguchi-Hanson case where there is a self-duality condition on the fields. Here we will find a generalisation of the Weyl doubling formula  \eqref{CeqCF}, where the additional terms are also given in terms of the derivative field strength $F$.  The Gibbons-Hawking metric is given by
\begin{equation}
\label{GHmetric}
ds^2 = {1\over V} \left( dx^4 + A_i dx^i\right)^2 + V dx^i dx_i \, ,
\end{equation}
where the fields $V, A^i$ are functions of the spatial coordinates  $x^i, i=1,2,3$, and are related by
\begin{equation}
\label{SDVA}
\nabla V = \nabla\times A\, .
\end{equation}
This equation implies $V$ is harmonic. It may thus be solved by a superposition of harmonic functions with arbitrary centres. The two centre solution can be shown to be equivalent to the Eguchi-Hanson solution, after a coordinate transformation.
From the Killing vector $K^\mu=(0,0,0,1)$ we can form the anti-self-dual  field strength
\begin{equation}
\label{SDVA2}
F_{\mu\nu}  = 2\partial_{[\mu}K_{\nu]} \, ,
\end{equation}
satisfying $F_{\mu\nu}^*=\frac{1}{2}\epsilon_{\mu\nu\rho\sigma}F^{\rho\sigma}=-F_{\mu\nu}$ with $\sigma=1/V$ via \eqref{sigma}. 
We find in this case that a simple Weyl doubling formula of the form of equation \eqref{CeqCF} does not hold.
To explore this further, note that as the Weyl curvature is anti-self-dual, it has only five independent components, which may be taken to be the components of the symmetric traceless matrix $3\times 3$ matrix $C_{i4j4}$. One can express this using the following three anti-self-dual two forms. First note that the vierbein one-forms are given by
\begin{align}
\label{vierbeinGH}
e^4 &={1\over \sqrt{V} }\big(dx^4 + A_i dx^i\big)   \, , \\ \nonumber
e^i  &= \sqrt{V} dx^i \, .
\end{align}
Then the two forms are given by
\begin{equation}
\label{SDtwoformsGH}
\Sigma^i = e^4 e^i + {1\over 2} \epsilon^{ijk}e^j e^k \, .
\end{equation}
We now solve the following equation for $\hat \Omega^i$:
\begin{equation}
\label{hatAGH}
d \Sigma^i +\epsilon^{ijk}\hat \Omega^j \Sigma^k  =0\, .
\end{equation}
This implies that $\hat \Omega^i $ is given by the anti-self-dual part of the spin connection, $\omega^\mu{}_\nu$ and thus
\begin{equation}
\hat{\Omega}^i= \omega^{i4}- \frac{1}{2} \epsilon^i {}_{jk} \omega^{jk} \, .
\end{equation} 
The  curvature can then be constructed directly from the curvature of $\hat \Omega^i$ and is given by 
\begin{equation}
\label{curvGH}
\hat C^i =d\hat \Omega^i+{1\over 2} \epsilon^{ijk}\hat \Omega^j \hat \Omega^k  =  \hat C^{ij}\Sigma^j \, ,
\end{equation}
where we have introduced $\hat C^{ij}$ encoding the Weyl tensor.
We  now use Cartan's first structure equation to calculate the connection one forms for the vierbeins as follows
\begin{equation}
\omega^{4i}= V^{-3/2}(- \frac{1}{2} \partial^i V e^4 + \partial^{[i} A^{j]} e_j) \, , \qquad \omega^{ij} =V^{-3/2} ( \frac{1}{2} \partial^j V  e^i - \partial^{[i} A^{j]} e^4)\, .
\end{equation}
Then, expanding both the Maxwell  field strength and the  curvature in terms of the two forms $\Sigma^i$ we find the following doubling formula
\begin{align}
\label{SDCGH}
F &=\alpha_i  \Sigma^i   \, , \\ \nonumber
\hat C^i  &= \big(V \alpha_i\alpha_j - \partial_i\alpha_j \big)^T  \, \Sigma^j \, ,
\end{align}
where the superscript $T$ means to take the traceless part of the expression within the brackets and
\begin{equation}
\label{alphaGH}
\alpha_i = {1\over V^2} \, \partial_i V  \, .
\end{equation}
In terms of the Weyl tensor we have the relation $\hat C_{ij} = -2 C_{i4j4}$, with the other components of the Weyl tensor  related to these by the anti-self-duality condition. Equation \eqref{SDCGH} may be viewed a a generalised form of Weyl doubling, where the quadratic, algebraic terms involving the gauge field are supplemented by terms depending on the derivatives of the gauge field. Thus we see that the general class of gravitational instantons in four dimensions satisfies a generalised Weyl doubling formula. 

An explicitly covariant version of this formula can be found as follows. There is the identity
\begin{equation}
\label{GHKVeqn}
C_{\mu\nu\rho\sigma} = -2V\Big( K_{\mu}K^\lambda C_{\nu\lambda\rho\sigma} + K_{\rho}K^\lambda C_{\sigma\lambda\mu\nu}
-2 g_{\rho\mu}C_{\nu\lambda\pi\sigma}K^\lambda K^\pi\Big)_{[\mu\nu][\rho\sigma]}  
\, 
\end{equation}
where the notation $[\mu\nu][\rho\sigma]$ means to antisymmetrise the expression within the preceding brackets in $\mu,\nu$ and separately in $\rho,\sigma$.  The above relation follows from the fact that $C^*=-C$ and $K^2=-\sigma=\frac{1}{V}$. To see this, define the three expressions on the right-hand side of \eqref{GHKVeqn}, including the antisymmetrisations, as $C_L+C_R+C_g$. Then, taking the (left) dual of each by contracting with  $\frac{1}{2}\epsilon^{\alpha\beta\mu\nu}$, one finds that $(C_L+C_R+C_g)^*=(-C-C_L)+(C_R)+(C_L-C_R)=-C=C^*$.

Equation  \eqref{GHKVeqn} is  equivalent to
\begin{equation}
\label{GHfinal}
C_{\mu\nu\rho\sigma} = V\Big( K_{[\mu}\nabla_{\nu]}F_{\rho\sigma} + K_{[\rho}\nabla_{\sigma]}F_{\mu\nu}\Big)^T\, ,
\end{equation}
where the notation $(...)^T$ means to subtract all traces. This can be re-expressed as the doubling formula
\begin{equation}
\label{GHfinal2}
C_{\mu\nu\rho\sigma} = V C_{\mu\nu\rho\sigma} [F]  +V\Big( 2\nabla_\rho(K_\mu F_{\nu\sigma}) -   \nabla_\mu(K_\nu F_{\rho\sigma})  -   \nabla_\rho(K_\sigma F_{\mu\nu}) - K_\mu\nabla_\nu F_{\rho\sigma} \Big)^T_{[\mu\nu][\rho\sigma]} 
\, .
\end{equation}
where the notation $(...)^T$ means to subtract all traces.

As the two-centred solution for the Gibbons-Hawking (GH) metric \eqref{GHmetric} is equivalent to the Eguchi-Hanson (EH) metric \eqref{EHdoubleKS} via a coordinate transformation one may wonder why there is a simple doubling formula for the EH metric described earlier, but not for the more general multi-centred GH metrics. This  special case can be understood by noting that for spherical polar coordinates the map from the EH metric to the two-centred GH metric (see \cite{Ghezelbash:2009we} for example) interchanges the periodic ``time'' coordinate and the 
azimuthal angle $\phi$. Mapping to the  two-centred GH case, the Killing vector which gives a simple Weyl doubling formula via the anti-self-dual part of the Maxwell field $2\partial_{[\mu}K_{\nu]}$ is then $K^\mu=\frac{\partial}{\partial\phi}$.
If one constructs the Maxwell field via the Killing vector $K^\mu=\frac{\partial}{\partial t}$ then one obtains the formul\ae\ in the analysis above for this particular potential. In the generic multi-centre case there is only the latter Killing vector, leading to the above analysis. (In the multi-centre case where the centres are all at different sites along the $z$-axis, the additional Killing vector is present of course - here we find evidence from a numerical analysis that the simpler Weyl doubling formula continues to hold.)


\section{An eight-dimensional example with Spin(7) holonomy}
\label{sec:SH}

The discussion above used features of the Gibbons-Hawking metrics which arise from the underlying self duality relation \eqref{SDVA}  which expresses the anti-self-duality of the field strength of the four vector gauge field $(V,A^i)$.  
In higher dimensions, manifolds of special holonomy are examples  where there are more novel duality conditions satisfied by the curvature, and two forms on the manifold can similarly be restricted to have duality properties determined by the canonical four-form defined by the special holonomy group (a pioneering paper on this is \cite{Corrigan:1982th}; see \cite{Gubser:2002mz} for a review relevant to string theory).
However, in general the formula \eqref{Dhigher} does not preserve duality properties. This means that if a two-form $F$ transforms in a certain representation of the holonomy group then this does not imply that $C[F]$ will as well (in each pair of indices). Furthermore,  projecting $C[F]$ onto appropriate representations does not in general preserve the algebraic symmetries needed to  relate it to a Weyl tensor. While these conditions may appear quite restrictive, we have found an example of an eight-dimensional manifold with Spin(7) holonomy  for which  Weyl doubling does work. We will use the conventions of \cite{Acharya:1997gp} in what follows. 
 
Begin with the metric \cite{Salur:2008pi}
\begin{equation}
\label{S7metric}
ds^2 = V^{-3/2} \big( dx^8 + A_i dx^i\big)^2 + V^{1/2} dx^i dx_i \, ,
\end{equation}
where here the indices $i,j,...$  run from $1$ to $7$,  and and $a,b,...$ and $\mu, \nu,...$ run from $1$ to $8$.  The fields $(A_i, V)$  are functions of the spatial coordinates  $x^i$ only. The spin(7) four-form $\phi_{abcd}$ has the following non-zero orthonormal frame components
\begin{align}
&[1256] =  [1278] =  [3456] =  [3478] =  [1357] =  
[2468] =  [1234] =  [5678] =  1    \, , \\ \nonumber
&[1368] =  [2457] =  [1458] =  [1467] =  [2358] =  [2367] = 
-1  \, ,
\end{align}
where $[abcd]$ means ${\phi}_{abcd}$. The four-form satisfies $\phi^2=4\phi+12$  and the projectors of a two form onto the
 \lq\lq self-dual" ${\bf 7}$ and  \lq\lq anti self-dual" ${\bf 21}$ representations of Spin(7) are given by
\begin{equation}
\label{projectors}
P_7 =  {1\over 4}(1+ {1\over 2}\phi), \qquad 
P_{21} =  {3\over 4}(1 - {1\over 6}\phi).
 \end{equation}
Define the acht-beins
\begin{align}
\label{achtbeinS7}
e^a &= (e^i,e^8)  \, , \nonumber \\ 
e^i  &=V^{1/4} \, dx^i    \, , \nonumber  \\ 
e^8 &=V^{-3/4} \, (dx^8 + A_i dx^i)\, .
\end{align}
Then the four form 
\begin{equation}
\label{fourform}
\phi =  {1\over 24} \phi_{abcd}e^ae^be^ce^d
 \end{equation}
 is closed provided that the fields $(A_i, V)$ satisfy the constraints  ($F_{ij} :=\partial_iA_j - \partial_j A_i$ and $V_i:=\partial_i V$)
\begin{equation}
\label{constraint}
V_{[i}\phi_{jklm]} - 2 F_{[ij}\phi_{klm]8} = 0\, .
 \end{equation}
These 21 equations can be solved in terms of 7 independent quantities. For example, choosing the independent variables to be $(F_{37}, F_{45}, F_{46}, F_{47}, F_{56}, F_{57}, F_{67})$ the other expressions are given by
\begin{align}
\label{Identities}
F_{12} &= F_{56}, F_{13} = F_{57}, F_{14} = -F_{67}, F_{15} = -F_{37}, F_{16} = F_{47}, F_{17} = -F_{46}, F_{23} = -F_{67}   \, , \nonumber \\ 
F_{24} &= -F_{57}, F_{25} = F_{47}, F_{26} = F_{37}, F_{27} = -F_{45}, F_{34} = F_{56}, F_{35} = -F_{46}, F_{36} = F_{45}   \, , \nonumber \\ 
V_1 &= -2F_{45}, V_2 = 2F_{46}, V_3 = 2F_{47}, V_4 = -2F_{37}, V_5 = 2F_{67}, V_6 = -2F_{57}, V_7 = 2F_{56}  \, .
\end{align}
It can be checked that $\phi$ is also covariantly constant when these conditions are met.  In fact, taking various linear combinations of the constraints
\eqref{constraint} one finds that $\partial_i \partial_j V=0$ for all $i,j$ so that $V$ is linear in the coordinates $x^i$ and the correction term is trivial. This also means that all the electric $V_i$ and magnetic $F_{ij}$ components of the gravitational field are constants.
Now consider the construction of the Maxwell field.  From the Killing vector $K^\mu=\delta^{\mu 8}$ we can again form the gauge  field strength
\begin{equation}
\label{KMax}
F_{\mu\nu}   = 2\partial_{[\mu}K_{\nu]} \, .
\end{equation}
This field is anti self-dual {\it i.e.,} in the {\bf 21} representation, as $P_7 F =0$. Define the anti self-dual two forms 
\begin{equation}
\label{ASDtwoformsS7}
\Sigma_-^{ij} =  P_{21} e^a \wedge e^b \, .
\end{equation}
These satisfy the orthonormality conditions 
\begin{equation}
\label{ASDortho}
\Sigma_-^{ab\mu\nu}\Sigma_{-cd\mu\nu} = 4{(P_{21})^{ab}}_{cd}, \qquad   \Sigma_{-}^{ab\mu\nu} \Sigma_{-ab\rho\sigma} =  {(P_{21})^{\mu\nu}}_{\rho\sigma}  \, .
\end{equation}
We can expand the gauge field two form in the basis of these as
\begin{equation}
\label{SCexpS7}
F_{\mu\nu}=  {1\over 2}\alpha_{ab}\, \Sigma^{ab}_{-\mu\nu}   \, ,
\end{equation}
with $\alpha_{ab}={1\over 2}\Sigma_{-ab}^{\mu\nu} F_{\mu\nu}$ which can be written as
\begin{equation}
\label{alphaS7}
  \alpha_{ij} = -{1\over 2V^2}\phi_{8ijk}V_k\, ,\qquad   \alpha_{i8} = -{3\over 2V^2} \, V_i  \, .
\end{equation}
The metric is Ricci-flat and it can be checked that the Weyl tensor $C_{\mu\nu\rho\sigma}$ satisfies $(P_7 C)_{\mu\nu\rho\sigma}=0$, with $P_7$ acting either on the first or second pair of indices of the Weyl tensor. The tangent space components of the Weyl tensor can be expressed purely in terms of the constants $V_i$. Let us then consider if these can be given as a doubling formula based on a  two form such as \eqref{SCexpS7} above. We will generalise this expression slightly and consider the ``twisted'' two form with components
\begin{equation}
\label{alphaS7ab}
  \alpha_{ij} =-{b\over 2V^2} \phi_{8ijk}V_k\, ,\qquad   \alpha_{i8} =  -{3a\over 2V^2}\, V_i  \, ,
\end{equation}
for some constants $a,b$. This two form will be anti self-dual if $a=b$. The natural expression to consider for a doubling formula is one based on 
\eqref{Dhigher} in eight dimensions, $C_{abcd}[\alpha]$ with $\alpha$ given by \eqref{alphaS7ab}. In four dimensions the formula \eqref{Dhigher} preserves duality - if $F_{\mu\nu}$ is self-dual or anti self-dual then $C_{\mu\nu\rho\sigma}[F]$ is also (in both pairs of indices). But this is not true in higher dimensions and  in this eight-dimensional case, in order to seek to match the Weyl tensor this expression must be projected onto the {\bf 21} on the left and right pairs of indices. Upon doing this we find that it is not possible to both preserve the duality conditions and the symmetry and trace properties of a Weyl-type tensor unless $a$ and $b$ satisfy the condition $23a^2-42 ab+27b^2=0$. We will write a solution of this as $a = \gamma b$ with $\gamma$ a particular complex number.
 In this case we find that there is a doubling  formula for the Weyl tensor
\begin{equation}
\label{weyl}
C_{abcd}=\lambda\, C^{21}_{abcd}[\alpha]   \, ,
\end{equation}
where the superscript $21$ on $C$ indicates that both pairs of indices are to be projected into the ${\bf 21}$ representation. The proportionality constant is $\lambda=-23 V^{3/2}/(32b^2(\gamma+1))$. The field $\alpha$ here is not anti self-dual as we have noted, but it can be checked that its anti self-dual part is proportional to the Maxwell tensor \eqref{SCexpS7}, and that the corresponding field strength $F_{\mu\nu}$ satisfies the Maxwell field equations.


\section{Reissner-Nordstrom and Born-Infeld}

In the sections above, we studied examples where the gauge field arises from intrinsic properties of the spacetime. We now turn to study situations where there is an {\it independent} gauge field defined on the spacetime, in order to see if there is a doubling relationship whereby 
the gauge fields in the gravity-gauge system give formul\ae\ for the full spacetime curvature - determining the Ricci tensor via the Einstein equations as usual but in addition fixing the Weyl tensor via  a doubling formula. We will call this ``self-doubling" for simplicity.

\subsection{Reissner-Nordstr\"om}
As a  first case, consider the Reissner-Nordstr\"om theory in $D$ dimensions, with metric
\begin{equation}
\label{RN}
ds^2 = -f(r) dt^2 + f^{-1}(r) dr^2 + r^2 d\Omega_{D-2}^2\, ,
\end{equation}
with 
\begin{equation}
\label{RN2}
f(r) = 1 - \frac{2M}{r^{D-3}} + \frac{Q^2}{r^{2(D-3)}}\, ,
\end{equation}
$d\Omega_{D-2}^2$ the metric on the $(D-2)$-dimensional sphere, and $M$ and $Q$ the mass and charge respectively. 

This theory is self-doubling in the sense that the gauge field already present in the theory provides the basis for the doubling. This can be seen as follows - the gauge field one form is
\begin{equation}
\label{RNsinglecopy}
A = -\frac{Q}{ r^{D-3}}\, dt \, ,
\end{equation}
satisfying the field equations ($F_{\mu\nu}=2\partial_{[\mu}A_{\nu]}$)
\begin{align}
\label{RNeqnsmotion}
\nabla^\mu F_{\mu\nu} & = 0\, , \nonumber  \\ 
R_{\mu\nu} -{1\over2}g_{\mu\nu}R &= \frac{D-2}{D-3} T_{\mu\nu} \, ,
\end{align}
with the usual stress tensor
\begin{equation}
\label{RNstresstensor}
T_{\mu\nu} = F_\mu\cdot F_\nu - \frac{1}{4}g_{\mu\nu}F\cdot F \, .
\end{equation}
We find the  Weyl doubling formula
\begin{equation}
\label{RNDC}
C_{\mu\nu\rho\sigma}  = \Lambda(r)\, C_{\mu\nu\rho\sigma}[F] \, ,
\end{equation}
where on the right-hand side of this  equation $C_{\mu\nu\rho\sigma}[F] $ is given by the formula \eqref{Dhigher} and the coefficient is
\begin{equation}
\label{RNcoeff}
\Lambda(r) = \frac{2(D-2)}{3(D-3)^2}\Bigg( (2D-5) - (D-1)\frac{M r^{D-3}}{Q^2}\Bigg)  \, .
\end{equation}
This result has been checked up to $D=10$ but there are no reasons why this would not hold for all dimensions given the structure of the metric.  Notice that one can take the extremal limit $Q\rightarrow M$ directly in the equations above. 


\subsection{Born-Infeld}
The discussion above may be generalised to the Born-Infeld theory in any dimension. The Lagrangian  is
\begin{equation}
\label{BILag}
L=   \sqrt{g} R + {1\over\lambda^2}\Big( \sqrt{g}-\sqrt{\vert \det{(g+\lambda F)} \vert}\Big) \, 
\end{equation}
with $F_{\mu\nu}$ the Maxwell field and $\lambda$ a constant. The solution with electric charge $Q$ in four dimension is given by (e.g.\cite{Rasheed:1997ns})
\begin{align}
\label{BI4D}
ds^2 & =   -\Big(1-{2 m(r)\over r^2} \Big)dt^2 + \Big(1-{2 m(r)\over r^2}\Big)^{-1}dr^2 +  r^2d\Omega_2^2 \, , \nonumber  \\
F_{tr} &= -F_{rt}= {Q\over\sqrt{r^4+\lambda^2 Q^2} }\, ,
\end{align}
with the other components of $F$ vanishing. The function $m(r)$ is fixed by the metric equation of motion, and satisfies
\begin{equation}
\label{4Dm}
m'(r) =  \, {1\over \lambda^2}\Big(\sqrt{r^4+\lambda^2 Q^2} -r^2\Big)\, ,
\end{equation}
the solution of which is given by equation \eqref{DDmSol} below with $D=4$. We also define the tensor
\begin{equation}
\label{Gtensor}
G^{\mu\nu} =  -{2\over \sqrt{g}}{\partial L\over \partial F_{\mu\nu}}
\end{equation}
which satisfies the (non-linear in $F$) field equations $\nabla^\mu G_{\mu\nu}=0$. For the solution $F$ in \eqref{BI4D} this is given simply by
$G_{tr} = - G_{rt} ={ Q\over r^2}$ with other components vanishing. It is straightforward to confirm that the Weyl tensor satisfies the equation
\begin{equation}
\label{FWeylBI}
C_{\mu\nu\rho\sigma} = \Lambda_4(r) C_{\mu\nu\rho\sigma}[G]\, 
\end{equation}
with
\begin{equation}
\label{FWeylBILambda}
\Lambda_4(r) =-{2 r\over 3Q^2}\Big( 6 m(r) - 4 r m'(r) + r^2 m''(r)\Big)\, .
\end{equation}

This discussion is easily generalised to $D$ dimensions. The metric is as in \eqref{BI4D} with $d\Omega_2^2\rightarrow d\Omega_{D-2}^2$ with the function $m(r)$ satisfying 
\begin{equation}
\label{DDm}
m'(r) =  \, {1\over \lambda^2}\Big(\sqrt{r^{2(D-2)}+\lambda^2 Q^2} -r^{D-2}\Big)\, .
\end{equation}
This is solved by
\begin{align}
\label{DDmSol}
\lambda^2\, m(r) =  \lambda^2\, m & + \frac{r^{D-1}}{D-1} - \frac{r \sqrt{\lambda^2Q^2+r^{2(D-2)}}}{D-1}\nonumber \\  & \quad + \frac{\lambda^2Q^2(D-2)}{(D-1)(D-3)r^{D-3}}\, \,{_2}F_1\Big[\frac{1}{2},\frac{D-3}{2(D-2)},1+\frac {D-3}{2(D-2)}, -\frac{\lambda^2Q^2}{r^{2(D-2)}}\Big] \, .
\end{align}
for constant $m$. Note that $m(r)\rightarrow m$ in the limit  $r\rightarrow0$.

One has the result 
\begin{equation}
\label{FWeylBID}
C_{\mu\nu\rho\sigma} = \Lambda(r) C_{\mu\nu\rho\sigma}[G]\, 
\end{equation}
with 
\begin{equation}
\label{FWeylBILambdaD}
\Lambda(r) =-{2 r^{D-3}\over 3 Q^2}\Big( (D-1)(D-2) m(r) - 2(D-2) r m'(r) + r^2 m''(r)\Big)\, 
\end{equation}
and the field $G$ in $D$ dimensions given by
\begin{equation}
\label{GD}
G_{tr} = - G_{rt} ={Q\over r^{D-2}}\, ,
\end{equation}
with other components vanishing. Notice that $G_{\mu\nu}$ is a function of the Maxwell field $F_{\mu\nu}$ via the relation \eqref{Gtensor}.
The result \eqref{FWeylBID} leads to curvature singularities as $r\rightarrow 0$ which are not present in the field $F_{\mu\nu}$.

This Born-Infeld solution reduces to the Reissner-Nordstr\"om model in the section above in the limit as $\lambda\rightarrow 0$, using the expansion
\begin{equation}
\label{BIexp}
\det{({{\delta^\mu}_\nu}+\lambda {{F^\mu}_\nu})}=  1-{1\over 2}\lambda^2(F^2) -{1\over 4}\lambda^4 \Big( (F^4) -{1\over2}(F^2)^2\Big) + o(\lambda^6)     \, ,
\end{equation}
where the brackets in $(F^2), (F^4)$ indicate matrix traces of powers of ${F^\mu}_\nu$. It can be checked that the coefficient in \eqref{FWeylBILambdaD} reduces in this limit to that in  \eqref{RNcoeff}.

The formula \eqref{FWeylBID} enables one to easily find invariants, for example
\begin{equation}
\label{CsqBI}
C^{\mu\nu\rho\sigma} C_{\mu\nu\rho\sigma} = \frac{4\,(D-3)}{(D-1)\,r^{2(D-1)}} \Big( (D-1)(D-2) m(r) - 2(D-2) r m'(r) + r^2 m''(r)\Big)^2 \, .
\end{equation}
Note that this diverges as $1/r^{2(D-1)}$ as $r\rightarrow 0$, although the non-zero Maxwell field strength component $F_{tr}= Q/\sqrt{r^{2D-4} +\lambda^2Q^2}$ does not. The divergence comes from the traces of the field $G_ {\mu\nu}$ which appear when squaring \eqref{FWeylBID}. (A discussion of the four dimensional Born-Infeld theory appeared recently \cite{Pasarin:2020qoa} and the $D=4$ version of \eqref{CsqBI} appeared there, although their equivalent of $m(r)$ satisfies a different equation.)


\section{Brane solutions}
\label{branes}

It is natural  to conjecture that BPS brane solutions in supergravity (we will follow the conventions of \cite{Stelle:1999ljt} in this section) might satisfy Weyl doubling.  The $(p+1)$-forms that minimally couple to the $p$-brane provide a potential from which one can construct a $(p+2)$-form field strength. This field strength can then be used in the formula for the Weyl curvature, as well as determining the Ricci curvature via the field equations and stress tensor. We  consider here the cases of $p$-branes where the scalar fields play no role.


\subsection{String in five dimensions}
A simple example is the string in five dimensions. The metric and two-form are given by
\begin{align}
\label{string5d}
ds^2 &= H^{-1}(r) \big(-dx_1^2 + dx_2^2 \big) + H^2(r) (dr^2 + r^2 \big(d\theta^2+{\rm sin}^2 \theta d\phi^2) \big) \, , \nonumber \\ 
B_{\mu\nu} &= \sqrt{3}\,\epsilon_{\mu\nu}H^{-1}(r) \, ,
\end{align}
with $H(r) = 1+k/r$ and indices $\mu,\nu=1,2$. The three-form field strength $F=dB$, obeys $\nabla^M F_{MNP}=0$ and the metric field equation
\begin{equation}
\label{stresstensor}
G_{MN} =\frac{1}{4}\Big( F_M\cdot F_N - \frac{1}{6}g_{MN}F\cdot F\Big) \, .
\end{equation}
For the $p$-forms $F$ discussed in the subsections below the equivalent equation is 
\begin{equation}
\label{stresstensorP}
G_{MN} =\frac{1}{2(p-1)!}\Big( F_M\cdot F_N - \frac{1}{2p}g_{MN}F\cdot F\Big) \, .
\end{equation}

We will now define the two-forms ${(F_\mu)}_{NP}={(K_\mu)}^M F_{MNP}$ ($\mu=1,2$) where the Killing vectors  $K_\mu$ correspond to translations in the $x^\mu$ directions in the string world sheet. Note that $K_1^2=H^{-1}(r)=-K_2^2$. We then find that the following Weyl doubling formula may be constructed from the three-form field strength $F$ and the two-form field strengths $F_\mu$
\begin{equation}
C_{MNPQ} = -\frac{(k+4r)}{6k} C_{MNPQ}[F] - \frac{r H(r)}{3k}\Big( -C_{MNPQ}[F_1] + C_{MNPQ}[F_2] \Big)\, .
\end{equation}
This can also be written  as
\begin{equation}
\label{5dstringDC}
{C_{MN}}^{PQ} = -\frac{1}{6}{C_{MN}}^{PQ}[F]  + \Sigma_1(r)  {T_{MN}}^{PQ}   \, ,
\end{equation}
where $ \Sigma_1(r)  = \frac{k }{r^3H^4}$. 
We note that $\Sigma_1(r)$ is proportional to the inverse of the volume of the transverse sphere. 
In this coordinate system all of the non-zero components of the Weyl tensor  take the form (up to sign)
${C_{MN}}^{MN}$.  The only non-zero components of ${C_{MN}}^{PQ}[F] $ and the tensor ${T_{MN}}^{PQ}$  are similarly  when $(P,Q)=(M,N)$ (or $(N,M)$). In this case the ${T_{MN}}^{PQ}$  are given by $(0,1,-1/2,-1,2)$ for $(MN)=(12,\mu 3,\mu \bar m, 3\bar m,45)$ respectively ($\bar m = 4,5$), and components related to these by the antisymmetry in $M,N$. 
Notice thus that the tensor $T$ vanishes when all its components are along the world-volume, and so from \eqref{5dstringDC} we see that on the world-volume there is a simple Weyl doubling formula, and this is corrected off the world volume by a tensor that takes a simple form.

In \cite{Lee:2018gxc} the Kerr-Schild formulation was investigated in the case where there is both a metric and a Kalb-Ramond field,  using doubled geometry. It was found that this involved two null vectors with the metric and $B$ field involving the symmetric and anti-symmetric product of these vectors. It is natural then to expect that a single copy in this case should involve two Maxwell gauge fields $A$ and $\bar A$. For the five-dimensional case under consideration here, one can see that the $B$ field
in \eqref{string5d} is given by
\begin{equation}
\label{BfieldDC}
B_{MN} = 2 \sqrt{3}\,H A_{[M}\bar A_{N]}   \, ,
\end{equation}
with 
\begin{align}
\label{BfieldSC2}
A_M &= (H^{-1},0,0,0,0)\, ,    \nonumber\\
\bar A_M &= (0,H^{-1},0,0,0) \,  .
\end{align}
This generates a formula for $F_{MNP}$ in terms of $A_M$ and $\bar A_M$. 

One avenue suggested by this work is to develop a Weyl doubling formula for DFT. Along these lines, one may view an expression of the formula \eqref{5dstringDC} as a Weyl ``doubling'' in terms of the two Maxwell fields in the formalism of \cite{Lee:2018gxc}. Inserting \eqref{BfieldDC} into \eqref{5dstringDC} would then give an expression for the Weyl tensor in terms of an expression quartic in fields and two derivatives. (Note that the expression \eqref{BfieldDC} is in the usual double copy where the vector fields in that formalism live in flat space.)


\subsection{M2 brane}

For the M2 brane in eleven dimensions, the metric and non-zero three-form potential components are given by
\begin{align}
\label{M2}
ds^2 &= H^{-2/3}(r) (-dx_1^2 + dx_2^2+ dx_3^2) + H^{1/3}(r) (dr^2 + r^2 d\Omega_{7}^2), \\ \nonumber
A_{\mu\nu\rho} &= \epsilon_{\mu\nu\rho}H^{-1}(r) \, ,
\end{align}
with $H(r) = 1+k/r^6$ and indices $\mu,\nu,...=1,2,3$, and $m,n,...=4,...,,11$ (with  $M,N,...=1,...,11)$. $r$ is the radial coordinate in
the eight-dimensional transverse space.
The four-form field strength $F=dA$ has non-zero components $F_{1234} = -6k H^{-2}/r^7$ and obeys $\nabla^M F_{MNPQ}=0$ and \eqref{stresstensorP} with $p=4$.

Again we define the tensors  ${(F_\mu)}_{MNP}={(K_\mu)}^Q F_{QMNP}$ where the Killing vectors  $K_\mu$ correspond to translations in the $x^\mu$ directions in the world-volume. The Weyl doubling formula can then be constructed as follows:
\begin{equation}
C_{MNPQ} = \frac{(k+4r^6)}{36k} C_{MNPQ}[F] - \frac{r^6 H(r)^{2/3}}{3k}\Big( -C_{MNPQ}[F_1] + C_{MNPQ}[F_2] + C_{MNPQ}[F_3] \Big) \, .
\end{equation}
This simplifies along the world-volume as in the string case above, with the analogue of \eqref{5dstringDC} being
\begin{equation}
\label{MDCfull}
{C_{MN}}^{PQ} = \frac{1}{36}{C_{MN}}^{PQ}[F]  + \Sigma_2(r)  {T_{MN}}^{PQ}   \, ,
\end{equation}
where here $ \Sigma_2(r)  = \frac{2k H^{-7/3}}{r^8}$ and the object ${T_{MN}}^{PQ} $ has non-zero components $(0,7,-1,-3,1)$ for $(PQ)=(MN) = (\mu\nu,\mu 4,\mu \bar m,4\bar m,\bar m\bar n)$ respectively, with $\bar m,\bar n=5,...,11$. Again we note that $\Sigma_2(r)$ is proportional to the inverse of the volume of the transverse sphere. 

Since the publication of the results in this section, a formulation of Kerr-Schild double copy for exceptional field theory has been obtained in \cite{Berman:2020xvs,Berman:2022bgy}; See \cite{Berman:2020tqn} for a review on exceptional field theory.


\subsection{D3 brane}

For the D3 brane in ten dimensions, the metric and non-zero four-form potential components are given by
\begin{align}
\label{D3}
ds^2 &= H^{-1/2}(r) (-dx_1^2 + dx_2^2+ dx_3^2+ dx_4^2) + H^{1/2}(r) (dr^2 + r^2 d\Omega_{5}^2), \\ \nonumber
A_{\mu\nu\rho\sigma} &= \epsilon_{\mu\nu\rho\sigma}H^{-1}(r) \, ,
\end{align}
with $H(r) = 1+k/r^4$ and indices $\mu,\nu...=1,2,3,4$, and $m,n,...=5...,,10$ (and  $M,N,...=1,...,10)$. $r$ is the radial coordinate in
the six-dimensional transverse space.
We will need the self-dual five-form field strength $F=\frac{1}{2}(dA+(dA)^*)$.  As above, we define  the tensors  ${(F_\mu)}_{MNPQ}={(K_\mu)}^R F_{RMNPQ}$ where the Killing vectors  $K_\mu$ correspond to translations in the $x^\mu$ directions in the world-volume. The Weyl doubling formula  is then found to be
\begin{equation}
C_{MNPQ} = \frac{r^4}{6k} C_{MNPQ}[F] - \frac{2r^4 H(r)^{1/2}}{3k}\Big( -C_{MNPQ}[F_1] + C_{MNPQ}[F_2]+ C_{MNPQ}[F_3] + C_{MNPQ}[F_4]  \Big)
\end{equation}
This also shows that the Weyl tensor vanishes for components along the world-volume. To see this, one has the equivalent expression 
\begin{equation}
\label{D3old}
{C_{MN}}^{PQ} = \Sigma_3(r)  {T_{MN}}^{PQ}   \, ,
\end{equation}
where  $ \Sigma_3(r)  = \frac{k H^{-5/2}}{r^6}$ and the object ${T_{MN}}^{PQ} $ has non-zero components $(0,5,-1,-4,2)$ for $(PQ)=(MN) = (\mu\nu,\mu 5,\mu \bar m,5\bar m,\bar m\bar n)$ respectively, with $\bar m,\bar n=6,...,10$. As in the cases above, $\Sigma_3(r)$ is proportional to the inverse of the volume of the transverse sphere. 
\\


\subsection{M5 brane}

There is a similar story for the M5 brane in eleven dimensions. The metric and non-zero 4-form field strength components are given by
\begin{align}
\label{M5}
ds^2 &= H^{-1/3}(r) \eta_{\mu\nu}dx^\mu dx^\nu + H^{2/3}(r) ( dr^2 + r^2 d\Omega_{4}^2 ),   \nonumber \\ 
F^{(4)}_{8\,9\,10\,11} &= 3k  \sin^3(\theta) \sin^2(\phi) \sin(\psi_1)\, ,
\end{align}
with $F^{(4)}$ antisymmetrised, $H(r) = 1+k/r^3$, world-volume coordinates $x^\mu$ ($\mu=1,...6)$, and transverse coordinates $r$ and spherical polars $(\theta,\phi,\psi_1,\psi_2)$. 

The fivebrane magnetically couples to the three form $C_3$ which means we will need to use the magnetic dual field strength given by the seven-form, $F={^*F^{(4)}}$.
Then define ${(F_\mu)}_{NPQRST}={(K_\mu)}^M F_{MNPQRST}$ ($\mu=1,2,...,6$) with the Killing vectors  $K_\mu$ corresponding to translations in the $x^\mu$ directions in the world-volume.  The Weyl doubling formula is then
\begin{align}
36 \, C_{MNPQ} = &-\frac{1}{60} C_{MNPQ}[F] +  \frac{r^3}{3k} C_{MNPQ}[F] \nonumber \\
  &-  \frac{r^3}{2k}H(r)^{1/3}\Big( -C_{MNPQ}[F_1] + C_{MNPQ}[F_2] + \dots + C_{MNPQ}[F_6] \Big) \, .
\end{align}

This is equivalent to
\begin{equation}
\label{M5DCfull}
36\,{C_{MN}}^{PQ} =- \frac{1}{60}{C_{MN}}^{PQ}[F]  + \Sigma_5(r) {T_{MN}}^{PQ}  \, ,
\end{equation}
where here $ \Sigma_5(r)  = \frac{18 k H^{-8/3}}{r^5}$ and the object ${T_{MN}}^{PQ} $ has non-zero components $(0,4,-1,-6,4)$ for $(PQ)=(MN) = (\mu\nu,\mu 7, \mu\, \bar m,7\bar m,\bar m\bar n)$ respectively, with $\bar m,\bar n=8,9,10,11$. We see again there is a simple doubling formula for the Weyl components along the world-volume directions and that $\Sigma_5(r)$ is proportional to the inverse of the volume of the transverse sphere. 

In all of  the cases above, for a brane with $V$-dimensional world-volume in $D$ dimensions, and $T=D-V$ transverse dimensions, the components of the tensor  ${T_{MN}}^{MN}$ 
are proportional to $(0,T-1,-1,-V,2V/(T-2))$ for $(PQ)=(MN) = (\mu\nu,\mu r, \mu\, \bar m,r\bar m,\bar m\bar n)$ respectively, in the notation used above. Given that the components along the world-volume vanish, this follows from the tracelessness condition on the Weyl tensor. The powers or $r$ and $H$ in the coefficients $\Sigma$ are  equal to $-T$ and that appearing in the inverse of the volume of the transverse sphere respectively. Similarly, the  powers of $r$ and $H$
in the doubling formul\ae\ above have common expressions: $r^{T-2}$ and the inverse of the power of $H$ which appears in the world-volume metric.

\section{Discussion}
\label{sec:disc}
We described Weyl doubling as the writing of the Weyl tensor (up to a scalar factor) in terms of a quadratic expression in an Abelian field strength obeying Maxwell's equations in a curved background. The curved background distinguishes these results from the usual double copy originating in scattering amplitudes and more recently classical solutions using the Kerr-Schild form. 
This phenomenon of Weyl doubling was found in a variety of solutions in different dimensions. The purely intrinsic case is where a Killing vector on the manifold is used to define a potential from which the field strength is derived. The Weyl tensor is given by the formula \eqref{Dhigher}. We showed that the  metrics for which the Weyl doubling formula \eqref{Dhigher}  applies fall into the Type D class in the general dimensional classification, so that this is a necessary condition.
It is also a sufficient condition in four dimensions \cite{Walker:1970abc,Hughston:1972qf} (see also \cite{Luna:2018dpt}). But it does not appear to be a sufficient condition in higher dimensions - the  five-dimensional  Myers-Perry metrics \cite{Myers:2011yc}  are Type D \cite{DeSmet:2003kt} and while we showed that the singly-rotating solution has a Weyl doubling formula,  the general solution with two rotation parameters  does not appear to satisfy such a formula. We also investigated the BPS solution in six-dimensions studied in \cite{Chen:2006xh}, which is Type D but  does not appear to satisfy a doubling formula.

A second way to generate a two-form gauge field strength is if the spacetime admits a closed, covariantly constant four-form which may be used to define a self-dual (or anti-self-dual)  two form. These manifolds have special holonomy and the Maxwell field equations then follow from the duality and closure conditions.  In the four-dimensional case these are manifolds with $SU(2)$ holonomy. The Gibbons-Hawking metrics provide  a broad class of such metrics. We found that in this case the Weyl doubling formula has a correction term which is linear in the gauge field strength - equation \eqref{SDCGH} or equivalently \eqref{GHfinal2}. This generalisation of doubling could be studied further to see if it applies in other cases. It may also suggest generalisations of the double copy construction. 
A reader might wonder that  such a doubling formula is inevitable given the symmetries of the Weyl tensor. The Gibbons-Hawking case where there is a derivative correction provides a good counter example that demonstrates the non-triviality of the algebraic relation in the algebraic Weyl doubling formula.

In studying a generalisation to higher dimensions an issue arises in that a Weyl tensor constructed from \eqref{Dhigher} using a self-dual two-form is not in general  self-dual, unlike in four dimensions.  One might have expected that this might be resolved by using self-dual projection operators, but this will not preserve the algebraic symmetries of the tensor in general. Nevertheless we found an example in eight dimensions where a sort of twisted  doubling construction exists which expresses the Weyl tensor in terms of a spin(7) self-dual  two form. It would be interesting to explore if other examples exist.

A different mechanism for Weyl doubling is if the gauge field strength is an {\it additional} field defined on the spacetime, rather than being expressed using the metric and/or Killing vectors. The most natural example of this is the  Reissner-Nordstr\"om metric in $D$ dimensions. This works in the general case as well as the BPS limit, although perhaps one might have expected that such a construction, requiring the Weyl and Maxwell curvatures to be related, would require the BPS constraint linking the charge and mass. We showed that this discussion generalises to the charged Born-Infeld solution in $D$ dimensions.  A recent paper \cite{Pasarin:2020qoa} discussed how the study of the Born-Infeld electrically charged solution might illuminate the investigation of stringy corrections of the double copy. It would be interesting to see if the doubling approach may provide insights into this. 

We then turned to study cases where the gauge field strength is a higher degree $p$-form, and analysed the associated brane solutions. Here it was found that a simple quadratic Weyl doubling formula holds using the $p$-form field strength and contractions of it with the world volume Killing vectors. Evaluating this gives a particularly simple Weyl doubling formula for the components of the Weyl tensor projected on to the world volume.
There is a variety of directions for further research, such as how does the inclusion of scalars, Kaluza-Klein reductions and supersymmetry impact the doubling construction. We have studied the Weyl tensor here but there will also be spinor analogues of our formul\ae\ in each dimension (c.f \cite{Monteiro:2018xev}). 

The central question that this chapter implicitly poses is, what is the relation to the usual double copy? Can the formul\ae\ for brane solutions be applied to the usual single and double copy? Does the derivative corrected Gibbon-Hawking expression have a conventional double copy interpretation in terms of Maxwell fields in flat space? It was shown in  \cite{Kim:2019jwm} that doubled geometry clarifies the double copy construction for the point charge, relating this to the JNW solution \cite{Janis:1968zz}. As we noted in section \ref{branes}, the Kerr-Schild form in double field theory involves more than one gauge field - for example, two Maxwell fields in the case where there is a Kalb-Ramond field as well as the metric. For the string in five dimensions the $B$ field can indeed be simply constructed from two Maxwell gauge fields; See \cite{Berman:2020xvs,Berman:2022bgy} for a exceptional field theoretic formulation of the Kerr-Schild double copy.
One approach following these ideas is to use a DFT or EFT generalised Killing vector as the basis for generating gauge field strengths in the extended space to express the DFT equations. 

The phenomenon of Weyl doubling that we have explored here in numerous examples, relating gravity and Abelian gauge theory, reveals a  structure  that would be interesting to develop further, and in particular to investigate its relationship to the double copy.

%% file: chaps/chap05.tex
\chapter{Holonomy Groups and the Single Copy}
\label{chap:Holonomy}
\section{Global Aspects of the Double Copy}
In the previous chapters we have dealt with local objects, namely solutions to equations of motions. We have been using the double copy procedure as a way to relate local solutions of gravitational and gauge theory. These solutions are not necessarily valid globally, as was the case with the Dirac monopole and Taub-NUT solutions in Chapter \ref{chap:s-duality}. The reason for this lack of global validity is simple, because generally a global section of the bundle need not exist. For example, if one takes the case of the tangent bundle of a 2-sphere, a global section, i.e., a tangent vector cannot be defined. There will always be a point where the vector is zero. One is forced to introduce an appropriate number of charts, and glue them together in order to describe the vector field globally over the entire bundle. However, one can use local objects to construct quantities that characterises bundles regardless of their local differential structure. For example, the Chern class, defined in terms of the curvature form, characterises vector bundles and provides an answer to whether two vector bundles are the same or not. The Chern class is an example of a characteristic class, topological invariants that do not depend on the local structure of the manifold. The question we ask in this chapter is whether one can construct such topological quantities with the aid of the double copy map. We will refer to these topological properties as global, since they do not depend on the local structure of the underlying object.

Studying non-perturbative  (topological or not) aspects in the context of the double copy has been attempted before. In \cite{Chacon:2020fmr} it was shown that the self-dual Yang-Mills equations map to that of self-dual gravity, and in \cite{White:2020sfn,Chacon:2021wbr} using twistor methods, shedding light the origins of the Weyl double copy of \cite{Luna:2018dpt}. Topological information, namely patching conditions of bundles have been studied in \cite{Alfonsi:2020lub} using the Kerr-Schild double copy. Indeed, our own study of solution generating transformations, and their mapping under the double copy in Chapter \ref{chap:s-duality} is another example of a non-perturbative effect. Solution generating transformations have also been explored in \cite{Huang:2019cja}.

In this chapter, we will study yet another non-perturbative, and in fact topological aspect of the double copy. Using the Kerr-Schild formulation of Chapter \ref{chap:s-duality}, we will study the notion of holonomy of a Riemannian manifold and its map under the double copy.
\section{Holonomy in Mathematics and Physics}
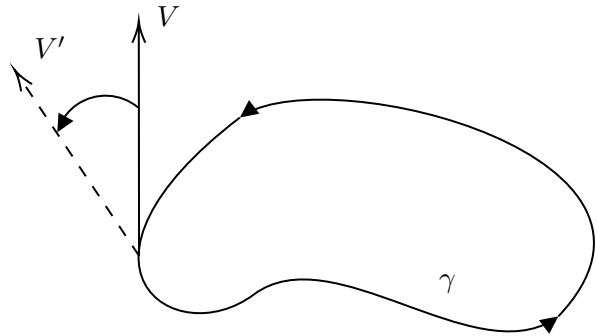
\begin{figure}[t]
    \centering
        \begin{tikzpicture}[x=0.75pt,y=0.75pt,yscale=-1,xscale=1]
            
            \draw    (249.68,181.52) -- (249.55,65.43) ;
            \draw [shift={(249.55,63.43)}, rotate = 89.93] [color={rgb, 255:red, 0; green, 0; blue, 0 }  ][line width=0.75]    (10.93,-3.29) .. controls (6.95,-1.4) and (3.31,-0.3) .. (0,0) .. controls (3.31,0.3) and (6.95,1.4) .. (10.93,3.29)   ;
            \draw  [dash pattern={on 4.5pt off 4.5pt}]  (249.68,181.52) -- (188.6,89.59) ;
            \draw [shift={(187.49,87.92)}, rotate = 56.4] [color={rgb, 255:red, 0; green, 0; blue, 0 }  ][line width=0.75]    (10.93,-3.29) .. controls (6.95,-1.4) and (3.31,-0.3) .. (0,0) .. controls (3.31,0.3) and (6.95,1.4) .. (10.93,3.29)   ;
            \draw    (307,201) .. controls (346.2,171.6) and (416.13,239.2) .. (456.56,213.68) ;
            \draw [shift={(459,212)}, rotate = 143.13] [fill={rgb, 255:red, 0; green, 0; blue, 0 }  ][line width=0.08]  [draw opacity=0] (8.93,-4.29) -- (0,0) -- (8.93,4.29) -- cycle    ;
            \draw    (302.58,110.29) .. controls (347.73,83.93) and (534.46,131.64) .. (459,212) ;
            \draw [shift={(300,112)}, rotate = 323.13] [fill={rgb, 255:red, 0; green, 0; blue, 0 }  ][line width=0.08]  [draw opacity=0] (8.93,-4.29) -- (0,0) -- (8.93,4.29) -- cycle    ;
            \draw    (300,112) .. controls (204,186) and (267,231) .. (307,201) ;
            \draw  [draw opacity=0] (208.88,119.42) .. controls (210.35,114.01) and (213.65,109.06) .. (218.59,105.57) .. controls (228.11,98.85) and (240.71,99.73) .. (249.19,106.95) -- (233,126) -- cycle ; \draw    (209.85,116.55) .. controls (211.59,112.27) and (214.54,108.43) .. (218.59,105.57) .. controls (228.11,98.85) and (240.71,99.73) .. (249.19,106.95) ;  \draw [shift={(208.88,119.42)}, rotate = 299.16] [fill={rgb, 255:red, 0; green, 0; blue, 0 }  ][line width=0.08]  [draw opacity=0] (8.93,-4.29) -- (0,0) -- (8.93,4.29) -- cycle    ;
            
            \draw (257,54.4) node [anchor=north west][inner sep=0.75pt]    {$V$};
            \draw (196,69.4) node [anchor=north west][inner sep=0.75pt]    {$V'$};
            \draw (398,189.4) node [anchor=north west][inner sep=0.75pt]    {$\gamma $};

            \end{tikzpicture}
    \caption{Parallel trasporting the vector field $V$ around the loop $\gamma$.}
    \label{fig:vecHolo}
    \end{figure}

The word holonomy comes from the greek \textgreek{ὅλος}, meaning \textit{whole}, and \textgreek{νόμος}, meaning \textit{law, custom, knowledge}. Therefore, holonomy can have the meaning of `entire law'. Indeed, the notion of holonomy of a Riemannian manifold is a topological property which can be used to classify Riemannian manifolds. Informally, the holonomy of a Riemannian manifold measures the failure for parallel transport to be the identity. Take a familiar example from physics, in particular from General Relativity where one parallel transports a tangent vector around a \emph{loop} $\gamma$ based at a point $p\in \mc M$. This induces an action on the tangent space at $p$, meaning
\begin{equation}
\begin{split}
    \tau_{(\gamma)} : \,&T_p\mc{M} \rightarrow T_p\mc{M}\\
     &V_p \mapsto \tau_{(\gamma)} (V)_p,
\end{split}
\end{equation}
If the parallel transported vector returns to the initial position unchanged, one says that the action of the parallel transport map is the identity; otherwise, it is some transformation that rotates the components of the vector field(see figure \ref{fig:vecHolo}). We will see that we can represent such a transformation by a matrix acting on the vector components. The matrix, as we shall see later, is an element of the holonomy group of the respective Riemannian manifold.

In \cite{MR2206889}, R. L. Bryant gives a historical account of the notion of holonomy. It seems that the notion of holonomy first appeared in the study of mechanics. One of the first people to study such a notion was Heinrich Hertz in his work  \textit{Die Prinzipien der Mechanik in neuem Zusammenhange dargestellt} published in 1894. Hertz coinded the terms `holonomic' and `non-holonomic' to distinguish between two types of constraints in configuration spaces. Bryant gives the examples of a flying object in space. If the object's velocity is always perpendicular to the radial vector from the origin, then the object will stay at a constant distance from the origin, and hence trace out a sphere. On the other hand, if the object's velocity is perpendicular to some other vector, one finds that the object can move to any other point in space, and not be confined like the previous case, while maintaining said perpendicularity. The first case is a said to be a holonomic constraint, while the latter, non-holonomic. In the 20th, thanks to the publications of Einstein's General Relativity, the notion of holonomy saw a considerable amount of work being done on it. In 1918, Schouten studied how frame fields change under parallel translation. He was trying to see what possible orientations the frame field can come back into after parallel transporting it around some Riemannian manifold. The notion of holonomy in its modern mathematical form first appeared in 1925 in the work of Elie Cartan \cite{BSMF_1926__54__214_0}, \textit{Sur une classe remarquable d'espaces de Riemann}. Around that time, Cartan was thinking about the mechanics of holonomic systems as he did work in Mathematics. His definition of holonomy was not exactly the same as Schouten's. Cartan took a tangent plane at a point on the manifold and rolled it along without twisting or slipping. He showed when the tangent plane is rolled back to the initial point, it comes back not only rotated, but also translated. Using that he gave interpretations for what now we call curvature and torsion. According to Bryant, Cartan defined the holonomy of a Riemannian manifold $(\mc M, g)$ as a subset of rotations of tangent spaces at each point:
\begin{equation}
    \text{H}_p \subseteq \text{SO}(T_p\mc{M}),\quad p\in \mc{M},
\end{equation}
and made four crucial remarks:
\begin{itemize}
    \item $\text{H}_p$ is a group. $\text{H}_p$ is connected if $\mc M$ is simply connected. And for some other point $q\in \mc{M}$, the holonomy groups are conjugate to each other, i.e., $\text{H}_p \simeq \text{H}_q$. Hence, holonomy as a conjugate class is independent of the point at which it is defined.
    \item $\text{H}_p$ is trivial if and only if the metric $g$ is flat.
    \item If $\text{H}_p$ preserves splitting, $T_p\mc{M}=E \oplus E^{\perp}$, then $g$ is locally a product of those two metric on $E$ and $E^{\perp}$.
    \item When $n=4$, $\text{H}_p = \text{SU}(2)$ is possible.
\end{itemize}
In fact we will see an example of the fourth remark later on when we discuss self-dual solutions in section \ref{sec:results}. In the period between 1950 and 1970, more results were established. In 1952, it was proved by Borel and Lichnerowicz that the holonomy group is in fact always a Lie group \cite{zbMATH03072651}. A link between holonomy and curvature was established by Ambrose and Singer in 1953 \cite{MR63739}. In 1954, Berger gave a classification of irreducibly acting holonomy groups.
\begin{theorem}\label{BergerTable}
    Suppose $(\mc{M}, g)$ is a connected Riemannian manifold such that $\text{H}_p$ acts irreducibly on $T_p\mc{M}$, $p\in\mc{M}$. Then either $(\mc{M}, g)$ is locally isometric to a symmetric space or $H_p$ os conjugate to one of the following subgroups of $\text{SO}(n)$:
\begin{center}
        \begin{tabular}{|c |c| c| c|} 
        \hline
        \rowcolor{BergerColor}
        Group & dim$(\mc{M})$ & Type of Manifold & Property \\ [0.5ex] 
        \hline\
        $\text{SO}(n)$ & $n$ & Orientable & - \\ 
        \hline
        $\text{U}(n)$ & 2$n$ & Kähler & - \\
        \hline
        $\text{SU}(n)$ & 2$n$ & Calabi-Yau & Ricci-flat, Kähler \\
        \hline
        $\text{Sp}(n)\cdot \text{Sp}(1)$ & 4$n$ & Quaternion-Kähler & Einstein \\
        \hline
        $\text{Sp}(n)$ & 4$n$ & Hyperkähler & Ricci-flat \\
        \hline
        $\text{G}_2$ & 7 & $G_2$ manifold & Ricci-flat \\
        \hline
        $\text{Spin}(7)$ & 8 & $\text{Spin}(7)$ manifold & Ricci-flat\\
        \hline
       \end{tabular}
    \end{center}
\end{theorem}
Indeed, the $\text{SU}(n)$, $G_2$ and $\text{Spin}(7)$ cases are of special importance in physics. The first group is the holonomy group of a Calabi-Yau manifold, which is used in compactification to obtain the right amount of supersymmetry. In fact, the holonomy group of the compactified space determines the number of supersymmetry one sees in the resulting theory \cite{GukovHolo}. Essentially, there is a link between the holonomy of the manifold one is compactifying on, and the fraction of supersymmetry preserved. The larger(more generic) the holonomy group, the fewer number of supersymmetry preserved. For example, if one considers toroidal compactifications of string theory, he will find that the original number of supersymmetry is preserved, since the holonomy group of a torus is trivial. Vectors come back to themselves unrotated upon parallel transportation. Another example is when the holonomy group is $\text{SU}(3)$, corresponding to Calabi-Yau manifolds of six real dimensions. Compactifying over such manifolds preserves one-fourth of the original supersymmetry. For instance, compactifying the heterotic string on said Calabi-Yau manifolds results in a $3+1$ dimensions effective field theory with $\mc{N} = 1$ supersymmetric charges. The following table, taken from \cite{GukovHolo}, summarises the relation between holonomy and supersymmetry.
\begin{center}
    \begin{tabular}{|c|c|c|c|c|}
        \hline
        Manifold & $T^n$ & $CY_3$ & $\mc{M}_{G_2}$ & $\mc{M}_{\text{Spin}(7)}$\\
        \hline
        \hline
        $\text{dim}_{\mathbb{R}}$ & $n$ & 6 & 7 & 8\\
        \hline $\text{Hol}$ & $\mathds{1}\subset$ & $\text{SU}(3)\subset$ & $G_2\subset$ & $\text{Spin}(7)$ \\
        \hline
        SUSY & 1 & 1/4  & 1/8  & 1/16\\
        \hline
    \end{tabular}
\end{center}
Another place where the notion of holonomy arises is in gauge theories, where it plays a key rule as a gauge invariant observable. In fact, in this case one defines what is known as the Wilson loop which is gauge invariant, and hence is considered to be a true observable in gauge theories. There has been a long standing confusion in the literature as to what one should define the Wilson loop for the theory at hand. We will shed some light on the issue and provide an answer in this chapter. Wilson loops first appeared in the physics literature with the work of K. Wilson in 1974 \cite{PhysRevD.10.2445}, where he used Wilson loops in the context of lattice QCD to describe the phenomenon of quark confinement. In turned out that one can use Wilson loops, or lines, to derive what is called as the area law, which characterises the phase that the quark is in. For more on this topic, the papers by 't Hooft are a good source \cite{tHooft:1977nqb,tHooft:1979rtg,tHooft:1981bkw}. More directly related to our work is the appearance of the holonomy, or Wilson lines, as the phase acquired by particles traversing through spacetime and other fields. The phase experienced by a particle depends on the coupling of the particle to whatever field it is travelling through, and whether the particle has charge, spin, or more generally, some kind of symmetry. We will discuss these ideas more explicitly after we introduce the mathematical formulation of holonomy in the next section.
\section{Mathematical Formulation of Holonomy}
As mention in the previous section, the idea of the notion of holonomy is to examine how the action of parallel transport map acts on a vector, or more generally sections of a fibre bundle. For example, parallel transporting a section of the tangent bundle $T\mc{M}$ means moving a tangent vector along some curve in $\mc M$, such that at each point the vector is tangent to the curve. The significance of this is that on Euclidean space $\mathbb{E}^n$, there is an obvious way to talk about vectors at different points. Two vectors, $V$ and $W$, at different points belong to the same tangent space. This is because at each point of $\mathbb{E}^n$, the tangent space is isomorphic to it, i.e., $T_p\mc{M} \simeq \mathbb{E}^n$. On the other hand, we do not have such an isomorphism for a general manifold $\mc M$. Therefore, a new notion of parallelism needs to be introduced. This will allow us to define the covariant derivative of a section, which is important since there exists a correspondence between parallel sections and the holonomy group of the manifold. Here we review the basics of fibre bundles and introduce the notion of the Horizontal lift, which will allow us to define the parallel transport map.

\subsection{Fibre bundles}
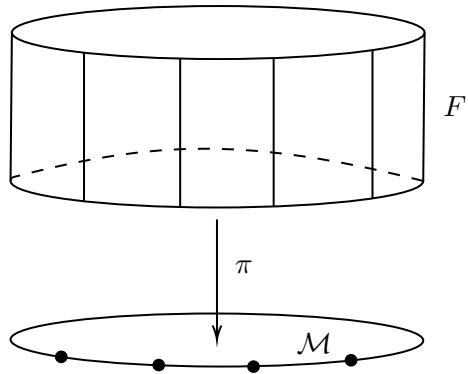
\begin{figure}[t]
    \centering
    \begin{tikzpicture}[x=0.5pt,y=0.5pt,yscale=-1,xscale=1]
        
        \draw   (485,260.4) .. controls (484.98,271.45) and (415.78,280.22) .. (330.45,280) .. controls (245.13,279.78) and (175.98,270.64) .. (176.01,259.6) .. controls (176.03,248.55) and (245.23,239.78) .. (330.56,240) .. controls (415.89,240.22) and (485.03,249.36) .. (485,260.4) -- cycle ;
        \draw   (177,28) .. controls (177,16.95) and (246.17,8) .. (331.5,8) .. controls (416.83,8) and (486,16.95) .. (486,28) .. controls (486,39.05) and (416.83,48) .. (331.5,48) .. controls (246.17,48) and (177,39.05) .. (177,28) -- cycle ;
        \draw  [draw opacity=0] (484.89,139.25) .. controls (484.96,139.5) and (485,139.75) .. (485,140) .. controls (485,151.05) and (415.83,160) .. (330.5,160) .. controls (245.17,160) and (176,151.05) .. (176,140) .. controls (176,139.06) and (176.5,138.13) .. (177.48,137.22) -- (330.5,140) -- cycle ; \draw   (484.89,139.25) .. controls (484.96,139.5) and (485,139.75) .. (485,140) .. controls (485,151.05) and (415.83,160) .. (330.5,160) .. controls (245.17,160) and (176,151.05) .. (176,140) .. controls (176,139.06) and (176.5,138.13) .. (177.48,137.22) ;  
        \draw    (486,28) -- (485,140) ;
        \draw    (177,28) -- (176,140) ;
        \draw  [dash pattern={on 4.5pt off 4.5pt}]  (485,140) .. controls (348,105) and (278,111) .. (177.48,137.22) ;
        \draw    (231,43) -- (231,155) ;
        \draw    (303,48) -- (303,160) ;
        \draw    (373,47) -- (373,159) ;
        \draw    (447,41) -- (447,153) ;
        \draw  [fill={rgb, 255:red, 0; green, 0; blue, 0 }  ,fill opacity=1 ] (434.96,275.13) .. controls (435.04,277.33) and (433.31,279.19) .. (431.1,279.26) .. controls (428.89,279.34) and (427.04,277.61) .. (426.97,275.4) .. controls (426.89,273.19) and (428.62,271.34) .. (430.83,271.27) .. controls (433.04,271.19) and (434.89,272.92) .. (434.96,275.13) -- cycle ;
        \draw  [fill={rgb, 255:red, 0; green, 0; blue, 0 }  ,fill opacity=1 ] (361.95,279.94) .. controls (362.03,282.14) and (360.3,283.99) .. (358.09,284.07) .. controls (355.88,284.14) and (354.03,282.42) .. (353.96,280.21) .. controls (353.88,278) and (355.61,276.15) .. (357.82,276.07) .. controls (360.02,276) and (361.88,277.73) .. (361.95,279.94) -- cycle ;
        \draw  [fill={rgb, 255:red, 0; green, 0; blue, 0 }  ,fill opacity=1 ] (290.95,278.75) .. controls (291.03,280.96) and (289.3,282.81) .. (287.09,282.88) .. controls (284.88,282.96) and (283.03,281.23) .. (282.96,279.02) .. controls (282.88,276.81) and (284.61,274.96) .. (286.82,274.89) .. controls (289.03,274.81) and (290.88,276.54) .. (290.95,278.75) -- cycle ;
        \draw  [fill={rgb, 255:red, 0; green, 0; blue, 0 }  ,fill opacity=1 ] (217.97,272.56) .. controls (218.04,274.77) and (216.32,276.62) .. (214.11,276.69) .. controls (211.9,276.77) and (210.05,275.04) .. (209.97,272.83) .. controls (209.9,270.62) and (211.63,268.77) .. (213.84,268.7) .. controls (216.04,268.62) and (217.89,270.35) .. (217.97,272.56) -- cycle ;
        \draw    (330.56,169) -- (330.56,258) ;
        \draw [shift={(330.56,260)}, rotate = 270] [color={rgb, 255:red, 0; green, 0; blue, 0 }  ][line width=0.75]    (10.93,-3.29) .. controls (6.95,-1.4) and (3.31,-0.3) .. (0,0) .. controls (3.31,0.3) and (6.95,1.4) .. (10.93,3.29)   ;
        
        \draw (342,197.4) node [anchor=north west][inner sep=0.75pt]    {$\pi $};
        \draw (498,73.4) node [anchor=north west][inner sep=0.75pt]    {$F$};
        \draw (388,252.4) node [anchor=north west][inner sep=0.75pt]    {$\mathcal{M}$};

        \end{tikzpicture}
        \caption{The cylinder as a product bundle}
        \label{fig:CylinBundle}
\end{figure}
\begin{figure}[t]
    \centering
    \begin{tikzpicture}[x=0.75pt,y=0.75pt,yscale=-1,xscale=1]
        
        \draw   (440.38,221.06) .. controls (440.36,230.04) and (384.08,237.18) .. (314.68,237) .. controls (245.28,236.81) and (189.03,229.38) .. (189.06,220.4) .. controls (189.08,211.42) and (245.36,204.28) .. (314.76,204.46) .. controls (384.16,204.64) and (440.4,212.07) .. (440.38,221.06) -- cycle ;
        \draw    (314.72,146.72) -- (314.72,218.73) ;
        \draw [shift={(314.72,220.73)}, rotate = 270] [color={rgb, 255:red, 0; green, 0; blue, 0 }  ][line width=0.75]    (10.93,-3.29) .. controls (6.95,-1.4) and (3.31,-0.3) .. (0,0) .. controls (3.31,0.3) and (6.95,1.4) .. (10.93,3.29)   ;
        \draw    (419.93,153.56) .. controls (363.34,167.18) and (216,104) .. (198.8,61.34) ;
        \draw    (420.98,84.39) -- (419.93,153.56) ;
        \draw    (272.16,117.93) .. controls (272.16,134.7) and (274.26,135.74) .. (288.93,139.94) ;
        \draw    (383.25,117.93) .. controls (390.58,133.65) and (391.63,132.6) .. (387.44,151.46) ;
        \draw    (408.4,104.3) .. controls (415.74,120.02) and (415.74,134.7) .. (411.54,153.56) ;
        \draw    (198.8,61.34) -- (198.8,143.08) ;
        \draw    (219.76,87.54) .. controls (211.38,122.12) and (219.76,133.65) .. (222.91,145.18) ;
        \draw    (244.91,102.21) .. controls (236.53,136.79) and (244.91,134.7) .. (248.06,146.22) ;
        \draw    (244.91,102.21) .. controls (310.94,83.34) and (319.32,87.54) .. (391.63,111.64) ;
        \draw    (257.49,54) -- (257.49,98.02) ;
        \draw    (301.51,56.1) -- (301.51,91.73) ;
        \draw    (360.19,61.34) -- (360.19,102.21) ;
        \draw    (198.8,143.08) .. controls (261,160) and (413,119) .. (420.98,84.39) ;
        \draw    (210.33,54) .. controls (316,45) and (372,56) .. (420.98,84.39) ;
        \draw  [line width=0.75] [line join = round][line cap = round] (210,54) .. controls (205.83,54) and (199,58.24) .. (199,62) ;
        
        \draw (322.81,168.4) node [anchor=north west][inner sep=0.75pt]    {$\pi $};
        \draw (359.57,213.13) node [anchor=north west][inner sep=0.75pt]    {$\mathcal{M}$};

        \end{tikzpicture}

        \caption{Möbius band as a non-trivial bundle}
        \label{fig:Mobius}
\end{figure}
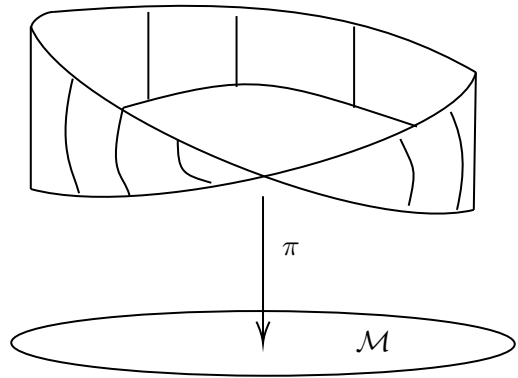

We shall start with giving the basic definition of a general fibre bundle. We closely follow \cite{Isham:1999qu}.
\begin{definition}\,\newline
    \begin{enumerate}
        \item A bundle is a triple $(E, \pi, \mc M)$ where $E$ and $\mc M$ are topological spaces and $\pi : E \rightarrow \mc M$ is a continuous map.
        \item The space $E$ is called the total space, and $\mc M$ is the base space. The map $\pi$ is called the projection, and its inverse image $\pi^{-1}(\{x\})$ is the fibre over $x\in \mc M$.
    \end{enumerate}
\end{definition}
A small comment is in order. In most applications that arise in physics, we require the fibres $\pi^{-1}(\{x\})$, $x\in \mc M$, to be homeomorphic to a common space $F$ called the fibre of the bundle. A bundle with fibre space $F$ will be denoted by $(E,\pi,\mc M, F)$, or
\begin{equation}
    \begin{tikzcd}
        F \arrow{r} & E \arrow{d}{\pi} \\%
        & \mc M
        \end{tikzcd}
\end{equation}
depending on ones preference. A typical example iof a fibre bundle is the trivial product bundle over base space $\mc M$ with fibre $F$, $(E = \mc{M}\times F, \pi, M)$. If we take the $\mc M = S^1$ and $F = \mathbb{R}$, then the product bundle is nothing but a cylinder(see figure \ref{fig:CylinBundle}). Another typical example of a fibre bundle is the Möbius band(figure \ref{fig:Mobius}). In this case the fibre can be taken to be the closed interval $[-1,1]$. However, note that the total space is not simply the product $S^1\times [-1,1]$; there is non-trivial twisting happening. We will come back to these two examples once we introduce the concepts of principal and its vector associated bundles.
\subsubsection{Cross-sections of fibre bundles}

From the perspective of application to theoretical physics, the idea of a cross-section is of great importance. Vectors, fields, frames, etc, can all be regarded as cross section of various different fibre bundles. Therefore, the local or global existence of sections is directly related to whether, for example, a field of charged particle can be defined locally or globally. Here we give the basic definition of the idea of a cross-section:
\begin{definition}
    A cross-section of a bundle $(E,\pi,\mc M)$ is a map $\sigma:\mc{M}\rightarrow E$ such that the image of each point $x\in \mc M$ lies in the fibre $\pi^{-1}(\{x\})$ over $x$,
    \begin{equation}
        \pi \circ \sigma = \text{id}_{\mc M}.
    \end{equation}
\end{definition}
As mentioned earlier, a cross section can be viewed as normal function in the case of the product bundle, $(\mc{M}\times F, \pi, \mc{M})$, since a global such cross section can be defined. This is the case in our cylinder example, where cross section is nothing but a function $\sigma(x): x\in\mc{M}\rightarrow F = \mathbb{R}$. However, the Möbius band as a bundle does not allow for such an interpretation. This is because there is non-trivial twisting happening, where the cross section changes sign depending on the point which it is taken. Next, we review the basics of principal fibre bundles.
\subsection{Principal fibre bundle}
The concept of the principal bundle is of utmost importance for both physics and mathematics. Principal bundles, whose fibre is some Lie group $G$, have the remarkable property that all non-principal bundles are associated with an underlying principal bundle. In fact, the twisting of these non-principal bundles associated to some principal bundle are uniquely determined by the latter.We have already seen some of the applications ofthe  theory of principal bundles in the case of the Dirac in \ref{sec:HodgeMag} and Taub-NUT monopoles in \ref{sec:DC}. Now we give the definition of a principal bundle:
\begin{definition}\,
    \begin{enumerate}
        \item A bundle $(E,\pi,\mc{M})$ is a G-principal bundle, $G$-bundle for short, if $E$ is a right $G$-space, and
        \item If $(E,\pi,\mc{M})$ is isomorphic(via some map $u$) to the bundle $(E,\rho,E/G)$ where $E/G$ is the orbit space of the $G$-action on $E$ and $\rho$ is the projection map:
        \begin{equation}
            \begin{tikzcd}
                E \arrow{r}{u}\arrow{d}{\pi} & E \arrow{d}{\rho} \\%
                \mc{M} \arrow{r}{\simeq}&  E/G
            \end{tikzcd}
        \end{equation}
        \item If $G$ acts \emph{freely} on $E$, then $(E,\pi,\mc M)$ is called a principal $G$-bundle and $G$ is its \emph{structure} group. The freedom of the $G$ on $E$ implies that each orbit of $G$ is homeomorphic to $G$ itself. Hence, we have a fibre bundle with fibre $G$.
    \end{enumerate}
    
\end{definition}
For the benefit of the reader, we shall give two important examples of principal bundles. In 1931, Heinz Hopf gave a construction relating the 3-sphere to the 2-sphere via a map from the former to the latter \cite{HopfHeinz}. It was an influential early example of a fibre bundle. One can understand the construction as a principal bundle in the following way. The $\text{U}(1)$ action on $\text{SU}(2)\simeq S^3$ gives rise to a bundle $\text{SU}(2)\rightarrow \text{SU}(2)/\text{U}(1)$, which is
\begin{equation}
    \begin{tikzcd}
        S^1 \arrow{r}& S^3 \arrow{d}{\pi} \\%
        &  S^2
    \end{tikzcd}
\end{equation}
This is indeed the famous fibration of $S^3$ by the circle, now known as Hopf fibration.

The second example is undoubtedly one of the most important principal bundles in physics, the frame bundle $(LM, \pi, \mc{M}, \text{GL}(m,\mathbb{R}))$. The idea is that for each point $x$ in an $m$-dimensional differentiable manifold $\mc{M}$, we attach a linear frame. A linear frame $e$ at $x\in\mc{M}$ is a an ordered set $(e_1, e_2,\ldots,e_m)$ of basis vectors for the tangent space $T_x\mc{M}$. The projection map is then simply $\pi : e_x \mapsto x$. As mentioned above, in a principal bundle the group has a right action on the total space $LM$ given by
\begin{equation}
    (e_1,e_2,\ldots,e_m)g := \big(   \sum^{m}_{j_1=1}e_{j_1}g_{j_11},\sum^{m}_{j_2=1}e_{j_2}g_{j_22},\ldots, \sum^{m}_{j_m=1}e_{j_m}g_{j_mm}   \big)
\end{equation}
for all $g\in \text{GL}(m, \mathbb{R})$. Now since the base manifold is a differentiable manifold, we can give the total space $LM$ a differentiable structure. In fact, the total space always inherits the differentiable structure from its base manifold \cite{fecko_2006}. Let $U\subset \mc{M}$ be a coordinate neighbourhood with coordinate functions $(x^1, x^2,\ldots, x^m)$ defined on it. Any basis for the vector space $T_x\mc{M}$ can be expanded uniquely as
\begin{equation}
    e_i = \sum^m_{j=1}e_i^j \big(\frac{\pd}{\pd x^j}\big)_x,\quad i = 1,\ldots,m
\end{equation}
for some non-singular $b^j_i\in \text{GL}(m,\mathbb{R})$. Thus we can define a map $h$ as follows
\begin{equation}
    \begin{split}
    h :\, &U \times \text{GL}(m,\mathbb{R}) \rightarrow \pi^{-1}(U)\\
    &(x,g) \mapsto \big(   \sum^{m}_{j_1=1}g^{j_1}_1(\pd_{j_1})_x, \sum^{m}_{j_2=1}g^{j_2}_2(\pd_{j_2})_x, \ldots, \sum^{m}_{j_m=1}g^{j_m}_m(\pd_{j_m})_x  \big)
    \end{split}
\end{equation}
and use it as a differentiable structure on $LM$, which becomes an $m+m^2$-dimensional manifold. Note that the section of the frame bundle is nothing but a choice of frame at some point in $\mc{M}$.
\subsection{Associated bundles}
Now we introduce the concept of associated bundles. General Relativity and gauge theories such as Maxwell's $\text{U}(1)$ theory are described by appropriate fibre bundles that are associated to some underlying principal bundle. The idea is that one can construct different kinds of fibre bundles that are associated to some principal bundle. The basic plan is to start with some principal bundle $(P,\pi,\mc{M})$ with structure group $G$, and form a fibre bundle with fibre $F$ by acting on on each space $F$ with $G$ as the group of transformations. First however, we must introduce the concept of a `$G$-product' of a pair of spaces on which $G$ acts.
\begin{definition}
    Let $X$ and $Y$ be any right $G$-spaces. Then the $G$-product of $X$ and $Y$, denoted by $X\times_G Y$, is the space of orbits of the action of $G$ on the Cartesian product $X\times Y$. The equivalence class is defined by $(x,y) := (x',y')$ if there exists $g\in G$ such that $x' = xg$ and $y' = yg$.
\end{definition}
Now for the definition of an associated fibre bundle
\begin{definition}
    Let $\xi=(P,\pi,\mc{M})$ be a principal G-bundle and let $F$ be a left $G$-space. Define $P_F:=P\times_G F$ by $(p,v):=(pg, g^{-1}v)$, and define a new projection map $\pi_F:P_F \rightarrow \mc{M}$ such that $\pi_F([p,v]):=\pi(p)$. Then $\xi_F=(P_F,\pi_F,\mc{M})$ is a fibre bundle with fibre $F$ and base space $\mc{M}$ associated to the principal bundle $\xi$.
\end{definition}
Indeed, for $\xi_F$ to really be a fibre bundle; each $\pi^{-1}(\{x\})$ must be homeomorphic to $F$. Therefore, we quote the following theorem:
\begin{theorem}
    For each $x\in\mc{M}$, the space $\pi_F^{-1}(\{x\})$ is homeomorphic to F.
\end{theorem}
An example of great importance is the tangent bundle. In this case we have the frame bundle $LM$ as the underlying principal bundle on an $m$-dimensional manifold $\mc{M}$, and the general linear group $\text{GL}(m,\mathbb{R})$ as the structure group. One can construct the associated bundle by choosing the left $G$-space to be $F=\mathbb{R}^m$ which is acted upon by the usual linear group of transformations. This associated bundle is denoted by $LM\times_{\text{GL}(m,\mathbb{R})}\mathbb{R}^m$ which can be identified with the tangent bundle via the map
\begin{equation}
    [e,\vec{r}]\mapsto \sum^m_{i=1}e_i r^i\in T_x\mc{M},
\end{equation}
where, as explained earlier, $e$ is a base for the tangent space at a point $x\in\mc{M}$ and $\vec{r}=(r^1,r^2,\ldots,r^m)$. More generally, one can have $\rho : \text{GL}(m,\mathbb{R})\rightarrow \text{Aut}(V)$ as any representation of $\text{GL}(m,\mathbb{R})$ on a real vector space $V$ acting as the group of linear transformations.

Next, we review the notion of a connection in the fibre bundle. This will allow us to talk about the tangent space of the total space, and its vertical and horizontal tangent spaces. In turn, this will allow us to define the notion of the horizontal lift of a curve, which is of central importance in the associated concepts of parallel transport and covariant differentiation. The latter is the result of being able to compare vectors, or sections in general, lying in different fibres on the manifold.
\subsection{Connections on fibre bundles}
A crucial notion, when discussing connections, is that of vertical and horizontal vectors; living in the vertical and horizontal parts of the tangent space of the total space respectively. In anticipation, let us quote the following definition:
\begin{definition}
    Let $G$ be a Lie group that has a right action $g\rightarrow \delta_g$ on a differentiable manifold $\mc{M}$. Then the vector field $X^A$ on $\mc{M}$ induced by the action of the  one parameter subgroup $t\mapsto \exp [tA]$, $A\in T_eG$, is defined as
    \begin{equation}
        X^A_x(f):= \frac{d}{dt}f(x\exp [tA])\Big{|}_{t=0}
    \end{equation}
    where $x\in \mc{M}$, $f\in C^{\infty}(\mc{M})$, and $\delta_g(p)$ is abbreviated to $pg$. Now we quote a theorem which states that the map that associates to $A\in T_eG$ a vector field $X^A$ on $\mc{M}$ is a homomorphism. This means that the Lie algebra of $G$ is represented by vector fields on $\mc{M}$ on which $G$ acts.
\end{definition}
\begin{theorem}
    Let $\mc{M}$ be a manifold on which a Lie group $G$ has a right action. Then the map $A\mapsto X^A$, which associates to each $A\in T_eG$ the induced vector field $X^A$ on $\mc{M}$, is a homomorphism of the Lie algebra $L(G)\simeq T_eG$ into the infinite dimensional Lie algebra of all vector fields on $\mc{M}$,i.e.,
    \begin{equation}
        [X^A,X^B] = X^{[A,B]}
    \end{equation}
    for all $A,B \in T_eG\simeq L(G)$.
\end{theorem}
Now we are ready to define the connection
\begin{definition}
    A connection $\omega$ is a $L(G)$-valued one-form on $P$ which is a projection of $T_pP$ onto the vertical component $V_pP \simeq L(G)$. It satisfies
    \begin{enumerate}
        \item $\omega_p(X^A) = A \,\forall p\in P, A\in L(G)$
        \item $\delta_g^*\omega = \text{Ad}_{g^{-1}}\omega$\,,i.e., $(\delta^*_g\omega)_p(\tau) = \text{Ad}_{g^{-1}}(\omega_p(\tau)), \forall \tau\in T_pP$
    \end{enumerate}
    where $\delta_g$ is the right action of $G$ on $P$ and $\tau$ is a vector in the tangent space of the total space $P$.
\end{definition}
One defines the horizontal subspace by the kernel of the connection on $TP$:
\begin{equation}\label{HoriDef}
    H_pP := \{ \tau\in T_pP | \omega(\tau) = 0 \}.
\end{equation}
The horizontal subspace satisfy
\begin{equation}
    \delta_{g*}(H_pP) = H_{pg}\,, \forall g\in G, \forall p\in P
\end{equation}
Therefore we have the decomposition of the tangent space of the total space, 
\begin{equation}
    T_pP \simeq V_pP \oplus H_pP,\, \forall p\in P.
\end{equation}
The connection $\omega\in \Omega^1(P, L(G))$ is defined on the total space. However, for many consideration in physics it is helpful to have a local form of the connection for doing concrete calculations. Indeed, in this chapter we shall make use of such local form of the connection, which we will call the Yang-Mills connection. Note that this does not specifically mean it is the usual Yang-Mills field, although that is indeed the local form of the connection. For example, in our terminology, the Christofell symbol is the local Yang-Mills connection of a $\text{GL}(m,\mathbb{R})$ frame bundle. The local Yang-Mills field can be regarded as a Lie algebra valued one-form on the base space $\mc{M}$, $A^a_\mu$, where $\mu = 1,\ldots, \text{dim}\,\mc{M}$, and $a = 1,\ldots, \text{dim}\, G$. Thus one can write
\begin{equation}\label{GValConn}
    \Gamma(x):= \sum^m_{\mu=1}\sum^{\text{dim}\, G}_{a = 1}\Gamma^a_\mu (x)E_a (dx^\mu)_x,
\end{equation}
where $\{E_1, E_2,\ldots\, E_{\text{dim}\,G}\}$ is a basis set for the Lie algebra $L(G)$. The relation between this local Yang-Mills field and the connection form on the total space $P$ is given by the following theorem \cite{Isham:1999qu,Nakahara:2003nw}
\begin{theorem}
    Given a $L(G)$-valued one-form $\Gamma_i$ on an open subset $U_i\subset \mc{M}$ and a local section $\sigma_i: U_i \rightarrow \pi^{-1}(U_i)$, there exists a connection one-form $\omega$ on $P$ such that $\Gamma_i = \sigma_i^*(\omega)$.
\end{theorem}
and if $G$ is a matrix group, $\Omega_{ij}: U_i \cap U_j \neq \emptyset \rightarrow G$ such that $\sigma_i(x) =\Omega_{ij}\sigma_j $, then two local representatives of $\omega$ are related by
\begin{equation}
    \Gamma_i = \Omega(x)^{-1}\Gamma_j\,\Omega(x) + \Omega(x)^{-1}d\Omega(x),
\end{equation}
where $\Omega$ is the transition function of the bundle relating two open sets of $\mc{M}$. This is nothing but the well-known gauge transformation from physics. However, some care should be taken as this transformation is induced by relating two different intersecting sets of the base space; relating two different local Yang-Mills fields. The relation two a gauge transformation at a fixed point in $\mc{M}$ is given by noting that if $\sigma:U\rightarrow P$ is a local section of $(P,\pi,\mc{M},G)$ with $\Gamma = \sigma^*(\omega)$, then an active gauge transformation $\phi: P\rightarrow P$ induces a transformation $\Gamma\mapsto \sigma^*(\phi^*\omega)= (\phi\circ \sigma)^*\omega$. But there exists a function $\Omega: U\rightarrow G$ such that $\sigma(x) = \phi\circ \sigma(x)\Omega(x)  $ for all $x\in \mc{M}$, then one can show that the transformation of the local representative $A$ on $U$ can be written as \cite{Isham:1999qu}
\begin{equation}
    \Gamma' = \Omega(x)^{-1}\Gamma\,\Omega(x) + \Omega^{-1}d\Omega(x),
\end{equation}
which is the usual gauge transformation of a single Yang-Mills field.
\subsection{Parallel transport}
Now we come to the pivotal concept of parallel transportation. Traditionally, the idea is to compare, say, vectors at some  point $x$ with another one at a different point $x'$. Now on $\mathbb{R}^m$, one can easily just transport the vector at $x$ to $x'$ while keeping the vector parallel to itself, i.e., the vector does not rotate. Thus, we can easily compare the two vectors, because they are in the same vector(tangent) space. However, the situation is rather different on a general curved manifold. How does one `parallel' transport a vector in this case? Even if one transports the vector in some way, the initial and final vectors would be living in different tangent spaces, thus comparing them would not be possible. In our case, we are dealing with the more general case of a principal bundle. The set of characters are the base space, total space, and the fibres over each point in the base space. How can we compare sections along different fibres in a well-defined way? With the aid of the connection and the notion of horizontal lifting, parallel translation of section of principal bundles can be defined. A classic resource for this topic is \cite{zbMATH03194988}.

Let us start with a principal bundle $(P,\pi, \mc{M})$ equipped with a connection $\omega$. An important concept for the idea of horizontal lifting are horizontal vector fields; fields whose flow lines move from one fibre to another without acquiring any vertical components along the way.
\begin{definition}
    Since $\pi_*: H_pP \rightarrow T_{\pi(p)}\mc{M}$ is an isomorphism, to each vector field on $\mc{M}$ there is a unique vector field , $X^{\uparrow}$, on $P$ such that, for all $p\in P$,
    \begin{enumerate}
        \item $\pi_*(X^{\uparrow}) = X|_{\pi(p)}$
        \item $\text{ver}(X^{\uparrow}) = 0$.
    \end{enumerate}
    $X^{\uparrow}$ is called the horizontal lift of $X$ onto $P$.
\end{definition}
In fact one can also have horizontal lifts of curves. That is, given a curve $\gamma$ in $\mc{M}$, one can lift it to get a curve $\gamma^\uparrow$ in $P$. But before we introduce that, here are some nice properties that horizontal lifting of vector fields have
\begin{enumerate}
    \item Horizontal lifting is $G$-equivariant; meaning $\delta_{g*}(X^\uparrow_p) = X^\uparrow_{pg}$.
    \item A sufficient condition for a vector field $Y$ on $P$ to be the horizontal lift of a field on $\mc{M}$ is that
    \begin{itemize}
        \item $\text{ver}(Y) = 0$; and
        \item $\delta_{g*}(Y_p) = Y_{pg}$ for all $p\in P$, and $g\in G$.
    \end{itemize}
    \item Algebraic properties:
    \begin{itemize}
        \item $(X^\uparrow + Y^\uparrow) = (X+Y)^\uparrow$
        \item If $f\in C^{\infty}(\mc{M})$, then $(fX^{\uparrow}) = f\circ \pi X^\uparrow$
        \item $[X,Y]^\uparrow = \text{hor}([X^\uparrow, Y^\uparrow])$.
    \end{itemize}
\end{enumerate}
Now we define the horizontal lift of a curve on the base space to the total space.
\begin{definition}
    Let $\gamma$ be a smooth curve $[a,b]\rightarrow \mc{M}$. A horizontal lift of $\gamma$ is a curve $\gamma^\uparrow : [a,b]\rightarrow P$ which is horizontal and such that $\pi(\gamma^\uparrow(t))=\gamma(t)$ for all $t\in [a,b]$,
\end{definition}
where the lift $\gamma^\uparrow$ being horizontal means $\text{ver}(\gamma^\uparrow) = 0$. In fact, there is a key theorem which guarantees the existence of the horizontal lift of a curve on $\mc{M}$ for each point $p$ lying in the fibre at $x\in \mc{M}$, or more precisely,
\begin{theorem}
For each point $p\in \pi^{-1}\{\gamma(a)\}$, there exists a unique horizontal lift of $\gamma$ such that $\gamma^\uparrow(a) = p$.
\end{theorem}
Now, for the purposes of calculation in theoretical physics, one would like to have as explicit an expression as possible for the horizontal lift of a curve $\gamma$. This can be done in practice by using the connection form and comparing two lifts of the curve $\gamma$; one horizontal while the other is some non-horizontal lift. Obtaining an explicit expression is very useful as it allows us to manipulate and do local calculations as the ones we are used to in physics, and also make contact with other concepts like angles or phases acquired by particles moving through spacetime.

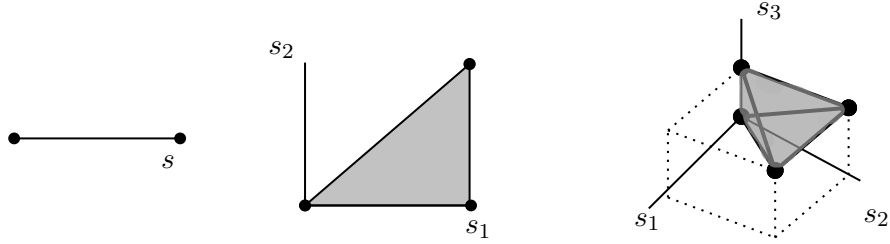
\begin{figure}[t]
    \centering
    \begin{tikzpicture}[x=0.55pt,y=0.55pt,yscale=-1,xscale=1]
        
        \draw    (60,115) -- (173,115) ;
        \draw [shift={(173,115)}, rotate = 0] [color={rgb, 255:red, 0; green, 0; blue, 0 }  ][fill={rgb, 255:red, 0; green, 0; blue, 0 }  ][line width=0.75]      (0, 0) circle [x radius= 3.35, y radius= 3.35]   ;
        \draw [shift={(60,115)}, rotate = 0] [color={rgb, 255:red, 0; green, 0; blue, 0 }  ][fill={rgb, 255:red, 0; green, 0; blue, 0 }  ][line width=0.75]      (0, 0) circle [x radius= 3.35, y radius= 3.35]   ;
        \draw    (258,161) -- (371,161) ;
        \draw [shift={(371,161)}, rotate = 0] [color={rgb, 255:red, 0; green, 0; blue, 0 }  ][fill={rgb, 255:red, 0; green, 0; blue, 0 }  ][line width=0.75]      (0, 0) circle [x radius= 3.35, y radius= 3.35]   ;
        \draw [shift={(258,161)}, rotate = 0] [color={rgb, 255:red, 0; green, 0; blue, 0 }  ][fill={rgb, 255:red, 0; green, 0; blue, 0 }  ][line width=0.75]      (0, 0) circle [x radius= 3.35, y radius= 3.35]   ;
        \draw    (258,63) -- (258,161) ;
        \draw [shift={(258,161)}, rotate = 90] [color={rgb, 255:red, 0; green, 0; blue, 0 }  ][fill={rgb, 255:red, 0; green, 0; blue, 0 }  ][line width=0.75]      (0, 0) circle [x radius= 3.35, y radius= 3.35]   ;
        \draw  [fill={rgb, 255:red, 0; green, 0; blue, 0 }  ,fill opacity=0.24 ] (370,64) -- (258,161) -- (370,161) -- cycle ;
        \draw    (555,33) -- (555,100) ;
        \draw [shift={(555,100)}, rotate = 90] [color={rgb, 255:red, 0; green, 0; blue, 0 }  ][fill={rgb, 255:red, 0; green, 0; blue, 0 }  ][line width=0.75]      (0, 0) circle [x radius= 3.35, y radius= 3.35]   ;
        \draw    (555,100) -- (492,163) ;
        \draw    (555,100) -- (636,144) ;
        \draw [shift={(555,100)}, rotate = 28.51] [color={rgb, 255:red, 0; green, 0; blue, 0 }  ][fill={rgb, 255:red, 0; green, 0; blue, 0 }  ][line width=0.75]      (0, 0) circle [x radius= 3.35, y radius= 3.35]   ;
        \draw  [dash pattern={on 0.84pt off 2.51pt}]  (555,66.5) -- (628,94) ;
        \draw  [dash pattern={on 0.84pt off 2.51pt}]  (505,109.5) -- (578,137) ;
        \draw  [dash pattern={on 0.84pt off 2.51pt}]  (578,137) -- (628,94) ;
        \draw  [dash pattern={on 0.84pt off 2.51pt}]  (505,109.5) -- (555,66.5) ;
        \draw  [dash pattern={on 0.84pt off 2.51pt}]  (628,140) -- (628,94) ;
        \draw  [dash pattern={on 0.84pt off 2.51pt}]  (578,183) -- (578,137) ;
        \draw  [dash pattern={on 0.84pt off 2.51pt}]  (505,155.5) -- (505,109.5) ;
        \draw  [dash pattern={on 0.84pt off 2.51pt}]  (505,155.5) -- (578,183) ;
        \draw  [dash pattern={on 0.84pt off 2.51pt}]  (578,183) -- (628,140) ;
        \draw [fill={rgb, 255:red, 155; green, 155; blue, 155 }  ,fill opacity=1 ][line width=1.5]    (555,66.5) -- (578,137) ;
        \draw [shift={(578,137)}, rotate = 71.93] [color={rgb, 255:red, 0; green, 0; blue, 0 }  ][fill={rgb, 255:red, 0; green, 0; blue, 0 }  ][line width=1.5]      (0, 0) circle [x radius= 4.36, y radius= 4.36]   ;
        \draw [shift={(555,66.5)}, rotate = 71.93] [color={rgb, 255:red, 0; green, 0; blue, 0 }  ][fill={rgb, 255:red, 0; green, 0; blue, 0 }  ][line width=1.5]      (0, 0) circle [x radius= 4.36, y radius= 4.36]   ;
        \draw [fill={rgb, 255:red, 155; green, 155; blue, 155 }  ,fill opacity=1 ][line width=1.5]    (555,100) -- (578,137) ;
        \draw [shift={(578,137)}, rotate = 58.13] [color={rgb, 255:red, 0; green, 0; blue, 0 }  ][fill={rgb, 255:red, 0; green, 0; blue, 0 }  ][line width=1.5]      (0, 0) circle [x radius= 4.36, y radius= 4.36]   ;
        \draw [shift={(555,100)}, rotate = 58.13] [color={rgb, 255:red, 0; green, 0; blue, 0 }  ][fill={rgb, 255:red, 0; green, 0; blue, 0 }  ][line width=1.5]      (0, 0) circle [x radius= 4.36, y radius= 4.36]   ;
        \draw [fill={rgb, 255:red, 155; green, 155; blue, 155 }  ,fill opacity=1 ][line width=1.5]    (628,94) -- (555,100) ;
        \draw [shift={(555,100)}, rotate = 175.3] [color={rgb, 255:red, 0; green, 0; blue, 0 }  ][fill={rgb, 255:red, 0; green, 0; blue, 0 }  ][line width=1.5]      (0, 0) circle [x radius= 4.36, y radius= 4.36]   ;
        \draw [shift={(628,94)}, rotate = 175.3] [color={rgb, 255:red, 0; green, 0; blue, 0 }  ][fill={rgb, 255:red, 0; green, 0; blue, 0 }  ][line width=1.5]      (0, 0) circle [x radius= 4.36, y radius= 4.36]   ;
        \draw [fill={rgb, 255:red, 155; green, 155; blue, 155 }  ,fill opacity=1 ][line width=1.5]    (628,94) -- (578,137) ;
        \draw [shift={(578,137)}, rotate = 139.3] [color={rgb, 255:red, 0; green, 0; blue, 0 }  ][fill={rgb, 255:red, 0; green, 0; blue, 0 }  ][line width=1.5]      (0, 0) circle [x radius= 4.36, y radius= 4.36]   ;
        \draw [shift={(628,94)}, rotate = 139.3] [color={rgb, 255:red, 0; green, 0; blue, 0 }  ][fill={rgb, 255:red, 0; green, 0; blue, 0 }  ][line width=1.5]      (0, 0) circle [x radius= 4.36, y radius= 4.36]   ;
        \draw [fill={rgb, 255:red, 155; green, 155; blue, 155 }  ,fill opacity=1 ][line width=1.5]    (555,66.5) -- (628,94) ;
        \draw [shift={(628,94)}, rotate = 20.64] [color={rgb, 255:red, 0; green, 0; blue, 0 }  ][fill={rgb, 255:red, 0; green, 0; blue, 0 }  ][line width=1.5]      (0, 0) circle [x radius= 4.36, y radius= 4.36]   ;
        \draw [shift={(555,66.5)}, rotate = 20.64] [color={rgb, 255:red, 0; green, 0; blue, 0 }  ][fill={rgb, 255:red, 0; green, 0; blue, 0 }  ][line width=1.5]      (0, 0) circle [x radius= 4.36, y radius= 4.36]   ;
        
        \draw  [color={rgb, 255:red, 155; green, 155; blue, 155 }  ,draw opacity=0.67 ][line width=6] [line join = round][line cap = round] (586,95.5) .. controls (577.27,95.5) and (581.21,91.42) .. (586,89.5) .. controls (591.6,87.26) and (598.1,102.12) .. (597,106.5) .. controls (596.15,109.91) and (587.88,109.38) .. (586,107.5) .. controls (583.06,104.56) and (582.41,99.77) .. (583,94.5) .. controls (583.29,91.89) and (591.62,90.47) .. (592,93.5) .. controls (592.45,97.09) and (589.45,110.23) .. (584,107.5) .. controls (577.08,104.04) and (578.61,91.87) .. (590,93.5) .. controls (598.6,94.73) and (592.71,116.19) .. (581,109.5) .. controls (572.62,104.71) and (579.56,90.28) .. (590,95.5) .. controls (598.52,99.76) and (587.33,113.16) .. (584,111.5) .. controls (575.84,107.42) and (582.14,92.55) .. (591,95.5) .. controls (595.13,96.88) and (592.33,105.52) .. (592,108.5) .. controls (591.16,116.06) and (590.6,117.3) .. (587,124.5) .. controls (586.73,125.04) and (578,129.5) .. (578,129.5) .. controls (578,129.5) and (581,129.5) .. (581,129.5) .. controls (581,129.5) and (580.37,122.31) .. (584,120.5) .. controls (586.85,119.08) and (589.28,119.41) .. (592,118.5) .. controls (592.83,118.22) and (607.88,104.62) .. (610,102.5) .. controls (612.26,100.24) and (614.64,97.62) .. (618,96.5) .. controls (618.71,96.26) and (620,95.5) .. (620,95.5) .. controls (620,95.5) and (617.15,94.57) .. (617,94.5) .. controls (612.1,92.05) and (600.84,91.34) .. (597,87.5) .. controls (596.19,86.69) and (591.46,84.73) .. (591,84.5) .. controls (590.06,84.03) and (586.95,83.5) .. (588,83.5) .. controls (592.86,83.5) and (600.07,88.03) .. (603,89.5) .. controls (603.92,89.96) and (611.34,92.46) .. (611,92.5) .. controls (607.67,92.87) and (604,93) .. (601,91.5) .. controls (600.7,91.35) and (600.04,91.17) .. (600,91.5) .. controls (599.24,97.59) and (598.35,101.83) .. (605,100.5) .. controls (605.73,100.35) and (606.67,100.17) .. (607,99.5) .. controls (608.37,96.75) and (604.1,95.6) .. (603,94.5) .. controls (598.97,90.47) and (594.4,88.3) .. (589,86.5) .. controls (588.49,86.33) and (585.32,86.13) .. (585,85.5) .. controls (584.59,84.69) and (582,81.5) .. (582,81.5) .. controls (582,81.5) and (588,83.5) .. (588,83.5) .. controls (588,83.5) and (583.92,83.14) .. (582,82.5) .. controls (578.14,81.21) and (573.39,78.96) .. (569,77.5) .. controls (568.08,77.19) and (560.21,71.29) .. (559,72.5) .. controls (558.92,72.58) and (560.89,79.99) .. (561,81.5) .. controls (561.31,85.82) and (561.33,90.18) .. (561,94.5) .. controls (560.84,96.6) and (559.94,94.41) .. (560,93.5) .. controls (560.33,88.18) and (560,82.83) .. (560,77.5) .. controls (560,76.75) and (559.05,78.76) .. (559,79.5) .. controls (558.69,84.49) and (559,89.5) .. (559,94.5) .. controls (559,95.17) and (559.31,89.89) .. (561,86.5) .. controls (563.36,81.78) and (559,79.52) .. (559,73.5) .. controls (559,72.5) and (558.76,75.53) .. (559,76.5) .. controls (559.48,78.41) and (560.82,83.41) .. (563,84.5) .. controls (568.48,87.24) and (576.45,90.5) .. (584,90.5) .. controls (586.24,90.5) and (580.09,88.29) .. (578,87.5) .. controls (574.63,86.24) and (570.63,82.59) .. (567,83.5) .. controls (566.28,83.68) and (567.72,84.81) .. (568,85.5) .. controls (570.18,90.96) and (572.75,93.99) .. (575,98.5) .. controls (575.8,100.11) and (578.27,104.77) .. (577,103.5) .. controls (576.74,103.24) and (570.62,93.03) .. (570,95.5) .. controls (568.4,101.91) and (574.31,112.81) .. (578,116.5) .. controls (579.58,118.08) and (581,124.74) .. (581,122.5) .. controls (581,119.74) and (582.2,115.3) .. (584,113.5) .. controls (584.23,113.27) and (587.34,112.16) .. (586,113.5) .. controls (583.69,115.81) and (576.41,125.32) .. (575,122.5) .. controls (572.26,117.02) and (568.99,106.47) .. (567,100.5) .. controls (566.66,99.48) and (565.04,89.46) .. (564,90.5) .. controls (552.87,101.63) and (578,117.21) .. (578,126.5) .. controls (578,127.55) and (577.47,124.44) .. (577,123.5) .. controls (576.8,123.11) and (574.04,122.56) .. (574,122.5) .. controls (573.45,121.67) and (574.32,120.45) .. (574,119.5) .. controls (573.99,119.47) and (566.6,109.1) .. (566,108.5) .. controls (565.67,108.17) and (564.55,107.65) .. (565,107.5) .. controls (565.34,107.39) and (584.35,112.17) .. (585,112.5) .. controls (588.13,114.06) and (588.31,118.81) .. (590,120.5) .. controls (590.77,121.27) and (593.5,116) .. (594,115.5) .. controls (594.03,115.47) and (602.3,111.8) .. (601,110.5) .. controls (599.6,109.1) and (592.96,114.43) .. (594,116.5) .. controls (594.4,117.3) and (595.4,116.3) .. (595,115.5) .. controls (593.99,113.49) and (590.33,113.92) .. (589,114.5) .. controls (584.76,116.36) and (581.14,119.43) .. (577,121.5) .. controls (576.61,121.7) and (574.22,125.38) .. (574,124.5) .. controls (572.3,117.7) and (568.39,108.62) .. (571,99.5) .. controls (571.37,98.22) and (573.71,99.18) .. (575,99.5) .. controls (577.79,100.2) and (586.85,104.27) .. (590,104.5) .. controls (594.32,104.81) and (598.68,104.83) .. (603,104.5) .. controls (604.05,104.42) and (605,103.17) .. (606,103.5) .. controls (607.55,104.02) and (607.17,107.02) .. (607,106.5) .. controls (605.87,103.1) and (618,96.5) .. (618,96.5) .. controls (618,96.5) and (617.19,99.33) .. (616,99.5) .. controls (606.14,100.91) and (599.86,104.53) .. (590,106.5) .. controls (580.06,108.49) and (585.65,98.5) .. (582,98.5) ;
        \draw  [color={rgb, 255:red, 155; green, 155; blue, 155 }  ,draw opacity=0.16 ][line width=6] [line join = round][line cap = round] (572,116.5) .. controls (566.36,116.5) and (564.85,102.5) .. (562,102.5) ;
        \draw  [color={rgb, 255:red, 155; green, 155; blue, 155 }  ,draw opacity=0.1 ][line width=6] [line join = round][line cap = round] (578,79.5) .. controls (574.16,79.5) and (572.78,76.5) .. (570,76.5) ;
        \draw  [color={rgb, 255:red, 155; green, 155; blue, 155 }  ,draw opacity=0.05 ][line width=6] [line join = round][line cap = round] (612,101.5) .. controls (612,101.83) and (612,102.17) .. (612,102.5) ;
        \draw  [fill={rgb, 255:red, 0; green, 0; blue, 0 }  ,fill opacity=1 ] (366.5,64) .. controls (366.5,62.07) and (368.07,60.5) .. (370,60.5) .. controls (371.93,60.5) and (373.5,62.07) .. (373.5,64) .. controls (373.5,65.93) and (371.93,67.5) .. (370,67.5) .. controls (368.07,67.5) and (366.5,65.93) .. (366.5,64) -- cycle ;
        
        \draw (158,124.4) node [anchor=north west][inner sep=0.75pt]    {$s$};
        \draw (365,169.4) node [anchor=north west][inner sep=0.75pt]    {$s_{1}$};
        \draw (231,46.4) node [anchor=north west][inner sep=0.75pt]    {$s_{2}$};
        \draw (480,164.4) node [anchor=north west][inner sep=0.75pt]    {$s_{1}$};
        \draw (636,161.4) node [anchor=north west][inner sep=0.75pt]    {$s_{2}$};
        \draw (563,19.4) node [anchor=north west][inner sep=0.75pt]    {$s_{3}$};

        \end{tikzpicture}
        
    \caption{n-simplices for $n = 1,2,3$}
    \label{fig:HoloSimplex}
\end{figure}

Let us begin by providing a relation between the global connection form $\omega$ living in the total space $P$ and the Yang-Mills connection defined via a local section on an open set in $\mc{M}$. First, as mentioned above, suppose there is a curve $\gamma: [a, b]\rightarrow M$ and two of its lifts onto the total space $P$; $\,\alpha^\uparrow, \beta : [a,b] \rightarrow
 P$, such that $\pi(\alpha^\uparrow) = \pi(\beta) = \gamma$. However, as indicated by the arrow, only $\alpha^\uparrow$ is a horizontal lift of $\gamma$ from $\mc{M}$ onto $P$. There is some unique function such that for all $t\in [a,b]$,
 \begin{equation}
     \alpha^{\uparrow}(t) = \beta(t)g(t), g \in G,
 \end{equation}
 then quoting the result from \cite{Isham:1999qu}, and specialising to a matrix group $G$ we have the expression for the global connection
 \begin{equation}
     \omega([\alpha^\uparrow]) = g(t)^{-1}\omega_{\beta(t)}([\beta])g(t) + g(t)^{-1}\frac{dg}{dt}(t),
 \end{equation}
but from \eqref{HoriDef} we know $\omega[\alpha^\uparrow] = 0$, i.e., the connection is a vertical form. Additionally, if $\sigma : U \rightarrow P$ is a section associated with a coordinate chart on an open set $U$, we can write the lift $\beta(t) = \sigma(\gamma(t))$. This then allows us to write $\omega(\beta(t)) = \omega(\sigma(\gamma(t))) = \sigma^*(\omega(\gamma(t)))$, which on a coordinate chart can be written as $\sigma^*(\omega) = \Gamma_\mu dx^\mu$. Thus we have reduced the problem of finding a function $g(t)$ that takes a lift $\beta(t)$ of $\gamma(t)$ to the horizontal lift $\alpha(t)^\uparrow$ to solving an ODE of the form
 \begin{equation}\label{ODEConn}
    \dot{g}(t) = -\Gamma_{\mu}(\gamma(t))\,\dot{\gamma}(t)^\mu g(t),
 \end{equation}
 where $\dot{\gamma} (t) = d\gamma (t)/dt$ and similarly for $\dot{g}(t)$. This equation is slightly tricky to solve, since from eq. \eqref{GValConn} we know the connection is a Lie algebra valued form and also the dependence on the curve parameter $t$ is also in the connection. A way to obtain a formal solution is to start with the initial condition $g(0) = g_0$ and write down the ansatz,
 \begin{equation}
     g(t) = g_0 - \int^t_a ds \,\Gamma_\mu(\gamma(s))\dot{\gamma}(s)g(s)\,.
 \end{equation}
This solution does indeed satisfy the initial condition set above, but unfortunately we have just shifted the dependence on the curve parameter to the second term. However, notice that we can plug the solution $g(t)$ into its second term's $g(s)$. This is exactly the trick needed to obtain an explicit solution of eq. \eqref{ODEConn}. Doing so repeatedly gets us
\begin{equation}\label{solPara}\begin{split}
    g(t) = g_0 &- \int^t_a ds\, \Gamma_\mu(\gamma(s))\dot{\gamma}(s)g(0)\\
     &+ \int^t_a ds_1\int^{s_1}_a ds_2\, \Gamma_{\mu_1}(\gamma(s_1))\Gamma_{\mu_2}(\gamma(s_2))\dot{\gamma}(s_1)\dot{\gamma}(s_2)g(s_1) + \cdots,
\end{split}
\end{equation}
which in fact can be interpreted as integral over $n$-simplices(see figure \ref{fig:HoloSimplex}). We can also integrate over $n$-cubes instead of simplices; however, dividing each $n$th term by $1/n!$ to take into account the extra volume of the $n$-cube. This will make the solution look like a formal expansion of an exponential function. Additionally, since each term in eq. \eqref{solPara} is Lie algebra valued, one has to be careful with ordering the integrals. This is done by doing each integral in increasing parameter value such that $s_n \geq s_{n-1} \geq \cdots \geq s_1$.
Taking the above steps into account, we write the following solution to eq. \eqref{ODEConn},
\begin{equation}
    g(t) = \mc{P}\exp \Big[   - \int_a^t ds\,\Gamma_\mu (\gamma(s))\dot{\gamma}(s) \Big]g_0,
\end{equation}
where the symbol $\mc{P}$ denotes the path ordered exponential as defined above in terms of the curve parameter value. Therefore, we can write the horizontal lift of the curve $\gamma$ in terms of its natural lift times the function which lifts it to its horizontal lift
\begin{equation}
    \begin{split}
    \gamma^\uparrow(t) &= \sigma(\gamma(t))\mc{P}\exp \Big[   - \int_a^t ds\,\Gamma_\mu (\gamma(s))\dot{\gamma}(s) \Big]g_0\\
    &= \sigma(\gamma(t))\Big( \mathds{1}  + \sum^\infty_{n = 1}  \frac{1}{n!} \int^t_a d^ns \mc{P} [\Gamma(s_n)\Gamma(_{s_{n-1}})\cdots \Gamma_{s_1}]  \Big)g_0,
    \end{split}
\end{equation}
where the second equality denotes the path-ordered exponential. Now we can give precise meaning to parallel transportation:
\begin{definition}
    Let $\alpha : [a,b]\rightarrow \mc{M}$ be a curved in $\mc{M}$. The parallel transportation along $\alpha$ is the map $\tau : \pi^{-1}(\{ \alpha(a) \}) \rightarrow \pi^{-1}(\{\alpha(b)\})$ obtained by associating with each point $p\in \pi^{-1}(\{ \alpha(a)\})$ the point $\alpha^\uparrow(b)\in \pi^{-1}(\{\alpha(b)\})$ where $\alpha^\uparrow$ is the unique horizontal lift pf $\alpha$ that passes through $p$ at $t=a$.
\end{definition}
An important case, as we shall see in the next section, is when $\gamma(a) = \gamma(b)$, i.e., a loop based at $\gamma(a)$. The parallel transport map then induces a map on the fibre above the point $\gamma(a)\in \mc{M}$:
\begin{equation}\label{NotClose}
    \begin{split}
        \tau &: \pi^{-1}(\{ \alpha(a) \}) \rightarrow \pi^{-1}(\{\alpha(a)\})\\
        p &\mapsto p \Big( \mc{P}\exp \Big[   - \oint_\gamma ds\,\Gamma_\mu (\gamma(s))\dot{\gamma}(s) \Big]\Big)
    \end{split}
\end{equation}
This is called the holonomy of the connection. We have a natural map from the loop space of $\mc{M}$ into the structure group $G$. The subgroup of all elements of $G$ that can be obtained in this way is called the \emph{holonomy group} of the bundle at $\gamma(a)\in \mc{M}$. Notice the fact that despite having a $\gamma$ loop on the base manifold $\mc{M}$, its horizontal lift $\gamma^{\uparrow}$ does not necessarily close. This difference is precisely given by eq. \eqref{NotClose}(see figure \ref{fig:noClose} ). More on this in the next section when we dive deeper into the concept of holonomy. First, let us briefly discuss how one extends the notion of parallel transportation to associated bundles, since as mentioned earlier they are used in physics to describe matter, such as the charged particles in a $\text{U}(1)$ gauge theory. Then we shall end this section by defining the covariant derivative using the notion of parallel transportation.
\begin{figure}[t]
    \centering
    \begin{tikzpicture}[x=0.75pt,y=0.75pt,yscale=-1,xscale=1]
        
        \draw   (170,198.5) .. controls (188,179.5) and (377,182.5) .. (404,207.5) .. controls (431,232.5) and (425,277.5) .. (392,274.5) .. controls (359,271.5) and (363.67,271.67) .. (314,264.5) .. controls (264.33,257.33) and (207.81,269.6) .. (195,265.5) .. controls (182.19,261.4) and (152,217.5) .. (170,198.5) -- cycle ;
        \draw    (227,205.5) .. controls (227,169) and (245,149.5) .. (255,114.5) ;
        \draw    (286,213.5) .. controls (286,177) and (304,157.5) .. (314,122.5) ;
        \draw    (294,232.5) .. controls (294,196) and (312,176.5) .. (322,141.5) ;
        \draw    (239,239.5) .. controls (239,203) and (269,166.5) .. (279,131.5) ;
        \draw    (227,205.5) .. controls (259.81,199.71) and (270.27,204.16) .. (284.44,212.57) ;
        \draw [shift={(286,213.5)}, rotate = 210.96] [color={rgb, 255:red, 0; green, 0; blue, 0 }  ][line width=0.75]    (10.93,-3.29) .. controls (6.95,-1.4) and (3.31,-0.3) .. (0,0) .. controls (3.31,0.3) and (6.95,1.4) .. (10.93,3.29)   ;
        \draw    (294,232.5) .. controls (295,221.5) and (291,218.5) .. (286,213.5) ;
        \draw    (239,239.5) .. controls (255,246) and (277,247) .. (294,232.5) ;
        \draw    (239,239.5) .. controls (228,236) and (194,219) .. (227,205.5) ;
        \draw [shift={(227,205.5)}, rotate = 337.75] [color={rgb, 255:red, 0; green, 0; blue, 0 }  ][fill={rgb, 255:red, 0; green, 0; blue, 0 }  ][line width=0.75]      (0, 0) circle [x radius= 3.35, y radius= 3.35]   ;
        \draw   (260.17,247.79) -- (250.27,242.97) -- (261,240.5) ;
        \draw  [fill={rgb, 255:red, 15; green, 107; blue, 224 }  ,fill opacity=1 ] (233.5,161.25) .. controls (233.5,159.46) and (234.96,158) .. (236.75,158) .. controls (238.54,158) and (240,159.46) .. (240,161.25) .. controls (240,163.04) and (238.54,164.5) .. (236.75,164.5) .. controls (234.96,164.5) and (233.5,163.04) .. (233.5,161.25) -- cycle ;
        \draw  [fill={rgb, 255:red, 255; green, 255; blue, 255 }  ,fill opacity=1 ] (247.5,126.25) .. controls (247.5,124.46) and (248.96,123) .. (250.75,123) .. controls (252.54,123) and (254,124.46) .. (254,126.25) .. controls (254,128.04) and (252.54,129.5) .. (250.75,129.5) .. controls (248.96,129.5) and (247.5,128.04) .. (247.5,126.25) -- cycle ;
        \draw [color={rgb, 255:red, 15; green, 107; blue, 224 }  ,draw opacity=1 ]   (236.75,161.25) .. controls (253,140) and (278,144) .. (303,151) ;
        \draw [color={rgb, 255:red, 15; green, 107; blue, 224 }  ,draw opacity=1 ]   (303,151) .. controls (312,155.5) and (313,161) .. (313,168) ;
        \draw [color={rgb, 255:red, 15; green, 107; blue, 224 }  ,draw opacity=1 ]   (313,168) .. controls (311,179.5) and (276,177.5) .. (264,160) ;
        \draw [color={rgb, 255:red, 15; green, 107; blue, 224 }  ,draw opacity=1 ]   (264,160) .. controls (259,152.5) and (252,144.75) .. (250.75,129.5) ;
        \draw  [color={rgb, 255:red, 15; green, 107; blue, 224 }  ,draw opacity=1 ] (282.45,140.66) -- (291.79,148.01) -- (279.98,149.32) ;
        \draw  [color={rgb, 255:red, 15; green, 107; blue, 224 }  ,draw opacity=1 ] (260.07,161.59) -- (257.28,150.04) -- (267.37,156.32) ;
        
        \draw (216,153.4) node [anchor=north west][inner sep=0.75pt]    {$p$};
        \draw (227,112.4) node [anchor=north west][inner sep=0.75pt]    {$p^{\uparrow }$};
        \draw (209,193.4) node [anchor=north west][inner sep=0.75pt]    {$x$};
        \draw (385,239.4) node [anchor=north west][inner sep=0.75pt]    {$\mathcal{M}$};
        \draw (296,235.9) node [anchor=north west][inner sep=0.75pt]    {$\gamma $};
        \draw (289,117.4) node [anchor=north west][inner sep=0.75pt]  [color={rgb, 255:red, 15; green, 107; blue, 224 }  ,opacity=1 ]  {$\gamma ^{\uparrow }$};

        \end{tikzpicture}
        
        \caption{Local view of the fibre bundle where the lines represent the fibre. $\gamma^\uparrow$ is the horizontal curve of the loop $\gamma$ on $\mc{M}$.}
        \label{fig:noClose}
\end{figure}
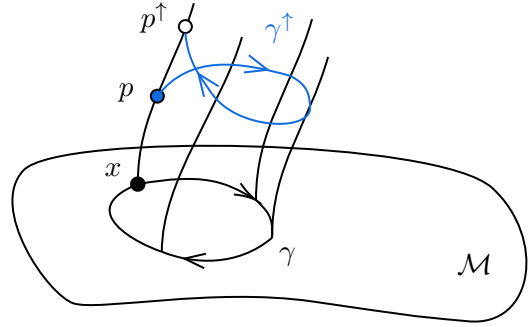
\subsubsection{Parallel transport in associated bundles}
\begin{figure}[h]
    \centering
    \begin{tikzpicture}[x=0.75pt,y=0.75pt,yscale=-1,xscale=1]
        
        \draw   (208.44,245.75) .. controls (227.53,230.98) and (254.98,219) .. (269.75,219) -- (490.12,219) .. controls (490.12,219) and (490.12,219) .. (490.12,219) -- (455.56,245.75) .. controls (436.47,260.52) and (409.02,272.5) .. (394.25,272.5) -- (173.88,272.5) .. controls (173.88,272.5) and (173.88,272.5) .. (173.88,272.5) -- cycle ;
        \draw    (268,136.5) -- (284,242) ;
        \draw    (375,129.5) -- (362,239.5) ;
        \draw    (289,273) -- (297,313.5) ;
        \draw    (359,272) -- (356,319.5) ;
        \draw [color={rgb, 255:red, 65; green, 117; blue, 5 }  ,draw opacity=1 ]   (269,246) .. controls (298,237) and (341,230) .. (373,243.5) ;
        \draw  [color={rgb, 255:red, 65; green, 117; blue, 5 }  ,draw opacity=1 ] (323,233) -- (334,236.25) -- (323,239.5) ;
        \draw [line width=0.75]    (284,242) -- (311.26,226.49) ;
        \draw [shift={(313,225.5)}, rotate = 150.36] [color={rgb, 255:red, 0; green, 0; blue, 0 }  ][line width=0.75]    (6.56,-1.97) .. controls (4.17,-0.84) and (1.99,-0.18) .. (0,0) .. controls (1.99,0.18) and (4.17,0.84) .. (6.56,1.97)   ;
        \draw  [fill={rgb, 255:red, 0; green, 0; blue, 0 }  ,fill opacity=1 ] (275.5,203.75) .. controls (275.5,202.23) and (276.73,201) .. (278.25,201) .. controls (279.77,201) and (281,202.23) .. (281,203.75) .. controls (281,205.27) and (279.77,206.5) .. (278.25,206.5) .. controls (276.73,206.5) and (275.5,205.27) .. (275.5,203.75) -- cycle ;
        \draw  [fill={rgb, 255:red, 0; green, 0; blue, 0 }  ,fill opacity=1 ] (370.5,142.75) .. controls (370.5,141.23) and (371.73,140) .. (373.25,140) .. controls (374.77,140) and (376,141.23) .. (376,142.75) .. controls (376,144.27) and (374.77,145.5) .. (373.25,145.5) .. controls (371.73,145.5) and (370.5,144.27) .. (370.5,142.75) -- cycle ;
        \draw  [fill={rgb, 255:red, 255; green, 255; blue, 255 }  ,fill opacity=1 ] (367.5,168.75) .. controls (367.5,167.23) and (368.73,166) .. (370.25,166) .. controls (371.77,166) and (373,167.23) .. (373,168.75) .. controls (373,170.27) and (371.77,171.5) .. (370.25,171.5) .. controls (368.73,171.5) and (367.5,170.27) .. (367.5,168.75) -- cycle ;
        \draw [color={rgb, 255:red, 15; green, 107; blue, 224 }  ,draw opacity=1 ]   (278.25,203.75) .. controls (318.25,173.75) and (338,221.5) .. (370.25,171.5) ;
        \draw   (319.13,191.79) -- (330,195.46) -- (318.88,198.29) ;
        
        \draw (304,244.4) node [anchor=north west][inner sep=0.75pt]  [color={rgb, 255:red, 65; green, 117; blue, 5 }  ,opacity=1 ]  {$\gamma $};
        \draw (293,167.4) node [anchor=north west][inner sep=0.75pt]  [color={rgb, 255:red, 15; green, 107; blue, 224 }  ,opacity=1 ]  {$\gamma ^{\uparrow _{F}}$};
        \draw (410,237.4) node [anchor=north west][inner sep=0.75pt]    {$\mathcal{M}$};
        \draw (250,128.4) node [anchor=north west][inner sep=0.75pt]    {$F$};
        \draw (382,118.4) node [anchor=north west][inner sep=0.75pt]    {$F$};
        \draw (285,219.4) node [anchor=north west][inner sep=0.75pt]    {$X$};
        \draw (195,179.4) node [anchor=north west][inner sep=0.75pt]    {$P_{F}$};
        \draw (339,247.4) node [anchor=north west][inner sep=0.75pt]  [font=\scriptsize]  {$\pi ^{-1}(\{\gamma ( b)\})$};
        \draw (380.94,139.62) node [anchor=north west][inner sep=0.75pt]  [font=\LARGE,rotate=-5.48]  {$\}$};
        \draw (395,150.4) node [anchor=north west][inner sep=0.75pt]    {$\nabla _{X} \psi $};

        \end{tikzpicture}
        \caption{An illustration of the covariant derivative along the tangent vector $X$ along $\gamma$. Note how the sections can be subtracted since $F$ is a vector space.}
        \label{fig:CovVec}
\end{figure}
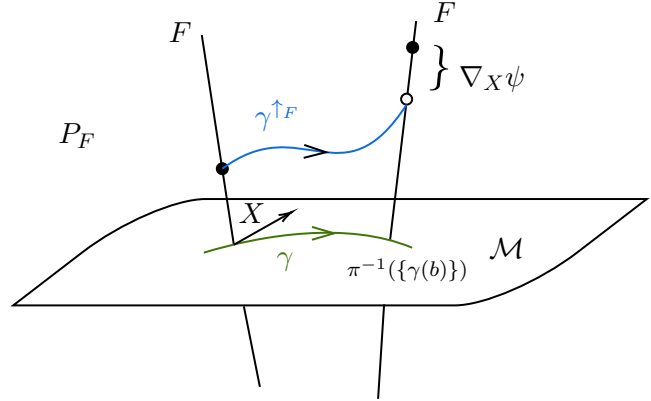

Fortunately, it is straightforward to transfer the previous results to associated bundles. They rest on the following important definition:
\begin{definition}
    Let $P \xrightarrow{\pi} \mc{M}$ be a $G$-principal bundle equipped with a connection one-form on the total space $P$. Let further $P_F \xrightarrow{\pi_F} \mc{M}$ be the associated bundle with fibre $F$ upon which $G$ has a left action.

    Now let $\gamma: [a,b]\rightarrow \mc{M}$ be a curve on $\mc{M}$ and let $\gamma^\uparrow : [a,b] \rightarrow P$ be the horizontal lift of $\gamma$ through $p\in \pi^{-1}\{(\gamma(a))\}$ on P. Then the horizontal lift of $\gamma$ to the associated bundle that passes through $[p, v] \in P_f$ is the curve $\gamma^{\uparrow_f} : [a,b]\rightarrow P_f, \gamma^{\uparrow_f}(t):= [\gamma^{\uparrow}(t), v]$.
\end{definition}
Now we can define the parallel transport map for associated bundles
\begin{definition}
    The parallel transport map on $P_f$ is the map
    \begin{equation}
    \begin{split}
        \tau_F: \pi^{-1}(\{\gamma(a) \}) &\rightarrow  \pi^{-1}(\{\gamma(b) \})\\
        [p,v] &\mapsto \gamma^{\uparrow_F}(b)
    \end{split}
    \end{equation}
\end{definition}
Now, as mentioned before, matter fields are described by associated bundles, but that is not the whole story. One needs to specialise to associated bundles that are vector bundles. This means that the fibres now carry the structure of a vector space upon which the structure group has a linear action.
\begin{definition}
    If $F$ is a vector space $(F,+,\cdot)$ and the action $G\times F\rightarrow F$ is linear, then $P_F$ is called a vector bundle.
\end{definition}
This allows us to define the notion of a covariant derivative; the same one that appears in general relativity and gauge theories as the gauge covariant derivative. Parallel transportation helps us compare sections, such as matter fields, living in different fibres. Using the connection, we can pull back the fibre over the second point to the fibre over the first point, and then subtract the result. More precisely
\begin{definition}
    Let $(P,\pi,\mc{M})$ be a principal $G$-bundle and let $F$ be a vector space that carries a linear representation of $G$. Let $\gamma: [0,\epsilon] \rightarrow \mc{M}, \epsilon > 0 $, be a curve in $\mc{M}$ such that $\gamma(0) = x_0\in \mc{M}$, and let $\psi: \mc{M}\rightarrow P_F$ be a section of the associated \emph{vector} bundle. Then the covariant derivative of $\psi$ in the direction $\gamma$ at $x_0$, is
    \begin{equation}
        \nabla_\gamma\psi:= \lim_{t\mapsto 0}\bigg(       \frac{\tau_F\psi(\gamma(t))-\psi(x_0)}{t}           \bigg) \in\pi^{-1}(\{x_0\})
    \end{equation}
    where $\tau_F$ is the linear parallel transport map from the vector space $\pi_F^{-1}(\{\gamma(t)\})$ to the vector space $\pi_F^{-1}(\{ x_0 \})$.
\end{definition}
Let us obtain an explicit expression for the covariant derivative of a local section of a vector bundle. Let $\psi(x)$ be the function representing a local section $U\subset \mc{M}\rightarrow F$. Now the element of $F$ that represents $\tau_F\psi(\gamma(t))$ is $g(t)^{-1}\psi(\gamma(t))$, and the local representative of $\nabla_\gamma\psi$ is given by
\begin{equation}
    \begin{split}
    \frac{d}{dt}\big(    g(t)^{-1}\psi(\gamma(t))   \big)\Big|_{t=0} &= \Bigg(       \frac{d}{dt}(g^{-1}(t))\psi(x_0) + g^{-1}(t)\frac{d\psi(\gamma(t))}{dt}         \Bigg)|_{t=0}\\
    &=\sum^m_{\mu=1}(\pd_\mu\psi(x_0) + \Gamma_{\mu}(x_0))\dot{\gamma}^\mu\Big|_{t=0}
    \end{split}
\end{equation}
with $g(0) = \mathds{1}$ and $\gamma(0) = x_0$, and in the last equality with have used the identity $d/dt (g(t)g(t)^{-1})=0$ and eq. \eqref{ODEConn} which relates $g(t)$ to the local Yang-Mills connection. Indeed, this is the familiar covariant derivative which satisfy the usual conditions of linearity, distributivity, and Leibniz. A special case when we one takes the covariant derivative along a tangent vector $X\in T_x\mc{M}$ to $\gamma$ as illustrated in figure \ref{fig:CovVec}.
\subsection{Holonomy groups}
We have mentioned the holonomy group under eq. \eqref{NotClose} in the context of a loop on the base manifold that does not close once lifted onto the total space. Now we give a formal definition of the holonomy group, and then specialise to the case of Riemannian holonomy groups, since this is the type of manifold we use to model spacetime in this thesis.
\begin{definition}
    Let $(P,\pi,\mc{M})$ be a $G$-principal bundle. Take a point $p\in P$ such that $\pi(p) = x\in \mc{M}$ and consider the set of loop $C_x(\mc{M})$ at $x$; $C_x(\mc{M}):=\{ \gamma: [0,1]\rightarrow \mc{M}| \gamma(0)=\gamma(1)=x \}$. The set of elements
    \begin{equation}
        \hol_p := \{ g\in G | \tau_{(\gamma)}(p) = pg, \gamma\in C_x(\mc{M}) \},
    \end{equation}
    where $\tau_{(\gamma)}$ denotes the parallel transport map associated to the loop $\gamma$.
\end{definition}
Sometime we shall drop the dependence on the path when it is clear to do so. Note that one can show that the parallel transport map satisfies the following. If $\alpha$ and $\beta$ are loops based at $x\in \mc{M}$ then so are $\alpha^{-1}$ and $\beta\alpha$. Moreover, the parallel transport map is in $G$ and satisfy $\tau_{(\alpha^{-1})} = \tau_{\alpha}^{-1}$ and $\tau_{(\beta\alpha)} = \tau_{(\beta)}\circ \tau_{(\alpha)}$; hence justifying calling $\hol_x$ a group \cite{joyce2000CompactMW}. Moreover, if $p,q\in P$ and $p \sim q=pg$ the the holonomy group at $p$ and the other one at $q$ are related by conjugacy\cite{joyce2000CompactMW,Nakahara:2003nw},
\begin{equation}\label{PHolConj}
    \hol_q = \hol_{pg} = g^{-1}\hol_p \,g.
\end{equation}
Therefore, the holonomy groups at different point are isomorphic in the sense that they are in the same conjugacy class
\begin{equation}
    \hol_q \sim g^{-1}\hol_p \,g.
\end{equation}
Therefore, it does not depend on the base point, and hence provides some topological information about the manifold. One can also define the notion of the \emph{restricted} holonomy group by
\begin{equation}
    \hol_0 := \{g\in G| \tau_{(\gamma)}(p) = pg, \gamma\in C^{0}_x(\mc{M}) \},
\end{equation}
where $C^{0}_x(\mc{M})$ is the set of loops at $x\in\mc{M}$ which are holomorphic to the constant loop, i.e., they can be contracted to a point. Until know, we have discussed holonomy groups of the point of view of principal bundle' however, as mentioned in previous section, in physics we often deal with theories that include matter fields, and these are described by vector bundles. Therefore, it is natural to understand holonomy groups in terms of vector bundles. Let us start with the definition the holonomy group
\begin{definition}
    Let $(E,\pi,\mc{M})$ be a vector bundle over $\mc{M}$
 equipped with a connection $\nabla^{E}$. Fix a point $x\in \mc{M}$ and let $\gamma: [0,1] \rightarrow \mc{M}$ be a loop based at $x$. The parallel transport map $\tau_{(\gamma)}:E_x\rightarrow E_x$ is an invertible linear map on $E_x$, thus $\tau_{(\gamma)}$ lies in $\text{GL}(E_X)$, the group of invertible linear transformations on $E_x$. The holonomy group $\hol_x(\nabla^E)$ at $x\in\mc{M}$ is defined by
 \begin{equation}
     \hol_x(\nabla^E):= \{  \tau_{(\gamma)}|\gamma\in C_x(\mc{M})   \}\in \text{GL}(E_x).
 \end{equation}
\end{definition}
We also have an analogue of eq. \eqref{PHolConj} in the vector bundle case. For a connected manifold $\mc{M}$, let $\gamma: [0,1]\rightarrow \mc{M}$ such that $\gamma(0)=x$ and $\gamma(1)=y$. Now if $\alpha$ is a \emph{loop} at x, then $\gamma\alpha\gamma^{-1}$ is a loop at $y$; and $\tau_{(\alpha)}$ and $\tau_{(\gamma\alpha\gamma^{-1})} = \tau_{(\gamma)}\tau_{(\alpha)}\tau_{(\gamma)}^{-1}$. Therefore, if $\tau_{(\alpha)}\in \hol_x(\nabla^E)$, then $\tau_{(\gamma)}\circ\tau_{(\alpha)}\circ\tau_{(\gamma)}^{-1}\in\hol_y(\nabla^E)$. Thus
\begin{equation}\label{Matconj}
    \tau_{(\gamma)}\,\hol_x(\nabla^E)\,\tau_{(\gamma)}^{-1} = \hol_y(\nabla^E).
\end{equation}
For example of the bundle has fibre $\mathbb{R}^m$ and we make the identification $E_x\simeq \mathbb{R}^m$; this induces the isomorphism $\text{GL}(E_x)\simeq \text{GL}(m,\mathbb{R})$, the general linear group of matrices. Therefore, eq. \eqref{Matconj} becomes matrix conjugation. Then the holonomy groups at different point in $\mc{M}$ are isomorphic to each other in the sense that they are related by conjugation. Hence, we can drop the point $x$ label from $\hol_x(\nabla^E)$. The restricted holonomy group is defined similarly with null-homotopic loops. Note that in the case of a simply-connected manifold $\mc{M}$, we have the canonical homomorphism $\pi_1(\mc{M}) \rightarrow \hol/\hol_0$, where $\pi_1(\mc{M})$ is the fundamental homotopy group of $\mc{M}$. Let us also introduce the Lie algebra of the holonomy group, since it plays a role in the relation between the curvature and the holonomy group, as given by Ambrose-Singer theorem \cite{MR63739},
\begin{definition}
    Let $\mc{M}$ be a manifold, $E$ a vector bundle over $\mc{M}$ with a connection $\nabla^E$ over $E$. Then $\hol_0(\nabla^E)$ is the restricted holonomy group defined up to conjugation. Define the \emph{holonomy algebra} $\ghol(\nabla^E)$ to be the lie algebra of $\hol(\nabla^E)$ as a Lie subalgebra of $\mathfrak{gl}(k,\mathbb{R})$ defined up to the adjoint action of $\text{GL}(m,\mathbb{R})$.
\end{definition}
In \cite{MR63739}, Ambrose and Singer proved a relation between the curvature of the connection and its holonomy algebra. This is not surprising since the geometric definition of curvature considers taking a vector around a parallelogram, which can be done by parallel transporting the same vector around the same parallelogram in the small loop limit. Here we state the relevant theory for vector bundles
\begin{theorem}\label{Ambrose-Singer}
    Let $\mc{M}$ be a manifold, $E$ a vector bundle over $\mc{M}$, and $\nabla^E$ a connection on $E$. Fix $x\in \mc{M}$ so that $\ghol_x(\nabla^E)$ is a Lie subalgebra of the endomorphisms of the fibre $E_x$, $\text{End}(E_x)$; then $\ghol_x(\nabla^E)$ is the vector subspace of $\text{End}(E_x)$ spanned by all elements of $\text{End}(E_x)$ of the form $\tau^{-1}_{(\gamma)}\,R(\nabla^{E})_y(X,Y)\, \tau_{(\gamma)}$, where $x\in \mc{M}$, $\gamma:[0,1]\rightarrow \mc{M}$ is a piecewise connected curve such that $\gamma(0)=x$ and $\gamma(1)= y$, and $X,Y\in T_y\mc{M}$.
\end{theorem}
This theorem shows that the curvature associated to the connection, $R(\nabla^E)$, determines $\ghol(\nabla^E)$, and hence $\hol^0(\nabla)$. Just to showcase how the curvature is related to the holonomy; compare the situation in \ref{fig:noClose} where the horizontal lift of a \emph{loop} $\gamma$ is not closing. Now recall the fact that curvature measure the difference of a vector upon being taken around a small loop. Let us see what happens from the perspective of principal bundles. The definition of curvature of a principal $G$-bundle $(P,\pi,\mc{M})$ with connection one-form $\omega$ is
\begin{equation}
    \Omega := d\omega \circ \text{hor},
\end{equation}
where $d$ is the usual exterior derivative of forms and $\text{hor}$ is the operation representing the horizontal component of the arguments of $d\omega$. Geometrically, this measures the amount of the failure of two horizontal arguments, say vectors on $\mc{M}$, to stay horizontal once commuted (see figure \ref{fig:CurvaHol} below). If one chooses a local section, and pull-back the curvature onto the base manifold $\mc{M}$; in local coordinate the curvature takes the familiar form,
\begin{equation}
    R^a_{\mu\nu} = \frac{1}{2}(\pd_\mu\Gamma^a_\nu - \pd_\nu\Gamma^a_\mu + \sum^{\text{dim G}}_{b,c = 1}C^a{}_{bc}\Gamma^b_\mu\Gamma^c_\nu),
\end{equation}
where $C^a_{bc}$ is the structure constants of the structure group $G$, and $\Gamma^a_{\mu}$ is the local Yang-Mills connection on $U\subset \mc{M}$.
\begin{figure}[t]
    \centering
    \begin{tikzpicture}[x=0.75pt,y=0.75pt,yscale=-1,xscale=1]
        
        \draw   (306.9,258) -- (470,258) -- (400.1,306) -- (237,306) -- cycle ;
        \draw  [dash pattern={on 0.84pt off 2.51pt}]  (237,89) -- (237,306) ;
        \draw  [dash pattern={on 0.84pt off 2.51pt}]  (306.9,93) -- (306.9,258) ;
        \draw  [dash pattern={on 0.84pt off 2.51pt}]  (400.1,91) -- (400.1,306) ;
        \draw  [dash pattern={on 0.84pt off 2.51pt}]  (470,93) -- (470,258) ;
        \draw [color={rgb, 255:red, 15; green, 107; blue, 224 }  ,draw opacity=1 ]   (237,220) .. controls (336,218) and (327,222) .. (399,237) ;
        \draw [shift={(324.05,221.6)}, rotate = 185.98] [fill={rgb, 255:red, 15; green, 107; blue, 224 }  ,fill opacity=1 ][line width=0.08]  [draw opacity=0] (10.72,-5.15) -- (0,0) -- (10.72,5.15) -- (7.12,0) -- cycle    ;
        \draw [color={rgb, 255:red, 15; green, 107; blue, 224 }  ,draw opacity=1 ]   (399,237) .. controls (423,194) and (433,186) .. (470,159) ;
        \draw [shift={(432.4,189.14)}, rotate = 133.77] [fill={rgb, 255:red, 15; green, 107; blue, 224 }  ,fill opacity=1 ][line width=0.08]  [draw opacity=0] (10.72,-5.15) -- (0,0) -- (10.72,5.15) -- (7.12,0) -- cycle    ;
        \draw [color={rgb, 255:red, 15; green, 107; blue, 224 }  ,draw opacity=1 ]   (307,103) .. controls (380,117) and (380,117) .. (470,159) ;
        \draw [shift={(385.2,121.58)}, rotate = 20.24] [fill={rgb, 255:red, 15; green, 107; blue, 224 }  ,fill opacity=1 ][line width=0.08]  [draw opacity=0] (10.72,-5.15) -- (0,0) -- (10.72,5.15) -- (7.12,0) -- cycle    ;
        \draw [color={rgb, 255:red, 15; green, 107; blue, 224 }  ,draw opacity=1 ]   (237,136) .. controls (269,112) and (270,112) .. (307,103) ;
        \draw [shift={(264,117.07)}, rotate = 332.02] [fill={rgb, 255:red, 15; green, 107; blue, 224 }  ,fill opacity=1 ][line width=0.08]  [draw opacity=0] (10.72,-5.15) -- (0,0) -- (10.72,5.15) -- (7.12,0) -- cycle    ;
        \draw    (237,220) -- (237,136) ;
        \draw [shift={(237,136)}, rotate = 270] [color={rgb, 255:red, 0; green, 0; blue, 0 }  ][fill={rgb, 255:red, 0; green, 0; blue, 0 }  ][line width=0.75]      (0, 0) circle [x radius= 3.35, y radius= 3.35]   ;
        \draw [shift={(237,220)}, rotate = 270] [color={rgb, 255:red, 0; green, 0; blue, 0 }  ][fill={rgb, 255:red, 0; green, 0; blue, 0 }  ][line width=0.75]      (0, 0) circle [x radius= 3.35, y radius= 3.35]   ;
        \draw    (237,306) -- (326.01,296.71) ;
        \draw [shift={(328,296.5)}, rotate = 174.04] [color={rgb, 255:red, 0; green, 0; blue, 0 }  ][line width=0.75]    (10.93,-3.29) .. controls (6.95,-1.4) and (3.31,-0.3) .. (0,0) .. controls (3.31,0.3) and (6.95,1.4) .. (10.93,3.29)   ;
        \draw    (237,306) -- (288.09,254.91) ;
        \draw [shift={(289.5,253.5)}, rotate = 135] [color={rgb, 255:red, 0; green, 0; blue, 0 }  ][line width=0.75]    (10.93,-3.29) .. controls (6.95,-1.4) and (3.31,-0.3) .. (0,0) .. controls (3.31,0.3) and (6.95,1.4) .. (10.93,3.29)   ;
        \draw   (338,301) -- (349,305.5) -- (338,310) ;
        \draw   (429.42,279.91) -- (441.03,277.38) -- (434.52,287.32) ;
        \draw   (379.96,263.05) -- (369,258.45) -- (380.04,254.05) ;
        \draw   (283.58,277.84) -- (272.58,282.36) -- (277.26,271.43) ;
        \draw  [fill={rgb, 255:red, 15; green, 107; blue, 224 }  ,fill opacity=1 ] (398,236) .. controls (398,235.17) and (398.67,234.5) .. (399.5,234.5) .. controls (400.33,234.5) and (401,235.17) .. (401,236) .. controls (401,236.83) and (400.33,237.5) .. (399.5,237.5) .. controls (398.67,237.5) and (398,236.83) .. (398,236) -- cycle ;
        \draw  [fill={rgb, 255:red, 15; green, 107; blue, 224 }  ,fill opacity=1 ] (468.5,158.5) .. controls (468.5,157.67) and (469.17,157) .. (470,157) .. controls (470.83,157) and (471.5,157.67) .. (471.5,158.5) .. controls (471.5,159.33) and (470.83,160) .. (470,160) .. controls (469.17,160) and (468.5,159.33) .. (468.5,158.5) -- cycle ;
        \draw  [fill={rgb, 255:red, 15; green, 107; blue, 224 }  ,fill opacity=1 ] (305.5,103) .. controls (305.5,102.17) and (306.17,101.5) .. (307,101.5) .. controls (307.83,101.5) and (308.5,102.17) .. (308.5,103) .. controls (308.5,103.83) and (307.83,104.5) .. (307,104.5) .. controls (306.17,104.5) and (305.5,103.83) .. (305.5,103) -- cycle ;
        
        \draw (333,283.4) node [anchor=north west][inner sep=0.75pt]    {$X$};
        \draw (262,240.4) node [anchor=north west][inner sep=0.75pt]    {$Y$};
        \draw (376,310.4) node [anchor=north west][inner sep=0.75pt]    {$\gamma $};
        \draw (440,114.4) node [anchor=north west][inner sep=0.75pt]    {$\gamma ^{\uparrow }$};
        \draw (161,171.4) node [anchor=north west][inner sep=0.75pt]    {$\Omega ( \tilde{X},\tilde{Y})$};
        \draw (486,106.4) node [anchor=north west][inner sep=0.75pt]    {$P$};

        \end{tikzpicture}
        
    \caption{The vertical component acquired by commuting two horizontal vectors is measured by the curvature, where $\tilde{X},\tilde{Y}$ are general lifts of $X,Y$} onto the total space $P$.
    \label{fig:CurvaHol}      
\end{figure}
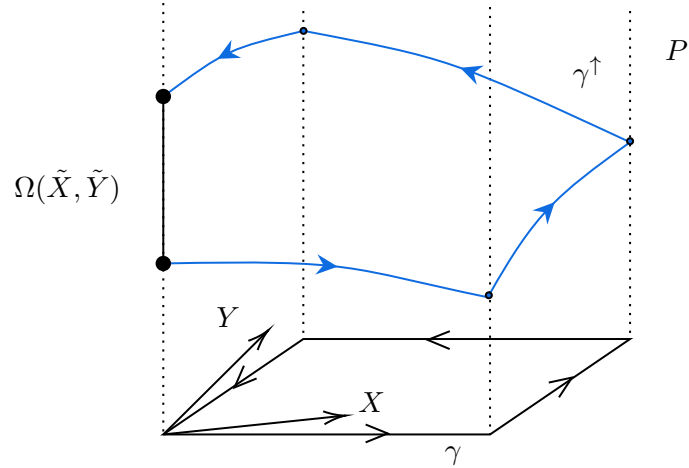

Now, there is a another way of determining the holonomy group of a manifold; by the notion of parallel sections of bundles\cite{Galaev:2015mxf,Besse:1987pua}. Let $E$ be a vector bundle over a base
manifold $M$, and $\nabla$ a connection induced on $M$. This then defines
parallel transport as above. For any piece-wise connected curve
$\gamma:[a,b]\subset \mathbb{R} \rightarrow M$ an isomorphism of the
vector spaces in the fibre of $E$ is defined as
\begin{equation}
    \tau_{(\gamma)}: E_{\gamma(a)}\rightarrow E_{\gamma(b)}.
\end{equation}
To introduce the holonomy group $\text{Hol}(\nabla)$, we fix a point $p\in M$ and parallel transport a section of the bundle $X_p\in\Gamma(E_p)$ along all piece-wise connected \emph{loops} at $p$. This defines the holonomy group of the connection $\text{Hol}_p(\nabla)$. If we restrict ourselves to null-homotopic loops, then we find the restricted holonomy group $\text{Hol}^0_p(\nabla)$. Since the holonomy groups of a connected manifold at different points are isomorphic, we can talk about the holonomy group of the connection $\text{Hol}(\nabla)\subseteq \text{GL}(d,E)$. \\

A section $X\in \Gamma(E)$ is called parallel if it is constant with respect to the connection, i.e., $\nabla X = 0$. This is equivalent to saying that $X$ is invariant under the parallel transport map
\begin{equation}
    \tau_{(\gamma)} : X_{\gamma(a)} \rightarrow X_{\gamma(b)},
\end{equation}
for any piece-wise connected path $\gamma: [a,b] \rightarrow \mc{M}$. Equivalently, we may write $\tau_{(\gamma)}(X_{\gamma(a)}) = X_{\gamma(b)}$, which is the non-coordinate basis expression of eq.~(\ref{Vchange}). Let us state the fundamental principle\cite{Galaev:2015mxf}:
\begin{theorem}
\textit{There exists a one-to-one correspondence between parallel sections $X$ of the bundle $E$ and vectors $X_p$ in the fibre $E_p$ invariant under $\rm{Hol}_p(\nabla)$}.
\end{theorem}
In other words, finding on a given (pseudo-)Riemannian manifold geometric objects whose covariant derivative vanishes is equivalent to finding the invariants of the holonomy group\cite{Besse:1987pua}. To elucidate the above statements, we now consider some examples of common Riemannian manifolds.

\begin{exmp}\upshape
On \textbf{orientable Riemannian manifolds} one can define the metric tensor $g$ as a section on the tensor product of the dual tangent bundle with itself $g\in\Gamma(T\mc{M}^*\otimes T\mc{M}^*)$, so that the metric tensor at a point $x\in \mc{M}$ is a map $g_x: T_x\mc{M} \otimes T_x\mc{M} \rightarrow \mathbb{R} $. There is a unique torsion-less connection called the Levi-Civita connection such that $\nabla g = 0$. Therefore, $g$ is called parallel. To find $\text{Hol}_x(\nabla)$, denote the group of linear transformations preserving $g$ by $\text{O}(T_x\mc{M},g_x)$. Since $g$ is parallel, we find $\text{Hol}_x(\nabla)\subset \text{O}(T_x\mc{M},g_x)$. Furthermore, as $T_x\mc{M}\cong \mathbb{R}^d$ we identify the holonomy group with a subgroup in ${\rm SO}(d)$. This corresponds to the first case in Berger's classification table \ref{BergerTable}. From now on we only consider Riemannian manifold equipped with a Levi-Civita connection.
\end{exmp}
\begin{exmp}\upshape
\textbf{K\"ahler Manifolds} are complex manifolds of dimension $d=2n$ and have a closed symplectic form $\omega=g(J,\,\cdot)\in\Omega ^2(\mc{M}_\mathbb{C})$, where $J$ is the almost complex structure and $\nabla\omega=0$. Now not only the metric tensor has vanishing exterior covariant derivative; the symplectic form $\omega$, which is invariant under the symplectic group ${\rm Sp}(2n,\mathbb{R})$, also has vanishing covariant derivative. Going by our philosophy, the holonomy group must both preserve lengths \textit{and} the symplectic form. The sought after group is then ${\rm U}(n)$. This can be seen as the intersection
\begin{equation}
    U(n)=SO(2n)\cap Sp(2n,\mathbb{R}).
\end{equation}
\end{exmp}
\begin{exmp}\upshape
\textbf{Hyper-K\"ahler manifolds} of dimension $d=4n$ have three almost complex structures $(I,J,K)$ with vanishing covariant derivative and obey the quaternionic relations $I^2=J^2=K^2=Id$ and $IJ=JI=-K$. The subgroup of ${\rm SO}(4n)$ that preserves the quaternionic almost complex structure is $\text{Sp}(n)\subset \text{SO}(4n)$. Therefore $\text{Hol(g)}=\text{Sp}(n)$. An example of this is the holonomy group of the self-dual Taub-NUT metric with $\text{Hol}(\nabla)={\rm Sp}(1)\cong {\rm SU}(2)$.
\end{exmp}
This then takes us to $\text{Spin}(7)$ manifolds, which we have seen in Chapter \ref{chap:Weyl}. We briefly discuss such manifolds following \cite{joyce2000CompactMW}. First let us define the $\text{Spin}(7)$ group as a subgroup of $\text{GL}(8,\mathbb{R})$.
\begin{exmp}\upshape
    \textbf{Spin(7) manifolds}\label{Spin7Man}
    Let $\mathbb{R}^8$  have coordinates $(x_1,\ldots, x_8)$. Write $dx_{ijkl}$ for the 4-form on $\mathbb{R}^8$. Define a 4-form $\phi$ on $\mathbb{R}^8$ by 
    \begin{equation}\begin{split}
        \phi =\, &dx_{1234} + dx_{1256} + dx_{1278}+dx_{1357}-dx_{1368}\\
        &-dx_{1458}
        - dx_{1467}-dx_{2358}-dx_{2367}-dx_{2457}\\
        &+dx_{2468}+dx_{3456}+dx_{3478}+dx_{5678}.
    \end{split}
    \end{equation}
\end{exmp}
The subgroup of $\text{GL}(8,\mathbb{R})$ preserving $\phi$ is the holonomy group $\text{Spin}(7)$ \cite{joyce2000CompactMW}. We also have that $\star \phi = \phi$, $\star$ is the hodge dual on $\mathbb{R}^8$ with the standard Euclidean metric, so that $\phi$ is a self-dual 4-form. An 8-dimensional manifold $\mc{M}$ equipped with a Levi-Civita connection $\nabla$ has a $\text{Spin}(7)$ structure if there exists a subset of 4-forms defined on it such that they can be identified with $\phi\in{\Omega^4(\mc{M})}$; moreover, $\nabla\phi = 0$. Then we say that $\mc{M}$ is a $\text{Spin}(7)$ manifold \cite{joyce2000CompactMW}. Indeed, by the fundamental principal, the condition $\nabla\phi = 0$ tells us of the existence of a parallel structure, which means that the holonomy group is $\hol(\nabla)=\text{Spin}(7)$.
\section{Holonomy and Wilson lines}
\label{sec:review}

As discussed above, holonomy refers, in general, to the change in
certain mathematical objects as they are transported around a closed
loop. The particular description depends on which theory we are in, as
well as which quantities are being transported. In this section, we
review the relevant ideas that we will need for the rest of the paper.

\subsection{Riemannian holonomy}
\label{sec:riemann}

Given a (pseudo-)Riemannian manifold ${\cal M}$, one may consider the
tangent space $T_p({\cal M})$ of all vectors at a point $p$. The
tangent spaces $T_p({\cal M})$ and $T_q({\cal M})$ associated with
points $p,q\in{\cal M}$ will be different in general, such that one
must define a prescription for comparing vectors at different
points. As is well-known, the solution is to consider a curve
$\gamma(t)$ from $p$ to $q$, and to say that a given vector $V^\mu$
undergoes {\it parallel transport} along the curve if it satisfies the
equation
\begin{equation}
\frac{d}{dt}V^\mu+\Gamma^\mu_{\sigma\rho}\frac{dx^\rho}{dt}V^\sigma=0,
\label{parallel}
\end{equation}
where $\Gamma^\mu_{\sigma\rho}$ is the Christoffel symbol. Vectors
may be compared after they are parallel-transported to the same point (or
tangent space), and one may indeed solve eq.~(\ref{parallel}) to find
the total change in $V^\mu$ after it has been transported from $p$ to $q$:
\begin{equation}
V_q^\mu =[\Phi_\Gamma(\gamma)]^\mu {}_\sigma\, V_p^\sigma,\quad
[\Phi_\Gamma(\gamma)]^\mu {}_\sigma={\cal P}\exp\left[ -\int_\gamma dx^\rho
  \Gamma^{\mu}_{\rho\sigma} \right].
\label{Vchange}
\end{equation}
Here, in a slight abuse of notation, we have exponentiated the
Christoffel symbol considered as the matrix
$\tensor{[\Gamma_\rho]}{^\mu_\sigma}$. Furthermore, the path-ordering symbol
${\cal P}$ indicates that, in expanding the exponential, these
matrices are to be ordered according to increasing parameter value
$\tau$. A special case of eq.~(\ref{Vchange}) occurs if $q$ and $p$ are
taken to be the {\it same} point, on a closed curve $C$. The vectors
appearing on the left- and right-hand sides of eq.~(\ref{Vchange}) are
then living in the same tangent space, such that the change in $V^\mu$
upon being transported around $C$ is effected by the transformation
matrix
\begin{equation}
[\Phi_\Gamma(C)]^\mu {}_\sigma={\cal P}\exp\left[ -\oint_C dx^\rho
  \Gamma^{\mu}_{\rho\sigma} \right],
\label{PhiGammadef}
\end{equation}
which we will call the {\it Riemannian holonomy operator}. The set of
all such transformations forms the {\it Riemannian holonomy
  group}. For a generic $d$-dimensional Riemannian manifold, one
expects this group to be the maximal set of possible transformations
on vectors in the tangent space, namely O($d$). If the manifold is
orientable this group will reduce to SO($d$). Further reductions occur
in other special cases~\cite{BSMF_1955__83__279_0}, the details of
which have been of much study in the mathematical literature. For
example, the holonomy group reduces to $\text{SU}(d/2)$ for Calabi-Yau
manifolds, to $\text{Sp}(d/4)$ for hyper-K\"ahler manifolds, and even
to $\text{G}_2$ for exceptional seven dimensional manifolds.
\subsection{The spin connection holonomy}
\label{sec:spinconnection}

In eq.~\eqref{parallel} we have defined parallel transport explicitly
with the Christoffel connection. One may also use an alternative
formalism involving the {\it spin connection}, which would in any
case be necessary if one were to consider the parallel transport of
spinors. Let us now describe the role of the spin connection.  As is
well known (see e.g. ref.~\cite{Carroll:2004st} for a pedagogical
summary), at each point in spacetime, one may introduce a set of
orthonormal basis vectors $\hat{e}_{(a)}$, related to the usual
tangent space vector basis $e_{(\mu)}\equiv\partial_\mu$ by
\begin{equation}
e_{(\mu)}= \tensor{e}{^a_\mu} \hat{e}_{(a)},\quad
\hat{e}_{(a)}= \tensor{e}{_a^{\mu}} e_{(\mu)},
\label{vierbein}
\end{equation}
which defines the {\it (inverse) vielbein} $\tensor{e}{^a_\mu}$
($\tensor{e}{_a^\mu}$), satisfying
\begin{equation}
\tensor{e}{^a_\mu} \tensor{e}{^b_\nu} \eta_{ab}=g_{\mu\nu},\quad \tensor{e}{_a^\mu} \tensor{e}{_b^\nu}g_{\mu\nu}=\eta_{ab}, \quad
\tensor{e}{^a_\mu}\tensor{e}{_b^\mu} = \delta^a_b, \quad \tensor{e}{^a_\mu}\tensor{e}{_a^\nu} = \delta^{\nu}_{\mu},
\label{vierbein2}
\end{equation}
where $\eta_{ab}$ is the Minkowski metric of the flat tangent space.\footnote{Like in the rest of the thesis, we use a $(-,+,+,+)$ metric signature throughout for Lorentzian spacetimes.} Thus, the Roman indices ($a,b,...$) are raised and lowered with this flat metric, while the Greek indices ($\mu, \nu,...$) are raised and lowered using the curved metric $g_{\mu \nu}$. \\

The vielbein may now be used to relate the components of an arbitrary vector $V^\mu$ at a point $p$ to the components of a vector in the tangent space at $p$ via
\begin{equation}
V^a=\tensor{e}{^a_\mu}(p) V^\mu, \quad V^{\mu} = \tensor{e}{_a^{\mu}}(p)V^a.
\label{Vadef}
\end{equation}
From the vielbein we may calculate the spin connection
$\tensor{\omega}{^a_b}$ using {\it Cartan's first structure
  equation}. In differential form language, with the torsion set to zero, this is:
\begin{equation}
d e^a + \omega^a{}_b \wedge e^b =0 \, .
\end{equation}
We may also invert this to write the components of the spin connection in terms of the vielbein:
\begin{equation}
(\omega_{\mu})^{a b}=\frac{1}{2} e^{a\nu}\left(\partial_{\mu} \tensor{e}{^b_\nu}-\partial_{\nu} \tensor{e}{^b_\mu}\right)-\frac{1}{2} e^{b\nu}\left(\partial_{\mu} \tensor{e}{^a_\nu}-\partial_{\nu} \tensor{e}{^a_\mu}\right)-\frac{1}{2} e^{a\rho} e^{b\sigma}\tensor{e}{^c_\mu}\left(\partial_{\rho} e_{c\sigma}-\partial_{\sigma} e_{c\rho}\right) \, .
\label{spinvierbein}
\end{equation}

A vector in the tangent space will then satisfy the parallel transport equation as in
eq.~(\ref{Vchange}), but now with the spin connection:
\begin{equation}
\frac{d}{dt}V^a+\tensor{(\omega_{\mu})}{^a_b}\frac{dx^\mu}{dt}V^b=0 \, .
\label{spinparallel}
\end{equation}

The solution of eq.~(\ref{spinparallel}), by direct analogy with
eq.~(\ref{Vchange}), is 
\begin{equation}
V_q^a=[\Phi_\omega(\gamma)]^a{}_b V_p^b,\quad
[\Phi_\omega(\gamma)]^a{}_b={\cal P}\exp\left[-\int_\gamma dx^\mu 
(\omega_\mu)^a{}_b\right],
\label{Vachange}
\end{equation}
where ${\cal P}$ denotes path-ordering of the matrices
$\tensor{(\omega_{\mu})}{^a_b}$ along the worldline. Choosing $p$ and $q$ to
correspond to the same point on a closed curve $C$, one obtains the
holonomy of the spin connection:
\begin{equation}
[\Phi_\omega(C)]^a{}_b={\cal P}\exp\left[-\oint_C dx^\mu (\omega_\mu)^a{}_b\right].
\label{spinholonomy}
\end{equation}
This form of the holonomy is straightforwardly related to the Riemannian
holonomy operator of eq.~(\ref{PhiGammadef}). Upon transforming both
sides of eq.~(\ref{Vchange}) to the orthonormal basis and rearranging, one
obtains~\cite{Modanese:1993zh}
\begin{equation}
[\Phi_\omega(p,q)]^a{}_b=\tensor{e}{^a_\mu}(p)\,[\Phi_\Gamma(p,q)]^\mu{}_\nu\, \tensor{e}{_b^\nu} (q),
\label{Phirel}
\end{equation}
so that for a closed curve one has
\begin{equation}
[\Phi_\omega(C)]^a{}_b= \tensor{e}{^a_\mu}\, [\Phi_\Gamma(C)]^\mu{}_\nu\, \tensor{e}{_b^\nu},
\label{Phirel2}
\end{equation}
where the two vielbeins on the right-hand side are evaluated at the same point. The physical interpretation of this
relation is straightforward. The Riemannian holonomy operator
tells us how the components of a vector transform after the vector has been transported around $C$. The expression in terms of 
the spin connection does the same, but in the orthonormal basis. Then, the two holonomy operators are related by a similarity transformation, which is the content of eq.~(\ref{Phirel2}).\\

Thus, in discussing the holonomy of a given manifold, one is free to use either. For our later
purposes, it is convenient to rewrite eq.~(\ref{spinholonomy}) yet
further. Noting that the spin connection is valued in the Lie algebra
of the Lorentz group, we may introduce explicit Lorentz generators
$M^{ab}$ via
\begin{equation}
(\omega_\mu)^c{}_d=\frac{i}{2}(\omega_\mu)_{ab}(M^{ab})^c{}_d,
\label{omegaM}
\end{equation}
where the normalisation factor arises from the components of the generators in the spin-1 representation:
\begin{equation}
(M^{ab})^c{}_d = i( \eta^{ac} \delta^b_d - \eta^{bc} \delta^a_d).
\label{Mspin1}
\end{equation}
The operator of eq.~(\ref{spinholonomy}) is then
\begin{equation}
[\Phi_{\omega}(C)]^c{}_d= \tensor{{\cal P}\exp\left[
-\frac{i}{2}\oint_C dx^\mu\, (\omega_\mu)_{ab}\,M^{ab}
\right]}{^c_d} \, .
\label{spinholonomy2}
\end{equation}
Here it is worth noting that this form allows us to easily extend the notion of holonomy to spinors. One replaces the generators in the spin-1 representation with the generators in the spin-(1/2) representation constructed from the associated Clifford algebra. Concretely, instead of $M^{ab}$ one uses $(\Gamma^{ab})^\alpha{}_\beta$ and the holonomy group will be generically $\text{Spin}(d)$ valued. Doing this relies on lifting the tangent bundle of the manifold to a spin bundle, which requires the existence of a spin structure on the manifold. There may of course be global obstructions to doing this which are given by the second Stiefel-Whitney class. In what follows we will not worry about such spinor-valued holonomies, however it is certainly worth understanding how the double copy works in this case and how different representations in the tangent bundle are related to representations in the single copy. Having described how holonomy works in gravity, let us now consider
gauge theory.

\subsection{Holonomy in gauge theory}
\label{sec:gauge}

Consider a gauge theory defined on a spacetime manifold ${\cal M}$
with structure group $G$ and a vector space fibre $V$ as the associated bundle $(P_G\times_{G} V,\pi_V,\mc{M})$. A field $\Psi^a$ transforming in a particular
representation of $G$ can then be defined as a section of
the associated bundle, where the gauge field itself is associated
with the connection 1-form on the total space. If we want to compare field values at different
points $p,q\in{\cal M}$, we must transform the gauge information
according to a suitable definition of parallel transport, leading to
an equation analogous to eq.~(\ref{Vchange}):
\begin{equation}
\Psi^a_q=[\Phi_A(\gamma)]^a{}_b\, \Psi^b_p,\quad 
[\Phi_A(\gamma)]^a{}_b= \tensor{{\cal P}
\exp\left[-g\int_\gamma dx^\mu {\bf A}_\mu\right]}{^a_b}.
\label{Phichange}
\end{equation}
Here ${\bf A}_\mu=A^a_\mu {\bf T}^a$ is the local matrix-valued gauge field,
${\bf T}^a$ are the generators of the Lie algebra in the representation
appropriate to the field $\Psi^a$, and $g$ is the coupling. If
we again take $p$ and $q$ to be the same spacetime point lying on a
closed curve $C$, the change in $\Psi^a$ after transport around
the loop is given by
\begin{equation}
\Phi_A(C)={\cal P}\exp\left[-g\oint_C dx^\mu {\bf A}_\mu \right].
\label{Phidef}
\end{equation}
The set of all such transformations forms the holonomy group
associated with gauge theory solutions, which will be a subgroup of the gauge group $G$. \\

The operator appearing in eq.~(\ref{Phichange}) is known as a {\it
  Wilson line} in the gauge theory literature.\footnote{Note that we
  have chosen anti-Hermitian colour generators, such that there is no
  explicit factor of $i$ in the exponent of
  eq.~(\ref{Phichange}).} It transforms covariantly under gauge
transformations according to
\begin{equation}
\Phi_A(\gamma)\rightarrow {\bf U}_p\Phi_A(\gamma) {\bf U}_q^{-1},
\label{Phitrans}
\end{equation}
where ${\bf U}_p$ is an element of $G$ in the appropriate
representation, and represents a local gauge transformation at the
point $p$. Thus, Wilson lines are ubiquitous in the study of
scattering amplitudes, and typically crop up whenever some physical
behaviour has to be expressed in a gauge-covariant manner.  The
operator of eq.~(\ref{Phidef}) is known as a Wilson loop once the
trace is taken on the right-hand side. From eq.~(\ref{Phitrans}), this
renders the Wilson loop gauge-invariant. \\

Notably, the gravitational operator of eq.~(\ref{Vchange}) also
transforms similarly to eq.~(\ref{Phitrans}), but where the gauge
transformations are replaced by diffeomorphisms. That is, upon
changing coordinates according to $x^\alpha\rightarrow y^\alpha$, one
has~\cite{Modanese:1993zh}
\begin{equation}
[\Phi_\Gamma(\gamma)]^\mu{}_\sigma\rightarrow [\Lambda_p]^\mu{}_\alpha
\,[\Phi_\Gamma(\gamma')]^\alpha{}_\beta\,[\Lambda_q^{-1}]^\beta{}_\sigma,\quad
[\Lambda_p]^\lambda{}_\delta=\left(\frac{\partial y^\lambda}
{\partial x^\delta}\right)_p,
\label{PhiGammatrans} 
\end{equation}
where the path $\gamma$ is transformed according to the diffeomorphism
to $\gamma'$. This property, together with the fact that there is a
common geometric interpretation of the operators $\Phi_\Gamma(\gamma)$
and $\Phi_A(\gamma)$ in gravity and gauge theory respectively, has led
to $\Phi_\Gamma$ also being referred to as a Wilson line in the
gravity
literature~\cite{Modanese:1991nh,Modanese:1993zh,Hamber:2007sz,Hamber:2009uz}. However,
there is a more sensible candidate for this, as we discuss in the
following section.\\

\subsection{The gravitational Wilson line}
\label{sec:gravwilson}

In an abelian gauge theory, the Wilson line operator of
eq.~(\ref{Phidef}) has a useful physical interpretation, in that it
represents the phase change experienced by a charged particle as it
traverses a closed loop. The non-abelian version is a generalisation
of this, once the trace is taken to form a gauge-invariant
quantity. The analogous operator in gravity is easy to write down. In gravity the equivalent of the charge is the mass of the particle and so  the phase will only depend on the (Lorentz-invariant) path length of the closed curve multiplied by the particle mass. For an arbitrary
curve $\gamma$, one may then define the {\it gravitational Wilson
  line}~\cite{Hamber:1994jh} as follows (this is also discussed in the much earlier work of
ref.~\cite{Dowker:1967zz}):
\begin{equation}
\Phi_g(\gamma)=\exp\left[-im\int_\gamma d\tau 
\sqrt{-g_{\mu\nu}\dot{x}^{\mu}
\dot{x}^{\nu}}\right],
\label{Phigdef}
\end{equation}
where $m$ is the mass of the particle, $\tau$ its proper time, and $\dot{x}^{\mu} \equiv dx^\mu/d\tau$. Throughout this section and the rest of the chapter, $\dot{x}^{\mu}$ will always denote differentiation with respect to the variable parameterising the curve. In
perturbation theory (as appropriate to the weak field limit), one may
introduce a graviton field $h_{\mu\nu}$ via
\begin{equation}
g_{\mu\nu}=\eta_{\mu\nu}+\kappa h_{\mu\nu},\quad \kappa=\sqrt{32\pi G_N}, 
\label{graviton}
\end{equation}
where $\eta_{\mu\nu}$ is the Minkowski metric, and $G_N$ Newton's
constant. Then the operator of eq.~(\ref{Phigdef}) simplifies, to first
non-trivial order in $\kappa$, to\footnote{We have ignored an overall
  multiplicative constant in eq.~(\ref{Phigdef2}), which will vanish
  in any vacuum expectation value of Wilson lines, once this is
  correctly normalised.}
\begin{equation}
\Phi_g(\gamma)=\exp\left[\frac{i\kappa}{2}\int_\gamma ds\, h_{\mu\nu}
\dot{x}^{\mu}\dot{x}^{\nu}\right],
\label{Phigdef2}
\end{equation}
where we have rescaled the integration variable to have mass dimension
$-2$. The square root in eq.~(\ref{Phigdef}) is cumbersome in general,
and one may further worry that the operator ceases to be defined for
massless particles. One may remove both problems by noting that the
exponent of eq.~(\ref{Phigdef}) contains the action for a point
particle. The latter can be replaced with the alternative
action\footnote{The use of this alternative point particle action in
a double copy context has been emphasised previously in
ref.~\cite{Plefka:2018dpa}.}
\begin{equation}
S_{\rm pp}= \frac{1}{2}\int d\tau \left[\frac{1}{e(\tau)}g_{\mu\nu}\dot{x}^{\mu}
\dot{x}^{\nu}-e(\tau)m^2 \right],
\label{Sdef}
\end{equation}
where $e(\tau)$ is an auxiliary parameter known as the {\it
  einbein}. Its field equation yields
\begin{equation}
\frac{\delta S}{\delta e}=-\frac{1}{2e^2}g_{\mu\nu}\dot{x}^{\mu}
\dot{x}^{\nu}-\frac{m^2}{2}=0,
\label{einbeineq}
\end{equation}
such that solving for $e(\tau)$ and substituting this into
eq.~(\ref{Sdef}) yields the original action of a point particle that
appears in eq.~(\ref{Phigdef}), in the massive case. The parameter
$e(\tau)$ plays the role of a ``metric'' on the worldline, and
transforms appropriately under reparametrizations. Choosing a value
for $e$ then amounts to fixing a gauge, and the choice $e=1$ in the
massless case immediately leads to the Wilson line of
eq.~(\ref{Phigdef2}). \\

It has recently been argued~\cite{Alfonsi:2020lub} that the operator
$\Phi_g(\gamma)$ is the double copy of the gauge theory Wilson line of
eq.~(\ref{Phichange}), which can be seen in a number of ways. It may be
related to scattering amplitudes, for example, by considering a
semi-infinite set of Wilson lines emanating from a common
point. Vacuum expectation values of such Wilson lines are known to
describe the infrared singularities of scattering amplitudes, where
the latter have been proven to formally double
copy~\cite{Oxburgh:2012zr}. That eq.~(\ref{Phigdef}) is the correct
Wilson line associated with IR singularities has been established in
refs.~\cite{Naculich:2011ry,White:2011yy}. Similar evidence comes from
the high energy (Regge) limit, where amplitudes are again known to
double-copy~\cite{Vera:2012ds,Johansson:2013aca}, and where there is
also a description in terms of the Wilson line operators of
eqs.~(\ref{Phichange})
and~(\ref{Phigdef2})~\cite{Korchemskaya:1994qp,Melville:2013qca,Luna:2016idw}. More directly, one may rewrite the operator in eq.~(\ref{Phichange}) as
\begin{equation}
\Phi(\gamma)={\cal P}
\exp\left[ig\tilde{\bf T}^a\int_\gamma ds  A^a_\mu\,\dot{x}^{\mu}\right],
\label{Phichange2}
\end{equation}
where we have temporarily adopted Hermitian colour generators defined
via $\tilde{\bf T}^a\equiv i {\bf T}^a$. If we consider a gauge field for which the double copy is known, then the gravitational Wilson line of
eq.~(\ref{Phigdef2}) is obtained by making the replacements
\begin{equation}
g\rightarrow\frac{\kappa}{2},\quad 
\tilde{\bf T}^a\rightarrow \dot{x}^{\mu},
\label{replacements}
\end{equation}
precisely mirroring the usual coupling and colour / kinematic
replacements associated with the BCJ double copy for
amplitudes. Further to the discussion in ref.~\cite{Alfonsi:2020lub},
it is interesting to note that the explicit double copy between the
gauge and gravity Wilson lines is manifest when using the alternative
point particle action of eq.~(\ref{Sdef}), for a particular choice of
einbein. This is perhaps not surprising: the original BCJ double copy
for amplitudes is known to be manifest only in certain {\it
  generalised gauges}, where conventional gauge transformations as
well as field redefinitions have potentially been applied. The same
property occurs also for classical solutions in general.  Thus, the
fact that a particular einbein is needed to get the double copy to
work -- itself a choice of gauge, as described above -- is entirely
consistent with previous instances of the double copy.\\

Aside from special kinematic limits of amplitudes, there are also
other situations in which $\Phi_g(\gamma)$ manifests itself as the
double copy of $\Phi_A(\gamma)$. It may be used, for example, to
quantify certain topological information that is the relevant
gravitational counterpart of that obtained in a gauge
theory~\cite{Alfonsi:2020lub}. Furthermore, a particular Wilson loop
involving $\Phi_A$ may be used to derive the Coulomb potential between
two static charges, such that replacing $\Phi_A$ with $\Phi_g$ instead
yields Newton's law of gravity~\cite{Hamber:1994jh}. That the two
potentials should indeed be related follows from the non-relativistic
limit of the classical double copy between the point charge and the
Schwarzschild black hole~\cite{Monteiro:2014cda}.\\

Given that the holonomy operator in gauge theory is related to the
Wilson line of eq.~(\ref{Phigdef2}) in gravity, it is therefore not
true that the holonomy operators in the two theories are related by
the double copy. The question then arises of whether one may find a
gauge theory single copy of the gravitational holonomy operator of
eq.~(\ref{PhiGammadef}). Indeed one can, as we explain in the
following section.

\section{The single copy of the gravitational holonomy}
\label{sec:singlecopy}

Above, we have seen that the holonomy operators in gauge and gravity
theory, whilst natural mathematical counterparts of each other, are
not physical counterparts in the sense of being related by the double/single copy. To find the correct single copy of the gravity result,
one must map the latter to a physical situation whose single copy is
already well-known. In the present case, the holonomy operator of
eq.~(\ref{spinholonomy2}) turns out to arise in the dynamics of
spinning particles, whose properties we review in the following
section.

\subsection{Relativistic spinning particles}
\label{sec:spinning}

In eq.~(\ref{Sdef}), we have seen the action for a spinless point
particle coupled to gravity. It is possible to generalise this to the
case in which a (possibly extended) object has an intrinsic angular
momentum (see ref.~\cite{Levi:2018nxp} for a modern pedagogical review, and also the classic works of refs.~\cite{Hanson:1974qy, Bailey:1975fe, Dixon:1970zza, Papapetrou:1951pa}). To this end, it is conventional to define a vierbein on the worldline, $\tensor{e}{^A_{\mu}}(\tau)$. The upper-case latin indices $(A,B,...)$ are those of a body-fixed frame; a frame fixed to the point-particle as it traverses the worldline. The vierbein $\tensor{e}{^A_{\mu}}$ therefore relates the body-fixed frame to the general coordinate frame. The angular velocity of the
object is then defined to be
\begin{equation}
\Omega_{\mu\nu}=e_{A\mu}\frac{D \tensor{e}{^A_\nu}}{D\tau},
\label{Omegadef}
\end{equation}
where 
\begin{equation}
\frac{D \tensor{e}{^A_\nu}}{D\tau} \equiv \dot{x}^\alpha D_\alpha \tensor{e}{^A_\nu}
=\dot{x}^\alpha\left(\partial_\alpha \tensor{e}{^A_\nu}-\Gamma^\lambda_{\alpha \nu}
\tensor{e}{^A_\lambda}\right)
\label{Dalpha} 
\end{equation}
is the spacetime covariant derivative of the body-fixed vierbein. That
eq.~(\ref{Omegadef}) satisfies the expected antisymmetry,
$\Omega_{\mu\nu}=-\Omega_{\nu\mu}$, follows from eq.~(\ref{vierbein2})
and the vanishing of the covariant derivative of the metric tensor
$g_{\mu\nu}$. The total action for the object can be written as
\begin{equation}
S_{\rm tot}=S_{\rm pp}+S_{\rm spin},
\label{Spp2}
\end{equation}
where $S_{\rm pp}$ is the spinless action of eq.~(\ref{Sdef}),
reflecting the fact that an extended object looks pointlike from a
sufficient distance. The correction term due to the spin
is\footnote{In writing eq.~(\ref{Sspin}), we have ignored additional
  gauge-fixing terms which are needed to eliminate residual arbitrary
  degrees of freedom in the precise definition of the spin
  tensor. Such terms will not matter for the arguments presented
  here.}
\begin{equation}
S_{\rm spin}=\int d\tau \left[\frac12 
\Omega_{\mu\nu}S^{\mu\nu}\right],
\label{Sspin} 
\end{equation}
where $S^{\mu\nu}$ is the {\it spin tensor} of the object, namely the
dynamical variable conjugate to the angular velocity. Physically this represents the intrinsic angular momentum of the object (in
either a classical or quantum setting). \\

In general, the vierbein $\tensor{e}{^a_\mu}$ that we choose for a given spacetime
will not correspond to the body-fixed vierbein $\tensor{e}{^A_\mu}$, and there will
therefore be a Lorentz transformation that relates the two:
\begin{equation}
{e^A}_\mu={\Lambda^A}_a {e^a}_\mu.
\label{etrans} 
\end{equation} 
The combination of terms appearing in eq.~(\ref{Sspin}) can then be
decomposed as~\cite{Vines:2016unv}
\begin{align}
S_{\mu\nu}\Omega^{\mu\nu}&=S_{\mu\nu}{\Lambda_A}^a {e_a}^\mu
\frac{D\Lambda^{Ab}{e_b}^\nu}{D\tau}\notag\\ &=S_{ab}\left({\Lambda_A}^a
\dot{\Lambda}^{Ab} - (\omega_\mu)^{ab}\dot{x}^\mu\right),
\label{omegadecompose}
\end{align}
where in the second line we have used eq.~(\ref{Dalpha}) together with
the known relation between the Christoffel symbol and spin connection
(see e.g.~\cite{Carroll:2004st})
\begin{equation}
\Gamma^\sigma_{\mu\nu}=\tensor{e}{_a^\sigma}\tensor{e}{^b_\nu}\tensor{(\omega_\mu)}{^a_b}
+\tensor{e}{_a^\sigma}\partial_\mu \tensor{e}{^a_\nu}.
\label{gamomega} 
\end{equation}

\subsection{The holonomy from a spinning particle}
\label{sec:holspin}

We now argue that the dynamics of spinning particles can be used to
construct a physical manifestation of the holonomy operator of
eq.~(\ref{spinholonomy2}). To this end, note that the two terms in
eq.~(\ref{omegadecompose}) have a straightforward physical
interpretation: the action of eq.~(\ref{Sspin}) dictates the dynamics
of the internal spin of the object under consideration, namely how the
body-fixed vierbein changes as one proceeds along the worldline. Put
another way, the action governs how a vector fixed to the moving
object will be modified, and there are clearly two distinct effects
causing such a vector to change: (i) the rotation of the object; (ii)
the fact that the body-fixed frame is changing due to the underlying
spacetime. These two effects are captured by the first and second terms
in the second line of eq.~(\ref{omegadecompose}) respectively, where
the first (rotation) term would be present even if the object were
moving in flat space.\\

In the previous section, we noted that the point particle action of
eq.~(\ref{Sdef}) could be used to form a Wilson line operator
representing the phase experienced by a particle as it traverses a
given contour. To do this, one forms the combination
\begin{equation}
e^{iS_{\rm pp}},
\label{Wilsonaction}
\end{equation}
and discards terms associated with flat space only i.e. that do not
involve the gravitational field. Such terms amount to an overall
multiplicative factor, that vanishes upon normalising vacuum
expectation values of Wilson lines. It is straightforward to repeat
this procedure for the action of eq.~(\ref{Spp2}), where the
corresponding Wilson line now represents the phase experienced by a
spinning particle. A given spin tensor will have the form
\begin{equation}
  S^{ab}(\tau)=Q^{ab}_{cd}(\tau) M^{cd}.
  \label{Sform}
\end{equation}
The quantity $Q^{ab}_{cd}$ denotes how much of each spin generator
is ``turned on'', and may in general depend on the parameter $\tau$
along the worldline. We wish to examine how all possible vectors are
transported around all possible loops. Thus, we may choose a spin
tensor such that
\begin{equation}
  Q^{ab}_{cd}=\frac12\left(\delta^{a}_{c}\delta^b_{d}
  -\delta^b_c\delta^a_d\right),
  \label{Qchoice}
\end{equation}
which physically amounts to a democratic assignment of unit spin along all axes. This discussion holds for a classical
particle. For a quantum particle in state $|\psi\rangle$, the spin
tensor will be given by a normalised expectation value
\begin{equation}
  S^{ab}=\frac{\langle\psi |Q^{ab}_{cd}\,M^{cd}
    |\psi\rangle}{\langle \psi|
\psi\rangle}.
\label{quantumspin}
\end{equation}
However, one may again make the choice of eq.~(\ref{Qchoice}), and for
concreteness we focus on a spin-1 particle with arbitrary
orientation. Furthermore, we will take our generalised Wilson line to
be matrix-valued in spin space, such that it describes how the spin
state of a test particle changes as it moves along its worldline. This
is directly analogous to how the gauge theory Wilson line of
eq.~(\ref{Phidef}) is matrix-valued in colour space, and in practical
terms amounts to the replacement
\begin{equation}
S^{ab}\rightarrow M^{ab}
\label{genspin}
\end{equation}
whilst defining the Wilson line according to the appropriate
generalisation of eq.~(\ref{Wilsonaction}). The result is
\begin{equation}
\Phi_g^{\rm spin}(\gamma)={\cal P}\exp\left[\frac{i\kappa}{2} \int_\gamma
ds \left(h_{\mu\nu}
\dot{x}^{\mu}\dot{x}^{\nu} - \dot{x}^{\mu}(\omega_\mu)_{cd}
M^{cd}\right)\right],
\label{Wilsonspin}
\end{equation}
where the path ordering is now necessary due to the matrix-valued
nature of the spin generators in the second term (note that there is also an
implicit identity matrix in spin space in the first term).  The first
term of eq.~(\ref{Wilsonspin}) represents how the mass of the test
particle couples to gravity, while the second term describes how its spin degrees of freedom couple to gravity. \\

As discussed above, a number of previous studies have attempted to
identify the operator of eq.~(\ref{PhiGammadef}) (and, by association,
the operator of eq.~(\ref{spinholonomy2})) as a gravitational Wilson
line, due to its mathematical similarity to the Wilson line in gauge
theory. However, the double copy tells us that
eq.~(\ref{spinholonomy2}) is not the gravitational counterpart to the
gauge theory Wilson line of eq.~(\ref{Phidef}). Rather, it is a
spin-dependent correction\footnote{Physically, the effects of the spin
term are suppressed by a power of the emitted graviton momentum, as we
discuss in section~\ref{sec:amplitudes}. So it is in this sense a
small correction to the spinless term in an appropriate kinematic
limit.} to the gravitational Wilson line of eq.~(\ref{Phigdef2}), and
represents the additional phase change that a particle experiences if
it happens to be spinning. It is interesting to note that a related
observation was made as early as the 1960s~\cite{Dowker:1967zz} (see
also ref.~\cite{Mandelstam:1962us}), predating the introduction of
Wilson lines!

\subsection{Single copy of the holonomy}
\label{sec:gaugetheory}

In the previous section, we have seen that a Wilson line constructed
from the action for a spinning particle coupled to gravity contains
the holonomy operator of eq.~(\ref{spinholonomy2}). This immediately
tells us how to take the single copy of the holonomy: we can simply
write down the action for a spinning particle coupled to a gauge
field, and use this to create a generalised Wilson line that contains
a spin correction to the phase. The relevant action for a spinning
particle coupled to a gauge field is (see e.g. ref.~\cite{Li:2018qap})
\begin{equation}
S_{\rm gauge}=\int d\tau \left[\frac{1}{2e(\tau)}\eta_{\mu\nu}
\dot{x}^{\mu}\dot{x}^{\nu}
-\frac{e(\tau)m^2}{2} +\frac12 \Omega_{\mu\nu} S^{\mu\nu}
+g c^a(\tau)
\left(\dot{x}^{\mu}{A}^a_\mu - \frac{e(\tau)}{2} F^a_{\mu\nu}S^{\mu\nu} 
\right)\right].
\label{gaugeaction}
\end{equation}
Here the first two terms are the usual point particle action of
eq.~(\ref{Sdef}) considered in Minkowski space. Furthermore,
$\Omega_{\mu\nu}$ is the flat space version of the angular velocity of
eq.~(\ref{Omegadef}), such that its contraction with the spin tensor
$S^{\mu\nu}$ matches the first term in the second line of
eq.~(\ref{omegadecompose}). There are then two terms involving the
gauge field, where $c^a(\tau)$ is a colour vector obtained by
evaluating the expectation value of the colour generator $\tilde{\bf T}^a$
at a given position on the worldline. The first of these terms gives
rise to the Wilson line operator of eq.~(\ref{Phidef}), once one
replaces the expectation value of the colour generator by the
generator itself. The remaining
term couples the field strength $F_{\mu\nu}^a$ to the spin tensor, and
thus represents the spin-dependent correction to the vacuum dynamics
of the spinning particle due to the presence of a gauge field. This is
the precise gauge theory analogue of the spin connection term in
eq.~(\ref{omegadecompose}), which amounts to the extra contribution to
the spin dynamics of the object arising from the gravitational
field.\\

Indeed, the double copy relationship between gauge theory and gravity
actions for spinning particles has been addressed in detail in
refs.~\cite{Goldberger:2017ogt,Li:2018qap}, which considered radiation
from such a particle interacting with a Yang-Mills field. The authors
calculated the effects of this radiation perturbatively, before
double-copying the results order-by-order in perturbation theory. The
resulting system is that of a spinning particle interacting with a
graviton, axion and dilaton, which is the usual spectrum arising from
the double copy of pure Yang-Mills theory. Thus, roughly speaking, the
final term in eq.~(\ref{gaugeaction}) double copies to multiple
operators, representing the coupling of the spin to the full field
spectrum in the gravity theory. This one-to-many nature of the double
copy does not affect our arguments here: given we are taking the {\it
  single copy}, which is many-to-one, we can unambiguously identify
the single copy of the graviton spin coupling as the final term
appearing in eq.~(\ref{gaugeaction}). Interestingly,
refs.~\cite{Goldberger:2017ogt,Li:2018qap} found that the spinning
particle actions in gauge theory and gravity were only strict physical
double copies of each other (in the sense that a conserved
energy-momentum tensor was found in the gravity theory) if a certain
fixed numerical coefficient was placed in front of the spin coupling
in the gauge theory. We shall ignore this complication here, given
that such a coefficient is irrelevant in elucidating the group of
transformations induced by the generalised Wilson line in the gauge
theory. Replacing the spin tensor according to eq.~(\ref{genspin}) as
before, as well as fixing the einbein $e=1$, this Wilson line is found
to be
\begin{equation}
\Phi_{\rm spin}(\gamma)={\cal P}\exp\left[ig\tilde{\bf T}^a\int_\gamma
ds \left( A_\mu^a \dot{x}^{\mu}-\frac{1}{2}F_{\mu\nu}^a M^{\mu\nu}
\right)\right]
\label{Phispin}
\end{equation}
Evaluating the spin correction over a closed curve $C$ then gives the
single copy of the gravitational holonomy:
\begin{equation}
\Phi_{F}(C)={\cal P}\exp\left[-\frac{ig}{2}\tilde{\bf T}^a\oint_C
ds F_{\mu\nu}^a M^{\mu\nu}\right].
\label{PhiFdef}
\end{equation}
As in the gravity theory, we can furnish the operator appearing in the
action of eq.~(\ref{gaugeaction}) with a physical interpretation,
where we may focus on the case of an abelian gauge theory for
simplicity. Identifying the electric and magnetic fields by 
\begin{equation}
F_{0i}=E_i,\quad F_{ij}=\epsilon_{ijk}B_k,
\label{EBfields}
\end{equation}
one finds
\begin{align}
F_{\mu\nu}S^{\mu\nu}&=F_{0i}S^{0i}+F_{ij}S^{ij}\notag\\
&=-\vec{E}\cdot \vec{d}-\vec{B}\cdot \vec{\mu},
\label{moments}
\end{align}
where we have defined
\begin{equation}
d_i=-S_{0i},\quad \mu_i=-\epsilon_{ijk}S^{jk}.
\label{moments2}
\end{equation}
Equation~(\ref{moments}) constitutes the electromagnetic coupling of
an electric dipole moment $\vec{d}$ and magnetic dipole moment
$\vec{\mu}$, and we may consider our test particle to have both of
these turned on in general.

\subsection{Relation to scattering amplitudes}
\label{sec:amplitudes}

In the previous section, we have used known results from the classical
double copy to identify the single copy of the gravitational
holonomy. We may also note, however, that our results can be linked to
known properties of scattering amplitudes. For example,
ref.~\cite{Arkani-Hamed:2019ymq} recently argued that the Kerr
(spinning) black hole and its single copy can be obtained from 3-point
amplitudes that are double-copies of each other. The implications of
this were explored further in
refs.~\cite{Emond:2020lwi,Guevara:2020xjx}. Similar conclusions have
been obtained independently in
refs.~\cite{Guevara:2018wpp,Guevara:2019fsj,Bautista:2019tdr}, which
also linked the double copy for spinning particles with the well-known
{\it (next-to-)soft theorems} for emission of low-momentum
radiation~\cite{Weinberg:1965nx,Gross:1968in,Low:1958sn,Burnett:1967km,DelDuca:1990gz,White:2011yy,Cachazo:2014fwa,Casali:2014xpa}. We
can see this directly from the above results as follows. Starting with
the Riemannian parallel transport operator of eq.~(\ref{Vchange}), we
can use the well-known relation between the Christoffel symbol and the
metric,
\begin{equation}
\Gamma^\mu_{\rho\sigma}=\frac12
g^{\mu\alpha}\left(\partial_\rho g_{\alpha\sigma }
+\partial_{\sigma}g_{\alpha\rho}-\partial_\alpha g_{\rho\sigma}\right),
\label{christoffelg}
\end{equation}
as well as the graviton definition of eq.~(\ref{graviton}), to obtain
\begin{equation}
\tensor{[\Phi_\Gamma(\gamma)]}{^\mu_\sigma}={\cal P}\exp\left[ -
\frac{\kappa}{2}\int_\gamma
  dx^\rho \left(\partial_\rho h^\mu_\sigma+\partial_\sigma h^\mu_\rho
  -\partial^\mu h_{\rho\sigma}+\ldots\right)\right],
\label{PhiGammaexp}
\end{equation}
where the ellipsis denotes terms of higher order in $\kappa$. Now
let us choose the case of a straight-line contour emanating from the
origin, as would be appropriate for a fast-moving particle with
momentum $p^\mu$ emerging from a scattering process:
\begin{equation}
x^\mu=s p^\mu,\quad 0\leq s < \infty.
\label{xparam}
\end{equation}
The first term in eq.~(\ref{PhiGammaexp}) is a total derivative, and
integrates to give a gauge-dependent artifact associated with the
endpoints of the contour. It will vanish for physical processes
e.g. in forming gauge-invariant amplitudes, or squaring amplitudes to
form a cross-section, which involves closing Wilson line contours to
make a closed loop. The remaining terms take the form
\begin{align}
-\frac{\kappa}{2} \int_0^\infty ds p^\rho \left(
\partial_\sigma h^\mu_\rho-\partial^\mu h_{\rho\sigma}\right)
&=\frac{i\kappa}{2}p^\rho \int\frac{d^dk}{(2\pi)^d}
\int_0^\infty ds \left(k_\sigma \tilde{h}^\mu_\rho 
-k^\mu\tilde{h}_{\rho\sigma}\right)e^{-isk\cdot p},
\label{NE1}
\end{align}
where we have introduced the Fourier components of the graviton field
via
\begin{equation}
h_{\mu\nu}=\int\frac{d^d k}{(2\pi)^d}\tilde{h}_{\mu\nu}(k)e^{-ik\cdot x}.
\label{Fourier}
\end{equation}
Carrying out the $s$ integral in eq.~(\ref{NE1})
yields\footnote{The upper limit of the $s$ integral in
  eq.~(\ref{NE1}) will vanish upon careful implementation of the
  Feynman $i\varepsilon$ prescription.}
\begin{equation}
\ln(\Phi_g)\sim \int\frac{d^d k}{(2\pi)^d}\tilde{h}_{\beta\rho}(k)
\left[\frac{\kappa}{2}\frac{p^\rho k_\alpha\tensor{(M^{\alpha\beta})}{^\mu_\sigma}}
{p\cdot k}\right],
\label{NE2}
\end{equation}
where we have written the second line in terms of the spin-1 Lorentz
generators of eq.~(\ref{Mspin1}). The square bracketed factor can be
recognised as the appropriate contribution to the next-to-soft theorem
for emission of a graviton~\cite{Cachazo:2014fwa}. Its appearance
in this context arises given that the spin-dependent coupling to the
worldline is suppressed by a single power of the momentum of emitted
radiation compared to the leading Wilson line operator of
eq.~(\ref{Phigdef}). Furthermore, one may perform an analogous calculation
for the gauge theory operator of eq.~(\ref{PhiFdef}), finding
\begin{equation}
\ln(\Phi_F)\sim \int\frac{d^d k}{(2\pi)^d}\tilde{A}^a_\mu(k)\left[
g\tilde{\bf T}^a\frac{k_\nu M^{\mu\nu}}{p\cdot k}\right],
\label{NE3}
\end{equation}
again in agreement with the appropriate next-to-soft
theorem~\cite{Casali:2014xpa}. The known double copy properties of
these results in the study of scattering amplitudes again corroborates
the fact that the single copy of the gravitational holonomy is the
operator of eq.~(\ref{PhiFdef}). It is also worth noting that these
generalised Wilson lines were derived before in the context of
next-to-soft physics, before the next-to-soft theorems were more
widely recognised~\cite{Laenen:2008gt,White:2011yy}. \\

Having presented a variety of arguments in favour of our single copy
holonomy operator, let us now note that further useful insights can be
obtained by focusing on a particular class of solutions, namely the
{\it Kerr-Schild} solutions in terms of which the first classical
double copy was formulated~\cite{Monteiro:2014cda}.

\subsection{Insights from Kerr-Schild solutions}

Kerr-Schild solutions of General Relativity (GR) are those for which the
metric assumes a particularly simple form, namely
\begin{equation}
    g_{\mu\nu}= \bar{g}_{\mu\nu}+\phi(x)k_\mu k_\nu.
\label{KSdef}
\end{equation}
Here $\bar{g}_{\mu\nu}$ is a background metric, which we will take to
be Minkowski throughout ($\bar{g}_{\mu\nu}=\eta_{\mu\nu}$), albeit not
necessarily in Cartesian coordinates. Furthermore, $\phi(x)$ is a
scalar field and $k_\mu$ a 4-vector field which is both null and
geodesic:
\begin{equation}
    \bar{g}^{\mu\nu}k_{\mu}k_{\nu}=g^{\mu\nu}k_\mu k_{\nu}=0,\quad k\cdot \nabla k_{\mu}=0.
\label{KSconditions}
\end{equation}
Comparing with eq.~(\ref{graviton}), we see that Kerr-Schild solutions
have a graviton field given by
\begin{displaymath}
h_{\mu\nu}=\phi k_\mu k_\nu.
\end{displaymath}
This ansatz turns out to greatly simplify the Einstein equations,
which then have a linear dependence on the graviton only. This allows
for an infinite family of exact solutions to be obtained, which incudes
e.g. known black holes. Also, the ``factorised'' form of the graviton
(i.e. involving an outer product of a 4-vector with itself) allows a
single copy to be straightforwardly
obtained. Reference~\cite{Monteiro:2014cda} proved that for static
solutions, the gauge field
\begin{equation}
{\bf A}_\mu=A_\mu^a{\bf T}^a,\quad A_\mu^a=c^a \phi k_\mu,
\label{AKS}
\end{equation}
where $c^a$ is an arbitrary colour vector, solves the Yang-Mills
equations, which again simplify to a linear form. Consequently, the field strength tensor for the gauge field takes an abelian-like form:
\begin{equation}
    F^a_{\mu\nu}(x)=\pd_\mu A^a_{\nu}(x)-\pd_\nu A^a_{\mu}(x).
\label{fieldstrength}
\end{equation}
In an orthonormal basis, one has
\begin{equation}
    g_{\mu\nu}=\eta_{ab}\tensor{e}{^a_\mu}\tensor{e}{^b_\nu},
\label{gee}
\end{equation}
which in turn implies the following form for the Kerr-Schild vierbein:
\begin{equation}
        \tensor{e}{^a_{\mu}} = \tensor{\bar{e}}{^{\, a}_{\mu}} + \frac{1}{2}\phi k^ak_{\mu}, \quad
        \tensor{e}{_a^{\mu}} = \tensor{\bar{e}}{_{a}^{\mu}} - \frac{1}{2}\phi k_ak^{\mu}.
\label{eKS}
\end{equation}
Here $\tensor{\bar{e}}{^{\, a}_\mu}$ is the vierbein associated with the background metric in eq.~\eqref{KSdef}, which for our purposes is the Minkowski metric $\eta_{\mu\nu}$. 
As we review in appendix~\ref{AppSpin:A}, the spin connection
associated with this particular vierbein (subject to the additional
conditions of eq.~(\ref{KSconditions})) assumes the form
    \begin{equation}
        (\omega_{\mu})_{ab} = \pd_b e_{a\mu} - \pd_a e_{b\mu}.
\label{omegaKS}
    \end{equation}
Unlike the general expression of eq.~(\ref{spinvierbein}), this has
the pleasing property of being linear in the vierbein. Substituting
the results of eq.~(\ref{eKS}) (after lowering indices appropriately)
yields
    \begin{align}
        (\omega_{\mu})_{ab} &= \frac{1}{2}\left[\pd_b(\phi k_{\mu}k_a) - \pd_a(\phi k_{\mu}k_b)\right] \\
        &= \frac{1}{2}\left[\tensor{e}{_b^{\sigma}}\pd_{\sigma}(\phi k_{\mu}k_a) - \tensor{e}{_a^{\sigma}}\pd_{\sigma}(\phi k_{\mu}k_b)\right]. \label{scTemp1}
    \end{align}
Note that due to the null property of the Kerr-Schild vectors, $k^{\mu}k_{\mu} = 0$, conversion between coordinate and orthonormal bases is done simply with the background vierbein:
    \begin{equation}
        k_a = \tensor{e}{_a^{\mu}}k_{\mu} = \tensor{\bar{e}}{_{a}^{\mu}}k_{\mu} - \frac{1}{2}\phi k_ak^{\mu}k_{\mu}
        = \tensor{\bar{e}}{_{a}^{\mu}}k_{\mu}.
    \end{equation}
Thus, the spin connection in eq.~\eqref{scTemp1} can be written as
    \begin{equation}\label{scTemp2}
        (\omega_{\mu})_{ab} = \frac{1}{2}\left[\tensor{\bar{e}}{_{a}^{\nu}}\tensor{e}{_b^{\sigma}} - \tensor{\bar{e}}{_{b}^{\nu}}\tensor{e}{_a^{\sigma}}\right]\pd_{\sigma}(\phi k_{\mu}k_{\nu}).
    \end{equation}
If we now write the remaining vierbeins explicitly, a great deal of
simplification occurs. To see this, consider the first term in the
above expression:
    \begin{align}
        \tensor{\bar{e}}{_{a}^{\nu}}\tensor{e}{_b^{\sigma}}\pd_{\sigma}(\phi k_{\mu}k_{\nu}) 
        &= \tensor{\bar{e}}{_{a}^{\nu}}\left[\tensor{\bar{e}}{_{b}^{\sigma}} - \frac{1}{2}\phi k_bk^{\sigma}\right]\pd_{\sigma}(\phi k_{\mu}k_{\nu}) \\
        &= \tensor{\bar{e}}{_{a}^{\nu}}\tensor{\bar{e}}{_{b}^{\sigma}}\pd_{\sigma}(\phi k_{\mu}k_{\nu}) - \frac{1}{2}\phi k_ak_bk_{\mu}k^{\sigma}\pd_{\sigma}\phi.
    \end{align}
In the second equality, we have expanded $\pd_{\sigma}(\phi
k_{\mu}k_{\nu})$ using the product rule, from which two of the three
resulting terms vanish due to the geodesic condition
$k^{\sigma}\pd_{\sigma}k_{\mu}=0$. Performing the same procedure for
the second term in eq.~\eqref{scTemp2}, we find that the terms which
contain only a derivative of the scalar field $\phi$ cancel, such that
eq.~\eqref{scTemp2} is simply
    \begin{equation}
        (\omega_{\mu})_{ab} =
        \frac{1}{2}\left[\tensor{\bar{e}}{_{a}^{\nu}}\tensor{\bar{e}}{_{b}^{\sigma}} - \tensor{\bar{e}}{_{b}^{\nu}}\tensor{\bar{e}}{_{a}^{\sigma}}\right]\pd_{\sigma}(\phi k_{\mu}k_{\nu}).
    \end{equation}
Finally, if we expand the spin connection in terms of the Lorentz
generators, we obtain
    \begin{equation}
        \frac{i}{2}(\omega_{\mu})_{cd}M^{cd}
        = -\frac{i}{2}\pd_{\sigma}(\phi k_{\mu}k_{\nu})M^{\nu\sigma},
      \label{KSintegrand}
    \end{equation}
where we have identified the spin-1 Lorentz generators as
    \begin{equation}
        (M^{\nu\sigma})_{ab}
        = i\left[\tensor{\bar{e}}{_{a}^{\nu}}\tensor{\bar{e}}{_{b}^{\sigma}} - \tensor{\bar{e}}{_{b}^{\nu}}\tensor{\bar{e}}{_{a}^{\sigma}}\right].
    \end{equation}
Thus, we are left with a simple expression
in which the exponent appearing in the Kerr-Schild gravitational
holonomy operator is written directly in terms of the graviton:
    \begin{equation}\label{gravHol}
        \oint dx^{\mu}(\omega_{\mu})_{ab}M^{ab} = -\oint dx^{\mu}\pd_{\sigma}(h_{\mu\nu})M^{\nu\sigma}.
    \end{equation}
Despite the name `graviton', we emphasise again that the expression above is exact due to the Kerr-Schild form of the metric tensor. The Kerr-Schild single copy of eq.~(\ref{AKS}) implies that we should
single copy eq.~(\ref{gravHol}) by replacing
\begin{equation}
  \dot{x}^\mu\rightarrow \tilde{{\bf T}}^a,\quad k_\mu\rightarrow c^a,
  \label{KSreplace}
\end{equation}
such that one obtains
    \begin{align}
        \oint dx^{\mu}(\omega_{\mu})_{ab}M^{ab} \to
        -\tilde{\bf T}^a \oint ds \, \pd_{\sigma}(\phi k_{\nu}c^a)M^{\nu\sigma}
        &= -\tilde{\bf T}^a
\oint ds \,\pd_{\sigma}(A^a_{\nu})M^{\nu\sigma} \label{SC}\\
        &= \frac{1}{2} \tilde{\bf T}^a\oint ds \, 
F^a_{\nu\sigma}M^{\nu\sigma}.
    \end{align}
This agrees with the conclusion reached above, namely that the single
copy of the gravitational holonomy is the operator of
eq.~(\ref{PhiFdef}).\\
\renewcommand{\arraystretch}{2}
\begin{table}[t]
\begin{center}
\begin{tabular}{c|c}
 Gauge Theory & Gravity\\
\hline
${\cal P}\exp\left[-g\oint_C dx^\mu {\bf A}_\mu \right]$
 & $\exp\left[\frac{i\kappa}{2}\oint_C ds \dot{x}^\mu \dot{x}^\nu
h_{\mu\nu}\right]$\\
 ${\cal P}\exp\left[\frac{g}{2}\oint_C ds {\bf F}_{\mu\nu} M^{\mu\nu}\right]$
& ${\cal P}\exp\left[-\frac{i\kappa}{2}\oint_C dx^\mu (\omega_\mu)_{ab}M^{ab}\right]$
\end{tabular}
\caption{Holonomy operators in gauge and gravity theories, and their
  single / double copies.}
\label{tab:operators}
\end{center}
\end{table}

By way of summarising the results of this section, we collect all of
the operators we have discussed in table~\ref{tab:operators}. The
gauge theory and gravity holonomies appear in the top-left and
bottom-right respectively, and the gravitational Wilson line appears
in the top-right. The operator in the bottom-left corner completes the
square, and has not been considered before as an analogue of the
gravity holonomy, due to its not having the appropriate differential
geometric definition. Indeed, its role is entirely different to the
gauge theory holonomy. Considering a gauge field via a connection on a
principal fibre bundle, the usual holonomy describes how vectors in
the internal colour space (associated with the fibres) are transported
as one moves along a curve. By contrast, the single copy of the
gravity holonomy instead describes how spacetime vectors are
transported, thus linking the gauge field with vectors living in the
tangent space of the base manifold. Our hope is that this operator
might also prove useful in classifying properties of different
Yang-Mills solutions, and we develop this notion in the following
section.

\section{Results for the single copy holonomy operator}
\label{sec:results}

The gravitational holonomy is useful in that it allows us to classify
solutions of General Relativity (and arbitrary manifolds more
generally) into qualitatively different types. In general, one expects
the holonomy group of a given manifold to be the most general group
acting on vectors in the tangent space, namely $\text{SO}(d)$ for a
$d$-dimensional orientable spacetime in Euclidean signature, or
$\text{SO}(1,d-1)$ for Lorentzian signature. In some cases, however, the
holonomy group reduces to a subgroup, and the classification of
manifolds based on this idea has been widely studied (see
e.g. ref.~\cite{BSMF_1955__83__279_0} for the most well-known
incarnation). This suggests that our single copy holonomy operator
might have a similar purpose. Certainly, the operator of
eq.~(\ref{PhiFdef}) defines a group of transformations for a given
gauge theory solution. For ease of reference, we shall refer
to eq.~(\ref{PhiFdef}) as the SCH operator (short for
``single copy holonomy operator''), and the resulting group of
transformations as the SCH group. It may well be that the
SCH group reduces for certain gauge fields. There is also the
interesting possibility of taking gauge theory solutions that are
known single copies of particular gravity solutions, and asking
if the SCH and holonomy groups match up! We will investigate
this by considering particular solutions, of increasing complexity.

\subsection{The Schwarzschild black hole}
\label{sec:schwarzschild}

Arguably the simplest non-trivial gravity solution is the
Schwarzschild solution. It may be sourced by a point mass $M$ sitting
at the origin, and has a known Kerr-Schild form involving spherical
polar coordinates $(t,r,\theta,\varphi)$, where the functions entering
eq.~(\ref{KSdef}) may be chosen as
\begin{equation}
\phi(r)=\frac{M}{4\pi r},\quad k_\mu=(1,1,0,0).
\label{SWKS}
\end{equation}
We can then use eq.~(\ref{gravHol}) to ascertain the holonomy group,
for which we must choose a number of different closed contours, and
see what the various elements of the holonomy group are. Let us first
choose a circular orbit at constant time $t$, in the equatorial plane,
parametrised by
\begin{equation}
C:\quad x^\mu= (0,0,0,\varphi),\quad \varphi\in[0,2\pi).
\label{phiparam}
\end{equation}
The integral appearing in the holonomy operator then reduces to
\begin{equation}
\oint_C d\varphi \, \partial_\sigma (h_{\varphi \nu})M^{\nu\sigma}=0,
\label{intphi0}
\end{equation}
where we have used the fact that the Kerr-Schild graviton implied by
eq.~(\ref{SWKS}) has no non-zero $h_{\varphi \nu}$ components. The
element of the holonomy group associated with $C$ is thus the identity
element. Furthermore, spherical symmetry implies that similar constant
time loops that are tilted with respect to the equatorial plane will
also have a trivial holonomy.\\

\begin{figure}
\begin{center}
\scalebox{0.6}{\includegraphics{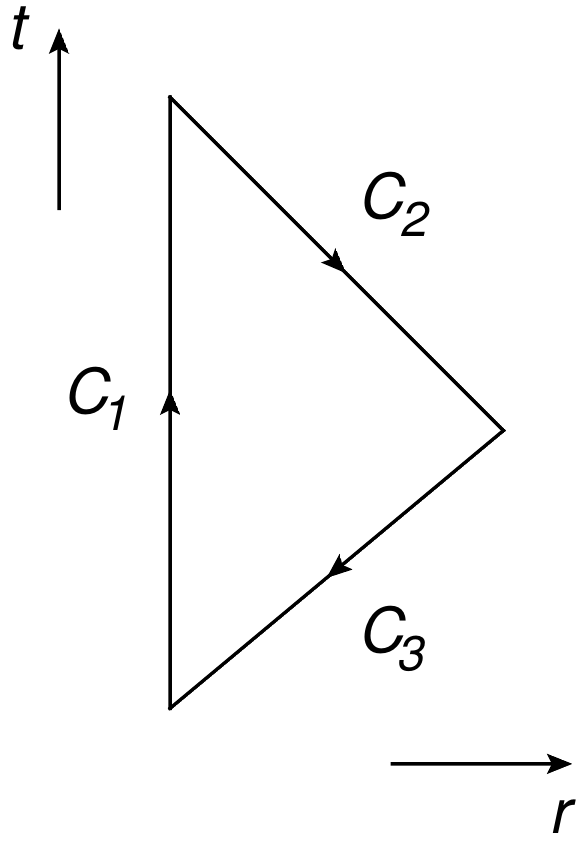}}
\caption{A loop consisting of three segments in the $(r,t)$ plane, where $C_2$ and $C_3$ are null lines.}
\label{fig:rtloop}
\end{center}
\end{figure}

To achieve a non-zero result, one may instead consider the curve shown
in figure~\ref{fig:rtloop}, consisting of three segments. 
The first segment $C_1$ is parallel to the time direction, and may be
parametrised by 
\begin{equation}
C_1:\quad x^\mu=(t,r_0,0,0),\quad 0\leq t\leq T,
\label{rtparam1}
\end{equation}
such that the curve is at a fixed radius $r=r_0$, and of total length
$T$. We have also chosen fixed values $\theta=\varphi=0$. From
eq.~(\ref{gravHol}), one finds
\begin{align}
\int_{C_1} dx^\mu \partial_\sigma (h_{\mu\nu})M^{\nu\sigma}=
\int_0^T dt \partial_r(h_{00})M^{0r}=
-\frac{MT}{4\pi r_0^2}M^{0r},
\label{C1int}
\end{align}
where we have noted the only non-zero contribution after contraction
of indices in the intermediate step. The remaining segments are
parametrised by
\begin{align}
C_2:\quad x^\mu&=(t,r_0+T-t,0,0),\quad T\geq t\geq T/2,\notag\\
C_3:\quad x^\mu&=\left(t,r_0+t,0,0\right),
\quad \frac{T}{2}\geq t\geq 0,
\label{rtparam2}
\end{align}
such that one finds
\begin{align}
\int_{C_2} dx^\mu \partial_\sigma (h_{\mu\nu})M^{\nu\sigma}=
\left.\int_T^{T/2}dt \partial_r(h_{00})M^{0r}\right|_{r=r_0+T-t} + 
\int_{r_0}^{r_0+T/2}
dr\partial_r(h_{r0})M^{0r};\notag\\
\int_{C_3} dx^\mu \partial_\sigma (h_{\mu\nu})M^{\nu\sigma}=
\left.\int_{T/2}^{0}dt \partial_r(h_{00})M^{0r}\right|_{r=r_0+t} + 
\int_{r_0+T/2}^{r_0}
dr\partial_r(h_{r0})M^{0r}.
\label{C23int}
\end{align}
One sees that the radial components cancel. Evaluating the remaining
integrals after using eq.~(\ref{SWKS}) gives
\begin{align}
\int_{C_2\cup C_3}dx^\mu \partial_\sigma (h_{\mu\nu})M^{\nu\sigma}
=\frac{M}{2\pi}\frac{T}{r_0(T+2r_0)}M^{0r},
\label{C23int2}
\end{align}
such that the total contribution from the entire loop is
\begin{align}
\oint dx^\mu \partial_\sigma (h_{\mu\nu})M^{\nu\sigma}=
\alpha M^{0r},\quad \alpha=-\frac{M}{4\pi}\frac{T^2}{r_0^2(T+2r_0)}.
\label{rtholonomy}
\end{align}
This constitutes an infinitesimal boost in the $(t,r)$ plane with
hyperbolic angle $\alpha$. To see this, we may recall the definition
of the boost generators $K_i$ and rotation generators $J_i$ in terms
of the $\{M^{\mu\nu}\}$:
\begin{equation}
K_i=M^{0i},\quad J_i=\frac12 \epsilon_{ijk}M^{jk},
\label{KJ}
\end{equation}
in terms of which the Lorentz algebra 
\begin{equation}
[M^{\mu\nu},M^{\rho\sigma}]=i\left(\eta^{\sigma\mu}M^{\rho\nu}
+\eta^{\nu\sigma}M^{\mu\rho}-\eta^{\rho\mu}M^{\sigma\nu}
-\eta^{\nu\rho}M^{\mu\sigma}\right)
\label{Lorentz}
\end{equation}
may be written as
\begin{equation}
[J_i,J_j]=i\epsilon_{ijk}J_k,\quad
[J_i,K_j]=i\epsilon_{ijk}K_k,\quad 
[K_i,K_j]=-i\epsilon_{ijk}J_k.
\label{KJalgebra}
\end{equation}
Equation~(\ref{rtholonomy}) is then clearly seen to contain the boost
generator $K_r$. If one considers loops with the same $r_0$ but
different fixed values of $\theta$ and $\varphi$, the full set of
boosts associated with arbitrary directions will be obtained. This in
turn implies that the holonomy group of the Schwarzschild spacetime is
SO$(1,d-1)$: from eq.~(\ref{KJalgebra}), one sees that the boosts do
not close upon themselves, such that exponentiating the boost
generators will produce transformations corresponding to combinations
of boosts and rotations in general. Note that our conclusions are in
qualitative agreement with those of e.g. ref.~\cite{Rothman:2000bz}
which also considered the holonomy in Schwarzschild spacetime. Our
explicit result for the boost angle differs due to having a loop with
a slightly different orientation, and also the use of Kerr-Schild
rather than conventional Schwarzschild coordinates.\\

Given this holonomy group, we may now consider the single copy, which
is well-known to be an abelian-like point charge in the gauge
theory~\cite{Monteiro:2014cda}. We may thus consider an abelian
exponent for the SCH operator:
\begin{equation}
\ln(\Phi_F)\rightarrow ig\oint_C ds F_{\mu\nu}M^{\mu\nu}.
\label{abelian}
\end{equation}
The only non-zero component of the field strength in this case
is
\begin{equation}
F_{0r}=\frac{Q}{4\pi r^2},
\label{pointchagre}
\end{equation}
where $Q$ is the charge. Plugging this into eq.~(\ref{abelian}), we
see immediately that an infinitesimal boost in the $(t,r)$ plane is
obtained, directly analogous to the Schwarzschild case. Thus, the
SCH group of the point charge is SO$(1,d-1)$. It is reasonable
to ponder at this point whether it is always the case that the
SCH and holonomy groups match up for gauge theory and gravity
solutions related by the double copy. That this is not the case will
be seen in the following example.

\subsection{Taub-NUT space}
\label{sec:NUT}

The Taub-NUT solution, as explained in Chapter \ref{chap:s-duality}, was first derived in refs.~\cite{Taub:1950ez,Newman:1963yy}, and is a
non-asymptotically flat solution of GR, that has a rotational
character to the gravitational field at infinity. This is due to a
so-called {\it NUT charge} $N$, that is present in addition to a
Schwarzschild-like mass $M$. With a suitable choice of coordinates,
the metric may be written in Lorentzian signature as
    \begin{equation}\label{TNmetric1}
        ds^2 = -A(r)\left[dt+B(\theta)d\phi\right]^2 + A^{-1}(r)dr^2 + C(r)\left[d\theta^2 + D(\theta)d\phi^2\right],
    \end{equation}
where for convenience we have defined 
    \begin{equation}\label{ABCD}
        A(r) = \frac{(r-r_+)(r-r_-)}{r^2 + N^2}, \quad
        B(\theta) = 2N\cos{\theta}, \quad
        C(r) = r^2+N^2, \quad
        D(\theta) = \sin^2{\theta},
    \end{equation}
and 
\begin{equation}
r_{\pm} = M \pm \sqrt{M^2 + N^2}. 
\label{rpm}
\end{equation}
The single copy of this solution was first considered in
ref.~\cite{Luna:2015paa}, and relied on the fact that coordinates
exist in which the Taub-NUT solution has a double Kerr-Schild
form~\cite{Chong:2004hw}. In the single copy, the mass $M$ maps to an
electric charge $Q$, as is familiar from the Schwarzschild case. The
NUT charge $N$, on the other hand, corresponds to a magnetic monopole
in the gauge theory, such that the single copy of the full Taub-NUT
solution is an electromagnetic dyon. This correspondence was
considered further in
refs.~\cite{Bahjat-Abbas:2020cyb,Alfonsi:2020lub}, which demonstrated how
magnetic monopoles in arbitrary non-abelian gauge theories can be
mapped to NUT charge in gravity. \\

For our present purposes, we want to examine the relationship (if
any) between the SCH group in gauge theory, and the holonomy
in gravity. Given that we have already seen this relationship for a
mass term $M$ in gravity, it is convenient to take this to zero, and
thus to consider the metric associated with a pure NUT charge $N$,
which maps to a magnetic monopole in gauge theory, whose magnetic
field may be written as
\begin{equation}
\vec{B}=\frac{\tilde{g}}{4\pi r^2}\hat{\vec{r}},
\label{Bmonopole}
\end{equation}
where $\tilde{g}$ is the magnetic charge, $r$ the spherical radius and
$\hat{\vec{r}}$ a unit vector in the radial direction. Note that we have
again chosen an abelian gauge group in the single copy for simplicity,
but the generalisation to a non-abelian context is
straightforward~\cite{Bahjat-Abbas:2020cyb,Alfonsi:2020lub}. From
eq.~(\ref{EBfields}), one sees that the only non-zero components of
the electromagnetic field strength are
\begin{equation}
F_{\theta\phi}=-F_{\phi\theta}=B_r=\frac{\tilde{g}}{4\pi r^2}.
\label{Fmonopole}
\end{equation}
Thus, the integral appearing in the SCH operator of
eq.~(\ref{PhiFdef}) reduces to
\begin{equation}
\oint_C ds \frac{\tilde{g}}{2\pi r} M^{\theta\phi},
\label{scholB}
\end{equation}
such that only the rotation generator in the $(\theta,\phi)$ plane is
turned on. This integral is indeed non-zero in general. Perhaps the
simplest case one may consider is a constant-time curve of radius
$r=r_0$ in the equatorial plane. All factors appearing in the
integrand in eq.~(\ref{scholB}) can then be taken outside the
integral, which then simply yields the length of the curve. Taking all
possible curves, the generator $M^{\theta\phi}$ will generate
rotations in all possible (Cartesian) directions, but not boosts. From
eq.~(\ref{KJalgebra}), one sees that the rotation algebra closes upon
itself. Thus, we straightforwardly obtain that the SCH group of
the magnetic monopole is SO(3) in four dimensions, and is therefore reduced
compared to the electric case of SO(1,3). \\

\begin{figure}
\begin{center}
\scalebox{0.6}{\includegraphics{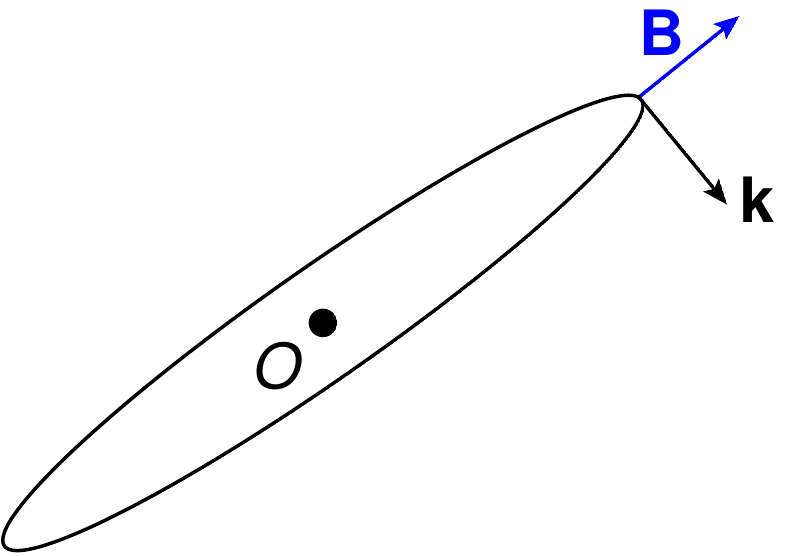}}
\caption{A closed spatial loop tilted with respect to the equatorial
  plane, where $O$ denotes the origin. A monopole magnetic field
  $\vec{B}$ generates a boost in the direction $\vec{k}$.}
\label{fig:boostloop}
\end{center}
\end{figure}

Let us now consider whether this matches up with the holonomy group of
the Taub-NUT solution in gravity. Given the metric of
eq.~(\ref{TNmetric1}), we may choose the vierbein
    \begin{equation}\label{viel}
        e^0 = A^{\frac{1}{2}}(dt + Bd\phi), \quad
        e^1 = A^{-\frac{1}{2}}dr, \quad
        e^2 = C^{\frac{1}{2}}d\theta, \quad 
        e^3 = (CD)^{\frac{1}{2}} d\phi.
    \end{equation}
It will also be useful to have the inverse of these expressions:
    \begin{equation}\label{vielInverse}
        dt = A^{-\frac{1}{2}}e^0 - B (CD)^{-\frac{1}{2}}e^3, \quad
        dr = A^{\frac{1}{2}}e^1, \quad
        d\theta = C^{-\frac{1}{2}}e^2, \quad
        d\phi = (CD)^{-\frac{1}{2}}e^3.
    \end{equation}
The spin connection can be obtained from the torsion-free form of Cartan's structure equations,
    \begin{equation}\label{cartan}
        \tensor{\omega}{^a_b} \wedge e^b = -de^a,
    \end{equation}
along with the metric compatibility condition
    \begin{equation}\label{metricComb}
        \omega_{ab} = -\omega_{ba}.
    \end{equation}
Thus, we first calculate the exterior derivatives of the basis in eq.~\eqref{viel}, and use eq.~\eqref{vielInverse} to write the results in terms of the vielbein basis:
    \begin{align}
        de^0 &= (\pd_r A^{\frac{1}{2}})e^1 \wedge e^0 + (\pd_{\theta} B)C^{-1}\left(\frac{A}{D}\right)^{\frac{1}{2}}e^2 \wedge e^3, \\
        de^1 &= 0, \\
        de^2 &= (\pd_r C^{\frac{1}{2}}) \left(\frac{A}{C}\right)^{\frac{1}{2}}e^1 \wedge e^2, \\
        de^3 &= (\pd_r C^{\frac{1}{2}}) \left(\frac{A}{C}\right)^{\frac{1}{2}}e^1 \wedge e^3 + (\pd_{\theta} D^{\frac{1}{2}})\left(\frac{1}{CD}\right)^{\frac{1}{2}}e^2 \wedge e^3.
    \end{align}
These can now be used in the Cartan structure equations of eq.~\eqref{cartan} which, along with the metric compatibility condition in eq.~\eqref{metricComb}, yields the following non-zero components of the spin connection: 
    \begin{align}
        \tensor{\omega}{^0_1} &= \tensor{\omega}{^1_0} 
        = (\pd_r A^{\frac{1}{2}})A^{\frac{1}{2}}(dt + Bd\phi), \label{sc01} \\
        \tensor{\omega}{^0_2} &= \tensor{\omega}{^2_0} 
        = \frac{1}{2}(\pd_{\theta}B)\left(\frac{A}{C}\right)^{\frac{1}{2}}d\phi, \\
        \tensor{\omega}{^0_3} &= \tensor{\omega}{^3_0}
        = -\frac{1}{2}(\pd_{\theta}B)\left(\frac{A}{CD}\right)^{\frac{1}{2}}d\theta, \\
        \tensor{\omega}{^1_2} &= -\tensor{\omega}{^2_1}
        = -(\pd_rC^{\frac{1}{2}})A^{\frac{1}{2}}d\theta, \\
        \tensor{\omega}{^1_3} &= -\tensor{\omega}{^3_1}
        = -(\pd_rC^{\frac{1}{2}})(AD)^{\frac{1}{2}}d\phi \\
        \tensor{\omega}{^2_3} &= -\tensor{\omega}{^3_2}
        = -\frac{1}{2}(\pd_{\theta}B)\frac{A}{CD^{\frac{1}{2}}}(dt + Bd\phi)
        -(\pd_{\theta}D^{\frac{1}{2}})d\phi. \label{sc23}
    \end{align}
Now by substituting eqs.~\eqref{ABCD} and performing the derivatives we find for Taub-NUT:
    \begin{align}
        \tensor{\omega}{^0_1} &= \tensor{\omega}{^1_0}
        = \frac{M(r^2-N^2)+2N^2r}{(r^2+N^2)^2}\left[dt+2N\cos{\theta}d\phi\right], \label{omegaTN01} \\
        \tensor{\omega}{^0_2} &= \tensor{\omega}{^2_0}
        = -\frac{N\sin{\theta}}{r^2+N^2}\sqrt{(r-r_+)(r-r_-)}d\phi, \\
        \tensor{\omega}{^0_3} &= \tensor{\omega}{^3_0}
        = \frac{N}{r^2+N^2}\sqrt{(r-r_+)(r-r_-)}d\theta, \\
        \tensor{\omega}{^1_2} &= -\tensor{\omega}{^2_1}
        = -\frac{r}{r^2+N^2}\sqrt{(r-r_+)(r-r_-)}d\theta, \\
        \tensor{\omega}{^1_3} &= -\tensor{\omega}{^3_1}
        = -\frac{r\sin{\theta}}{r^2+N^2}\sqrt{(r-r_+)(r-r_-)}d\phi, \\
        \tensor{\omega}{^2_3} &= -\tensor{\omega}{^3_2}
        = \frac{N(r-r_+)(r-r_-)}{(r^2+N^2)^2}dt
        + \left[\frac{2N^2(r-r_+)(r-r_-)}{(r^2+N^2)^2} - 1\right]\cos{\theta}d\phi. \label{omegaTN23}
    \end{align}
For the single copy solution of a magnetic monopole above, we
considered a loop at constant time and radius in the equatorial plane
$\theta = \pi/2$. The integral in the holonomy operator is then simply
    \begin{equation}
        \oint dx^{\mu} (\omega_{\mu})_{ab} M^{ab} 
        = 2\oint d\phi [(\omega_{\phi})_{02} M^{02} + (\omega_{\phi})_{13} M^{13}]
        = 4\pi[(\omega_{\phi})_{02} M^{02} + (\omega_{\phi})_{13} M^{13}].
        \label{hTN}
    \end{equation}
This yields a boost in the 0-2 plane and a rotation in the 1-3 plane,
with coefficients $(\omega_{\phi})_{02}$ and $(\omega_{\phi})_{13}$
respectively, where
    \begin{align}
        (\omega_{\phi})_{02} &=
      \frac{N}{r^2+N^2}\sqrt{(r-r_+)(r-r_-)}, \label{holonomyNUT02}\\ (\omega_{\phi})_{13}
      &= -\frac{r}{r^2+N^2}\sqrt{(r-r_+)(r-r_-)}.
\label{holonomyNUT13}
    \end{align}
Note that the boost and rotation planes are mutually orthogonal, and
such a transformation is conventionally referred to as a {\it Lorentz
  four-screw}. Furthermore, our results are in agreement with those of
ref.~\cite{Bini:2002wd}, despite the different choice of vierbein
adopted by that reference. Equations~(\ref{holonomyNUT02},~\ref{holonomyNUT13}) still
correspond to the general Taub-NUT solution. We wish to consider the
double copy of the pure magnetic monopole, i.e. a pure NUT charge,
such that we may set $M\rightarrow0$ in eqs.~(\ref{holonomyNUT02},~\ref{holonomyNUT13}). The
integral of eq.~(\ref{hTN}) then becomes
    \begin{equation}
        \oint dx^{\mu} (\omega_{\mu})_{ab} M^{ab}=
        \frac{4\pi\sqrt{r^2-N^2}}{r^2+N^2}\left[
          N M^{02}-r M^{13}\right].
        \label{hTN2}
    \end{equation}
We thus see that the boost generator survives even in the case of a
pure NUT charge. By the arguments of the previous section, this will
potentially lead to the holonomy group SO(1,3), unless the effect of
the boost can be removed by performing a similarity transformation on
all group elements. However, upon considering other loops, boosts in
different Cartesian directions are generated. To see this, we may use
the fact that the metric for a pure NUT charge has a single
Kerr-Schild form, and thus we may use the expression of
eq.~(\ref{KSintegrand}) for the integrand of the holonomy operator. The
coefficient of the boost generators, including the measure, is then
\begin{align}
dx^\mu \partial_\sigma(\phi k_\mu)M^{0\sigma}=
dx^\mu\left[\partial_\sigma(\phi k_\mu)-\partial_\mu(\phi k_\sigma)\right]
M^{0\sigma},
\label{hcalcKS}
\end{align}
where we have used the fact that $k_0=1$ for this
solution~\cite{LunaGodoy:2018tyq}, and in the second term we have introduced a total
derivative term that integrates to zero around a closed loop. From
eq.~(\ref{AKS}), we may recognise the expression in the closed brackets as the field strength tensor of a gauge field that is the single
copy of a Kerr-Schild graviton. One then finds

\begin{align}
dx^\mu \partial_\sigma(\phi k_\mu)M^{0\sigma}&=
dx^\mu F_{\sigma\mu}M^{0\sigma}\notag\\
&= dx_j \epsilon_{ijk} B_k K_i\notag\\
&=\vec{K}\cdot\left[d\vec{x}\times\vec{B}\right],
\label{hcalcKS2}
\end{align}
where we have used eqs.~(\ref{EBfields}, \ref{KJ}). The physical content of eq.~(\ref{hcalcKS2}) can be understood by considering a loop at constant time and
radius, that is tilted relative to the equatorial plane, as shown in
figure~\ref{fig:boostloop}.
The field of a magnetic monopole points radially outwards, whereas the
tangent 3-vector to the curve $d\vec{x}$ points into the page at the
point shown. This generates a boost in the direction $\vec{k}\propto
d\vec{x}\times\vec{B}$, which is easily seen to be in the increasing
$\theta$ direction. Thus, the conclusion reached above for the $\theta=\pi/2$ case, namely that there is a boost in the $(t,\theta)$
plane, turns out to be general for all such constant time loops. We
can therefore conclude that boosts in all Cartesian directions will be
turned on, such that the holonomy group of the pure NUT charge is
indeed SO(1,3), in contrast to the SCH group of the magnetic
monopole. Na\"{i}vely, one might have expected these groups to match
up, given that there are a well-known set of analogies between
electromagnetism and gravity known as {\it gravitomagnetism}. The pure
NUT charge is an extremal case of the Taub-NUT solution, that is most
like a magnetic monopole from a purely gravitational point of
view. However, the fact that its holonomy is sensitive to
gravitoelectric as well as gravitomagnetic effects is
well-documented~\cite{Bini:2002wd}, and arises from the fact that
spacelike hypersurfaces in the NUT solution have a nonzero extrinsic
curvature. This is not the case in the single copy gauge theory, which
lives in Minkowski space.

\subsection{Self-dual solutions}
\label{sec:self-dual}

The above examples probe different types of behaviour of the
SCH operator. For the Schwarzschild / point charge system, neither the
SCH nor the holonomy group reduce from their general form of
SO(1,3). For the NUT / monopole solutions, the SCH group does indeed
reduce, but the holonomy does not. In this section, we demonstrate
another possibility. Namely, that the SCH and holonomy groups
both reduce to the same subgroup of SO(4). We will
explicitly consider the case of {\it self-dual} solutions in gauge
theory and gravity. The following argument, in particular for gravity and non-Ableian gauge theories, relies on requiring the loops to be `small'. The reason is that one can not use Stoke's theorem if the underlying bundle is non-trivial; see \cite{schreiber2011smooth,Alvarez_1998} for nice reviews and theorems. Now in gravity, one may decompose the Riemann tensor
into self-dual and anti-self-dual parts, given respectively by
\begin{equation}
R^\pm_{\mu\nu\rho\lambda}={(P^\pm)^{\alpha\beta}}_{\mu\nu} 
R_{\alpha\beta\rho\lambda},
\label{Rpmdef}
\end{equation}
where we have defined the projectors
\begin{align}
{(P^\pm)^{\alpha\beta}}_{\mu\nu} = \frac{1}{2}\left(\delta^{\alpha}_\mu\,\delta^\beta_\nu-
\delta^{\alpha}_\nu\,\delta^{\beta}_\mu\pm\sqrt{g}
{\epsilon^{\alpha\beta}}_{\mu\nu}\right),
\label{proj}
\end{align}
with $g$ denoting the determinant of the metric.
Note also that we work in Euclidean signature throughout this section
only. We may use Stokes' theorem, while having the above point in mind, to rewrite the holonomy
operator of eq.~(\ref{spinholonomy2}) as~\cite{Dowker:1967zz}
\begin{align}
\Phi_\omega&={\cal P}\exp\left(-\frac{i}{2}\iint_\Sigma
d\Sigma^{\mu\nu}\,R_{cd\mu\nu}M^{cd}\right)\notag\\
&={\cal P}\exp\left(-\frac{i}{2}\iint_\Sigma
d\Sigma^{\mu\nu}\,(R^+_{\rho\lambda\mu\nu}
+R^-_{\rho\lambda\mu\nu})M^{\rho\lambda}\right).
\label{Stokes}
\end{align}
where $\Sigma$ is the area bounded by the curve $C$, with area element
$d\Sigma^{\mu\nu}$, and we have also converted vielbein indices to
coordinate frame indices in the second line. Note that $\Sigma$ is
simply connected, and therefore the holonomy group reduces to the
restricted holonomy group ${\rm Hol}^0(\nabla)$, which is only equal
to the full group $\text{Hol}(\nabla)$ when the fundamental homotopy
group $\pi_1$ is trivial. The second term is zero by definition for a
self-dual solution, in which case the holonomy operator becomes
\begin{align}
\Phi_\omega&={\cal P}\exp\left(-\frac{i}{2}\iint_\Sigma
d\Sigma^{\mu\nu}\, {(P^+)^{\alpha\beta}}_{\rho\lambda}R_{\alpha\beta\mu\nu}
M^{\rho\lambda}\right)\notag\\
&={\cal P}\exp\left(-\frac{i}{2}\iint_\Sigma
d\Sigma^{\mu\nu}\, R_{\alpha\beta\mu\nu}
(M^+)^{\alpha\beta}\right),
\label{SDhol2}
\end{align}
where we have defined two linearly independent sets of $\text{SO}(4)$
generators via
\begin{equation}
(M^\pm)^{\alpha\beta}={(P^\pm)^{\alpha\beta}}_{\rho\lambda}M^{\rho\lambda}.
\label{Mpm}
\end{equation}
This amounts to the known lie algebra isomorphism $\mathfrak{so}(4)\simeq \mathfrak{su}(2)\oplus \mathfrak{su}(2)$,
where each $\mathfrak{su}(2)$ subgroup corresponds to the generators
$(M^{\pm})^{\alpha\beta}$ respectively. It follows that for self-dual
solutions, the holonomy group reduces to ${\rm SU}(2)$, and
similar arguments can be applied to the case of anti-self-dual
solutions. The previous argument is really a demonstration of the Ambrose-Singer theorem  \cite{MR63739} from section \ref{Ambrose-Singer}.\\

The single copy of a self-dual gravity solution is also self-dual in
the gauge theory~\cite{Monteiro:2013rya}. One may define the
(anti-)self-dual parts of the field strength as follows:
\begin{equation}
F^{\pm}_{\mu\nu}={(P^\pm)_{\mu\nu}}^{\alpha\beta}F_{\alpha\beta},
\label{FSD} 
\end{equation}
where one uses the projectors of eq.~(\ref{proj}), but where now
the metric corresponds to that of Euclidean flat space (albeit
potentially in a curvilinear coordinate system, so that one must keep
the factor of $\sqrt{g}$). Self-dual solutions are defined by
$F^-_{\mu\nu}=0$, so that the SCH operator becomes
\begin{align}
\exp\left[ig\oint_C
ds F^+_{\mu\nu} M^{\mu\nu}\right]
&=\exp\left[ig\oint_C
ds {(P^+)_{\mu\nu}}^{\alpha\beta}F_{\alpha\beta} M^{\mu\nu}\right]\notag\\
&=\exp\left[ig\oint_C
ds F_{\alpha\beta}(M^+)^{\alpha\beta}
\right].
\label{PhiFSD}
\end{align}
Again only half of the generators are turned on, so that the SCH group
reduces to ${\rm SU}(2)$. The self-dual sector thus provides an interesting
example, in which the SCH and holonomy groups reduce, and are 
isomorphic. Furthermore, it is interesting to ponder whether the
arguments of this section generalise to manifolds of exotic holonomy
in higher dimensions, such as the well-known cases with holonomy
groups G$_2$ and Spin(7). It is not known how to explicitly single
copy such manifolds (see e.g. ref.~\cite{Alawadhi:2020jrv} for a
related discussion), but seeking Yang-Mills solutions with a suitable
SCH group might be a good place to start.

\section{Discussion}
\label{sec:conclude}

In this paper, we have considered the holonomy group in gravity, which
consists of the group of transformations acting on vectors that have
been parallel transported around a closed curve. The analogue of this in a (non)-abelian gauge theory is the
Wilson loop, which has a physical interpretation in terms of the phase
experienced by a scalar particle traversing a closed
contour. Attempts to match up the physical properties of the holonomy
and Wilson line -- or to interpret the holonomy operator itself as a
gravitational Wilson loop -- have been made
before~\cite{Modanese:1991nh,Modanese:1993zh,Hamber:2007sz,Hamber:2009uz,Brandhuber:2008tf},
with the conclusion that the gravitational holonomy should not be
thought as being the correct physical analogue of the gauge theory
Wilson line. Indeed, a different gravitational Wilson line exists,
which corresponds to the phase experienced by a scalar
particle~\cite{Hamber:1994jh,Dowker:1967zz,Naculich:2011ry,White:2011yy}. This
begs the question of what the correct gauge theory analogue of the
gravitational holonomy is. To investigate this we have
used the {\it single copy}. We showed that the gravitational
holonomy arises naturally in the description of a spinning particle
interacting with a gravitatonal field. The single copy of this
situation is well-known to be a spinning particle interacting with a
gauge field. This allowed us to construct a generalised Wilson line
operator in the gauge theory, which gives the phase (non-diagonal in
spin space) experienced by a spin-1 test particle having an electric
and magnetic dipole moment. \\

Having found the single copy of the holonomy operator -- which we
dubbed the SCH operator -- we then commenced an
exploration of its properties.  We looked at certain special cases in which the
SCH group reduces, which includes the case of a pure magnetic
monopole, and also solutions which are self-dual. For the former, the
SCH group reduces even though the gravitational holonomy of the
monopole's double copy counterpart (a pure NUT charge) does not. \\

There are a number of avenues for further work. Firstly, one could
apply the SCH operator to different Yang-Mills solutions, and
see what general conclusions can be reached about their possible
SCH groups. It would also be interesting to look at how to
match the holonomy and its single copy in gauge and gravity theories more
generally, which might help in extending the classical double copy to
more complicated cases than are currently possible. Thirdly, it would
be nice if the SCH operator could shed light on
non-perturbative aspects of the double copy. In particular, we note
that the SCH operator is matrix-valued both in colour and spin
space. It thus rotates vectors both in the internal space associated
with the colour degrees of freedom, and also in the tangent space of
the manifold, which is associated with kinematic information. Might the single copy of the holonomy then have something to do with BCJ
duality~\cite{Bern:2008qj}, which links colour and kinematics in an
intriguing way? 

%% file: chaps/chap06.tex
\chapter{The Ricci flow, RG flow, and the Single Copy}\label{chap:RicciFlow}
This chapter is based on \cite{Alawadhi:2022gwy}.
\section{Introduction}
The premise of this chapter is simple. We first notice that the double copy relates the metric tensor to the gauge connection, and the Ricci tensor to the gauge curvature, in the sense of eqs. \eqref{RicciSingle} and \eqref{RicciField}. Moreover, we have seen in Chapter \ref{chap:Holonomy} and the references within that the double copy could also preserve topological information between gravitational and gauge theories. In the previous chapter, the author of this thesis proposed to study the concept of holonomy in the context of the double copy for its own sake, as a mathematical object. As it turned out, both holonomy and its proposed single copy have physical interpretations which can prove useful to further understand physical theories. Similarly, for its own mathematical sake, the author asked the question of what is the single copy of the Ricci flow as we shall see soon. The topic of the Ricci flow is more commonly explored in the mathematical literature in particular in the fields of differential geometry and topology. Similarly to the concept of holonomy, the Ricci flow also appears in the physics literature as the beta function of the closed string. More specifically, the Ricci flow is said to be analogous to the heat equation for Riemannian metrics. Since it was first introduced in the mathematics literature, it has found a tremendous number of applications in the fields of geometry and topology, wherein Perelman used the Ricci flow to prove the Poincaré conjecture; then an open problem since 1904 \cite{Perelman:2006un,Perelman:2006up,Perelman:2003uq}. In the physics literature, the Ricci flow appeared in Friedan's PhD thesis \cite{Friedan:1980jm} as the beta function of the non-linear sigma model. Since then the Ricci flow has found many applications in physics including string theory, gravity, black hole thermodynamics, cosmology and more \cite{Woolgar:2007vz,Petropoulos:2010zz,DeBiasio:2020xkv,Headrick:2006ti}.
. In this chapter, we shall briefly review the concept of the Ricci flow in both the mathematical and physics literature. In particular, the Ricci flow appears as the beta function of the sigma model describing a closed string, and similarly its single copy describes the beta function of the open string, known as the Yang-Mills flow in the mathematical literature. The ability of the double copy in relating the beta functions of the open and closed string to each other is reminiscent of the well-known KLT relations in string theory \cite{Kawai:1985xq}.

In this chapter we use the Kerr-Schild double copy formalism on Minkowski background as given in \ref{KSReview}. We assume the single copy gauge field to be effectively abelian, or as a connection on a trivial $\text{U}(1)$-principal bundle.
\section{The Ricci flow}\label{sec:RicciFlow}

The Ricci flow is a particular PDE for the metric tensor, where the latter evolves according to its Ricci tensor. Given a $\text{d}$-dimensional Riemannian manifold endowed with a metric $(\mc M, g)$, the Ricci flow equation is 
\begin{equation}\label{RicciFloweq1}
    \frac{\pd}{\pd \lambda}g_{\mu\nu} = -2 R_{\mu\nu},\quad g_{\mu\nu}(\lambda = 0) = g^0_{\mu\nu},
\end{equation}
where both $g_{\mu\nu}$ and $R_{\mu\nu}$ are functions of the flow parameter $\lambda\in\mathbb{R}$. Eq.~\eqref{RicciFloweq1} describes the evolution of $g_{\mu\nu}(\lambda)$ in the space of metrics $\mathfrak{G}$. To illustrate this, let us see how a generic Einstein metric evolves. For Einstein spaces the Ricci tensor takes a particularly simple form given by $R_{\mu\nu} = k g_{\mu\nu}$ for some constant $k\in\mathbb{R}$. Plugging this form of the Ricci tensor into eq.~\eqref{RicciFloweq1} we find
\begin{equation}
    \frac{\pd}{\pd\lambda}g_{\mu\nu} = -2kg_{\mu\nu}.
\end{equation}
Let us take the $d$-sphere and see how it evolves under the flow. In this case we have $g_{\mu\nu}=r^2g^{S^d}_{\mu\nu}$ and $k = \frac{d-1}{r^2}$. Plugging into the the above equation we find
\begin{equation}
    r(\lambda) = \sqrt{r^2_0-2(d-1)\lambda},
\end{equation}
where $r(\lambda)$ is the radius of the $\text{d}$-sphere at a given value of the flow parameter $\lambda$. Therefore, we see that the sphere shrinks as the metric flows until it becomes singular in finite flow time given by $\lambda^* = r^2_0/(2(d-1))$. Another physically relevant example are dS and AdS spaces. In this case we have 
\begin{equation}
    R_{\mu\nu} = \frac{2\Lambda}{d-2}g^0_{\mu\nu},
\end{equation}
 and the flow gives us
 \begin{equation}
     g_{\mu\nu}(\lambda)=\bigg( 1-\frac{4\Lambda}{d-2}\lambda \bigg)g^0_{\mu\nu}.
 \end{equation}
 For $\Lambda > 0 $ the space shrinks to a point in finite flow time given by $\lambda^* = \frac{d-2}{4\Lambda}$. The case $\Lambda = 0$ is a critical point on the flow and therefore the space does not evolve. For $\Lambda<0$ the flow is immortal and the space expands indefinitely. See \cite{DeBiasio:2020xkv} for some examples and in more detail As mentioned earlier, the Ricci flow is analogous to the heat equation for the metric tensor. To make this analogy more clear, one can compute the Ricci tensor in harmonic coordinates. Harmonic coordinates were studies considerably in the context of Lorentzian geometry and relativity by Einstein and Cornelius Lanczos \cite{zbMATH01585275, zbMATH02603727}. In 1922, Lanczos discovered the local form of the Ricci tensor in harmonic coordinates which resembles the heat equation, an elliptic operator. For a Rimeannian manifold $(\mc{M},g)$, a coordinate chart with coordinate functions $(x^1,x^2,\ldots, x^m)$ is said to be harmonic if the coordinate functions satisfy
 \begin{equation}
     \Delta_g x^m = 0,
 \end{equation}
 where $\Delta_g$ is the Laplace-Beltrami operator of the metric $g$. In these coordinates, the Christoffel symbols satisfy the following identity
 \begin{equation}
     g^{\mu\nu}\Gamma^\alpha_{\mu\nu} = 0.
 \end{equation}
 In particular, the Ricci tensor takes a simple form,
  \begin{equation}
    R_{\mu\nu} = -\frac{1}{2}g^{\alpha\beta}\frac{\pd^2 g_{\mu\nu}}{\pd x^\alpha\pd x^\beta}+\cdots,
  \end{equation}
  where the dots denote lower order terms in $\pd$ \cite{topping_2006, ASENS_1981_4_14_3_249_0 }. The Ricci flow equation \eqref{RicciFloweq1} then gives
  \begin{equation}
    \frac{\pd}{\pd \lambda}g_{\mu\nu} = g^{\alpha\beta}\frac{\pd^2 g_{\mu\nu}}{\pd x^\alpha\pd x^\beta}+\cdots,
  \end{equation}
  which resembles the familiar heat equation. Essentially, as mention earlier, one evolves the geometry of the manifold according to the local value of the curvature at each point. While this is a parabolic equation(for a Riemannian manifold), harmonic coordinates need not remain harmonic along the flow at a later time and the flow might seize to be strongly parabolic. In \cite{DennisFlow}, DeTurck provides a way to restore strong parabolicity along the flow. We shall come to this issue in the next section.
  \section{Kerr-Schild flow}
  Authors in \cite{Coll:2000rm,Hildebrandt:2002qh, Hildebrandt:2002qg} take a different point of view on Kerr-Schild metrics. They view it as a local continuous transformation that acts on the metric. They show that there exists a continuous group of such transformations, putting it on equal footing with isometries and conformal transformations of the metric. In the following, we shall follow closely \cite{Coll:2000rm,Hildebrandt:2002qh, Hildebrandt:2002qg} and summarise the main points relevant to us. Let $\mathcal{M}$ be a Lorentzian manifold equipped with a metric tensor $g$ with signature $(-,+,\ldots,+)$. The indices of a coordinate system on $\mathcal{M}$ are to be written in greek and run from $0$ to $d-1$ where $d$ is the dimension of $\mathcal{M}$.
  \begin{definition}\label{Def:KSDiff}
  A one parameter group of transformations \{$\phi_s$\} of $\mathcal{M}$, where $s\in \mathbb{R}$, is called a \textit{Kerr-Schild diffeomorphism} if the transformed metric has the form
  \begin{equation}\label{KS_Diff}
      \phi_s^*(g) = g + 2H_s \ell\otimes \ell,
  \end{equation}
  where $H_s$ are functions on $\mathcal{M}$.
  \end{definition}
  Group homomorphism, $\phi_s\phi_r = \phi_{s+r}$, requires the null-forms to transform as
  \begin{equation}\label{null_Diff}
      \phi^*_s(\ell) = M_s\ell,
  \end{equation}
  where $M_s$ are functions on $\mathcal{M}$. If we differentiate equations \eqref{KS_Diff} and \eqref{null_Diff} with respect to the parameter $s$ around zero, we find that the transformations take the infinitesimal forms
  \begin{equation}\label{KS_rel1}
      \pounds_{\Vec{\xi}}\,g = 2h\ell\otimes \ell,
  \end{equation}
  and
  \begin{equation}\label{KS_rel2}
  \pounds_{\Vec{\xi}}\,\ell = m\ell,
  \end{equation}
  where we defined $h = dH_s/ds|_{s=0}$ and $m = dM_s/ds|_{s=0}$ and $\pounds_{\Vec{\xi}}$ is the lie derivative with respect to some vector $\Vec{\xi}$. Unlike the concept of isometries, the null 1-form $\ell$ plays a crucial role in Kerr-Schild groups. To this end, the authors give the following definitions.
  \begin{definition}
      Any solution $\Vec{\xi}$ of the Kerr-Schild equations \eqref{KS_rel1} and \eqref{KS_rel2} will be called a Kerr-Schild vector field(KSVF) with respect to a particular $\ell$. The functions $h$ and $m$ are called the gauges of the metric $g$ and the null 1-form $\ell$.
  \end{definition}
  Since any killing vector $\Vec{X}$ which leaves $\ell$ invariant is also a KSVF but with $h=0$ from $\pounds_{\Vec{X}}g=0$, one defines \emph{proper} Kerr-Schild vector fields,
  \begin{definition}
      A non-zero KSVF $\Vec{\xi}$ is called a proper KSVF if its metric deformation $\pounds_{\Vec{X}}g$ is non-vanishing.
  \end{definition}
  An important property of KSVF, which will help us find solutions, is given by the following proposition,
  \begin{proposition}
      Two metrics related by a Kerr-Schild transformation, $\Tilde{g} = g + 2 H \ell \otimes \ell$, admit the same KSVFs with respect to $\ell$.
  \end{proposition}
  An important corollary when the KSVF is a killing vector field of the Kerr-Schild transformed metric
  \begin{equation}
      \Tilde{g} = g + 2H\ell\otimes\ell,\quad \pounds_{\Vec{\xi}}\,\Tilde{g} = 2\Tilde{h}\ell\otimes\ell=0,
  \end{equation}
where $\Tilde{h} = h + \pounds_{\Vec{\xi}}H + 2mH$. Therefore, the equation $\Tilde{h}=0$ admits local solutions in $H$. For further discussion of the Kerr-Schild transformation as a diffeomorphism on the manifold, the interested reader is referred to the papers \cite{Coll:2000rm, Hildebrandt:2002qh, Hildebrandt:2002qg}; in particular, the last reference which acts as a mathematically detailed review for the topic.

Treating Kerr-Schild metrics as those obtained by a Kerr-Schild diffeomorphism, we turn to the issue of diffeomorphism invariance of the Ricci flow. This is important to show that the metric tensor remains of Kerr-Schild form along the flow, allowing us to apply the Kerr-Schild double copy prescription. To this end, we note that in \cite{DennisFlow} it was shown that one can make the Ricci flow equation strongly parabolic by adjusting it with a diffeomorphism along the flow. This can be done by relating the Ricci flows of two metrics related by a diffeomorphism. We shall sketch the idea without providing a rigorous proof. The reader who is looking for rigour should consult \cite{chow2004ricci}, and \cite{WanFlow} for a treatment in local coordinates for specific non-compact Riemannian manifolds. Indeed, the reader should be warned that the aforementioned and following cases assume non-Lorentzian and compact manifolds; therefore, care should be taken when applying such arguments to our Kerr-Schild flow which is generally defined for Lorentzian and non-compact manifold. Let us begin by defining the flow equations for two metrics $g(t)$ and $\bar{g}(t) = (\varphi_t)_*g(t)$ where $\varphi_t : \mc{M}\rightarrow\mc{M}$ is a 1-parameter family of diffeomorphism generated by some vector field $W(t)\in T\mc{M}$. The flow equation are
\begin{equation}
    \begin{split}
        \frac{\pd}{\pd t}g(t) &= -2\text{Ric}[g(t)]\\
        \frac{\pd}{\pd t}\bar{g}(t) &= -2\text{Ric}[\bar{g}(t)],\\
    \end{split}
\end{equation}
the idea is then to show that the two flows are related to each other by a diffeomorphism term involving the Lie derivative. To this end we differentiate and use the definition of the Lie derivative,
\begin{equation}
    \begin{split}
    \frac{\pd}{\pd t}g(t) &= \frac{\pd}{\pd t}\Big( (\varphi_t^{-1})^*\bar{g}(t) \Big)\\[5pt]
    &= (\varphi_t^{-1})^* \bigg( \frac{\pd}{\pd t}\bar{g}(t) \bigg) + \mc{L}_{(\varphi_t)_{*}(\frac{\pd}{\pd t}\varphi_t)}\Big[  (\varphi_t^{-1})^*\bar{g}(t) \Big]\\[5pt]
    &= (\varphi_t^{-1})^*(-2\,\text{Ric}[\bar{g}(t)]) + \mc{L}_{  (\varphi_t)_{*}(\varphi_t^{*}[W(t)])    }[g(t)]\\[5pt]
    &= -2\,\text{Ric}[g(t)] + \mc{L}_{W(t)}[g(t)].
    \end{split}
\end{equation}
Hence, the flow of the metrics $g(t)$ and $\bar{g}(t)$ are related by a diffeomorphism term generated by a vector field $W(t)$ on $\mc{M}$. In local coordinates it takes the form
\begin{equation}\label{DiffFlow}
    \frac{\pd}{\pd t}g_{\mu\nu} = -2R_{\mu\nu} + \nabla_\mu W_\nu + \nabla_\nu W_\mu.
\end{equation}
Now if we take the diffeomorphism $\varphi$ to be a Kerr-Schild diffeomorphism, as described above in \ref{Def:KSDiff}, generated by a vector field $W = \xi$, then equation \eqref{DiffFlow} relates the Ricci flow of two metrics related by a Kerr-Schild diffeomorphism.
\section{The Ricci flow as RG flow}
The Ricci flow also has relevance in the physics literature, first appearing in the context of the RG flow of the non-linear sigma model \cite{Friedan:1980jm}. Let us briefly illustrate how.

 Consider a map $f$ between two Riemannian manifolds with metrics $\gamma$ and $g$
\begin{equation}\label{mapf}
    f: (\mc B, \gamma)\rightarrow (\mc M, g),
\end{equation}
where $\mc B$ is 2-dimensional and $\mc M$ is $d$-dimensional. An action integral is given by
\begin{equation}
    S[f; \gamma, g] =\frac{1}{4\pi\alpha'} \int_{\mc V \subset \mc B}(\gamma, f^{*}g)_{\gamma}\omega_\gamma.
\end{equation}
where $\alpha'\in\mathbb{R}$ is some constant, $f^* g$ is the pullback of the metric $g$ of $\mc M$ to $\mc B$. The parenthesis $({\,},{\,})_{\gamma}$ denotes contraction with metric $\gamma_{ab}$. Finally, $\omega_\gamma = \sqrt{|\gamma|}d^2x$ denotes the volume-form on $\mc B$. Therefore, our action can be written more explicitly as

\begin{equation}
    S[f; \gamma, g] = \frac{1}{4\pi\alpha'}\int_{\mc V}\gamma^{ab}\frac{\pd y^\mu}{\pd x^a}\frac{\pd y^\nu}{\pd x^b} g_{\mu\nu}\sqrt{|\gamma|}d^2x ,
\end{equation}

where $x^a=(x^1,x^2)$ are coordinates on $\mc V\subset \mc B$ and $y^\mu=(y^1,\ldots,y^d)$ on $\mc M$. In the context of string theory one would call $\mc B$ the world-sheet, $\mc M$ spacetime, and $\alpha' \in \mathbb{R}$ the Regge slope. Writing the action for a 2D world-sheet in conformal gauge \cite{Green:2012oqa} where $\gamma_{\mu\nu}\rightarrow\eta_{\mu\nu}$ we get
\begin{equation}
      S[f; \gamma, g] = \frac{1}{4\pi\alpha'}\int_{\mc V} \pd_a y^\mu \pd^{a} y^\nu g_{\mu\nu} d^2x,
\end{equation}
where a contraction with the flat world-sheet metric is done. The non-linear sigma model action describes the propagation of a closed string on a $d$-dimensional spacetime $\mc (M,g)$. Now if we view the target space metric $g_{ab}$ as a coupling constant governing the strength of the interaction, expand around a classical background $y^\mu(x) = \bar{y}^\mu + \sqrt{\alpha'}Y^\mu (x)$, and compute the amplitude at one loop order we find that we have to renormalise to avoid divergences. The renormalisation results in a correction to the target space metric given by $g_{\mu\nu} \rightarrow g_{\mu\nu} + \alpha' R_{\mu\nu} $, where $R_{\mu\nu}$ is the Ricci tensor. The beta function is indeed given by
\begin{equation}\label{betaG}
    \beta_{\mu\nu}^g \coloneqq \frac{\pd}{\pd \lambda}g_{\mu\nu} = \alpha' R_{\mu\nu}
\end{equation}
for some energy scale $\lambda \in \mathbb{R}$. Indeed, eq.~\eqref{betaG} is the Ricci flow equation(with $\alpha' = -2 $).

In this section we reviewed the basics of the Ricci flow with some examples. We also gave a brief description of its relevance in the physics literature in the context of string theory. In the following section we introduce the main result of the chapter, the Yang-Mills flow as the single copy of the Ricci flow. Physically, the Yang-Mills flow is known to be the first order correction to the Yang-Mills field of an open string.

\section{Yang-Mills flow}
\label{sec:YMFlow}
To obtain the single copy of the Ricci flow of eq.~\eqref{RicciFloweq1}, and in accordance with our method of the double copy, we restrict our class of metrics to the algebraically special Kerr-Schild metrics that satisfy the properties in eqs. \eqref{KSFORM} and \eqref{KSrelations}, and a flat Minkowski spacetime as the background. Thus writing the Ricci flow equations as
\begin{equation}
    \frac{\pd}{\pd\lambda}\Big( \eta_{\mu\nu} + \phi(\lambda) k_{\mu}(\lambda)k_{\nu}(\lambda) \Big) = -2 R_{\mu\nu},
\end{equation}

and noting that the first term on the LHS does not depend on the flow parameter $\lambda$, hence $\pd_\lambda \eta_{\mu\nu} = 0$. Dropping the explicit dependence on $\lambda$ we now have
\begin{equation}
    \frac{\pd}{\pd\lambda}( \phi\, k_{\mu}k_{\nu} ) = -2 R_{\mu\nu}.
\end{equation}

Now the Ricci tensor and hence the vacuum Einstein equations only linearise in $\phi$ if one of the indices is contravariant \cite{Monteiro:2014cda}. Therefore we raise one index by contracting it with $g^{\mu\alpha}$, thus obtaining

\begin{equation}
    \frac{\pd}{\pd\lambda}(\phi\, k^{\alpha}k_{\nu})-\phi^2 k^{\alpha}k_{\mu}\frac{\pd k^{\mu}}{\pd\lambda}= -2 R^{\alpha}{}_{\nu},
\end{equation}
where we used the fact that $k_{\mu}$ is null as given by eq.~\eqref{KSrelations}. In fact, the second term on the LHS is zero since

\begin{equation}
   k_{\mu} \frac{\pd k^{\mu}}{\pd\lambda} = \frac{1}{2}\frac{\pd}{\pd\lambda}(k_\mu k^\mu) = 0.
\end{equation}
Therefore, the Ricci flow equation takes on a particularly simple form ready for single copying:
\begin{equation}\label{KSmixFlow}
    \frac{\pd}{\pd\lambda}(\phi\, k^{\alpha}k_{\nu}) = -2 R^{\alpha}{}_{\nu}.
\end{equation}
Now if we simply follow the argument underlined in Chapter \ref{chap:s-duality} and define a Yang-Mills connection form $A_{\nu}=\phi\, k_{\nu}$ with $\alpha = 0$, we find the Yang-Mills flow\footnote[1]{For a discussion on parabolicity, gauge, and diffeomorphism invariance for both the Ricci and Yang-Mills flow see \cite{Headrick:2006ti,Streets:2007kct}.}
\begin{equation}\label{YMFlow}
    \frac{\pd}{\pd\lambda}A_{\mu} = -\pd_{\nu}F_{\mu}{}^{\nu}
\end{equation}

Given our choice of applying the Kerr-Schild single copy on a flat Minkowski background, equation~\eqref{YMFlow} should be interpreted as the flow in the space of connections of a principal $\text{U}(1)$-bundle $(P,\mathbb{R}^{1,d-1},\text{U}(1))$ with curvature $F \in \Omega^{2}(\mathbb{R}^{1,d-1},\mathfrak{u}(1))$. Again, eq.~\eqref{YMFlow} is analogous to a diffusion equation for the gauge field $A_{\mu}$ since with $F_{\mu\nu} = 2\pd_{[\mu}A_{\nu]}$ we have
\begin{equation}
    \frac{\pd}{\pd \lambda}A_{\mu} = \pd^2A_{\mu} - \pd_\nu \pd_\mu A^{\nu}.
\end{equation}
Now that we have introduced the Yang-Mills flow, let us see where it appears in the physics literature.

In studying the interaction of closed and open strings in background fields \cite{Abouelsaood:1986gd,Fradkin:1985qd,Callan:1986bc}, divergences arise which upon cancelling them give rise to a set of loop corrected beta functions. One of those is the beta function for the open string coupled to the electromagnetic gauge field potential $A_{\mu}$. Fradkin and Tseytlin showed in \cite{Fradkin:1985qd} that one finds non-linear corrections to Maxwell's equations upon cancelling the divergences that appear at tree and loop level in the open string sigma model. Subsequently, it was further studied by others in \cite{Abouelsaood:1986gd,Callan:1986bc}. Let us briefly show how the above comes about. Starting from the sigma model functional for an open string coupled to a background gauge field $A_{\mu}$,
\begin{equation}
    \begin{split}\label{openStringSigma}
        S[f; \gamma, \eta]= \frac{1}{4\pi\alpha'}\int_{\mc{V} \in \mc{B}}\gamma^{a b}&\pd_{a}y^{\mu}\pd_{b}y_{\mu}\sqrt{\gamma}d^{2}x\\
    & - \oint_{\pd\mc{B}}A_{\mu}(y)\frac{\pd}{\pd s}y^{\mu}ds,
    \end{split}
\end{equation}
where the second integral is over the boundary induced by $\gamma_{ab}$. We have taken the target space to be Minkowski $\eta_{\mu\nu} = \text{diag}(-1,1,\ldots,1)$. The beta function for eq.~\eqref{openStringSigma} was exactly found for all orders in $\alpha'$ and \emph{lowest in derivatives} of the field strength $F_{\mu\nu}$(see \cite{Abouelsaood:1986gd} for a detailed derivation),
\begin{equation}\label{betaOpen}
    \beta_{\mu}^{A} \coloneqq \frac{\pd}{\pd \lambda}A_{\mu} = 2\pi\alpha' \pd^{\nu}F_{\mu}{}^{\delta}\Big[\mathds{1} - (2\pi\alpha' F)^2\Big]^{-1}_{\delta\nu},
\end{equation}
which is in fact obtainable from varying the Born-Infeld functional with respect to $A_{\mu}$ \cite{Callan:1986bc}
\begin{equation}
    S_\text{{B-I}} = \int d^dx\sqrt{\det(\mathds{1}+2\pi\alpha' F)}.
\end{equation}
For fields $F \ll 1/\alpha'$ we can expand in powers of $\alpha'$
\begin{equation}
    S_\text{{B-I}} = \int d^dx \Big(1 + \frac{(2\pi\alpha')^2}{4}F_{\mu\nu}F^{\mu\nu} + O(\alpha'^3)\Big),
\end{equation}
where the second term in the action is just the Yang-Mills Lagrangian. Indeed, the beta function \eqref{betaOpen} admits the same expansion in $\alpha'$
\begin{equation}\label{betaYM}
    \beta_{\mu}^{A} = 2\pi\alpha' \partial_{\nu}F_{\mu}{}^{\nu} + O(F^2,\pd^2 F),
\end{equation}
 giving us one of Maxwell's equations since conformal symmetry demands the beta function to be equal to zero. Treating the beta function more generally, say off-shell, eq.~\eqref{betaYM} is nothing but the Yang-Mills flow equation for a $\text{U}(1)$-connection. It is remarkable that the single copy relates the beta function of the closed string on a curved background to the beta function of the open string on a flat background that is coupled to a gauge field. This is in agreement with the spirit of the single copy and also reminiscent of the KLT relations that relate the amplitude of the closed string to the square of the open string \cite{Kawai:1985xq}. More precisely, the single copy relates the flow in the space of metrics $\mathfrak{G}$ to the flow in the space of $\text{U}(1)$-connections $\mc A$. Indeed, the space of metrics is restricted to a particular subset of metrics that are compatible with our single copy procedure.

 \section{The zeroth copy}
\label{sec:zerothFlow}
Indeed, in the context of the double copy, there is the notion of the zeroth copy where one goes from the gauge theory side to a scalar theory called the biadoint theory by applying the zeroth copy. Schematically,
\begin{equation}
    g_{\mu\nu} = \eta_{\mu\nu} + \phi\, k_{\mu}k_{\nu} \rightarrow A_{\mu} = \phi\, k_{\mu} \rightarrow \phi
\end{equation}
where the arrow indicates the stripping off of one Kerr-Schild null vector $k_{\mu}$. Furthermore, we can either take the zeroth copy of the Yang-Mills flow eq.~\eqref{YMFlow} or simply start with the Kerr-Schild Ricci flow eq.~\eqref{KSmixFlow} and set $\mu=\nu=0$ to obtain the flow equation for the scalar field $\phi$
\begin{equation}\label{phiFlow}
    \frac{\pd}{\pd\lambda}\phi = \pd^2 \phi,
\end{equation}
which is nothing but the familiar heat equation for $\phi(\lambda, x)$. As implied by the Kerr-Schild Einstein equations, at the critical point of eq.~\eqref{phiFlow}, the scalar field satisfies
\begin{equation}
    \pd^2\phi = 0,
\end{equation}
i.e. it is a harmonic function. The relation to the dilaton beta function of the non-linear sigma model is unclear since the zeroth copy corresponds to biajoint scalar theory \cite{Monteiro:2014cda}, which is unrelated to the dilaton appearing in the non-linear sigma model \cite{Polchinski:1998rq}.

\section{Discussion}
In this chapter we have reviewed the non-perturbative double copy in its Kerr-Schild form and gave a simple example of a Schwarzschild black hole single copying to a point-like Coulomb charge. In section two, we have introduced the Ricci flow with some examples and its appearance in the physics literature as the first order (in $\alpha'$) correction to the target space metric tensor. Afterwards, in section 3, we discussed Kerr-Schild diffeomorphisms as introduced in \cite{Coll:2000rm}, and argued how one could use the DeTurck trick to keep the metric in Kerr-Schild form along the flow. In the following section the Yang-Mills flow is obtained by single copying the Ricci flow equation. Physically, the Yang-Mills flow is the first order correction to the Yang-Mills gauge field arising from the beta function of the open string theory. This has been done to all orders in $\alpha'$ where the exact form of the beta function was found to be related to the Born-Infeld non-linear generalisation of Maxwell's electromagnetism.

Two obvious avenues to extend this work into are, firstly, generalising the classical double copy relation to higher orders in the Regge slope $\alpha'$. While mathematically the map from the Ricci flow eq.~\eqref{RicciFloweq1} to the Yang-Mills flow eq.~\eqref{YMFlow} is non-perturbative, physically the flows are perturbative corrections obtained by calculating amplitudes of the non-linear sigma model. Recently, Pasarin and Tseytlin \cite{Pasarin:2020qoa} obtained a metric tensor satisfying the Einstein equations by double copying a gauge field corrected by the Born-Infeld theory, hinting at a non-linear generalisation of the classical double copy. See \cite{Easson:2020esh} for a related work involving the double copy of non-singular black holes. Secondly, the Ricci flow has been used in studying the thermodynamics and phase transitions of black holes \cite{Headrick:2006ti,DeBiasio:2020xkv}. It would be interesting to see what are the parallel behaviours and quantities on the gauge theory side by applying the single copy procedure.

%% file: chaps/chap07.tex
\chapter{Outlook and Future Directions}\label{Outlook}
Here we shall lay out the avenues that one could explore given the work presented in this thesis. The majority of the work presented in this thesis uses the Kerr-Schild formulation of the double copy. Therefore, an obvious goal would be to extend the Kerr-Schild double copy formulation beyond its current capabilities. Indeed, as discussed in Chapter \ref{chap:Weyl}, the Weyl double copy is an example of such a formulation; however, one might want to work with solutions rather than their respective curvatures, like in the Kerr-Schild formulation. Another generalisation of the Kerr-Schild double copy, which is ongoing, is its Double and Exceptional Field theoretic formulation as seen in \cite{Lee:2018gxc,Kim:2019jwm,Berman:2020xvs}; extending the range of applicability to more general GR solutions, and making contact with its amplitudal formulation by agreeing on doubling pure Yang-Mills to gravity endowed with a Kalb-Ramon field, and a dilaton\cite{Johansson:2014zca,Bern:2019prr}. Recently, there has been a push towards generalising the Kerr-Schild formalism to include non-abelian single copy solutions; in particular, for gauge fields and biadjoint scalar related to the Eguchi Hanson instanton on the gravity side \cite{Armstrong-Williams:2022apo}; see \cite{Cheung:2022mix} for a 2-dimensional study.

The Kerr-Schild formulation is in a sense too restrictive, for it forces us to consider a very special class of solutions, although it does include a host of physically interesting spacetimes. Moreover, we are restricted to Abelian single copy solutions, since the Kerr-Schild ansatz linearised Eintein's equations. The linearisation causes the Ricci tensor to resemble the Abelian Yang-Mills equation, as seen in Chapter \ref{chap:s-duality}, which is the source of Abelian-like single copy gauge fields. One could either stay within the Kerr-Schild formalism, but somehow devise a procedure to obtain non-Abelian single copies, or find a new formulation that deals directly with solutions to the equations of motion, just like Kerr-Schild, but is able to deal with the non-linearity of Einstein equations and obtain non-Abelian single copy field. Given the above discussion on possible generalisations of the Kerr-Schild double copy, let us see how can these apply to the results of the previous chapter individually.

In Chapter \ref{chap:s-duality}, we saw how the double copy was used to find the equivalent of Electromagnetic duality transformation on the gravity side. This was done by the use of solution generating transformations in General Relativity; Ehlers transformation in our case. While initially this was done using the Kerr-Schild formalism, which restricts the range of applicability of said analysis; however, in Chapter \ref{chap:Weyl} we saw how this can be generalised to type D spacetimes using the Weyl formalism of the double copy. The latter provides a mean of freeing us from the restrictions brought by working in the Kerr-Schild formalism, and identified Ehlers transformation as an $\text{SL}(2,\mathbb{R})$ transformation acting on solutions of Einstein's equations. An area of exploration, in this case, would be to generalise to non-Abelian gauge fields which make up the Weyl tensor, and see transforming these non-Abelian gauge fields double copy to solutions of Einstein's equation.

In Chapter \ref{chap:Holonomy}, we applied the double copy formalism to both establish the appropriate Wilson line operator in gravity, and also find the single copy of the gravitational holonomy operator, which we dubbed the SCH. It is worth extending the work done in this chapter in two direction. First, by applying the single copy procedure to a bigger range of solutions, and examining how both the holonomy and SCH groups behave. This will allow us to further understand the relationship between solutions of Einstein's equations and their single copy gauge fields. Secondly,  it will be worthwhile to explore the possibility of some sort of classifications of Yang-Mills equations' solutions obtained as single copies of those of General Relativity. This will be in the same spirit of Berger's classification of possible holonomy groups of Riemannian manifolds as discussed in the beginning of Chapter \ref{chap:Holonomy}. Indeed, as discussed at the beginning of this chapter, a generalisation of this double copy formalism to more general spacetimes and non-Abelian gauge fields is a valuable research direction. This will allow us to obtain more SCH groups as single copies of holonomy groups of solutions of Einstein's equations that do not admit a Kerr-Schild form, allowing us to analyse more exotic manifolds such as Spin manifolds as seen in section \ref{sec:SH} and eq. \eqref{Spin7Man}.

Finally, in Chapter \ref{chap:RicciFlow}, we have discussed the relevance of the Ricci flow in theoretical physics. In particular, in the context of string theory where one obtains the Ricci flow as the beta function of the metric on the target manifold. Using the machinery of the double copy, we have obtain an analogous flow equation that is known as the Yang-Mills flow equation. The latter appears, again, as the beta function of the gauge field coupled to the open string. Essentially, the flow equation acts as a correction, in the $\text{U}(1)$ gauge group case, to the Maxwell equations in the string coupling $\alpha'$, the same way the Ricci flow is the leading order stringy correction to the Einstein's equation. A possible avenue to explore is to generalise this map from the Ricci to Yang-Mills flow to spacetimes that are not Kerr-Schild. This could or could not include a generalisation to non-Abelian gauge field, depending on the nature of the extension of the double copy. An interesting application utilising the result in this chapter is on black hole thermodynamics and cosmology. The former would be in the same spirit of \cite{Headrick:2006ti,DeBiasio:2020xkv}; however, in our case, it is the single copy of the results obtained in those papers that could be of novel nature. Works in the context of cosmology featuring the Ricci flow have been explored in \cite{Luo:2021zpi,SharmaCosmo}.

%% file: appendix.tex
\appendix
\chapter{Derivation of the Kerr-Schild spin connection}
\label{AppSpin:A}

In this appendix, we provide a derivation of the form of the spin
connection in Kerr-Schild coordinates, as reported in
eq.~(\ref{eKS}). The spin connection satisfies Cartan's first structure equation in the
absence of torsion,
\begin{equation}
    de^{a}+\omega^a_{\Indx c}\wedge e^c=0.
\end{equation}
In tensorial language this takes the form
\begin{equation}
    \pd_\mu \tensor{e}{^a_\nu} - \pd_\nu \tensor{e}{^a_\mu} + (\omega_\mu)^a_{\Indx \nu} - (\omega_\nu)^a_{\Indx \mu} = 0,
\end{equation}
were we have contracted the vierbein with the spin
connection. Multiplying by a factor of $\tensor{e}{_b^\mu}\tensor{e}{_c^\nu}$, one finds
\begin{equation}
     (\pd_b \tensor{e}{^a_\nu}) \tensor{e}{_c^\nu}
 - (\pd_c \tensor{e}{^a_\mu}) \tensor{e}{_b^\mu} + (\omega_b)^a_{\Indx c} - (\omega_c)^a_{\Indx b} = 0.
\end{equation}
Next, one can substitute the explicit forms of the Kerr-Schild
vierbein given in eq.~(\ref{eKS}), and use the null condition from
eq.~(\ref{KSconditions}), to obtain
\begin{equation}
     \pd_b \tensor{e}{^a_c} - \pd_c \tensor{e}{^a_b} + (\omega_b)^a_{\Indx c} - (\omega_c)^a_{\Indx b} - \frac{1}{4}\phi^2k^ak^{\mu} \left[ k_c \pd_b k_{\mu} -  k_b \pd_c k_{\mu} \right] = 0.
\end{equation}
Upon lowering the index $a$, one may cyclically permute the indices
$(a,b,c)$ and consider the combination $(a,b,c) - (b,c,a) - (c,a,b) = 0$, which yields 
\begin{equation}
    (\omega_{\mu})_{bc} = (\omega_a)_{bc}e^a_{\Indx \mu} = \left(\pd_c e_{ab} - \pd_b e_{ac} \right)e^a_{\Indx \mu} + \frac{1}{4}\phi^2k_{\mu}k^{\nu} \left( k_c\pd_b k_\nu - k_b \pd_c k_\nu  \right),
\end{equation}
where we have also multiplied the entire equation by $e^a_{\Indx \mu}$
to turn $a$ into a spacetime index. One may again use eqs.~(\ref{eKS},
\ref{KSconditions}) for the vierbein in the first term, after which
cancellations occur, leading to
\begin{equation}
    (\omega_\mu)_{ab} = \pd_b e_{a\mu} - \pd_a e_{b\mu}.
\end{equation}
This agrees with a similar result in ref.~\cite{Chakrabarti:1999mb}, and can also be obtained by plugging the Kerr-Schild vierbein of eq.~\eqref{eKS} into eq.~\eqref{spinvierbein}.

\chapter{Maxwell field strength from killing vectors}\label{Maxapp}

It is well known that one can obtain solutions to Maxwell's equations using killing vectors defined on generally curved geometries. We closely follow Wald \cite{Wald:1984rg}.

Suppose we have a Killing vector field $\xi$ defined on a Riemannian manifold $(\mc{M}, g)$; meaning, it satisfies
\begin{equation}
    \pounds_\xi g = 0,\quad \nabla_a\xi_b + \nabla_b\xi_a = 0,
\end{equation}
where $\nabla$ is the connection associated with the metric $g$, and the second equation, namely Killing's equation, follows from the first.

Now, by the definition of the Riemann tensor on a Riemannian manifold, we have
\begin{equation}
    \nabla_a\nabla_b \xi_c - \nabla_b\nabla_a \xi_c = R_{abc}{}^{d}\xi_d.
\end{equation}
Using Killing's equation, we can write
\begin{equation}
    \nabla_a\nabla_b \xi_c + \nabla_b\nabla_c \xi_a = R_{abc}{}^{d}\xi_d.
\end{equation}
Now if we denote this equation by its free indices, i.e. $(abc)$, then simply cycle through the indices adding the first permutation and then subtracting the last, $(abc) + (bca) - (cab)$, and using the symmetries of the Riemann tensor, we obtain
\begin{equation}
\begin{split}
    2\nabla_b\nabla_c\xi_a &= (R_{abc}{}^d + R_{bca}{}^{d} - R_{cab}{}^d)\xi_d\\
    &= -2R_{bca}{}^d\xi_d.
\end{split}
\end{equation}
Now if we contract the indices $b,c$ with $g^{bc}$ we obtain
\begin{equation}\label{WaldMAx}
    \nabla^a\nabla_a\xi_c = -R_c{}^d\xi_d.
\end{equation}
Therefore, we see that any Killing vector satisfies the source-free Maxwell's equations in Lorentz gauge in a vacuum spacetime $R_{c}{}^d = 0$. The Killing vectors satisfy the Lorentz gauge property by virtue of the Killing's equation. It is worth noting that  \eqref{WaldMAx} is not simply Maxwell's equation in curved spacetime as there is a sign difference \cite{Wald:1984rg}.

\chapter{Myers-Perry supplement}
The pentad used for the analysis of the Myers-Perry metric in subsection \ref{MPsubsection} is based on the two null vectors $l^\mu=L^\mu_\pm$ satisfying the equation $l^\nu l^\rho C_{\mu\nu\rho[\pi}l_{\omega]}=0$ \cite{Pravda:2005kp}
\begin{equation}
\label{pentadnl}
\begin{split}
L_\pm  = {1\over(x^2-1)(-1+L^2y)}&\Bigg( {L^2x y - y + L^2 x +1 -2 L^2y\over x-y} R\partial_t -L\partial_\psi\Bigg)\\
& \pm \sqrt{{L^2x-1\over(x-y)(y-1)}}\Bigg(\partial_x+{y^2-1\over x^2-1}\partial_y\Bigg)
 \, .
\end{split}
\end{equation}
We take $l^\mu = L_+^\mu, n^\mu=L_-^\mu$ and choose the three unit norm vectors $m^\mu_i$ ($i=1,2,3$) to be
\begin{align}
\label{pentadmi}
m_1  &= \sqrt{ {(1+x)(x-y)(1+y)(-1+L^2y)^2\over (1+L^2)R^2(x-1)}} \Bigg(0,-{(x-1)(-1+L^2x)\over (y-1)(-1+L^2y)},-1,0,0\Bigg)\, ,  \nonumber  \\
m_2&= \sqrt{- {(x-y)^2\over (1+x)(-1+L^2x)(y-1)^2R^2}} \Bigg(0,0,0,-1,0\Bigg) \, ,  \nonumber\\ 
m_3&=  \sqrt{- {(x-y)^2\over (-1+x)(1+L^2)(y^2-1)R^2}}    \Bigg( {LR(x-1)(y+1)\over x-y},0,0,0,-1\Bigg)\, .
\end{align}
These vectors satisfy the conditions in and beneath \eqref{nullpentad}. This choice is convenient in this case as the non-zero weight Weyl-NP (and Maxwell-NP) components vanish directly using them.